\documentclass[longauth]{aa}
\usepackage{graphicx}
\usepackage{txfonts}
\usepackage[colorlinks=true, allcolors=blue]{hyperref}
\usepackage{placeins}
\usepackage[rightcaption]{sidecap}
\usepackage[english]{babel}
\usepackage{array}
\usepackage{capt-of}
\usepackage[normalem]{ulem}
\usepackage{lscape}
\usepackage{multirow}
\usepackage{multicol}

\usepackage{bm}

\usepackage{xspace}
\newcommand{\kms}{km~s$^{-1}$\xspace}
\newcommand{\mum}{$\mu$m\xspace}

\begin{document}

   \title{JWST Observations of Young protoStars (JOYS)}

  \subtitle{Overview of gaseous molecular emission and absorption in low-mass protostars}

   \author{
          % MAIN CONTRIBUTORS
          M.~L. van Gelder\inst{1}
          \and
          L. Francis\inst{1}
          \and
          E.~F. van Dishoeck\inst{1,2}
          \and
          {\L}. Tychoniec\inst{1}
          \and
          % JOYS co-PIs
          T.~P. Ray\inst{3}
          \and
          % JOYS TEAM MEMBERS
          H. Beuther\inst{4}
          \and
          A. Caratti o Garatti\inst{5}
          \and \\
          Y. Chen\inst{1}
          \and
          R. Devaraj\inst{3}
          \and
          C. Gieser\inst{2}
          \and
          K. Justtanont\inst{6}
          \and
          P.~J. Kavanagh\inst{7}
          \and
          P. Nazari\inst{8}
          \and
          S. Reyes\inst{4}
          \and
          W.~R.~M. Rocha\inst{1,9}
          \and \\
          K. Slavicinska\inst{1,9}
          \and
          % MIRI co-PIs
          M. G\"udel\inst{10,11}
          \and
          Th. Henning\inst{4}
          \and
          P.~-O. Lagage\inst{12}
          \and
          G. Wright\inst{13}
          }

   \institute{
              % 1 
              Leiden Observatory, Leiden University, PO Box 9513, 2300RA Leiden, The Netherlands\\
              \email{vgelder@strw.leidenuniv.nl}
              \and
              % 2
              Max Planck Institut f\"ur Extraterrestrische Physik (MPE), Giessenbachstrasse 1, 85748 Garching, Germany
              \and
              % 3
              School of Cosmic Physics, Dublin Institute for Advanced Studies, 31 Fitzwilliam Place, D02 XF86, Dublin, Ireland
              \and
              % 4
              Max Planck Institute for Astronomy, K\"onigstuhl 17, 69117 Heidelberg, Germany
              \and
              % 5
              INAF-Osservatorio Astronomico di Capodimonte, Salita Moiariello 16, 80131 Napoli, Italy
              \and
              % 6
              Department of Space, Earth and Environment, Chalmers University of Technology, Onsala Space Observatory, 439 92 Onsala, Sweden
              \and
              % 7
              Department of Experimental Physics, Maynooth University, Maynooth, Co Kildare, Ireland
              \and
              % 8
              European Southern Observatory (ESO), Karl-Schwarzschild-Strasse 2, 1780 85748 Garching, Germany
              \and
              % 9
              Laboratory for Astrophysics, Leiden Observatory, Leiden University, P.O. Box 9513, 2300 RA Leiden, The Netherlands
              \and
              % 10
              Department of Astrophysics, University of Vienna, T\"urkenschanzstrasse 17, A-1180 Vienna, Austria
              \and
              % 11
              ETH Z\"urich, Institute for Particle Physics and Astrophysics, Wolfgang-Pauli-Strasse 27, 8093 Zürich, Switzerland
              \and
              % 12
              Universit\'e Paris-Saclay, Universit\'e Paris Cit\'e, CEA, CNRS, AIM, 91191 Gif-sur-Yvette, France
              \and
              % 13
              UK Astronomy Technology Centre, Royal Observatory Edinburgh, Blackford Hill, Edinburgh EH9 3HJ, UK
             }

   \date{Received 23 August 2024 / Accepted 2 October 2024}

% \abstract{}{}{}{}{}
% 5 {} token are mandatory
 
  \abstract
   {
   The Mid-InfraRed Instrument (MIRI) onboard the {\it James Webb} Space Telescope (JWST) allows for probing the molecular gas composition at mid-infrared (mid-IR) wavelengths at unprecedented resolution and sensitivity. It is important to study these features in low-mass embedded protostellar systems since the formation of planets is thought to start in this phase. Previous studies were sensitive primarily to high-mass protostars.
   }
   {
   The aim of this paper is to derive the physical conditions of all gas-phase molecules detected toward a sample of 18 low-mass protostars as part of the JWST Observations of Young protoStars (JOYS) program and to determine the origin of the molecular emission and absorption features. This includes molecules such as CO$_2$, C$_2$H$_2$, and CH$_4$ that cannot be studied at millimeter wavelengths.
   }
   {
   We present JWST/MIRI data taken with the Medium Resolution Spectrometer (MRS) of 18 low-mass protostellar systems, focusing on gas-phase molecular lines in spectra extracted from the central protostellar positions. The column densities and excitation temperatures are derived for each molecule using local thermodynamic equilibrium (LTE) slab models. Ratios of the column densities (absorption) or total number of molecules (emission) are taken with respect to H$_2$O in order to compare these to ratios derived in interstellar ices.
   }
   {
   Continuum emission is detected across the full MIRI-MRS wavelength toward 16/18 sources, the other two sources (NGC~1333~IRAS~4B and Ser-S68N-S) are too embedded to be detected. The MIRI-MRS spectra show a remarkable richness in molecular features across the full wavelength range, in particular toward B1-c (absorption) and L1448-mm (emission). Besides H$_2$, which is not considered here, water is the most commonly detected molecule (12/16) toward the central continuum positions followed by CO$_2$ (11/16), CO (8/16), and OH (7/16). Other molecules such as $^{13}$CO$_2$, C$_2$H$_2$, $^{13}$CCH$_2$, HCN, C$_4$H$_2$, CH$_4$, and SO$_2$ are detected only toward at most three of the sources, particularly toward B1-c and L1448-mm. The JOYS data also yield the surprising detection of SiO gas toward two sources (BHR71-IRS1, L1448-mm) and for the first time CS and NH$_3$ at mid-IR wavelengths toward a low-mass protostar (B1-c). The temperatures derived for the majority of the molecules are 100--300~K, much lower than what is typically derived toward more evolved Class~II sources ($\gtrsim500$~K). Toward three sources (e.g., TMC1-W), hot ($\sim1000-1200$~K) H$_2$O is detected, indicative of the presence of hot molecular gas in the embedded disks, but such warm emission from other molecules is absent. The agreement in abundance ratios with respect to H$_2$O between ice and gas point toward ice sublimation in a hot core for a few sources (e.g., B1-c) whereas their disagreement and velocity offsets hint at high-temperature (shocked) conditions toward other sources (e.g., L1448-mm, BHR71-IRS1). 
   }
   {
   Molecular emission and absorption features trace various warm components in young protostellar systems, from the hot core regions to shocks in the outflows and disk winds. The typical temperatures of the gas-phase molecules of $100-300$~K are consistent with both ice sublimation in hot cores as well as high-temperature gas phase chemistry. Molecular features originating from the inner embedded disks are not commonly detected, likely because they are too extincted even at mid-IR wavelengths by small not-settled dust grains in upper layers of the disk.
   }


   \maketitle
%
%________________________________________________________________

\section{Introduction}
Molecules play a crucial role in the formation of protostellar and planetary systems \citep[e.g.,][]{vanDishoeck1998,Caselli2012,Ceccarelli2023}. 
Not only is their evolution from molecular clouds to protoplanetary disks important for setting the initial composition of planetary bodies in the disks, but they also provide constraints on the physical conditions during all protostellar stages. 
It is especially relevant to study the molecular gas composition in the earliest phases of star formation since planet formation is suggested to start in these early Class~0 and I phases \citep[e.g.,][]{Harsono2018,Tychniec2018,Tychoniec2020}. 
% Recent observations suggest an inheritance of intersellar ices from molecular cloud to disk \citep[e.g.,][]{Booth2021,Tobin2023}, but observing the warm gas in the terresterial planet forming zone of young embdedded systems remains difficult. 
In this work, we present new observations with the {\it James Webb} Space Telescope (JWST) at mid-infrared (mid-IR) wavelengths tracing the molecular gas composition in young and embedded protostellar systems.

Molecular emission in embedded protostellar systems is commonly observed at (sub)millimeter wavelengths with interferometers such as the Atacama Large Millimeter/submillimeter Array (ALMA). These observations have shown that molecular emission is present on all scales of embedded protostellar systems \citep[see overviews of][]{Jorgensen2020,Tychoniec2021,Tobin2024}, from the large-scale envelope \citep[e.g.,][]{Jorgensen2002,Tobin2013}, to the inner envelope and hot core \citep[e.g.,][]{Bottinelli2004,Kristensen2012,Oya2019,vanGelder2020,Nazari2021}, embedded disks \citep[e.g.,][]{Harsono2014,van'thoff2023,Lee2024}, and outflows and jets \citep[][]{Arce2010,Codella2014,Lee2017,Tychoniec2019}. However, ALMA is mostly sensitive to the colder ($\lesssim500$~K) regions and not to the hotter material located in the strong outflow and jet shocks, disk winds, and the inner embedded disk. Moreover, several important and abundant molecules such as H$_2$, CO$_2$, C$_2$H$_2$, and CH$_4$ lack a permanent dipole moment and therefore do not show pure rotational lines at sub-millimeter wavelengths \citep[][]{vanDishoeck2004}. In order to probe the (hot) rovibrational transitions of such species one has to observe them at mid-IR wavelengths.

Prior to the launch of JWST, gaseous molecular features at mid-IR wavelengths were difficult to detect toward low-mass protostellar sources \citep[][]{Lahuis2010}. Water (H$_2$O) was the most common molecule (next to H$_2$ and CO) detected with the {\it Spitzer} Space Telescope with rather high temperatures of up to 1500~K that likely originates from embedded disks or shocks \citep[e.g.,][]{Watson2007,Lahuis2010}. Detections of other molecules such as CO$_2$, C$_2$H$_2$ were limited to a few sources and did not allow derivation of their physical conditions \citep[][]{Lahuis2010}. Toward high-mass sources, on the other hand, gaseous emission and absorption lines were more commonly observed at mid-IR wavelengths using first the Infrared Space Observatory Short Wavelength Spectrometer (ISO-SWS) \citep[e.g.,][]{Helmich1996,Lahuis2000,Boonman2003_Orion,Boonman2003_H2Oabs,vanDishoeck2004}, and then {\it Spitzer} as well as several ground-based telescopes such as the Very Large Telescope (VLT) and the Stratospheric Observatory For Infrared Astronomy (SOFIA) \citep[e.g.,][]{Lacy1989,Lacy1991,Evans1991,Sonnentrucker2006,Sonnentrucker2007,Barr2020}. Molecular absorption is typically observed toward the bright continuum sources and suggested to arise from the envelope or a disk surface layer above the accretion-heated midplane \citep[e.g.,][]{Knez2009,Barr2020,Barr2022}. Not only CO, H$_2$O, and CO$_2$ are commonly detected toward high-mass sources, but also species like C$_2$H$_2$, HCN, SO$_2$, CS, and NH$_3$ were detected in absorption \citep[][]{Keane2001,Boonman2003_H2Oabs,Dungee2018,Barr2020,Nickerson2023}. Furthermore, molecular emission was detected toward high-mass protostellar systems at positions where they could clearly be attributed to shocks \citep[][]{Boonman2003_Orion,Sonnentrucker2006,Sonnentrucker2007}. 
Nevertheless, previous mid-IR observatories either lacked the spatial and/or spectral resolution and sensitivity to detect gaseous molecular features at high signal-to-noise ratios (S/N) in a larger sample of sources or suffer from telluric absorption at crucial wavelengths. 

The Medium Resolution Spectrometer \citep[MRS;][]{Wells2015,Argyriou2023} of the Mid InfraRed Instrument \citep[MIRI;][]{Rieke2015,Wright2015,Wright2023} on JWST is excellent at detecting molecular emission and absorption features due to its unprecedented spatial and spectral resolution and sensitivity. This has led to the detection of various molecular emission and absorption features toward the distant high-mass protostellar system IRAS~23385+6053 \citep[][]{Beuther2023,Gieser2023,Francis2024}. Interestingly, they seem to trace gas at a temperature of $\sim150$~K toward this high-mass source, likely originating from either a disk surface layer or alternatively the outflow. For a couple of low-mass sources, clear features of CO and rovibrational H$_2$O lines are detected \citep[][]{Yang2022,Kospal2023,Salyk2024}. Moreover, \citet{vanGelder2024} recently observed SO$_2$ for the first time at mid-IR wavelengths toward a low-mass protostellar system, NGC~1333~IRAS~2A, finding a temperature of $\sim100$~K which is consistent with the SO$_2$ tracing the hot core. 

The mid-IR molecular lines allow for tracing molecules in the inner hot cores of low-mass protostellar systems that cannot be observed otherwise. At millimeter wavelengths, these hot cores show spectra dominated by emission of complex organics \citep[e.g.,][]{Jorgensen2016,Bianchi2020,vanGelder2020,Nazari2021,Nazari2024_Band3}. Given their similar gas-phase abundance ratios across protostellar luminosities \citep[e.g.,][]{Coletta2020,Nazari2022_N-COMs,Chen2023}, these are suggested to originate from thermal ice desorption. This was recently supported by JWST/MIRI-MRS observations of complex organics in the ices \citep[][]{Rocha2024,Chen2024}, showing similar ratios between ice and gas for several (but not all) complex organics. For dominant ice species such as H$_2$O, CO$_2$, CH$_4$, and NH$_3$, however, this is not yet clear. The dominant ice species have been studied in detail toward several low-mass sources with ground-based telescopes and {\it Spitzer} \citep[e.g.,][]{Boogert2008,Pontoppidan2008,Oberg2008}, showing typical abundance ratios with respect to H$_2$O of $10^{-2}-10^{-1}$ \citep[see overview by][]{Boogert2015}.
Gas-phase observations toward low-mass sources were limited to a few detections \citep[mostly H$_2$O; e.g.,][]{Watson2007,Lahuis2010} and could not constrain whether the emission or absorption was originating from thermal ice sublimation in the hot core. Moreover, (additional) high-temperature gas-phase chemistry in the hot cores is possible for these species and alter their abundances following ice sublimation \citep[e.g.,][]{Charnley1992,Garrod2022}. Nevertheless, similar abundance ratios between ice and gas will be a strong indication of the gas-phase molecules originating from thermal ice sublimation in a hot core.

Molecular emission at mid-IR wavelengths can also originate from shocks in outflows or disk winds \citep[e.g.,][]{Boonman2003_Orion,Sonnentrucker2006,Sonnentrucker2007,Francis2024}. Extended jets and outflows are frequently seen at mid-IR wavelengths through H$_2$ and atomic tracers \citep[e.g.,][]{Maret2009,Dionatos2009,Dionatos2014,CarattioGaratti2024,Narang2024}, but also the smaller-scale disk winds can now be resolved with JWST \citep[e.g.,][]{Harsono2023,Sturm2023,Federman2024,Tychoniec2024,Assani2024}. These outflows and disk winds are important for protostellar systems and disk evolution by carrying away angular momentum as well as disk dispersal \citep[e.g.,][]{Frank2014,Bally2016,Tabone2022a,Tabone2022b,Pascucci2023}. Velocity-shifted molecular emission is a good indication of the presence of outflows or small-scale disk winds, although disk winds are also often seen in absorption toward the bright IR continuum \citep[e.g.,][]{Thi2010,Herczeg2011}. Shocks in these outflows or disk winds can produce molecules such as H$_2$O, CO$_2$, SO$_2$, and SiO through high-temperature gas-phase chemistry or sputtering of the ices from the dust grains \citep[e.g.,][]{Caselli1997,Gusdorf2008_a,Gusdorf2008_b,vanDishoeck2021}. 

Toward Class~II protoplanetary disks, molecular emission at mid-IR wavelengths is commonly detected \citep[e.g. H$_2$O, CO$_2$, C$_2$H$_2$, HCN;][]{Carr2008,Salyk2011,Grant2023,Pontoppidan2024,Henning2024} and given the derived temperatures ($\gtrsim500$~K), it is suggested to originate from the inner disk or warm surface layers \citep[e.g.,][]{Blake2004,Banzatti2023_JWST,Gasman2023,Temmink2024_H2O,Schwarz2024}. Especially toward very low-mass stars a wealth of hydrocarbon molecules were identified recently \citep[e.g., CH$_4$, C$_4$H$_2$, C$_6$H$_6$;][]{Tabone2023,Arabhavi2024}. An interesting first comparison between embedded Class~0/I and more evolved Class~II sources suggests that the temperatures are lower in protostellar systems \citep[$\sim100-300$~K;][]{vanDishoeck2023,Francis2024,vanGelder2024,Salyk2024} compared with Class~II disks \citep[$\gtrsim500$~K;][]{Grant2023,Ramirez-Tannus2023,Banzatti2023_JWST,Gasman2023,Temmink2024_H2O,Temmink2024_CO}. This implies that MIRI-MRS observations of protostellar systems are not necessarily tracing the embedded disks. However, the number of low-mass embedded sources with accurate constraints on the molecular excitation conditions remains limited.

The most common tools adopted to analyze molecular emission and absorption features at mid-IR wavelengths are local thermodynamic equilibrium (LTE) slab models \citep[e.g.,][]{Salyk2011,Tabone2023,Francis2024}. 
% These LTE models assume that the level populations are fully thermalized and can be described by a single excitation temperature. 
However, while the rovibrational transitions may be (almost) fully thermalized for the high densities ($\gtrsim10^{10}$~cm$^{-3}$) in inner Class~II disks, the densities in the inner envelopes of protostellar systems ($\sim10^{6}-10^{8}$~cm$^{-3}$) are lower than their critical densities (typically $>10^{10}$~cm$^{-3}$).
% Whereas a full non-LTE analysis can be performed for some molecules that have their collisional rate coefficient calculated \citep[e.g., HCN and CO$_2$;][]{Bruderer2015,Bosman2017}, it is not possible for many other species (e.g., SO$_2$, CH$_4$). 
An important effect to take into account for mid-IR emission lines in protostellar systems is infrared pumping \citep[e.g.,][]{Boonman2003_Orion,Sonnentrucker2006,Sonnentrucker2007,vanGelder2024}, where the vibrationally excited levels get more strongly populated by a strong mid-IR radiation field than through just collisional excitation. This leads LTE models to overpredict the total number of molecules by several orders of magnitude \citep[][]{vanGelder2024}. 
% An important effect to take into account for mid-IR lines in emission in protostellar systems is infrared pumping \citep[e.g.,][]{Boonman2003_Orion,Sonnentrucker2006,Sonnentrucker2007,vanGelder2024}, where LTE models of mid-IR lines can overpredict the number of molecules by several orders of magnitude \citep[][]{vanGelder2024}. 
% The excitation temperature can still be accurately constrained from the LTE slab models \citep[][]{vanGelder2024}, but IR pumping remains something to take into account when deriving the number of emitting molecules. Fortunately, absorption models do not suffer from such non-LTE effects. 
Furthermore, it is important to note that even at the spectral resolution of MIRI-MRS \citep[$R=3500-1500$;][]{Labiano2021,Jones2023}, line blending and optical depth needs to be taken into account in order to derive accurate physical parameters \citep[e.g.,][]{Li2024}.

In this paper, we present an overview of the JWST/MIRI-MRS observations of 18 low-mass protostellar systems from the JWST Observations of Young protoStars (JOYS) program\footnote{\url{https://miri.strw.leidenuniv.nl/}}, focusing on the gaseous molecular emission and absorption features. This includes all molecules except H$_2$, which will be presented in a separate paper (Francis et al. in prep.). Moreover, the emphasis of this paper lies on the physical and chemical conditions in the inner envelope and embedded disks (i.e., $<100$~au scales). Molecular emission located on larger scales in the outflow will also be discussed in a separate paper (Francis et al. in prep.).
This paper is organized as follows. The sample, data reduction, and LTE analysis are described in Sect.~\ref{sec:obsservations_analysis}. The detection statistics and the results from the LTE analysis are presented in Sect.~\ref{sec:results}. We discuss these results in context of other evolutionary stages and what molecule is tracing what component in Sect.~\ref{sec:discussion}. Lastly, our main conclusions are summarized in Sect.~\ref{sec:conclusions}. 

\section{Observations and analysis}
\label{sec:obsservations_analysis}
\subsection{Sample}
The full sample of sources studied in this work and their properties are listed in Table~\ref{tab:sample}. All observations are part of the JOYS program. The sample was selected to cover both the earliest Class~0 phases as well as more evolved Class~I systems. It contains sources in three star-forming regions, Taurus, Perseus, Serpens, as well as two protostars from the BHR71 cloud (IRS~1 and IRS~2), spanning a large range of bolometric luminosities ($0.2-109$~L$_\odot$) and bolometric temperatures ($28-189$~K). Furthermore, several of the targeted protostellar systems contain confirmed embedded disks \citep[e.g., L1527, TMC1, TMC1A;][]{Tobin2012,Harsono2014,Tychoniec2021} or are part of small-scale ($<1000$~au) multiple systems \citep[B1-a, Ser-SMM1, Ser-S68N, TMC1;][le Gouellec et al. in prep.]{Choi2009,Tobin2016,van'tHoff2020_Taurus}. The Class~I source B1-a remains (partially) unresolved at mid-IR wavelengths and is therefore considered as a single source in the remaining analysis. Furthermore, the JOYS sample also includes SVS4-5, a low-mass Class~I/II source showing deep CH$_3$OH ice features that is located behind or inside the envelope and outflow of the Class~0 source Ser-SMM4 \citep[][]{Pontoppidan2004}. For Ser-emb8(N), the central protostellar position was not observed and only the blueshifted part of the outflow was covered. Similarly, only the blueshifted outflow was targeted for IRAS~4A, but the protostellar position is also covered in another JWST program (PID 1236; Ressler et al. in prep.) with no continuum detected. These two sources will therefore not be discussed further in this paper. This gives a total of 18 sources (with binaries explicitly counted) where the central protostellar position is covered by MIRI-MRS (11 Class~0, 1 Class 0/I, 5 Class~I, 1 Class~I/II; see Table~\ref{tab:sample}).

\begin{figure*}
    \centering
    \includegraphics[width=\linewidth]{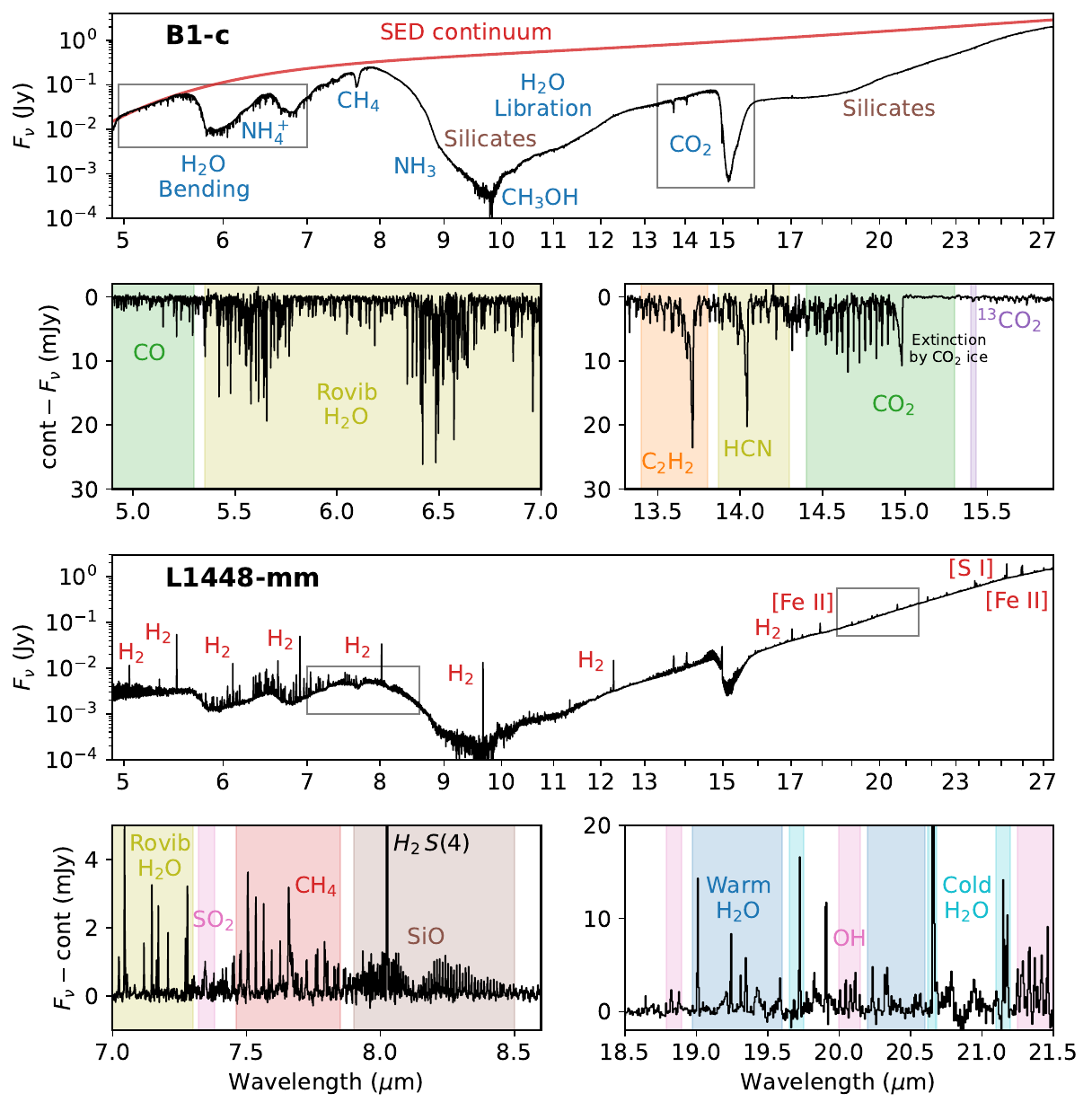}
    \caption{Spectra of B1-c (top row) and L1448-mm (third row) and the main spectral features detected among the JOYS data of these two sources (two panels per source below the full spectrum). In the top spectrum of B1-c, the SED continuum fit is shown in red and dominant ice and silicate absorption features are labeled in blue and brown, respectively. The two gray boxes mark the spectral range covered in the two insets in the second row. The {baseline-subtracted} absorption spectra of B1-c are presented in the second row, highlighting the CO and rovibrational H$_2$O features between $4.9-7.0$~\mum and the CO$_2$, $^{13}$CO$_2$, C$_2$H$_2$, and HCN features between $13.3-15.9$~\mum. The third row shows the spectrum of L1448-mm with dominant emission lines of H$_2$, {[S I]}, and {[Fe II]} labeled in red. The bottom row shows the {baseline-subtracted} emission spectra from the two gray boxes, focusing on the rovibrational H$_2$O, SO$_2$, CH$_4$, and SiO features between $7.0-8.6$~\mum and the pure rotational lines of H$_2$O and OH between $18.5-21.5$~\mum.}
    \label{fig:specoverview_B1-c_L1448-mm}
\end{figure*}

\subsection{Observations}
The MIRI-MRS data analyzed in this work were taken as part of guaranteed time observations (GTO) program 1290 (PI: E.F. van Dishoeck). The majority of the sources were observed using a single pointing centered on the protostar, often with a small offset to cover a portion of the blueshifted outflow. For a few selected sources (i.e., B1-c, L1448-mm, BHR71~IRS1), multiple pointings were used to cover the larger-scale blueshifted outflows. However, since the focus of this paper is on the molecular emission from the central protostellar regions, only the pointings centered on the protostars themselves are used. 

The observations were taken with a 2-point dither pattern that was optimized for extended sources except for B1-c and Ser-SMM1A for which a 4-point dither pattern was adopted. The pointing of TMC1A was not centered on the protostar but on the blueshifted disk wind which resulted in the source itself lying at the edge of the field of view (FOV) in channels 1 and 2. Only one dither position was therefore used for channels 1 and 2 in order to reduce instrumental artifacts created by combining two dithers when the point spread function (PSF) is not fully sampled. For each of the four star-forming regions, a dedicated background observation was performed to allow for a proper subtraction of the telescope background and detector artifacts. For Taurus, the background observations were carried out with a single dither position whereas for the other three dedicated backgrounds a 2-point dither pattern was used. All observations used the FASTR1 read-out mode in all three gratings (A, B, C), providing the full 4.9-27.9~\mum wavelength coverage of MIRI-MRS. 
For each source, the total on-source integration time in each grating is listed in Table~\ref{tab:sample}. The integration time was evenly divided between the three gratings except for B1-c where the integration time in grating B was twice as long as in the other gratings (4000~s in grating B and 2000~s in gratings A and C) to get a higher S/N in the silicate absorption feature around 10~\mum.

All data were processed through the JWST calibration pipeline version 1.13.4 \citep{Bushouse2024} using reference context {\tt jwst$\_$1188.pmap} of the JWST Calibration Reference Data System \citep[CRDS;][]{Greenfield2016}. First, the raw {\tt uncal} data was processed through the {\tt Detector1Pipeline} using the default settings. Second, the {\tt Spec2Pipeline} was carried out to produce calibrated detector images. The dedicated background was also subtracted on the detector level in this step. To circumvent subtracting astronomical features in the dedicated background observations from the science data, any clear emission lines (e.g., H$_2$ S(1) and S(2) lines) in the dedicated background {\tt rate} files were masked before subtraction. Additionally, the fringe flat for extended sources was applied, as well as the residual fringe correction (Kavanagh et al. in prep.). Following this step, an additional bad pixel map was created from the {\tt cal} files using the Vortex Imaging Processing (VIP) package \citep[][]{Christiaens2023}. Last, data cubes were created from the calibrated detector files in the {\tt Spec3Pipeline} for each band and each channel separately using the drizzle algorithm \citep{Law2023}. In this step, the master background and outlier rejection steps were switched off. The new wavelength calibration derived from H$_2$O lines was included in the data reduction \citep[][]{Pontoppidan2024}.

The spectra of all sources were extracted from apertures centered on the continuum sources, see Table~\ref{tab:sample}. Toward some sources (e.g., L1448-mm), extended molecular emission is present in the outflow, which will be presented in a separate paper (Navarro et al. in prep.). For the sources where the continuum emission is unresolved, spectra were extracted using an aperture that increases with wavelength following the size of the PSF \citep[${\rm FWHM_{PSF}} = 0.033 (\lambda/\mu{\rm m}) + 0.106 ''$;][]{Law2023}. By default, the diameter was set to $4\times{\rm FWHM_{PSF}}$ to encompass as much of the PSF as possible without including additional noise. However, for sources where extended molecular emission is present in the outflow (e.g., L1448-mm), a smaller aperture with a diameter of $2\times{\rm FWHM_{PSF}}$ was adopted to exclude the larger scale outflow in the spectra. Furthermore, the aperture for B1-a was set to a larger diameter of $5\times{\rm FWHM_{PSF}}$ to include both sources in a single aperture since the individual components of the binary cannot be resolved at longer wavelengths (see Fig.~\ref{fig:specext_B1-a-NS}). For the edge-on disk L1527, the source appears as extended in scattered light at the shortest wavelengths (see Fig.~\ref{fig:specext_L1527}). Hence, a circular aperture with a radius of 3$''$ is used to capture both sides of the disk in a single aperture. Following the spectral extraction, an additional 1D residual fringe correction was applied to the spectra, in particular to remove the high-frequency dichroic noise in channels 3 and 4 (Kavanagh et al. in prep.). Finally, the spectra of all 12 subbands were stitched together to allow for a single analysis of the full wavelength range that MIRI-MRS provides. {Channel 1A was used as the baseband since its photometric calibration is generally the most accurate \citep[][]{Argyriou2023}, but since the photometric calibration between the all 12 subbands matched down to at most a few \%, only minor offsets were needed to stitch all the 12 subbands to each other.} The final spectra of B1-c and L1448-mm are presented in Fig.~\ref{fig:specoverview_B1-c_L1448-mm}. For all sources, overview figures including spectra, continuum images, and the corresponding aperture are shown in Appendix~\ref{app:specextr} (Figs.~\ref{fig:specext_B1-a-NS}-\ref{fig:specext_BHR71-IRS2}).

% \begin{figure*}
%     \centering
%     \includegraphics[width=\linewidth]{{Figures/spec_extr/B1-c_conePSF_4.00_lambdaD}.png}
%     \caption{Extracted spectrum (top panel) and continuum images (bottom row) for B1-c. The 12 subbands in the spectrum are not stitched,  and each color represents one of the three MIRI-MRS gratings: A (red), B (black), and C (yellow). The bottom row shows from left to right the mid-IR continuum images around 5.3~\mum (channel 1 short), 8.1~\mum (channel 2 short), 12.5~\mum (channel 3 short), and 19.3~\mum (channel 4 short). The continuum images are scaled with a {\tt sqrt} stretch to enhance fainter features without saturating brighter emission. A 100~au scale bar is displayed in the bottom right and the FWHM of the PSF is shown in the bottom right of each panel.}
%     \label{fig:specext_B1-c}
% \end{figure*}

\subsection{Analysis methods}
This subsection describes in detail the specific the steps used in the analysis of the paper. The {baseline} subtraction is explained in Sect.~\ref{subsubsec:contsub}, {followed by a brief description of the extinction determination in Sect~\ref{subsubsec:extinction}.} The molecular spectroscopy used in this work is summarized in Sect.~\ref{subsubsec:mol_spec}. The LTE slab model fit procedure is described in Sect.~\ref{subsubsec:LTE_model_fitting} and the importance of including IR pumping in the derivation of the total number of molecules is stated in Sect.~\ref{subsubsec:IR_pumping}. Last, in Sect.~\ref{subsubsec:abundances}, the approach in taking abundance ratios is detailed. Readers interested in the results of this paper should proceed to Sect.~\ref{sec:results}.

\subsubsection{{Baseline} subtraction}
\label{subsubsec:contsub}
In order to fit molecular emission and absorption features, it is important to subtract {the baseline}. This was achieved by a combination of an automated fit to the {observed} continuum followed by a visual inspection and correction where necessary. First, the full 4.9--27.9~\mum range was fitted automatically with a univariate spline function to line-free {bins}. The latter were selected based on whether their measured flux was higher or lower than the 40\% and 60\% quantiles of 10 neighboring {bins}. If the {wavelength bin} had a lower or higher value, it was considered as either an emission or absorption line or a bad pixel, whereas if {its value} was within the 40-60\% quantile range, it was considered as a line-free {wavelength bin} in the fit with the univariate spline function. For spectra very rich in emission or absorption lines (e.g., L1448-mm, B1-c, TMC1-W), the range of quantiles was varied to 10-30\% and 70-90\%, respectively, in order to provide a better automated fit of the {baseline}. 

Second, the automated fit was checked by visual inspection and line-free {wavelength regions} were added or removed manually. This was especially important for very line rich sources (e.g., TMC1-W) where no clear line-free {regions} are present in some parts of the spectrum. Furthermore, the automated {baseline} estimate often did not provide a proper result in intermediately-broad ice absorption features such as those of CH$_4$ (around 7.7~\mum) and CO$_2$ (15~\mum). 

Following the {baseline} subtraction, the noise level $\sigma$ was determined per subband from line-free {wavelength regions}. The derived $\sigma$ (in mJy) are presented in Table~\ref{tab:sigma_data}. The noise level is $\sigma\sim0.1-0.5$~mJy in channels 1-3 for most sources, with $\sigma$ increasing toward values of up to $\sim$10~mJy in channel~4C. TMC1A shows higher noise levels of several mJy already in channels 1-3 because the source is located near the edge of the FOV.

% Besides determining the baseline of the spectra needed for fitting the molecular emission and absorption features, the global thermal spectral energy distribution (SED) continuum was also fitted in a similar manner as done by \citet{Rocha2024} and \citet{Chen2024} (see top panel of Fig.~\ref{fig:specoverview_B1-c_L1448-mm}). This is important for determining the amount of differential extinction on the molecular features by the various ice and silicate absorption features \citep[see e.g., Appendix C of][]{vanGelder2024}.

\subsubsection{Extinction}
\label{subsubsec:extinction}
The total extinction was determined in a similar way as was done by \citet[][see their Appendix~C]{vanGelder2024}.  In short, the total extinction is decomposed into two components: the differential extinction caused by various ice absorption bands and silicates ($\tau_{\rm ice,silicates}$) and the absolute extinction ($\tau_{\rm ext}$) based on the \citet{McClure2009} extinction law. In this paper, $\tau_{\rm ice,silicates}$ was determined for each source by using the baseline continuum fit (see Sect.~\ref{subsubsec:contsub}) via,
\begin{align}
    \tau_{\rm ice,silicates} = -\ln\left(\frac{F_{\rm baseline}}{F_{\rm SED}}\right),
\end{align}
where $F_{\rm baseline}$ is the local continuum {baseline} (in mJy) and $F_{\rm SED}$ is the thermal spectral energy distribution (SED) global continuum fit (in mJy). The latter was derived in a similar manner as done by \citet{Rocha2024} and \citet{Chen2024} (see top panel of Fig.~\ref{fig:specoverview_B1-c_L1448-mm}).

The absolute extinction was determined by fitting a powerlaw function to regions of the \citet{McClure2009} extinction law where ice and silicate absorption features are absent (see Appendix~C of \citealt{vanGelder2024} for the derivation) and scaling the optical depth with $A_{\rm K}$,
\begin{align}
    \tau_{\rm ext} = 0.085\lambda^{-0.25} A_{\rm K}/7.75,
\end{align}
where $A_{\rm K}$ was set to $7$~mag \citep[maximum $A_{\rm K}$ for which the][excintion law is valid]{McClure2009} for all sources, corresponding to $A_{\rm V} \approx 55$~mag. 
Since the interest of this paper lies primarily in column density ratios rather than in their absolute values, the precise value of $A_{\rm K}$ does not matter for our purposes.

\subsubsection{Molecular spectroscopy}
\label{subsubsec:mol_spec}
The spectroscopic information needed for fitting the emission and absorption of all targeted molecules was taken from the HITRAN database\footnote{\url{https://hitran.org/}} \citep{Gordon2022}. The full list of molecules included in this work contains H$_2$O, CO, OH, CO$_2$, $^{13}$CO$_2$, C$_2$H$_2$, $^{13}$CCH, HCN, C$_4$H$_2$, CH$_4$, SO$_2$, CS, SiO, and NH$_3$. The most relevant spectroscopic information is the rest wavelength, upper energy level $E_{\rm up}$, Einstein $A_{\rm ij}$ coefficients, and upper level degeneracy $g_{\rm up}$ for each transition. The HITRAN data were converted to the Leiden Atomic and Molecular Database \citep[LAMDA;][]{Schoier2005,vanderTak2020} format to make it compatible with the {\tt radexpy} slab model code \citep[e.g.,][]{Grant2023,Tabone2023,Francis2024}. The partition functions $Q$ as function of temperature were similarly obtained from the HITRAN database. 
The partition function of SiO was taken from the Cologne Database for Molecular Spectroscopy\footnote{\url{https://cdms.astro.uni-koeln.de/}} \citep[CDMS;][]{Muller2001,Muller2005_CDMS,Endres2016}.
% The spectroscopic information of CH$_3^+$ was obtained from \citet{Berne2023}.

\subsubsection{LTE model fitting}
\label{subsubsec:LTE_model_fitting}
For each molecule, the best fitting excitation temperature $T_{\rm ex}$ and column density $N$ were determined by setting a grid covering a large range of conditions \citep[see e.g.,][]{Francis2024}. Two different grids are explored in this work. One higher resolution grid covers $T_{\rm ex} = 50-800$~K in steps of 10~K and $\log_{10}(N) = 14-21$ in units of cm$^{-2}$ with steps of 0.125 on a $\log_{10}$ scale. The second grid has a lower resolution but covers a larger range in both $T_{\rm ex}$ and $N$: $50-2500$~K in steps of 25~K and $\log_{10}(N) = 14-23$ in units of cm$^{-2}$ with steps of 0.167 on a $\log_{10}$ scale. The higher resolution grid is applicable to most molecules since typical temperatures measured lie in the 100--500~K range. However, for CO, OH, and in some cases H$_2$O, higher temperatures (up to 2000~K) are measured, therefore requiring the need for exploring a broader parameter space. 

Given that H$_2$O could not be fitted by an LTE model with a single excitation temperature in many sources, multiple components were used that probe different reservoirs of H$_2$O in the protostellar system with different temperatures \citep[see e.g.,][]{Gasman2023,Temmink2024_H2O}. Here, three different components are adopted. All the rovibrational lines shortward of 9~\mum were fitted using a single LTE model (Rovib H$_2$O). The pure rotational lines longward of 11~\mum were fitted using a colder ($T\sim100-300$~K) and warm ($T>300$~K) component.
% What grid was used for which molecule is shown in Table~xx.

Absorption and emission model grids were computed separately. For each grid point, the optical depth as a function of wavelength was computed on a high spectral resolution of $R = 10^6$ taking into account line overlap producing optically thick lines \citep[][]{Tabone2023}. For emission models, the intrinsic line broadening $\Delta V$ was set to the commonly adopted value of 4.71~km~s$^{-1}$ based on the thermal line broadening of H$_2$ molecules at 500~K \citep{Salyk2011}. It is important to note that in case of optically thin emission, the derived column density is independent of $\Delta V$, whereas for optically thick emission, the column density scales with $\Delta V^{-1}$ \citep[e.g.,][]{Tabone2023}. For absorption models, on the other hand, the depth of absorption features is more dependent on the assumed $\Delta V$. Since determining $\Delta V$ accurately for the sample is not the main goal of this paper, only a few values of $\Delta V$ (2, 4.71, 10, and 20~km~s$^{-1}$) were explored in order to get good fits to the data. Larger $\Delta V$ were tested by visual inspection, but did not improve the fit for any of the sources. 

{The radial velocity offset ($v_{\rm line}$) with respect to the $v_{\rm lsr}$ was determined for CO$_2$, SiO, and H$_2$O in our three most line-rich sources, B1-c, L1448-mm, and BHR71-IRS1, through Gaussian fitting of selected unblended lines. For H$_2$O, the rovibrational lines and rotational lines were fitted separately. Since multiple lines of the same species were included, the uncertainty of the velocity could be derived at the level of $\sim5$~\kms. The $v_{\rm line}$ of all other species was determined by visual inspection with a typical uncertainty of $\sim$10-20~\kms. The aim of this paper is not to derive accurate velocity information for all sources and species, but only for the selected sources and species for which more accurate velocities are needed in interpreting the results. For all other sources and species, $v_{\rm line}$ will be derived in future works focusing on spatially extended molecular emission in the outflows.}

% The radial velocity offset ($v_{\rm line}$) with respect to the $v_{\rm lsr}$ was determined by visual inspection with an estimated uncertainty of about 10\% of the spectral resolution \citep[$R\sim3500-1000$;][]{Labiano2021,Jones2023}, giving a typical uncertainty on $v_{\rm line}$ of $\sim$10~\kms for species emitting at shorter wavelengths ($<12$~\mum) and $\sim$20--30~\kms at longer wavelengths ($>12$~\mum).

For absorption models, the $\chi^2$ was minimized for each grid point on the optical depth scale following the description of \citet{Francis2024} which is based on earlier work of \citet{Helmich1996_PhD},
\begin{align}
    \chi^2 = \sum_{i=1}^{N_{\rm bin}} \left(\frac{\tau_{\rm obs,i}-\tau_{\rm model,i}}{\tau_{\sigma}}\right)^2,
    \label{eq:chi2_abs}
\end{align}
where $N_{\rm bin}$ is the number of selected wavelength bins used in the fit, $\tau_{\rm obs}$ is the observed line optical depth defined as $\tau_{\rm obs} = -\ln(F_{\rm obs}/F_{\rm baseline})$ with $F_{\rm obs}$ the observed flux (in mJy) and $F_{\rm baseline}$ the local continuum {baseline} (in mJy) determined in Sect.~\ref{subsubsec:contsub}, $\tau_{\rm model}$ is the model optical depth, and $\tau_{\sigma}$ the uncertainty on the optical depth defined as $\tau_{\sigma} = \sigma/F_{\rm obs}$ {with $\sigma$ the noise level in mJy presented in Table~\ref{tab:sigma_data}}. In order to match the observations, the optical depth of the model in Eq.~\eqref{eq:chi2_abs} ($\tau_{\rm model}$) was scaled to the MIRI-MRS resolution as function of wavelength \citep[$R = 3500-1000$;][]{Labiano2021,Jones2023,Argyriou2023,Pontoppidan2024} 

In case of emission, the model optical depth {as function of wavelength} was first converted to the model flux ($F_{\rm model}$) {before calculating the $\chi^2$} via,
\begin{align}
    F_{\rm model} = \pi \left(\frac{R}{d}\right)^2 B_\nu(T_{\rm ex}) \left(1-e^{\tau_{\rm model}}\right),
    \label{eq:flux_em}
\end{align}
where $d$ is the distance to the source, $R$ is the radius of the emitting area, and $B_\nu(T_{\rm ex})$ is the Planck function at the excitation temperature of the corresponding grid point. In Eq.~\eqref{eq:flux_em}, the emitting area was parameterized as a circle with a radius $R$, but in reality the emitting area can have any shape with an area equal to $\pi R^2$. In contrast to absorption models, emission models thus have one additional parameter $R$ related to the size of the emitting area which was needed to scale the model to the observed flux. In case of optically thin emission, $N$ and $R$ are completely degenerate with each other, but for optically thick emission, $N$ and $R$ can both be constrained. Using the model flux, the $\chi^2$ for each grid point was calculated analogous to Eq.~\eqref{eq:chi2_abs},
\begin{align}
    \chi^2 = \sum_{i=1}^{N_{\rm bin}} \left(\frac{F_{\rm obs,i}-F_{\rm model,i}}{\sigma}\right)^2,
    \label{eq:chi2_em}
\end{align}
with $\sigma$ the noise level (in mJy) for each band listed in Table~\ref{tab:sigma_data}.

{In contrast to absorption models, it is important to take into account the extinction to provide a good fit to the data and derive accurate column densities. This is especially important for species with transitions in deep ice absorption features \citep[e.g., CO$_2$ $P$-branch lines in the CO$_2$ ice feature;][]{Francis2024,vanGelder2024}. The extinction as described in Sect.~\ref{subsubsec:extinction} is applied to the LTE model before computing the $\chi^2$. }

% In contrast to absorption models, it is important to take into account extinction by silicates and ices in order to provide a good fit to the data \citep[see e.g.,][]{Francis2024,vanGelder2024}. The total extinction was determined in a similar way as was done by \citet[][see their Appendix~C]{vanGelder2024} by computing it as a superposition of the differential extinction by ices and silicates and the absolute extinction following the \citet{McClure2009} extinction law. In this paper, the differential extinction was determined for each source by using the global continuum fit (see Sect.~\ref{subsubsec:contsub}). The absolute extinction was determined using (see Appendix~C of \citealt{vanGelder2024} for the derivation),
% \begin{align}
%     \tau_{\rm ext} = 0.085\lambda^{-0.25} A_{\rm K}/7.75,
% \end{align}
% where $A_{\rm K}$ was set to $7$~mag \citep[maximum $A_{\rm K}$ for which the][excintion law is valid]{McClure2009} for all sources, corresponding to $A_{\rm V} \approx 55$~mag. 
% Since the interest of this paper lies primarily in column density ratios rather than in their absolute values, the precise value of $A_{\rm K}$ does not matter for our purposes.

Only wavelengths that show emission or absorption features were included in the fit while wavelengths overlapping with strong atomic or H$_2$ lines were excluded. Furthermore, the shape of the $Q$-branch of molecular emission or absorption features is very sensitive to the excitation temperature but is also very degenerate with line optical depth. Hence, even though the line optical depth was taken into account in the LTE models, strongly blended lines such as $Q$-branches were omitted in the fit to not suffer from high line optical depths. This indeed improves the accuracy of the derived excitation temperatures and column densities \citep[see discussion in Appendix~\ref{app:line_optical_depth} and also][]{Li2024}. 
{However, $Q$-branches or other possibly optically thick lines were included if these were the only detected lines (i.e., when both $R$ and $P$-branches were not detected).}
% These wavelengths are listed per source in {Table~xx}. 
Following the procedure of \citet{Grant2023} and \citet{Francis2024}, an iterative fit of the molecules was performed in order of decreasing flux where the best-fit model for each molecule was subtracted before continuing to the next molecule.

The best-fit model was determined by the minimum $\chi^2$ of the grid. The confidence intervals of $T_{\rm ex}$, $N$, and $R$ (for emission models) were determined following \citet{Carr2008} and \citet{Salyk2011} by placing contours on the $\chi^2$ maps. Similarly, the best-fit number of molecules $\mathcal{N}_{\rm mol}$ of emission models was computed via,
\begin{align}
    \mathcal{N}_{\rm mol}  = N \pi R^2,
\end{align}
and its confidence intervals were calculated similarly as those of $T_{\rm ex}$ and $N$. The number of free parameters $K$ was set to 2 \citep[see discussion in][]{Avni1976}, and the 1,2,3$\sigma$ confidence intervals were computed from the reduced $\chi^2$ ($\chi^2_{\rm red} = \chi^2 /K$) by setting a $\Delta\chi^2_{\rm red}$ of 2.3, 6.2, and 11.8, respectively. 

A molecule (or component of H$_2$O) was considered detected when at least 3 wavelength bins showed emission or absorption above the 5$\sigma$ level. For the cases where no accurate constraints on the parameters could be achieved for a certain molecule (e.g., $^{13}$CO$_2$ and $^{13}$CCH$_2$ in L1448-mm), the temperature was fixed to that of other species in the same source and only the column density and emitting area were fitted for. 
% Similarly, $T_{\rm ex}$ was fixed to 200~K for $^{13}$CO$_2$ in L1448-mm to improve the quality of the fit. 
Moreover, in case of non-detections, the 3$\sigma$ upper limit to the column density was derived for a typical excitation temperature of 150~K and setting the emitting radius to a typical radius of 10~au. It is important to note that the emission (and absorption) was assumed to be optically thin and the resulting upper limit on $\mathcal{N}_{\rm mol}$ is therefore not dependent on the assumed radius. For the warm H$_2$O component, an excitation temperature of 450~K was set {when deriving the 3$\sigma$ upper limit to the column density}.

\subsubsection{Infrared pumping}
\label{subsubsec:IR_pumping}
Column densities derived from absorption lines are not susceptible to possible non-LTE effects such as infrared pumping. However, for column densities (or number of molecules) derived from emission line models, this has to be taken into account so that these parameters are not overestimated \citep[e.g.,][]{Boonman2003_Orion,Bruderer2015,Bosman2017,vanGelder2024}. The most direct way to take infrared pumping into account would be to run a grid of non-LTE {\tt RADEX} models \citep[][]{vanderTak2007,Bruderer2015,Bosman2017}, but this is beyond the scope of this work and only possible for a select number of molecules (i.e., H$_2$O, CO$_2$, and HCN) for which collisional rate coefficients are available. It would also require detailed physical and chemical model of the source with the infrared radiation field specified at each point.
An indirect method is through measurements of the same molecules at sub-millimeter wavelengths in their pure rotation transitions as was done for SO$_2$ in NGC~1333~IRAS~2A \citep[][]{vanGelder2024}, but this is not possible for molecules such as CH$_4$, CO$_2$, and C$_2$H$_2$ that lack pure rotational transitions at millimeter wavelengths due to the absence of a permanent dipole moment.

\renewcommand{\arraystretch}{1.2}
\begin{table}[t]
    \centering
    % \footnotesize
    \caption{Infrared pumping information for each molecule.}
    \begin{tabular}{lcc}
    \hline\hline
    Species & Mode\tablefootmark{(1)} & $\lambda_{\rm pump}$ \tablefootmark{(2)} \\
    & & \mum \\
    \hline
    Rovib H$_2$O & $\nu_2$ & 6.0 \\
    CO$_2$ & $\nu_3$ & $4.3$ \\
    $^{13}$CO$_2$ & $\nu_3$ & 4.3 \\
    C$_2$H$_2$ & $\nu_4+\nu_5$ & $7.7$ \\
    $^{13}$CCH$_2$ &  $\nu_4+\nu_5$ & $7.7$ \\
    HCN & $\nu_2$ & $7.0$ \\
    C$_4$H$_2$ & $\nu_6+\nu_8$ & 8.0 \\
    CH$_4$ & $\nu_4$ & 7.6 \\
    SO$_2$ & $\nu_3$ & 7.3 \\
    CS & $\nu_2$ & 8.0 \\
    SiO & $\nu_1$ & 8.1 \\
    NH$_3$ & $\nu_4$ & 6.1 \\ 
    % Others & -- & max\tablefootmark{(3)} \\
    \hline
    \end{tabular}
    \label{tab:IR_pump}
    \tablefoot{
    % The brightness temperature shortward of 4.9~\mum was obtained from GO NIRSpec data within the JOYS collaboration (private communication) or through interpolation of the SED to shorter wavelengths for sources where NIRSpec data is not available. 
    For species not listed in this table, the pumping wavelength was assumed to be equal to wavelength corresponding to the observed maximum flux.
    \tablefoottext{1}{Vibrational mode through which the IR pumping is assumed to occur.} \tablefoottext{2}{Wavelength of IR pumping that is applied (as frequency $\nu$) in Eqs.~\eqref{eq:T_IR} and \eqref{eq:IR_pump}.} 
    % \tablefoottext{3}{For all other species, the wavelength equal to the maximum flux was assumed.}
    }
\end{table}
\renewcommand{\arraystretch}{1.0}

Nevertheless, the effect of infrared pumping can be estimated when assuming the vibrational temperature ($T_{\rm vib}$) is set by the infrared radiation field \citep[as was the case for SO$_2$ in NGC~1333~IRAS~2A;][]{vanGelder2024}. The vibrational temperature is then approximately equal to the brightness temperature ($T_{\rm IR}$) at the frequency ($\nu$) of the vibrational mode through which the IR pumping occurs,
\begin{align}
    T_{\rm vib} \approx T_{\rm IR} = \frac{h \nu}{k_{\rm B}} \ln^{-1}\left(1 + \frac{2 h \nu^3}{I_\nu c^2}\right),
    \label{eq:T_IR}
\end{align}
where $I_\nu$ is the extinction corrected surface brightness (in Jy~sr$^{-1}$), $h$ is Planck's constant, $k_{\rm B}$ is the Boltzmann constant, and $c$ is the speed of light. The corrected number of molecules ($\mathcal{N}_{\rm mol,corr}$) can then be computed via \citep[][]{vanGelder2024},
\begin{align}
    \mathcal{N}_{\rm mol,corr} = \mathcal{N}_{\rm mol} \frac{e^{-h \nu/(k_{\rm B} T_{\rm rot})}}{e^{-h \nu/(k_{\rm B} T_{\rm vib})}}, 
    \label{eq:IR_pump}
\end{align}
where $\mathcal{N}_{\rm mol}$ is the number of molecules measured from the LTE model fits (Sect.~\ref{subsubsec:LTE_model_fitting}) and $T_{\rm rot}$ is the rotational temperature in the vibrational ground state \citep[assumed to be equal to $T_{\rm ex}$ measured from the LTE model fits;][]{vanGelder2024}. Statistical degeneracies $g$ have been neglected in Eq.~\eqref{eq:IR_pump} {for simplicity} but this does not significantly affect the derived $\mathcal{N}_{\rm mol,corr}$ {since the uncertainties on $T_{\rm ex}$ and $T_{\rm vib}$ dominate the uncertainty of $\mathcal{N}_{\rm mol,corr}$ (see discussion below)}. The correction was only applied to species that are detected and not to derived upper limits on $\mathcal{N}_{\rm mol}$. 

% Assuming that $T_{\rm vib}$ is set by the infrared radiation field, as was the case for SO$_2$ in NGC~1333~IRAS~2A \citep[][]{vanGelder2024}, it is approximately equal to the brightness temperature at the frequency $\nu$ of the vibrational mode through which the IR pumping occurs,

% \begin{align}
%     T_{\rm vib} \approx T_{\rm IR} = \frac{h \nu}{k_{\rm B}} \ln^{-1}\left(1 + \frac{2 h \nu^3}{W I_\nu c^2}\right),
% \end{align}
% where $I_\nu$ is the extinction corrected surface brightness (in Jy~sr$^{-1}$) and $W$ is the dilution factor. This dilution factor takes into account that the flux that the IR-pumped material receives is diluted due to the its distance from the infrared emitting source (i.e., the inner disk). A dilution factor of $W=1$ (i.e., no dilution) results in the largest correction factor on $\mathcal{N}$, but is likely not realistic. Typically, the dilution is assumed to be about 10\% ($W=0.1$), which in turn results in a factor 10 difference on the derived $\mathcal{N}_{\rm corr}$. 
It is important to take into account that the IR pumping does not necessarily go through the vibrational level of the observed emission. For CO$_2$, for example, pumping can occur through the $\nu_3$ mode around $4.3$~\mum followed by de-excitation through other vibrational bands such as the $\nu_2$ bending mode around 15~\mum \citep[][]{Bosman2017}. Many of the detected molecules have vibrational modes at shorter wavelengths (i.e., higher $T_{\rm IR}$) than those that are detected here (e.g., C$_2$H$_2$ with the $\nu_4+\nu_5$ mode around 7.7~\mum, which is not detected in our data). The bands through which the IR pumping is assumed to occur are presented in Table~\ref{tab:IR_pump}. For CO$_2$, the brightness temperature at 4.3~\mum (i.e., outside the MIRI-MRS range) was obtained from a NIRSpec program (PID: 1960) within the JOYS collaboration (private communication) or through interpolation of the SED to shorter wavelengths for sources where NIRSpec data is not available. 

% It is important to note that Eq.~\eqref{eq:IR_pump} assumes that the infrared pumping occurs from the ground state toward the first vibrational excited state from which the emission lines are detected. In reality, higher vibrational levels may get pumped and subsequently populate the first vibrational level through radiative decay. However, we assume that this effect negligible compared to the infrared pumping from the vibrational ground state to the first excited state. 

Computing $\mathcal{N}_{\rm corr}$ comes with some significant uncertainties (i.e., $T_{\rm vib} \approx T_{\rm IR}$, 
% dilution factor $W$, 
pumping through higher order excited states, neglecting $g$ factors), see discussion in Appendix.~\ref{app:IR_pumping}. Small differences of $5-10$~K in the derived $T_{\rm vib}$ can already lead to an order of magnitude difference in the derived $\mathcal{N}_{\rm corr}$, see Fig.~\ref{fig:IR_pump}.
The uncertainty on $\mathcal{N}_{\rm corr}$ can therefore be orders of magnitude even when the difference between $T_{\rm vib}$ and $T_{\rm rot}$ is small. Nevertheless, it is important to provide an estimate for the effect of infrared pumping on the derived number of molecules. Therefore, only the minimum and maximum values of of $\mathcal{N}_{\rm corr}$ are presented, where the maximum value of $\mathcal{N}_{\rm corr}$ corresponds to the uncorrected $\mathcal{N}_{\rm mol}$ and the minimum value of $\mathcal{N}_{\rm corr}$ to that computed using Eq.~\eqref{eq:IR_pump}.
% The uncertainty on $\mathcal{N}_{\rm corr}$ is therefore calculates by propagating the errors in Eq.~\eqref{eq:IR_pump} whilst adopting a factor 90\% uncertainty on the computed correction factor. 
A full non-LTE analysis, taking all these effects into account, is beyond the scope of this work. 

\subsubsection{Abundances}
\label{subsubsec:abundances}
Column densities or total number of molecules by themselves do not provide accurate constraints on the abundance of a molecule due to observational dependencies (i.e., emitting area). The ratio between two molecules, on the other hand, do provide such constraints, assuming that the two molecules are roughly located in the same region of the protostellar system. It is, however, important that a relevant reference species is selected. 

In this paper, all abundance ratios were taken with respect to H$_2$O since it is a dominant ice species and because it is abundantly detected in many sources. Water is therefore an excellent reference species for ice sublimation in hot cores. High-temperature gas-phase chemistry in either the hot core or in shocks can also produce H$_2$O, but as long as the H$_2$O is just recycled in the hot gas (i.e., destroyed and reformed), it is still a good reference species. The main issue would be if there is a significant amount oxygen not originally in H$_2$O ice or gas (e.g., atomic O, refractory dust) that is driven into H$_2$O by high-temperature gas-phase chemistry, but this cannot contribute to the amount of H$_2$O by more than a factor of a few \citep[see e.g.,][]{vanDishoeck2021}.

Water shows emission or absorption lines through both its rovibrational lines between $5-9$~\mum and its pure rotational lines longward of $13$~\mum. However, the ro-vibrational lines are susceptible to IR pumping effects (see Sect.~\ref{subsubsec:IR_pumping}) whereas the pure rotational lines are not. On the other hand, the pure rotational lines at mid-IR wavelengths originate from high energy levels of several thousands of K, needing warm gas to be excited, whereas rovibrational lines in absorption originate from the ground state with $E_{\rm low}$ as low as 0~K, therefore being more sensitive to the cold gas. Nevertheless, given that the derived temperatures of H$_2$O from the pure rotational lines are low ($100-500$~K, see Table~\ref{tab:LTE_Tex}) and in order to avoid the effect of infrared pumping, the total column density (absorption) or number of molecules (emission) of H$_2$O was derived from the pure rotational lines. The rotational lines were fitted using two components (cold, warm) and the total column density or number of molecules was calculated through the sum of these two components. In the majority of the sources, the total rotational H$_2$O component (hereafter H$_2$O-rot) is dominated by the cold component, although for some sources this is not detected (e.g., B1-c). 

An alternative to H$_2$O as a reference species would be CO$_2$, which is also a dominant ice species detected toward many of the sources with a rather constant CO$_2$/H$_2$O ice ratio \citep[e.g.,][]{Pontoppidan2008}. However, gaseous CO$_2$ appears to be more often associated with disk winds or outflowing material and its derived column density or number of molecules is often uncertain due to IR pumping. 
Molecular hydrogen is another option as it is the most abundant molecule and would provide absolute abundances. However, the emission of the low-$J$ lines, which are most sensitive to the warm ($\sim$few~100~K) gas, is dominated by the outflow or disk winds \citep[e.g.,][Francis et al. in prep.]{Tychoniec2024}, making it difficult to disentangle the contribution from the warm inner envelope and disk. 
Therefore, H$_2$O was selected as the reference species for abundance ratios.

\section{Results}
\label{sec:results}
\subsection{Continuum emission}
Within the JOYS low-mass sample, 18 continuum point sources are detected, see Appendix~\ref{app:specextr}. The binary B1-a is only very marginally resolved at the shortest wavelengths (see Fig.~\ref{fig:specext_B1-a-NS}) and fully unresolved from $\sim12$~\mum onwards and therefore analyzed as a single source B1-a-NS. On the other hand, the TMC1 binary is resolved up to $\sim16$~\mum (see Figs.~\ref{fig:specext_TMC1-E} and \ref{fig:specext_TMC1-W}) and therefore both components are analyzed individually. At longer wavelengths, the binary becomes unresolved, which could explain the similar temperature of the cold H$_2$O components (see Sect.~\ref{subsubsec:results_H2O}). However, this does not affect any of the derived conclusions. For L1527, the continuum traces the scattered light on both sides of the disk (see Fig.~\ref{fig:specext_L1527}). 
% From visual inspection, no clear discrepancies are visible in the molecular gas content between both sides. The spectrum was therefore extracted from a circular aperture (i.e., area that is constant in wavelength) with a radius of 1.5$''$ to encompass both sides of the disk. 

\begin{figure*}[p]
    \centering
    \includegraphics[width=\linewidth]{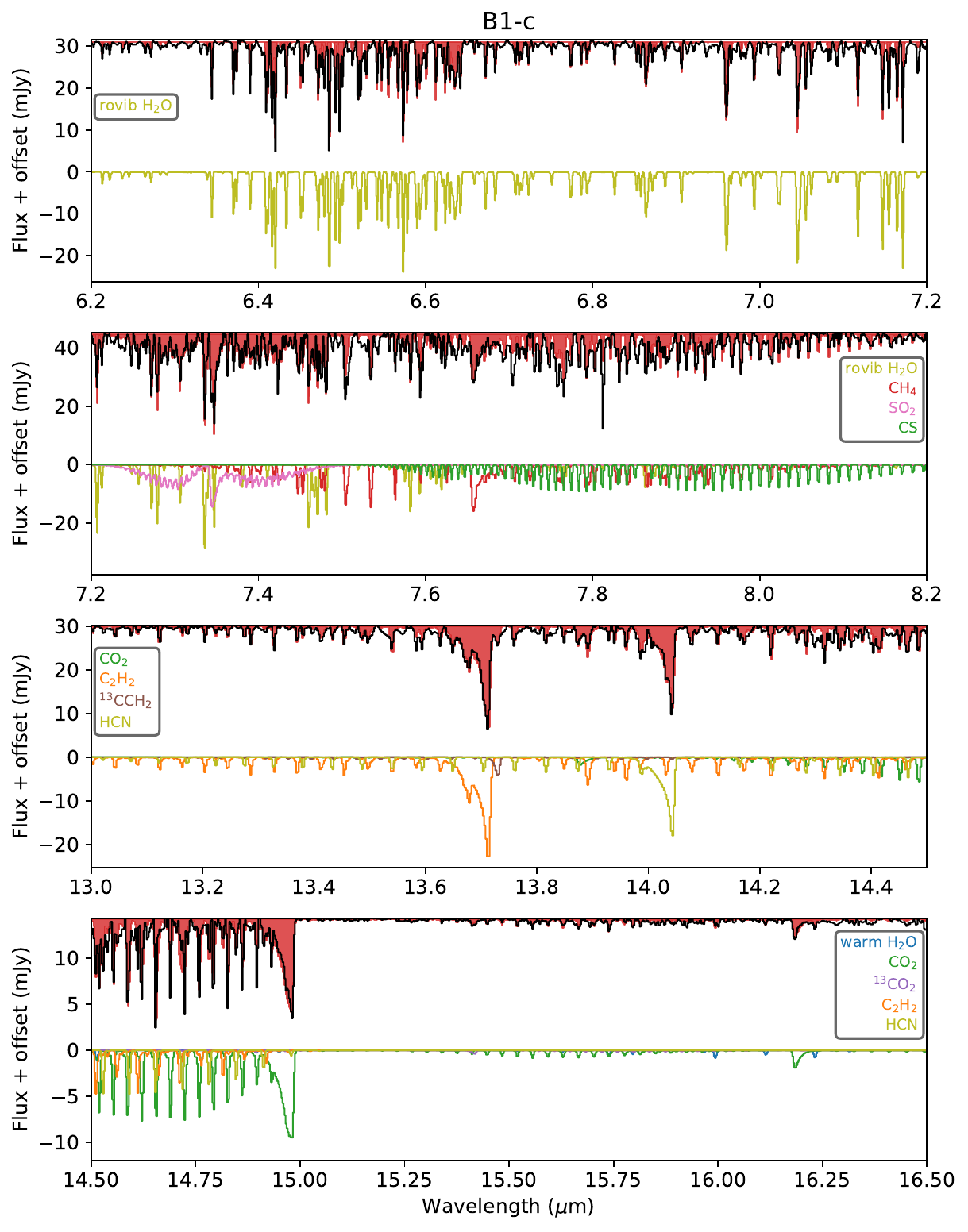}
    \caption{Overview of the main molecular absorption features detected toward B1-c in the MIRI-MRS wavelength range. In each panel, the {baseline-subtracted} spectrum is shown in black and best-fit LTE model including all molecules listed is overlaid as the red shaded area. In the bottom of each panel, the individual best-fit LTE models of all molecules contributing to the corresponding wavelength range are displayed at an arbitrary constant offset, with each color denoting a different species. 
    % Deep negative absorption features originating from detector artifacts are clipped for clarity. 
    The surprising detection of NH$_3$ toward B1-c is highlighted in Fig.~\ref{fig:specfit_B1-c_CS_NH3} .
    }
    \label{fig:specfit_B1-c_overview}
\end{figure*}

Both IRAS~4B and Ser-S68N-S are not detected at their source position \citep[based on ALMA data;][]{Tychoniec2021} in the continuum over the full MIRI-MRS wavelength range. For Ser-S68N-S, some continuum flux appears to be present longward of 12~\mum, but this likely originates from the wings of the PSF from Ser-S68N-N (see Fig.~\ref{fig:specext_Ser-S68N-S}). Both IRAS~4B and Ser-S68N-S are likely too embedded to be detectable even at MIRI-MRS wavelengths, similar to the case of HH~211 \citep[][]{Ray2023,CarattioGaratti2024}. Ser-S68N-S has an outflow oriented nearly in the plane of the sky \citep[i.e., edge-on disk;][]{Podio2021}, which could also explain the lack of continuum emission, but for IRAS~4B the orientation of the outflow is unknown \citep[][]{Podio2021}. Toward the source position of IRAS~4B no clear molecular features are present, but in particular strong emission of H$_2$O is detected in the jet about 4$''$ to the south for IRAS~4B, as was also seen by {\it Herschel} \citep[][]{Herczeg2012}, which will be discussed in a separate paper. Toward Ser-S68N-S, weak CO and CO$_2$ emission is detected around 5~\mum and 15~\mum, respectively, which is likely related to the strong molecular outflow \citep[][Francis et al. in prep.]{Tychoniec2019}. Additionally, H$_2$O and OH emission is detected at $>20$~\mum wavelengths, but this is likely related to Ser-S68N-N since its PSF starts overlapping with that of Ser-S68N-S at longer wavelengths. IRAS~4B and Ser-S68N-S are therefore excluded in the remainder of the analysis. However, it is important to note that the absence of molecular features in the spectra in these two sources does not mean that the molecules are absent in these systems but are possibly hidden due to the high extinction by their natal envelopes and/or dusty accretion disks.

\subsection{Molecular spectra}
\label{subsec:spectra}
An overview of the two most line-rich spectra of the JOYS sample, B1-c and L1448-mm, is presented in Fig.~\ref{fig:specoverview_B1-c_L1448-mm}. 
For B1-c, the {baseline-subtracted} spectrum around a few key wavelength ranges is also presented in Fig.~\ref{fig:specfit_B1-c_overview}. 
The spectra show an unprecedented richness in gas-phase molecular features across the full wavelength range, as well as multiple deep ice absorption features. These ice features include both simple (i.e., H$_2$O, CO$_2$, CH$_4$, NH$_3$) and complex (e.g., CH$_3$OH, C$_2$H$_5$OH, CH$_3$OCHO) molecules and are presented in separate papers \citep[e.g.,][]{Rocha2024,Chen2024,Brunken2024,Slavicinska2024}. Multiple sources also show emission of H$_2$ (not analyzed in this work) and atoms (i.e., {[S I]}, {[Fe II]}) tracing mostly the disk wind, outflow, or jet \citep[e.g.,][]{Tychoniec2024}. The molecular gas-phase lines are detected both in absorption (e.g., B1-c, BHR71~IRS1) and in emission (e.g., B1-a-NS, L1448-mm) toward the JOYS sources.

The shorter wavelengths (i.e., $<7$~\mum) are dominated by features from the first overtone of CO and the $\nu_2$ rovibrational lines of H$_2$O, see for B1-c in Fig.~\ref{fig:specoverview_B1-c_L1448-mm}. Only the $J>25$ $P$-branch lines of CO are covered in the MIRI-MRS wavelength range. 
% Due to the degeneracy between high excitation temperature and line optical depth predicted by LTE models \citep[e.g.,][see also discussion in \citealt{Rubinstein2024} for the NIRSpec range]{Herczeg2011,Francis2024}, these CO lines will not be analyzed in detail further. 
The rovibrational lines of H$_2$O are commonly detected toward the JOYS sources (see Sect.~\ref{subsec:mol_statistics}), both in emission and in absorption. In several sources (e.g., B1-c, TMC1-W), the forest of H$_2$O lines is so dense that almost no line-free {wavelength regions} are present. 

At longer wavelengths (i.e., $>7$~\mum), other molecular features start to appear such as those of the SO$_2$ $\nu_3$ band with the $Q$-branch at $7.35$~\mum \citep[see also][]{vanGelder2024} and the CH$_4$ $\nu_4$ mode with its $Q$-branch at $7.65$~\mum. Interestingly, clear emission (L1448-mm) and absorption (BHR71-IRS1) features of SiO are detected between $8-8.5$~\mum (see bottom left panel of Fig.~\ref{fig:specoverview_B1-c_L1448-mm}). This triples the number of SiO detections at mid-IR wavelengths toward protostellar systems \citep[see e.g.,][]{McClure2024}. Moreover, toward B1-c, clear absorption features associated with CS and NH$_3$ are detected (see e.g., Fig.~\ref{fig:specfit_B1-c_overview}), marking the first detection of these molecules at mid-IR wavelengths toward a low-mass protostellar system.

\renewcommand{\arraystretch}{1.2}
\begin{table*}[t]
    \centering
    \footnotesize
    \caption{Presence of molecules in absorption (A), emission (E), or a non-detection (--). 
    % {Will add detections in extended outflow as "O"}
    }
    \begin{tabular}{lccccccccccccccc}
    \hline\hline
    % Source & rot-H$_2$O & rovib-H$_2$O & CO$_2$ & $^{13}$CO$_2$ & C$_2$H$_2$ & $^{13}$CCH$_2$ & HCN & C$_4$H$_2$ & CH$_4$ & SO$_2$ & CS & SiO & NH$_3$ & CO & OH \\
    Source & \multicolumn{2}{c}{H$_2$O} & CO$_2$ & $^{13}$CO$_2$ & C$_2$H$_2$ & $^{13}$CCH$_2$ & HCN & C$_4$H$_2$ & CH$_4$ & SO$_2$ & CS & SiO & NH$_3$ & CO & OH \\ \cline{2-3}
    % & rot & rovib & $15$~\mum & $15.6$~\mum & $13.6$~\mum & $13.7$~\mum & $14.1$~\mum & $15.9$ & $7.66$~\mum & $7.35$~\mum & $8$~\mum & $8.3$~\mum & $9$~\mum \\
    % \hline
    & rot & rovib &  \\
    \hline
B1-a-NS & E & E & A & -- & -- & -- & -- & -- & -- & -- & -- &-- & -- & E & E  \\
B1-b & -- & -- & -- & -- & -- & -- & -- & -- & -- & -- & -- & -- & -- & -- & --  \\
B1-c & A & A & A & A & A & A & A & -- & A & A & A & -- & A & A & --  \\
L1448-mm & E & E & E & E & E & E & E & E & E & E & -- & E & -- & E & E  \\
IRAS~4B\tablefootmark{(1)} & -- & -- & -- & -- & -- & -- & -- & -- & -- & -- & -- & -- & -- & -- & --  \\
Per-emb8 & -- & -- & -- & -- & -- & -- & -- & -- & -- & -- & -- & -- & -- & -- & --  \\
Ser-S68N-N & E & -- & -- & -- & -- & -- & -- & -- & -- & -- & -- & -- & -- & -- & --  \\
Ser-S68N-S\tablefootmark{(1)} & E\tablefootmark{(2)} & -- & E\tablefootmark{(3)} & -- & -- & -- & -- & -- & -- & -- & -- & -- & -- & -- & --  \\
Ser-SMM1A & --& -- & -- & -- & -- & -- & -- & -- & -- & -- & -- & -- & -- & -- & --  \\
Ser-SMM1B & -- & A & E & -- & -- & -- & -- & -- & -- & -- & -- & -- & -- & A & --  \\
Ser-SMM3 & -- & -- & E & -- & -- & -- & -- & -- & -- & -- & -- & -- & -- & -- & --  \\
SVS4-5 & E & E & E & -- & -- & -- & -- & -- & E & -- & -- & -- & -- & E & E  \\
L1527 & -- & E & -- & -- & -- & -- & -- & -- & -- & -- & -- & -- & -- & -- & --  \\
TMC1A & -- & A & A & -- & -- & -- & -- & -- & -- & -- & -- & -- & -- & -- & E  \\
TMC1-E & E\tablefootmark{(4)} & A & E & -- & -- & -- & -- & -- & -- & -- & -- & -- & -- & E & E  \\
TMC1-W & E\tablefootmark{(4)} & E & E & -- & -- & -- & -- & -- & -- & -- & -- & -- & -- & E & E  \\
BHR71-IRS1 & E & A & A & -- & A & -- & -- & -- & -- & -- & -- & A & -- & A & E  \\
BHR71-IRS2 & -- & A & E & -- & -- & -- & -- & -- & -- & -- & -- & -- & -- & -- & --  \\
    \hline
    \end{tabular}
    \label{tab:detection_statistics}
    \tablefoot{\tablefoottext{1}{No continuum from protostar detected across the full MIRI wavelength range.} \tablefoottext{2}{Detected but likely originates from Ser-S68N-N.} \tablefoottext{3}{Detected but is clearly associated with the outflow rather than the central protostellar position.} \tablefoottext{4}{The binary components of TMC1 are indistinguishable at longer wavelengths.}}
\end{table*}
\renewcommand{\arraystretch}{1.0}

In channel 3B (i.e., $13<\lambda<15.5$~\mum), the bands of C$_2$H$_2$ ($\nu_5$, $Q$-branch at $13.6$~\mum), HCN ($\nu_2$, $Q$-branch at $14.1$~\mum), and CO$_2$ ($\nu_2$, $Q$-branch at $15.0$~\mum) are located, see the case of B1-c in Fig.~\ref{fig:specfit_B1-c_overview}. Next to the $Q$-branch of C$_2$H$_2$, also its $^{13}$CCH$_2$ isotopologue is detected toward both B1-c and L1448-mm. Similarly, the $Q$-branch of $^{13}$CO$_2$ at $15.45$~\mum is detected toward both of these sources (see Figs.~\ref{fig:specoverview_B1-c_L1448-mm} and \ref{fig:specfit_B1-c_overview}). Toward L1448-mm, also di-acetelyne (C$_4$H$_2$) is observed in emission through its $\nu_8$ bending mode at $15.92$~\mum.

The longest wavelengths (i.e., $>17$~\mum) are dominated by OH features and the pure rotational transitions of H$_2$O (see bottom right panel of Fig.~\ref{fig:specoverview_B1-c_L1448-mm}).
% In some sources (e.g., B1-a-NS, TMC1-E, BHR71-IRS1), prompt OH emission between $9-11$~\mum is also detected, but since neither the OH lines longward of $17$~\mum nor these prompt OH lines are originating from gas in LTE \citep[but from H$_2$O photodissociation or formation pumping;][]{Carr2014,Tabone2021,Tabone2024,Zannese2024,Neufeld2024}, they will not be discussed further. 
The rotational lines of H$_2$O originate from energy levels of $>1000$~K and are therefore mostly sensitive to warmer temperatures. However, some rotational transitions are clearly sensitive to colder temperatures of $\sim100-200$~K than the bulk of the lines which trace $T>300$~K gas.

All together, the JOYS data provide a wealth of molecular emission and absorption features of various molecules. In Sections \ref{subsec:mol_statistics} and \ref{subsec:LTE_fit_restults}, the detection statistics and LTE slab model fit results will be presented. The LTE models provide remarkably good fits to the data, see e.g., Fig.~\ref{fig:specfit_B1-c_overview} for B1-c. 
{In Appendix~\ref{app:B1-c_residuals}, the residuals of the full LTE fit to B1-c are presented. The residuals show remaining absorption lines which could either originate from either a different temperature component to the fitted molecular species or a species that is not considered in this paper (e.g., isotopologues). Nevertheless, the residuals are mostly below the level of 10\% with higher residuals in some line-rich wavelength regions (e.g., $7-8$~\mum, see Fig.~\ref{fig:B1-c_residuals_pt2}).}

\subsection{Molecular detection statistics}
\label{subsec:mol_statistics}
The molecular detections per source are listed in Table~\ref{tab:detection_statistics}. The most commonly observed molecule is H$_2$O, which is detected toward 12/16 of the studied sources. The rovibrational lines around 5--8~\mum are detected in three more sources (11/16) than the pure rotational lines longward of 13~\mum (8/16). 
% This likely originates from the decrease in sensitivity at longer wavelengths (see Table~\ref{tab:sigma_data}), making it more difficult to detect the pure rotational lines. 
% This could be related to the decrease in sensitivity at longer wavelengths (see Table~\ref{tab:sigma_data}), but the pure rotational lines appear to be more often detected than the rovibrational lines toward more evolved Class~II disks, likely due to subthermal excitation of the vibrational levels \citep[see e.g.,][]{Banzatti2023_preJWST,Banzatti2023_JWST}. In our protostellar sources, the higher detection rate of the rovibrational lines is therefore likely to originate from IR pumping of the vibrational levels.
This could be related to the decrease in sensitivity at longer wavelengths (see Table~\ref{tab:sigma_data}).
However, it is just opposite to what is observed toward more evolved Class~II disks where the pure rotational lines appear to be more often detected than the rovibrational lines, likely due to subthermal excitation of the vibrational levels \citep[see e.g.,][]{Banzatti2023_preJWST}.
% A more likely explanation for the higher detection rate of the rovibrational lines in protostellar sources is therefore that the vibrational levels are excited through IR pumping. 
Moreover, three of the four sources with rovibrational H$_2$O detected but an absence of rotational H$_2$O (Ser-SMM1B, TMC1A, BHR71-IRS2) show the rovibrational lines in absorption. These absorption lines originate from the vibrational ground state with $E_{\rm low}$ as low as 0~K whereas the pure rotational lines originate from much higher energy levels ($>1000$~K), which could explain their absence. The other source, L1527, shows only a few rovibrational lines of H$_2$O weakly in emission and only located in the western side of the disk or outflow (Devaraj et al. in prep.), which could mean that they are the result of IR pumping rather than collisional excitation.
B1-c is the only source that shows absorption of H$_2$O in the pure rotational lines arising from $E_{\rm low} > 1000$~K levels. Interestingly, the pure rotational lines of H$_2$O are in emission toward BHR71-IRS1 whereas the rovibrational lines are in absorption, indicating that they are likely tracing two different components (e.g., hot core, disk wind or outflow) within the protostellar system. Moreover, the rovibrational lines are seen in absorption toward TMC1-E whereas they are in emission toward TMC1-W.

Almost as commonly detected as H$_2$O is CO$_2$, which is seen toward 11/16 sources. Similar to H$_2$O, it is mostly observed in emission, but is in absorption toward four sources. In the case of Ser-S68N-S, the CO$_2$ emission is clearly spatially offset and located in the outflow rather than at the central position and is therefore excluded from the remaining analysis. Toward several other sources (e.g., L1448-mm, BHR71-IRS1, Ser-SMM3), CO$_2$ also shows an outflow component but hosts a bright central component as well. The $^{13}$CO$_2$ isotopologue is only detected toward the two sources that are most rich in molecular features, B1-c (in absorption) and L1448-mm (in emission). 

Other molecules are far less present than H$_2$O and CO$_2$. Besides CO and OH, which are discussed further below, C$_2$H$_2$ and CH$_4$ are the most detected species (3/16). Several other species such as $^{13}$CCH$_2$, HCN, and SO$_2$ are only detected toward B1-c and L1448-mm. The carbon-chain molecule C$_4$H$_2$, commonly observed in more evolved Class II disks around very low-mass stars \citep[e.g.,][]{Tabone2023,Arabhavi2024}, is detected in emission only toward L1448-mm. 
Silicon monoxide is detected in emission (L1448-mm) and absorption (BHR71-IRS1) around 8.5~\mum (see Sect.~\ref{subsubsec:results_SiO}). Toward B1-c, also CS ($8-8.5$~\mum) and NH$_3$ ($9-11$~\mum) are detected in absorption for the first time toward a low-mass protostellar system.
% Interestingly, we detect clear emission (L1448-mm) and absorption (BHR71-IRS1) features of SiO around 8.5~\mum (see Fig.~\ref{fig:specfit_L1448-mm_BHR71-IRS1_SiO}). This triples the number of SiO detections at mid-IR wavelengths toward protostellar systems (see e.g., McClure et al. 2024, submitted). Moreover, toward B1-c, clear absorption features associated with NH$_3$ are detected, marking the first detection of this molecule at mid-IR wavelengths toward a low-mass protostellar system.

Lastly, CO and OH are commonly detected toward the JOYS sources (8/16 and 7/16, respectively). Carbon monoxide is seen in absorption toward the sources that also show the rovibrational H$_2$O lines in absorption except for TMC1-E, where CO is in absorption but the rovibrational lines of H$_2$O are in emission. However, only the high-$J$ lines of the $P$-branch of CO are detectable with MIRI-MRS, hence a non-detection of CO here does not imply that it is not present in the protostellar systems. Furthermore, because only the high-$J$ lines are detectable, the LTE analysis is degenerate between high temperatures and high column densities \citep[e.g.,][see also discussion in \citealt{Rubinstein2024} for the NIRSpec range]{Herczeg2011,Francis2024} and will therefore not be further discussed in this paper. 
% Due to the degeneracy between high excitation temperature and line optical depth predicted by LTE models \citep[e.g.,][see also discussion in \citealt{Rubinstein2024} for the NIRSpec range]{Herczeg2011,Francis2024}, these CO lines will not be analyzed in detail further. 
The OH radical is only detected in emission. Toward several sources (e.g., B1-a-NS, TMC1-E, BHR71-IRS1), prompt OH emission is detected between 9--11~\mum, which is suggested to originate from H$_2$O photodissociation \citep[][]{Tabone2021,Tabone2024,Zannese2024,Neufeld2024}. In all sources with OH detections, also the lines at longer wavelengths ($>13$~\mum) are detected which likely result from formation pumping \citep[e.g.,][]{Carr2014,Zannese2024}. 
Since the OH emission is therefore likely not originating from regions that are in LTE, it will also not be discussed further in this paper. 

\renewcommand{\arraystretch}{1.1}
\begin{table*}[]
    \centering
    % \footnotesize
    \caption{Excitation temperatures in units of Kelvin derived from the LTE slab models.}
    \begin{tabular}{lcccccccccc}
\hline\hline
Source & cold H$_2$O & warm H$_2$O & rovib H$_2$O & CO$_2$ & C$_2$H$_2$ & HCN & CH$_4$ & SO$_2$ & SiO \\
\hline
B1-a-NS & $160 \pm 10$ & $370 \pm 10$ & $405 \pm 15$ & $115 \pm 25$ & -- & -- & -- & -- & --  \\
% B1-b & -- & -- & -- & -- & -- & -- & -- & -- & --  \\
B1-c & -- & $325 \pm 15$ & $300 \pm 10$ & $330 \pm 10$ & $285 \pm 25$ & $180 \pm 10$ & $200 \pm 10$ & $150 \pm 10$ & --  \\
L1448-mm & $130 \pm 10$ & $390 \pm 10$ & $180 \pm 10$ & $120 \pm 10$ & $110 \pm 10$ & $110 \pm 10$ & $130 \pm 10$ & $115 \pm 15$ & $315 \pm 15$  \\
% Per-emb8 & -- & -- & -- & -- & -- & -- & -- & -- & --  \\
Ser-S68N-N & $200 \pm 10$ & -- & -- & -- & -- & -- & -- & -- & --  \\
% Ser-SMM1A & -- & -- & -- & -- & -- & -- & -- & -- & --  \\
Ser-SMM1B & -- & -- & $980 \pm 20$ & $205 \pm 25$ & -- & -- & -- & -- & --  \\
Ser-SMM3 & -- & -- & -- & $90 \pm 10$ & -- & -- & -- & -- & --  \\
SVS4-5 & $210 \pm 10$ & $515 \pm 15$ & $1040 \pm 20$ & [200] & -- & -- & $460 \pm 80$ & -- & --  \\
L1527 & -- & -- & $90 \pm 10$ & -- & -- & -- & -- & -- & --  \\
TMC1A & -- & -- & $440 \pm 10$ & $60 \pm 10$ & -- & -- & -- & -- & --  \\
TMC1-E & $185 \pm 15$ & $405 \pm 15$ & $580 \pm 20$ & $315 \pm 25$ & -- & -- & -- & -- & --  \\
TMC1-W & $190 \pm 10$ & $485 \pm 15$ & $1180 \pm 20$ & $190 \pm 10$ & -- & -- & -- & -- & --  \\
BHR71-IRS1 & $140 \pm 10$ & -- & $560 \pm 10$ & $260 \pm 10$ & [150] & -- & -- & -- & $400 \pm 20$  \\
BHR71-IRS2 & -- & -- & $440 \pm 40$ & $75 \pm 15$ & -- & -- & -- & -- & --  \\
\hline
\end{tabular}
\tablefoot{Square brackets indicate that the excitation temperature was fixed during the fit. The excitation temperatures of C$_4$H$_2$, CS, and NH$_3$ are not shown since they are only detected toward one source (L1448-mm or B1-c). Sources for which no molecular emission or absorption features are detected (i.e., B1-b, Per-emb 8, Ser-SMM1A) are not shown.}
\label{tab:LTE_Tex}
\end{table*}
\renewcommand{\arraystretch}{1.0}

The JOYS data were also checked for other molecules including H$_2$S and hydrocarbons such as C$_6$H$_6$, C$_2$H$_4$, C$_2$H$_6$, and CH$_3$ that are often detected toward Class~II disks around very low-mass stars \citep[][]{Tabone2023,Arabhavi2024}. However, these molecules are not detected in any of the sources. Likewise, complex organics such as CH$_3$OH and CH$_3$CN have infrared bands in the MIRI-MRS wavelength range and are not detected toward any of the sources. Furthermore, the MIRI-MRS spectral range also covers cations such as N$_2$H$^+$, HCO$^+$, and CH$_3^+$, but these are also not detected toward any of the sources. 
% All these species will therefore not be discussed further in this paper.
% Toward the most line-rich sources, 

Three sources (i.e., B1-b, Per-emb~8, Ser-SMM1A) do not show any gas-phase molecular features in their spectra albeit a clear detection of the continuum across the full wavelength range (see Figs.~\ref{fig:specext_B1-b}, \ref{fig:specext_Per-emb8}, and \ref{fig:specext_Ser-SMM1A}). Here, the absence of molecular features can not be explained by envelope or cloud extinction as is likely the case for IRAS~4B and Ser-S68N-S.

\subsection{LTE model fit results}
\label{subsec:LTE_fit_restults}
As an example of our results, an overview of the main molecular features detected toward B1-c and the best-fit LTE model is shown in Fig.~\ref{fig:specfit_B1-c_overview}. The full {baseline-subtracted} spectra are presented in Appendix~\ref{app:specfit} for all sources. The overlaid best-fit model clearly presents a very good fit to the data, with the majority of the absorption features fitted by the molecules listed in Fig.~\ref{fig:specfit_B1-c_overview}. Moreover, the discrepancy between the $R$ and $P$-branches due to the extinction of the CO$_2$ ice band is evident: almost no absorption features are present between 15~\mum and 15.4~\mum which is nicely consistent with the LTE models (i.e., there is no continuum to absorb against). Similarly, for L1448-mm this effect is very evident for CO$_2$ in emission, as it was to a lesser extent also for the $P$ and $R$-branch lines in the high-mass source IRAS~23385+6053 \citep[][]{Francis2024}. 

In Table~\ref{tab:LTE_Tex}, the derived excitation temperatures are presented for all sources showing molecular emission or absorption features. The best-fit LTE results (i.e., column density $N$, excitation temperature $T_{\rm ex}$, radius of emitting area $R$, number of molecules $\mathcal{N}_{\rm mol}$, number of molecules corrected for infrared pumping $\mathcal{N}_{\rm mol,corr}$) are tabulated per source in Appendix~\ref{app:LTE_fit_results}.

It is important to note that the results from absorption modeling (i.e., B1-c, BHR71-IRS1) are very reliable since they are not dependent on an emitting area and do not suffer from non-LTE effects such as IR pumping. The results from emission models (i.e., L1448-mm) are therefore more uncertain. Nevertheless, the majority of the excitation conditions ($T_{\rm ex}$) are well constrained (typical error of $\sim10$~K) and are rather cold in the range of $T_{\rm ex}\sim100-300$~K for most species. Similarly to $T_{\rm ex}$, column densities ($N$) and number of molecules ($\mathcal{N}_{\rm mol}$) are within a few orders of magnitude for the majority of the sources. Below, the main findings per set of molecules are presented.

\subsubsection{H$_2$O}
\label{subsubsec:results_H2O}
\begin{figure*}
    \centering
     \includegraphics[width=\linewidth]{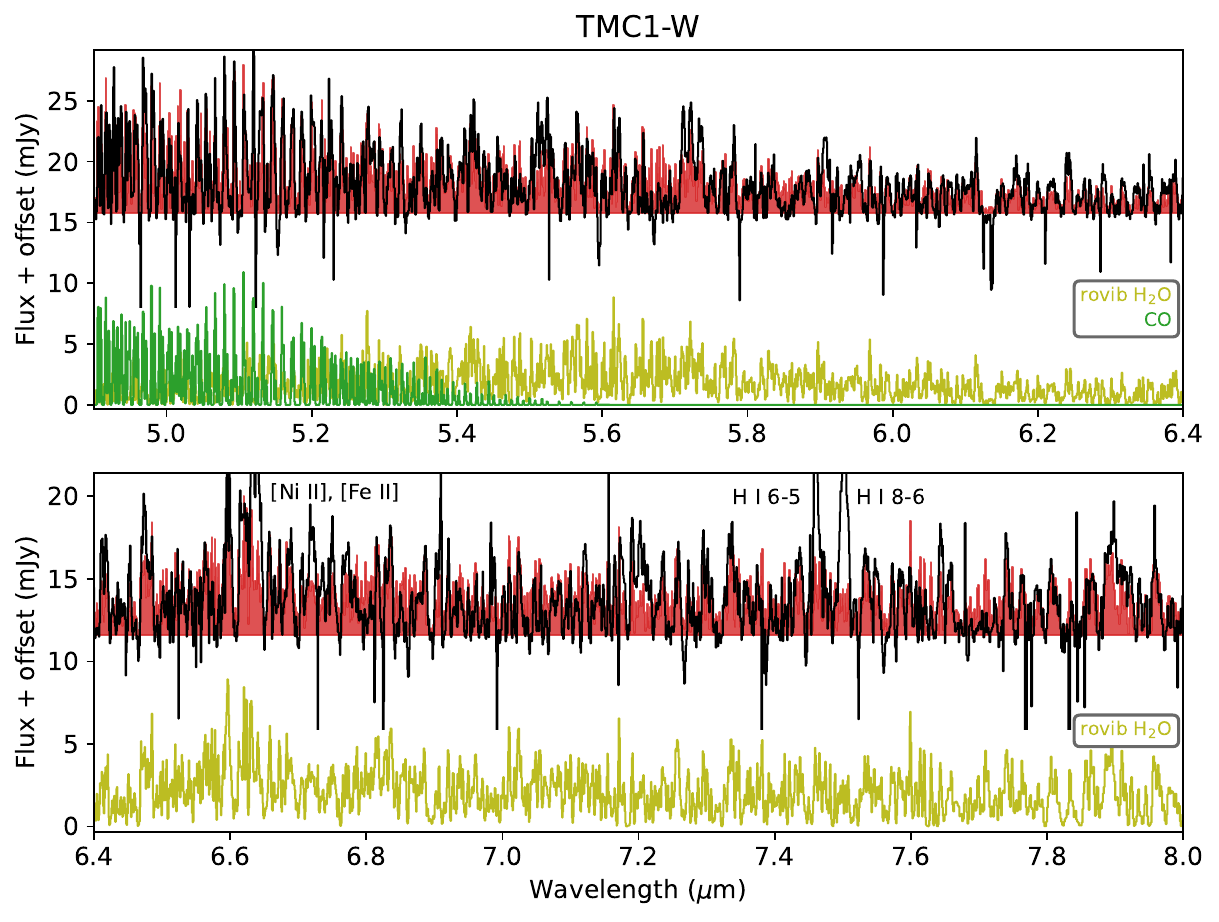}
    \caption{
     {Baseline-subtracted} spectrum (black) and best-fit LTE model (shaded red) for TMC1-W in the $4.9-8.0$~\mum range. In the bottom of each panel, the individual best-fit LTE models of rovibrational H$_2$O (yellow) and CO (green) are shown at an arbitrary constant offset. Deep negative absorption features originating from detector artifacts are clipped for clarity.
    }
    \label{fig:specfit_TMC1-W_H2O-rovib}
\end{figure*}

\begin{figure*}
    \centering
     \includegraphics[width=\linewidth]{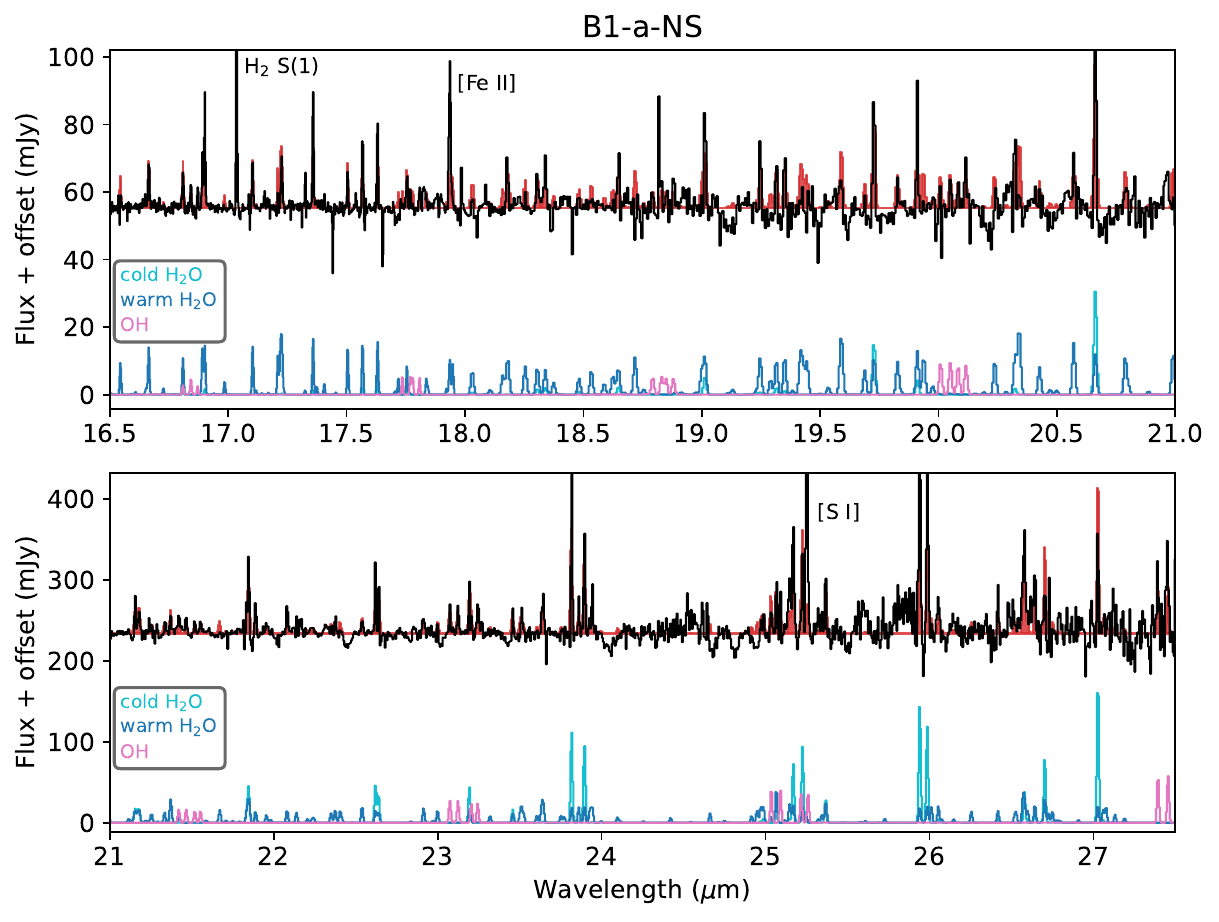}
    \caption{
     {Baseline-subtracted} spectrum (black) and best-fit LTE model (shaded red) for B1-a-NS in the $16.5-27.5$~\mum range. In the bottom of each panel, the individual best-fit LTE models of cold H$_2$O (cyan), warm H$_2$O (blue), and OH (pink) are shown at an arbitrary constant offset. Deep negative absorption features originating from detector artifacts are clipped for clarity. 
    }
    \label{fig:specfit_B1-a-NS_H2O_OH}
\end{figure*}

As discussed in Sect.~\ref{subsec:mol_statistics}, H$_2$O is the most commonly detected molecule in the JOYS data. In total, three different components were used to fit all the water lines across the full MIRI-MRS wavelength range. For the rovibrational lines shortward of 10~\mum, one warm component provided a good fit to the data, see e.g., Fig.~\ref{fig:specfit_TMC1-W_H2O-rovib} for TMC1-W. On the other hand, for the pure rotational lines longward of 10~\mum, two components with two different temperatures were needed to provide an accurate fit to the data with typical temperatures of $\sim150$~K and $\sim400$~K, called cold and warm respectively, see Table~\ref{tab:LTE_Tex} and Fig.~\ref{fig:specfit_B1-a-NS_H2O_OH}. In all sources where two components were needed to fit the pure rotational emission, the cold H$_2$O components traces the bulk of the H$_2$O (i.e., a higher $N$ or $\mathcal{N}_{\rm mol}$ than the warm component). 

In several sources, the rovibrational component appears to be tracing the same H$_2$O as the pure rotational lines. This is especially clear toward B1-c, the only source where both the rotational and rovibrational lines of H$_2$O (and those of other species) are detected solely in absorption. The excitation temperature derived from the pure rotational lines ($325\pm15$~K) is in perfect agreement with that derived from the rovibrational lines ($300\pm10$~K) and also the column densities are within errorbars in good agreement ($4.4\pm1.2\times10^{18}$~cm$^{-2}$ and $7.8\pm 2.2\times10^{18}$~cm$^{-2}$, respectively, see Table~\ref{tab:LTE_results_B1-c}). This shows that there is one warm component of H$_2$O that is absorbing against the bright continuum across the full wavelength range.

Other sources with agreement between the excitation conditions of the rovibrational lines and the pure rotational lines are B1-a-NS and L1448-mm, see Table~\ref{tab:LTE_Tex}. Toward both of these sources, H$_2$O is present in emission and two components were needed to fit the pure rotational lines, see Fig.~\ref{fig:specfit_B1-a-NS_H2O_OH}. The rovibrational lines in B1-a-NS seem to be tracing the same gas as the warm component of H$_2$O ($\sim400$~K) with an additional cold component of $160\pm10$~K traced by the lower $E_{\rm up}$ rotational lines. On the other hand, toward L1448-mm the temperature derived from the rovibrational lines ($180\pm10$~K) is in better agreement with the cold H$_2$O component of the rotational lines ($130\pm10$~K) than with the warm rotational H$_2$O component ($390\pm10$~K) which is clearly tracing warmer gas. 

Interestingly, toward the TMC1 binary, the rovibrational lines are in absorption toward TMC1-E (see Fig.~\ref{fig:specfitfull_TMC1-E_4.9-13.0}) whereas they are in emission toward TMC1-W (see Fig.~\ref{fig:specfit_TMC1-W_H2O-rovib}). The pure rotational lines are in emission (see Figs.~\ref{fig:specfitfull_TMC1-E_13.0-27.5} and \ref{fig:specfitfull_TMC1-W_13.0-27.5}), but at these wavelengths the two components of the binary are barely resolved so it is not possible to determine which of the two components (or both) is showing this H$_2$O in emission. The excitation temperature of the rovibrational lines between the two components is very different with TMC1-W showing much hotter H$_2$O than TMC1-E ($1280\pm20$~K and $580\pm20$~K, respectively). The radius of the emitting area derived for TMC1-W is $0.06\pm0.01$~au, consistent with a hot inner disk, whereas for TMC1-E no emitting area can be derived since the rovibrational lines are in absorption. In contrast, the radii of the emitting areas derived from the pure rotational lines are $0.3\pm0.1$~au (warm component) and $6.4\pm1.9$~au (cold component), suggesting that these are tracing more extended H$_2$O emission. Similarly to TMC1, the rovibrational lines of H$_2$O are in absorption toward BHR71-IRS1 whereas the rotational lines are in emission. The temperatures between these two components are significantly different ($560\pm10$~K and $140\pm10$~K, respectively). 
{The emitting area derived for the cold component is $24\pm1$~au, which is consistent with it being present on larger scales in the system (e.g., outflow), although it remains spatially unresolved at the continuum position. For the rovibrational lines, no emitting area can be derived since the lines are in absorption, but the velocity offset for both H$_2$O components ($-42\pm3$~\kms and $-44\pm9$~\kms, respectively) is consistent with outflowing material.}
% Likely, the cold H$_2$O is more extended whereas the warm H$_2$O traced by the rovibrational lines is more compact.

In multiple sources (e.g., SVS 4-5, TMC1-E, TMC1-W), all water components are tracing three distinct physical components, where the rovibrational lines trace the hottest component. 
% This shows that in these systems, H$_2$O is present in the envelope or disk at large range of radii from the central protostar, showing an onion-like temperature structure. Alternatively, some of the H$_2$O could also be located on small-scale shocks such as in the inner outflow or at the disk-envelope boundary. 
The majority of the H$_2$O in all sources resides at temperatures of $\lesssim600$~K. However, toward a few sources (i.e., Ser-SMM1B, SVS4-5, TMC1-W) very hot water with temperatures of up to 1200~K is detected. Toward TMC1-W and SVS4-5, this hot water is seen in emission (see Fig.~\ref{fig:specfit_TMC1-W_H2O-rovib}) whereas toward Ser-SMM1B it is seen in absorption. This hot water component is similar to that found in Class~II disks \citep[e.g.,][]{Banzatti2023_JWST,Perotti2023,Pontoppidan2024,Temmink2024_H2O} and likely originates from the inner disk close to the protostar where the temperatures are high enough to produce such hot water. 

\begin{figure*}
    \centering
    \includegraphics[width=\linewidth]{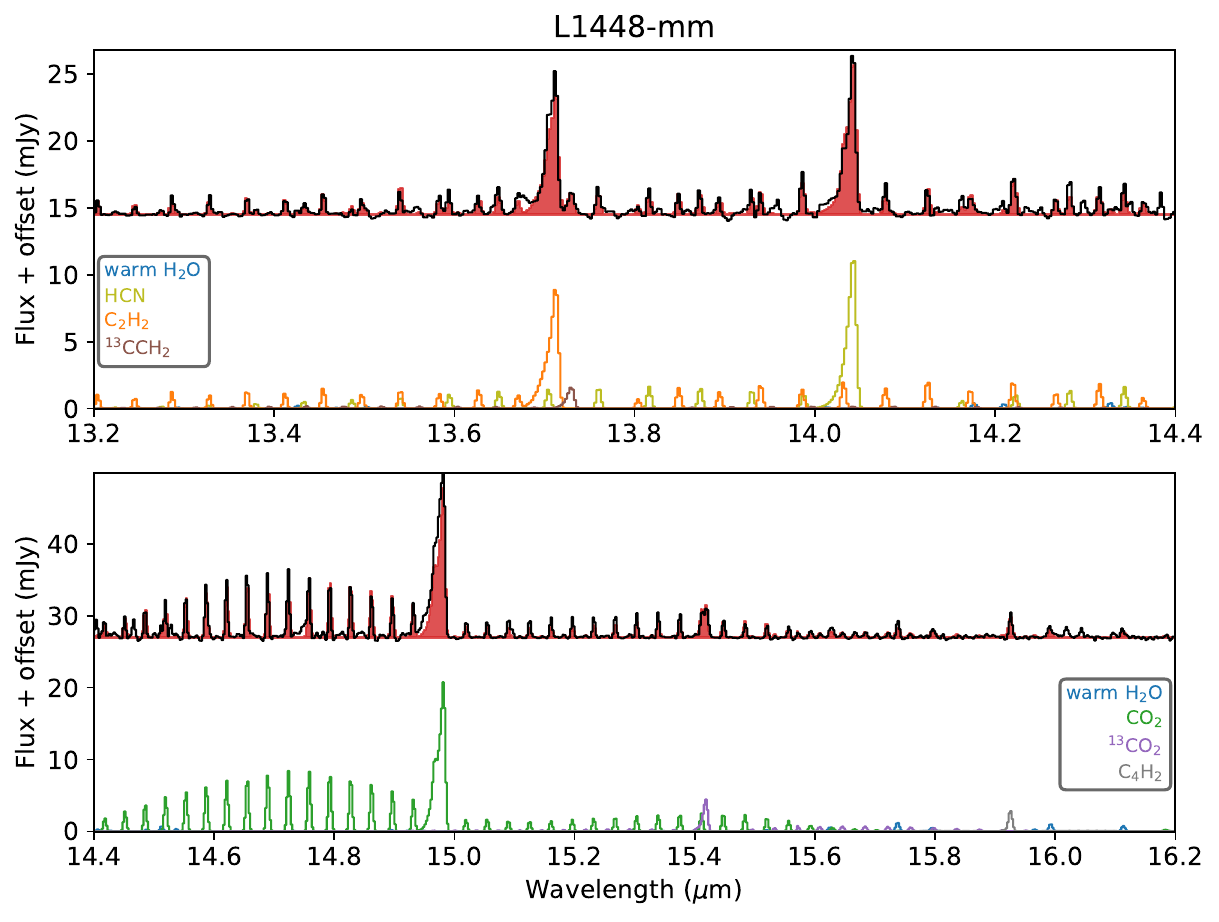}
    \caption{
     {Baseline-subtracted} spectrum (black) and best-fit LTE model (shaded red) for L1448-mm in the $13.2-16.2$~\mum range. In the bottom of each panel, the individual best-fit LTE models of warm H$_2$O (blue), C$_2$H$_2$ (orange), $^{13}$CCH$_2$ (brown), HCN (yellow), CO$_2$ (green), $^{13}$CO$_2$ (purple), and C$_4$H$_2$ are shown at an arbitrary constant offset. Deep negative absorption features originating from detector artifacts are clipped for clarity.
    }
    \label{fig:specfit_L1448-mm_13.2-16.2}
\end{figure*}

\subsubsection{CO$_2$}
\label{subsubsec:results_CO2}
For L1448-mm, the best-fit models of CO$_2$, C$_2$H$_2$, HCN, and C$_4$H$_2$ are presented in Fig.~\ref{fig:specfit_L1448-mm_13.2-16.2}. As mentioned in Sect.~\ref{subsec:mol_statistics}, CO$_2$ is the second most detected molecule next to H$_2$O. It seems to be tracing rather cold regions in embedded protostellar systems, see Table~\ref{tab:LTE_Tex}. Typical temperatures are about $\sim200$~K, with the highest temperature of $315\pm25$~K measured toward TMC1-E and the lowest temperature of $60\pm10$~K measured toward TMC1A. This is in strong contrast to what is seen toward more evolved Class II disks where temperatures are typically much higher \citep[$>500$~K;][]{Grant2023}, although also Class~II disks with lower temperature CO$_2$ exist \citep[e.g.,][]{Schwarz2024}. The abundance ratios with respect to H$_2$O (derived from the cold and warm components of the pure rotational lines) suggest a range of abundances from $10^{-5}-10^{-1}$ due to the effect of infrared pumping. 

In several cases, the lines of CO$_2$ are blueshifted (i.e., BHR71-IRS1, SVS4-5) or even redshifted (i.e., B1-a-NS, Ser-SMM1B) by up to several tens of \kms, suggesting that it is either located in the outflow or is infalling. A clear case where CO$_2$ (as well as other molecular species) is tracing outflowing material is BHR71-IRS1, where all lines {of CO$_2$ are shifted by $-10\pm4$~\kms with respect to the $v_{\rm lsr}$ and show some spatially extended emission along the outflow cavity walls. The velocity shift is similar but slightly lower than other species such as H$_2$O and SiO ($-45$~\kms and $-23\pm2$~\kms, respectively; see Table~\ref{tab:LTE_results_BHR71-IRS1})}. 
% Interestingly, CO$_2$ is moving at a slightly lower velocity of $-25$~\kms and shows some spatially extended emission along the outflow cavity walls. 
BHR71-IRS1 also shows strong and spatially extended H$_2$ and atomic lines that are blueshifted by similar velocities, which will be discussed in a separate paper (Tychoniec et al. in prep.). The CO$_2$ lines toward SVS4-5 are even shifted by -80~\kms, but these likely originate from the outflow of the nearby Class~0 source Ser-SMM4 \citep[][]{Pontoppidan2004}. On the other hand, the absorption lines toward B1-a-NS are redshifted by $35$~\kms, which is consistent with either tracing the redshifted outflow or with infalling material. Some weak extended emission of CO$_2$ is present in the southern direction, which is indicates that an outflow origin is more likely. However, the S/N of the CO$_2$ is low and the lines are strongly blended with H$_2$O lines (see Fig.~\ref{fig:specfitfull_B1-a-NS_13.0-27.5}), hampering a more precise analysis. 
% The upper part of this range is in agreement with interstellar ices \citep[$\sim$few$\times10^{-1}$][]{Boogert2015}. 
% These are all lower than what is expected from interstellar ices \citep[$\sim$few$\times10^{-1}$][]{Boogert2015}, which might suggests that either the amount of CO$_2$ is underestimated (see below) or that our observations are not probing full sublimation of ices within the hot cores.

Toward B1-c and L1448-mm, also the $^{13}$CO$_2$ isotopologue is detected, see Figs.~\ref{fig:specfit_B1-c_overview} and \ref{fig:specfit_L1448-mm_13.2-16.2}. The excitation temperature measured for B1-c ($220\pm30$~K) is very similar to that of CO$_2$ ($270\pm10$~K), suggesting that they are tracing the same component. For L1448-mm, the excitation temperature cannot be accurately constrained below 300~K, and was therefore fixed to that of CO$_2$ ($120$~K). 
Adopting a typical $^{12}$C/$^{13}$C ratio of 70 for the local ISM \citep[][]{Milam2005}, the derived abundance ratio of CO$_2$ with respect to H$_2$O in B1-c is a factor of $\sim5$ higher than derived directly from the main isotopologue, likely due to line optical depth effects of CO$_2$. For L1448-mm, the abundances of $^{13}$CO$_2$ multiplied by 70 and CO$_2$ itself agree perfectly, given that the $Q$-branch is excluded from the fit (see Appendix~\ref{app:line_optical_depth}).
% Toward B1-c, the abundance of CO$_2$ derived from $^{13}$CO$_2$ agrees very well with typical ice abundances \citep[][]{Boogert2015}, suggesting that our JWST observations are indeed probing its hot core. For L1448-mm, however, the abundance of CO$_2$ is only in agreement with ices when IR pumping is neglected and can be a factor 10\,000 lower than the ices when IR pumping is maximally efficient. This implies that we are not necessarily only probing ice sublimation but that gas-phase chemistry may also contribute.

% \subsubsection{C$_2$H$_2$, HCN, and C$_4$H$_2$}
\subsubsection{C$_2$H$_2$}
Acetylene is detected toward three sources, B1-c, L1448-mm, and BHR71-IRS1 through its $\nu_5$ bending mode around 13.6~\mum (see e.g., Figs.~\ref{fig:specfit_B1-c_overview} and \ref{fig:specfit_L1448-mm_13.2-16.2}). The derived excitation temperatures are $300\pm10$~K and $110\pm10$~K for B1-c and L1448-mm, respectively, see Table~\ref{tab:LTE_Tex}. For BHR71-IRS1, the excitation temperature could not be constrained since only the $Q$-branch is weakly detected and was fixed to 150~K. This is significantly lower than what is typically observed toward more evolved Class~II sources \citep[$\gtrsim500$~K; e.g.,][]{Salyk2011,Grant2023,Tabone2023,Arabhavi2024}. Interestingly, the excitation temperature of $^{13}$CCH$_2$ ( $140\pm30$~K) is lower than that of C$_2$H$_2$ for B1-c, suggesting that $^{13}$CCH$_2$ might be probing a different region than the main isotopologue. 
However, the temperature of the main isotopologue may also be overestimated due to line optical depth effects \citep[e.g., Appendix~\ref{app:line_optical_depth}, ][]{Li2024}, even though the optically thick $Q$-branch was excluded from the fit. 
Furthermore, the lines of $^{13}$CCH$_2$ are blended with those of C$_2$H$_2$ and HCN (see e.g., Fig.~\ref{fig:specfit_B1-c_overview}), which are much stronger, therefore hampering the analysis of $^{13}$CCH$_2$. 

The high optical depth of the C$_2$H$_2$ lines is further supported by the ratios of the column densities for B1-c. The derived column density ratio with respect to H$_2$O is about a factor of $\sim7$ higher than derived directly from C$_2$H$_2$ when adopting a $^{12}$C/$^{13}$C ratio of 70 \citep[][]{Milam2005}. 
For L1448-mm, the ratio of $^{13}$CCH$_2$ is in perfect agreement with C$_2$H$_2$ if the $Q$-branch of the latter excluded from the fit (see Appendix~\ref{app:line_optical_depth}). The effect of IR pumping is only a factor of a few given that the derived excitation temperature of C$_2$H$_2$ ($110\pm10$~K) is only slightly lower than the brightness temperature at 7.7~\mum (118~K).
% For L1448-mm, the ratio of $^{13}$CCH$_2$ is highly dependent on IR pumping since it's excitation temperature is much lower than the brightness temperature at 7.7~\mum (see Sect.~\ref{subsubsec:IR_pumping}), therefore making an accurate column density difficult to derive. Taking $T_{\rm IR}$ at 7.7~\mum, the ratio of $^{13}$CCH$_2$ (multiplied by 70) is in agreement with that derived directly from C$_2$H$_2$, but it may be underestimated due to the IR pumping correction. 
% In any case, the derived ratio with respect to H$_2$O is about an order of magnitude lower in L1448-mm than toward B1-c.
% However, the ratio of C$_2$H$_2$/CO$_2$ is similar between B1-c and L1448-mm. This suggests that the abundance of H$_2$O may be enhances toward L1448-mm rather than that the abundances of CO$_2$ and C$_2$H$_2$ are lower.
The ratio of C$_2$H$_2$/CO$_2$ is similar between B1-c and L1448-mm, but in BHR71-IRS1 C$_2$H$_2$ appears to be almost 2 orders of magnitude less abundant than CO$_2$. However, given both the lack of $^{13}$CO$_2$ and $^{13}$CCH$_2$ detections toward this source and the fact that only the possibly optically thick $Q$-branches are detected for CO$_2$ and C$_2$H$_2$, the effect of line optical depth cannot be excluded. The upper limit of $^{13}$CCH$_2$ is consistent with the ratio of C$_2$H$_2$/CO$_2$ in B1-c and L1448-mm.  

\begin{figure*}
    \centering
    \includegraphics[width=\linewidth]{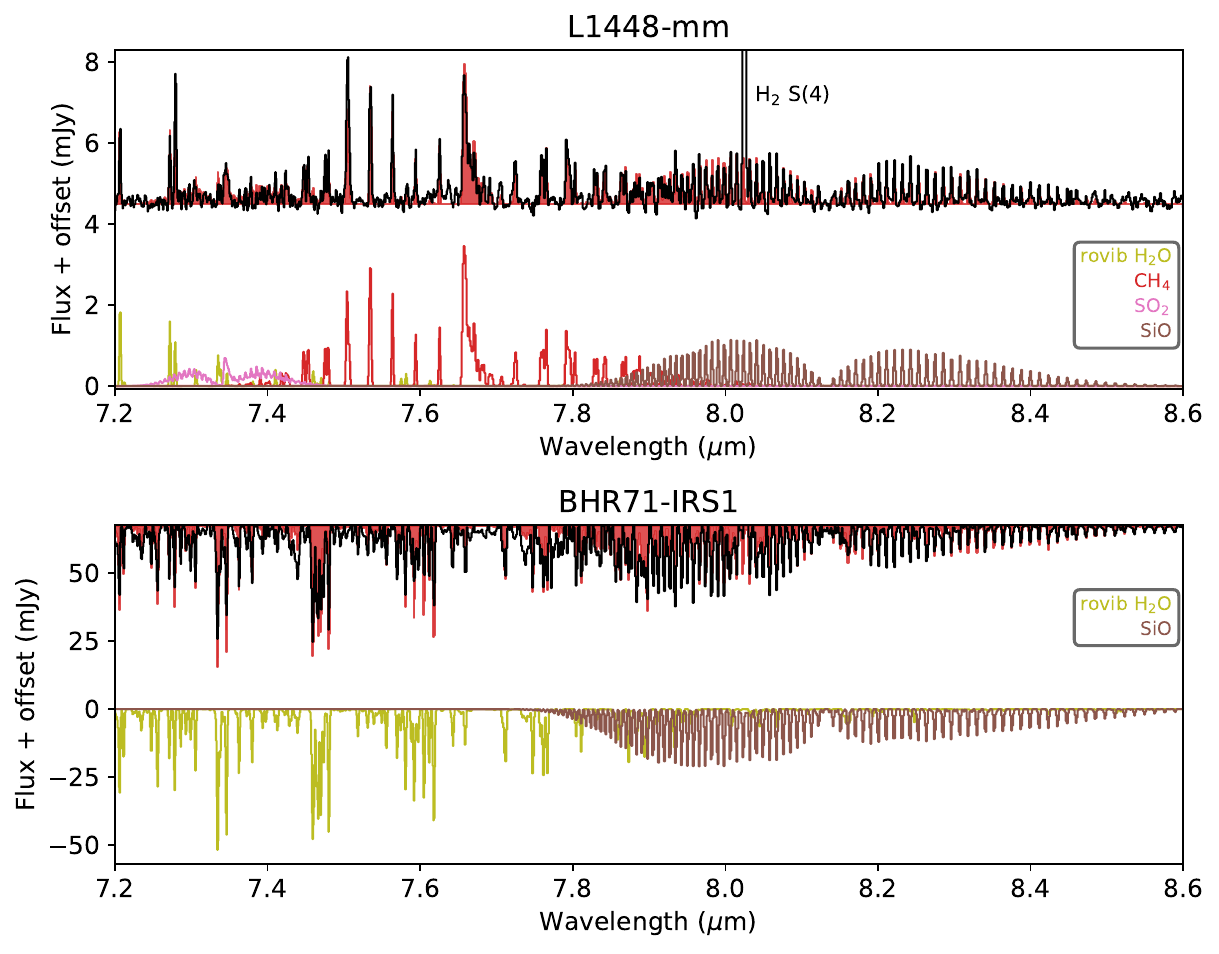}
    \caption{
    {Baseline-subtracted} spectrum (black) and best-fit LTE model (shaded red) for L1448-mm (top panel) and BHR71-IRS1 (bottom panel) in the $7.2-8.6$~\mum range. In the bottom of each panel, the individual best-fit LTE models of rovibrational H$_2$O (yellow), CH$_4$ (red), SO$_2$ (pink), and SiO (brown) are shown at an arbitrary constant offset. Deep negative absorption features originating from detector artifacts are clipped for clarity.
    % {Baseline-subtracted} spectrum (top) and best-fit LTE models per molecule (bottom) for L1448-mm in the 7.2--8.6~\mum range. In the top spectrum, the red shaded area is the combined best-fit LTE model to the data. Each molecule contributing to this wavelength range is shown in the bottom. The bright emission lines around 8.08~\mum is the H$_2$ S(4) transition.
    % The gray vertical bands indicate wavelengths that were excluded from the fit due to the common presence of strong H$_2$ or atomic emission lines (e.g., H$_2$ S(7) and [Fe II]~5.33~\mum.
    }
    \label{fig:specfit_L1448-mm_BHR71-IRS1_SiO}
\end{figure*}

\subsubsection{HCN}
Hydrogen cyanide is only detected in its $\nu_2$ mode around 14~\mum toward B1-c and L1448-mm (see Figs.~\ref{fig:specfit_B1-c_overview} and \ref{fig:specfit_L1448-mm_13.2-16.2}). Interestingly, similar temperatures are derived from its $\nu_2$ mode around 14~\mum as for $^{13}$CCH$_2$ in both B1-c ($180\pm10$~K) and C$_2$H$_2$ in L1448-mm ($110\pm10$~K). This likely means that they are located in the same component of the protostellar system. Similarly to C$_2$H$_2$, the derived excitation temperatures are significantly lower than commonly observed toward more evolved Class II sources. Furthermore, the abundance ratios with respect to H$_2$O indicates that HCN is equally abundant as C$_2$H$_2$ in both B1-c ($7.3\pm2.9\times10^{-2}$) and L1448-mm ($1\times10^{-3}-2\times10^{-2}$), although for the latter source the effect of IR pumping provides an order of magnitude uncertainty. The uncorrected abundance of HCN is $\sim2\times10^{-2}$, which would make it almost similarly abundant with respect to H$_2$O as C$_2$H$_2$. 

\subsubsection{C$_4$H$_2$}
Toward L1448-mm, also C$_4$H$_2$ is detected in emission through the $Q$-branch of its $\nu_8$ bending mode at 15.92~\mum (see Fig.~\ref{fig:specfit_L1448-mm_13.2-16.2}). Its excitation temperature cannot be accurately constrained and was therefore fixed to 100~K, but any temperatures higher than 150~K can be excluded from the shape of the $Q$-branch. Similarly to C$_2$H$_2$ and HCN, the derived upper limit on $T_{\rm ex}$ is significantly lower than in Class II sources. The abundance with respect to H$_2$O ($\sim10^{-4}-10^{-3}$) is about an order of magnitude lower than C$_2$H$_2$ and HCN, which is also consistent with the upper limit derived for B1-c ($<2\times10^{-4}$) and other sources.

% \begin{figure*}
%     \centering
%     % \includegraphics[width=\linewidth]{{Figures/spec_fit/B1-c_13.3-16.3}.pdf}
%     \includegraphics[width=\linewidth]{{Figures/spec_fit_selection/B1-c_6.2-8.2}.pdf}
%     \caption{
%      {Baseline-subtracted} spectrum (black) and best-fit LTE model (shaded red) for B1-c in the $6.2-8.2$~\mum range. In the bottom of each panel, the individual best-fit LTE models of rovibrational H$_2$O (yellow), CH$_4$ (red), SO$_2$ (pink), and CS (green) are shown. 
%     % {Baseline-subtracted} spectrum (top) and best-fit LTE models per molecule (bottom) for B1-c. In the top spectrum, the red shaded area is the combined best-fit LTE model to the data. Each molecule contributing to this wavelength range is shown in the right in its corresponding color.
%     % The gray vertical bands indicate wavelengths that were excluded from the fit due to the common presence of strong atomic emission lines (e.g., [Ne III] at 15.5~\mum).
%     }
%     \label{fig:specfit_B1-c_CH4_SO2_CS}
% \end{figure*}

% \subsubsection{CH$_4$, SO$_2$, and CS}
\subsubsection{CH$_4$}
The best-fit LTE model of CH$_4$ in L1448-mm is presented in the top panel of Fig.~\ref{fig:specfit_L1448-mm_BHR71-IRS1_SiO}. The $\nu_4$ mode of CH$_4$ is clearly detected through its $Q$-branch at 7.66~\mum, as well as through the $R$ and $P$-branch lines on either sides. 
The CH$_4$ features mildly suffer from extinction due to the (rather narrow) ice absorption band of CH$_4$ at 7.67~\mum which also hampered the continuum determination. This differential extinction is taken into account in the LTE slab model fits but has only minor effects on the results.
Besides B1-c and L1448-mm, CH$_4$ is also detected (weakly) in emission toward SVS4-5. The excitation temperatures derived for all three sources vary significantly, see Table~\ref{tab:LTE_Tex}, with L1448-mm and B1-c showing a much lower excitation temperatures ($130\pm10$~K and $200\pm10$~K, respectively) compared to SVS4-5 ($460\pm80$~K). However, as for CO$_2$, the absorption of CH$_4$ in SVS4-5 likely originates from the outflow of Ser-SMM4 in front of SVS4-5 given the high blueshifted velocity of $-40$~\kms. Methane thus shows similarly low temperatures between $100-200$~K as the other molecules toward the protostellar positions. For B1-c the derived ratio with respect to H$_2$O ($9.7\pm3.8\times10^{-2}$) is similar to C$_2$H$_2$ and HCN and about a factor of two lower than CO$_2$. On the other hand, the abundance with respect to H$_2$O is almost an order of magnitude higher ($0.62\pm0.09$) toward L1448-mm, making it only a factor of two less abundant than H$_2$O. 
% In contrast to C$_2$H$_2$, HCN, and C$_4$H$_2$, the derived temperature toward SVS4-5 is roughly in agreement with what is derived for Class~II sources ($>500$~K), although the number of sources where CH$_4$ is detected remains limited. 
% Interestingly, the abundance of CH$_4$ with respect to H$_2$O appears to be higher than those in ices in L1448-mm. 

% For B1-c the derived ratio with respect to H$_2$O ($1.3\pm0.5\times10^{-1}$) is a factor of a few higher than derived in the ice for the same source \citep[$3.7\pm1.9\times10^{-2}$;][]{Chen2024} but still in agreement with the typical range for low-mass sources \citep[$\sim10^{-2}-10^{-1}$;][]{Oberg2008,Boogert2015}. On the other hand, the gas-phase abundance in SVS4-5 ($4.3\pm2.4\times10^{-3}$) is a factor of a few lower than in the ices \citep[][]{Oberg2008,Boogert2015}. For L1448-mm the abundance with respect to H$_2$O is more an order of magnitude higher than ices, making CH$_4$ about equally abundant as H$_2$O itself. This is in strong contrast to CO$_2$, which is under-abundant by an order of magnitude. This suggests that additional gas-phase chemistry of CH$_4$, possibly in the small-scale outflow, must be present that enhances the amount of CH$_4$. This is further supported by the velocity offset of $-60$~\kms and $-20$~\kms with respect to the $v_{\rm lsr}$ of the lines.

\subsubsection{SO$_2$}
Sulfur-dioxide is only detected toward B1-c and L1448-mm through its $\nu_3$ asymmetric stretching mode around 7.35~\mum (see Figs.~\ref{fig:specfit_B1-c_overview} and \ref{fig:specfit_L1448-mm_BHR71-IRS1_SiO}), whereas its $\nu_1$ symmetric stretching mode around 9~\mum and $\nu_2$ bending mode around 19~\mum are both not detected. 
The absence of the $\nu_1$ and $\nu_2$ modes in B1-c is consistent with the best-fit LTE model, but according to the best-fit LTE model of L1448-mm the $\nu_1$ and $\nu_2$ modes should have been easily detected. This suggests that IR pumping through the $\nu_3$ band could be important in L1448-mm similarly to the low-mass sources NGC~1333~IRAS~2A \citep[][]{vanGelder2024}, consistent with other molecules such as CO$_2$ in this source (see e.g., Sect.~\ref{subsubsec:results_CO2}). The derived temperatures for B1-c ($135\pm25$~K) and L1448-mm ($115 \pm 15$~K) are very similar to that derived for IRAS~2A \citep[$95\pm10$~K;][]{vanGelder2024}. The abundance of SO$_2$ with respect to H$_2$O in B1-c is $1.3\pm0.5\times10^{-2}$, which is about one order of magnitude lower than CO$_2$, C$_2$H$_2$, HCN, and CH$_4$. For L1448-mm, the range of SO$_2$ abundances is $0.03-4$ when taking IR pumping into account.
% Interestingly, for B1-c, the abundance of SO$_2$ with respect to H$_2$O is about an order of magnitude higher than what is derived in the ices \citep[][]{Rocha2024,Chen2024}, suggesting that gas-phase chemistry is important. For L1448-mm, the abundance with respect to H$_2$O is even higher at $10^{-2}-10^{0}$, although also uncertain due to the effect of IR pumping.

\subsubsection{CS}
\label{subsubsec:results_CS}
\begin{figure*}
    \centering
    \includegraphics[width=\linewidth]{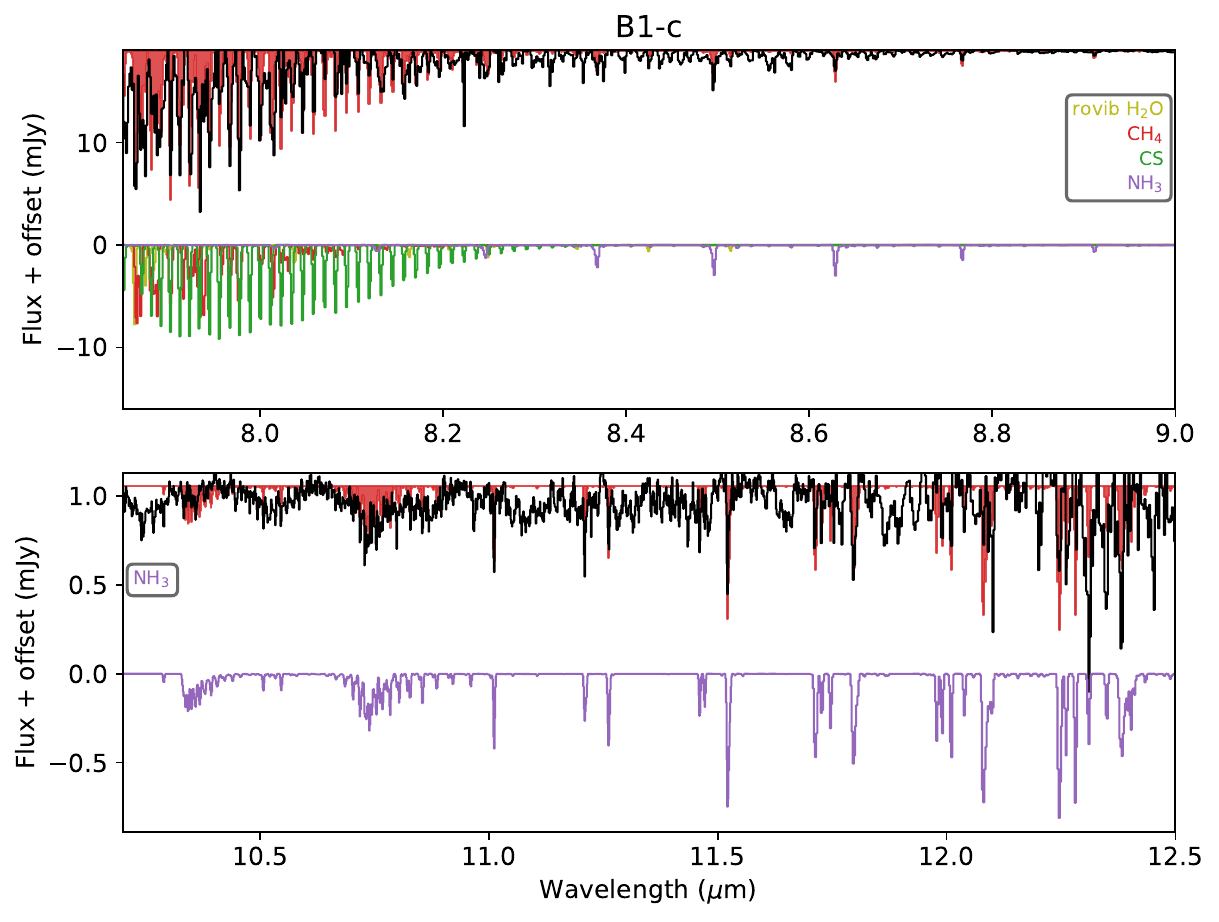}
    \caption{{Baseline-subtracted} spectrum (black) and best-fit LTE model (shaded red) for B1-c in the $8.2-12.5$~\mum range. In the bottom of each panel, the individual best-fit LTE models of rovibrational H$_2$O (yellow), CH$_4$ (red), CS (green), and NH$_3$ (purple) are shown at an arbitrary constant offset. Deep negative absorption features originating from detector artifacts are clipped for clarity. {Between $10-12$~\mum, some features of NH$_3$ are over or under reproduced, but the S/N at these wavelengths is low due to the strong absorption by the silicate and H$_2$O ice libration bands.}
    }
    \label{fig:specfit_B1-c_CS_NH3}
\end{figure*}

Toward B1-c, CS is seen in absorption through its $\nu_2$ mode between $7.5-8.5$~\mum, see Figs.~\ref{fig:specfit_B1-c_overview} and \ref{fig:specfit_B1-c_CS_NH3}. Carbon monosulfide has so far only been detected in absorption toward high-mass protostellar systems \citep[e.g.,][]{Knez2009,Barr2018,Nickerson2023} but never at mid-IR wavelengths toward a low-mass sources. The $R$-branch lines are severely blended with the $P$-branch lines of CH$_4$ and were therefore excluded in the fit, but $P$-branch of CS shows clear unblended transitions. The excitation temperature of $285\pm15$~K is in very good agreement with other molecules toward B1-c (see Table~\ref{tab:LTE_Tex}), suggesting it is tracing the same gas. The derived abundance ratio of $5.5\pm2.1\times10^{-2}$ is consistent with that of C$_2$H$_2$, HCN, and CH$_4$ and a factor of $\sim5$ higher than that of SO$_2$.

% \subsubsection{SiO and NH$_3$}
% The JOYS data contain two surprises, SiO and NH$_3$. 
\subsubsection{SiO}
\label{subsubsec:results_SiO}
Silicon monoxide is detected through its $\nu_1$ mode in emission toward L1448-mm and in absorption toward BHR71-IRS1, see Fig.~\ref{fig:specfit_L1448-mm_BHR71-IRS1_SiO}. Thus far, SiO has only been detected at mid-IR wavelengths in absorption toward low-mass protostellar systems in the disk around an oubursting protostar \citep[][]{McClure2024}. Here, the $R$ and $P$-branches are clearly detected between 7.8~\mum and 8.5~\mum and well-fitted with the LTE models. The derived excitation temperatures are similar for the two sources, $305\pm15$~K and $405\pm25$~K for L1448-mm and BHR71-IRS1, respectively. This is much lower than the temperature needed for thermal sublimation of silicon grains ($\sim1500$~K), implying that the SiO is not located within the dust sublimation radius. 
% A possible origin could be small grains that are lifted to higher layers of the disk and (partly) being destroyed by strong UV radiation. 
A likely origin is that the SiO is located in inner jet shocks where the silicon is sputtered off the dust grains in the shocks. This is further supported by blueshifted velocities of $-24\pm2$~\kms and $-23\pm2$~\kms with respect to the $v_{\rm lsr}$ of the SiO lines and the emitting area of $1.9\pm0.3$~au derived for L1448-mm, pointing toward shocked regions in the inner jet or outflow. Furthermore, both L1448-mm and BHR71-IRS1 show very prominent outflows \citep[e.g.,][]{Guilloteau1992,Tobin2018,Nazari2024_L1448mm,Gavino2024} and in particular L1448-mm hosts a collimated high-velocity jet traced by SiO at millimeter wavelengths \citep[][]{Jimenez-Serra2011,Podio2021,Tychoniec2021,Nazari2024_L1448mm}. 

% \begin{figure*}
%     \centering
%     \includegraphics[width=0.49\linewidth]{{Figures/B1-c_ratio=N_ref=H2O-rot}.pdf}
%     % \includegraphics[width=0.49\linewidth]{{Figures/SVS4-5_ratio=N_ref=H2O-rot}.pdf}
%     \includegraphics[width=0.49\linewidth]{{Figures/L1448-mm_ratio=N_ref=H2O-rot}.pdf}
%     % \includegraphics[width=0.49\linewidth]{{Figures/BHR71-IRS1_ratio=N_ref=H2O-rovib}.pdf}
%     \caption{
%     Abundance ratios of several detected molecules (colored datapoints) with respect to the total column density (absorption) or number of molecules (emission) of H$_2$O derived from the pure rotational lines (cold + warm components). 
%     % For BHR71-IRS1 (bottom right panel), the ratio is taken with respect to the rovibrational component since the rotational lines are in emission. 
%     The datapoints of $^{13}$CO$_2$ and $^{13}$CCH$_2$ are multiplied by 70 to take into account the $^{12}$C/$^{13}$C ratio of the local ISM \citep[][]{Milam2005}. For L1448-mm (right), the larger lighter errorbars indicate the range of the abundance ratios when IR pumping is taken into account. The range of column density ratios detected in the ices for all low-mass protostellar systems (i.e., not limited to four sources presented in this figure) are displayed as the colored shaded area for each species with a confirmed ice detection \citep[][]{Boogert2015,Rocha2024,Chen2024}.
%     }
%     \label{fig:B1-c_L1448-mm_column_ratio}
% \end{figure*}

\begin{figure*}
    \centering
    \includegraphics[width=\linewidth]{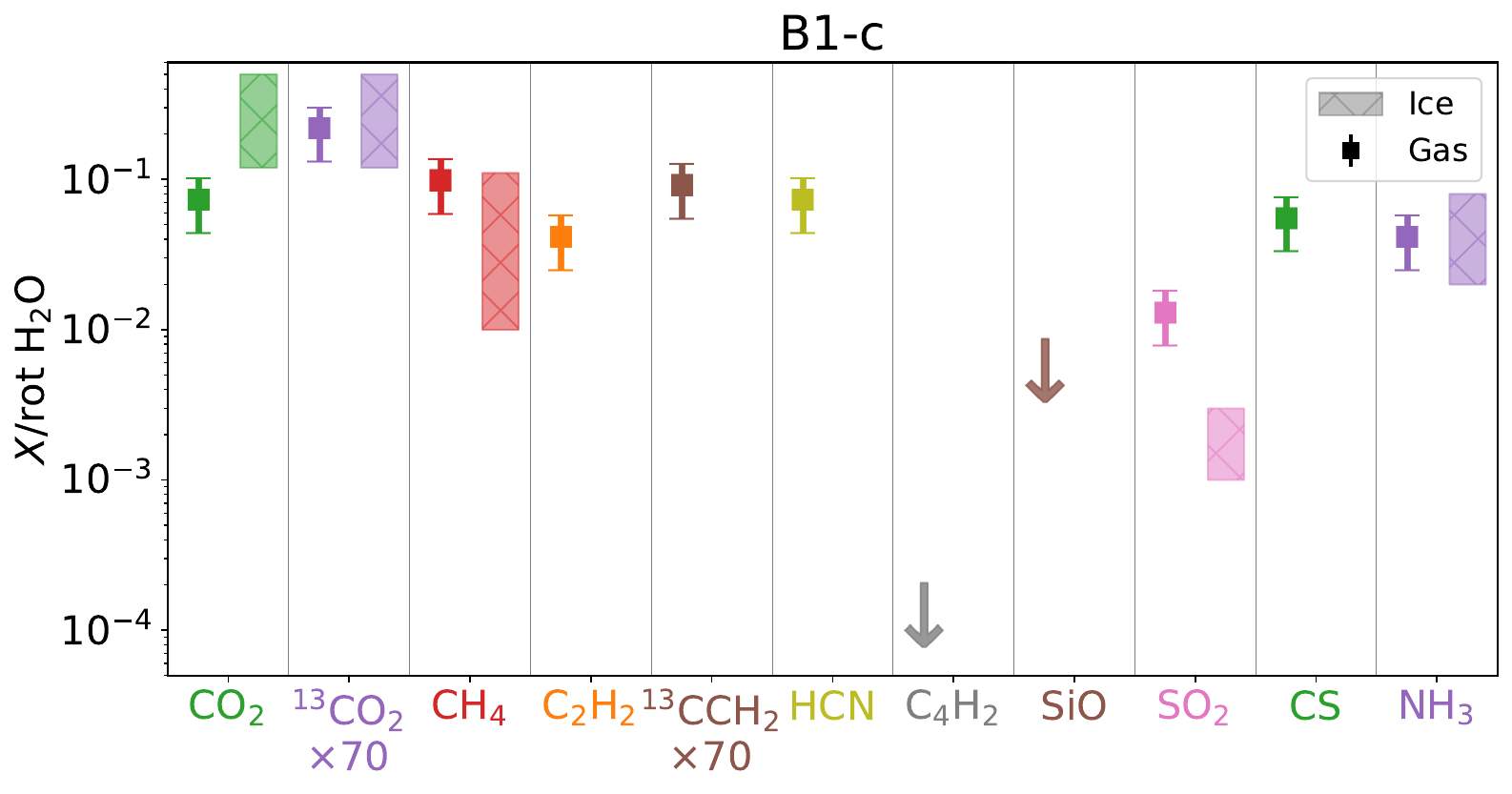}
    \caption{
    Abundance ratios of several detected molecules (colored datapoints) with respect to the total column density of H$_2$O derived from the pure rotational lines (cold + warm components) toward B1-c. 
    % For BHR71-IRS1 (bottom right panel), the ratio is taken with respect to the rovibrational component since the rotational lines are in emission. 
    The datapoints of $^{13}$CO$_2$ and $^{13}$CCH$_2$ are multiplied by 70 to take into account the $^{12}$C/$^{13}$C ratio of the local ISM \citep[][]{Milam2005}. 
    % For L1448-mm (right), the larger lighter errorbars indicate the range of the abundance ratios when IR pumping is taken into account. 
    The range of column density ratios detected in the ices for all low-mass protostellar systems (i.e., not limited to B1-c or other JOYS sources) are displayed as the colored shaded area for each species with a confirmed ice detection \citep[][]{Boogert2015,Rocha2024,Chen2024}.
    }
    \label{fig:B1-c_column_ratio}
\end{figure*}

The abundance ratios with respect to H$_2$O is $2.0\pm0.6\times10^{-4}$ in L1448-mm, which is three orders of magnitude lower than CH$_4$ and the minimum abundance of CO$_2$ and conistent with the range of C$_4$H$_2$. In contrast, the abundance ratio of SiO with respect to the rovibrational H$_2$O component in BHR71~IRS1 (the rotational lines are in emission) is much higher at $0.32\pm0.13$ which is two orders of magnitude higher than that of CO$_2$ ($5.6\pm2.2\times10^{-3}$) and three orders of magnitude higher than that of C$_2$H$_2$ ($1.8\pm0.7\times10^{-4}$). This suggests that especially in BHR71-IRS1, a lot of silicon is being released from the grains. The upper limit for B1-c is $<5.5\times10^{-3}$, which is consistent with L1448-mm but two orders of magnitude lower than BHR71-IRS1.

\subsubsection{NH$_3$}
\label{subsubsec:results_NH3}
Another surprising detection is that of the NH$_3$ $\nu_2$ symmetric bending mode in B1-c. This is, as to our best knowledge, the first detection of NH$_3$ at mid-IR wavelengths toward a low-mass embededded protostellar system, with only a few detections toward high-mass sources \citep[e.g.,][]{Evans1991,Barr2020, Nickerson2023}. The best-fit LTE model is shown in Fig.~\ref{fig:specfit_B1-c_CS_NH3}. Clear absorption features of NH$_3$ are detected both between $8.2$~\mum and $9.3$~\mum, as well as longward of $10.2$~\mum, which are very well reproduced by the best-fit LTE model. The derived excitation temperature is $330\pm10$~K, which is consistent with those of CO$_2$ and H$_2$O (warm and rovibrational lines, see Table~\ref{tab:LTE_Tex}). Furthermore, the derived column density ratio with respect to H$_2$O is $4.1\pm1.6\times10^{-2}$, which is very similar to almost all other molecules. However, the upper limits for several other sources are significantly lower (i.e., $<4\times10^{-3}$ for L1448-mm).
% in perfect agreement with that derived for ices \citep[][]{Boogert2015}, suggesting that we are probing the sublimation of NH$_3$ ices in the hot core of B1-c. 

\section{Discussion}
\label{sec:discussion}
\subsection{Hot core versus outflow}
\label{subsec:hotcore_outflow}
One key question is what the molecular emission and absorption features in the JOYS data are tracing. Most molecules have temperatures in the range of $100-300$~K, which means that they are likely associated with the hot core and warm inner envelope or alternatively with dense shocks in the outflow. A direct way to distinguish between ice desorption in the hot core or gas-phase chemistry is by comparing the abundance ratios with respect to H$_2$O between ice and gas. Assuming that the amount of H$_2$O is dominated by thermal ice desorption, similar abundances between gas and ice point toward an origin of the molecules in the ices whereas deviating abundances hint at (subsequent) gas-phase chemistry. 
However, H$_2$O can also have gas-phase formation routes in the hot core, but as long as H$_2$O is only recycled (i.e., destroyed and reformed) in the gas phase, the amount of H$_2$O will not be altered significantly. Only if large amounts of oxygen are driven from another form (e.g., atomic, refractory dust) into H$_2$O, its abundance (with respect to H$_2$) will be increased, but not by more than a factor of a few \cite[see e.g.,][]{vanDishoeck2021}.
Outflowing material is most directly recognized by a blueshifted velocity offset, such as for BHR71-IRS1 {($v_{\rm line}$ up to $-45$~\kms)}, or spatially extended emission such as for L1448-mm and B1-a. 

Only H$_2$O and CO$_2$ are detected toward a larger bulk of sources. The majority of the species are predominantly detected toward B1-c and L1448-mm, with some other detections toward other sources such as BHR71-IRS1 (C$_2$H$_2$, SiO). The discussion below on what the mid-IR molecular features are tracing is therefore biased toward B1-c and L1448-mm, but also information from (non)detections in the other sources is included.

\begin{figure*}
    \centering
    \includegraphics[width=\linewidth]{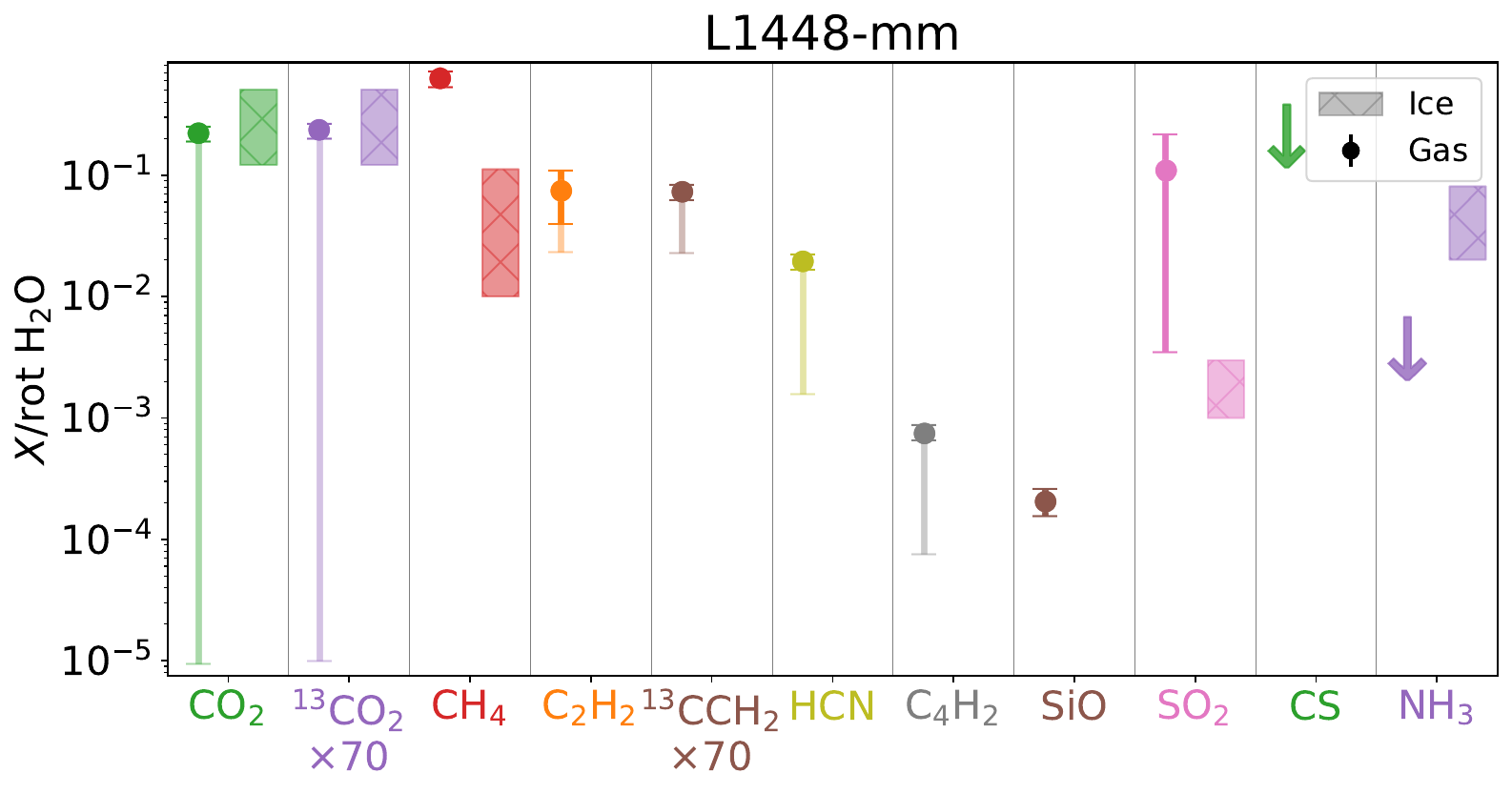}
    \caption{
    Abundance ratios in of several detected molecules (colored datapoints) with respect to the total number of molecules of H$_2$O derived from the pure rotational lines (cold + warm components) toward L1448-mm. 
    % For BHR71-IRS1 (bottom right panel), the ratio is taken with respect to the rovibrational component since the rotational lines are in emission. 
    The datapoints of $^{13}$CO$_2$ and $^{13}$CCH$_2$ are multiplied by 70 to take into account the $^{12}$C/$^{13}$C ratio of the local ISM \citep[][]{Milam2005}. 
    The larger lighter errorbars indicate the range of the abundance ratios when IR pumping is taken into account. 
    The range of column density ratios detected in the ices for all low-mass protostellar systems (i.e., not limited to L1448-mm or other JOYS sources) are displayed as the colored shaded area for each species with a confirmed ice detection \citep[][]{Boogert2015,Rocha2024,Chen2024}.
    }
    \label{fig:L1448-mm_column_ratio}
\end{figure*}

\subsubsection{Hot cores}
The strongest indication for the mid-IR gas-phase lines tracing ice sublimation in a hot core is the case of B1-c. Both the gas-phase and ice abundance ratios with respect to H$_2$O are shown in Fig.~\ref{fig:B1-c_column_ratio}. B1-c is one of the prototypical hot core sources that is rich in complex organics at millimeter wavelengths with no evidence of an embedded disk down to 10~au scales \citep[][]{vanGelder2020,Nazari2021}. Moreover, it is also a bright infrared source showing many deep ice absorption features \citep[e.g., Fig.~\ref{fig:specoverview_B1-c_L1448-mm};][]{Boogert2008,Pontoppidan2008,Oberg2008,Chen2024}. 
% The gas-phase abundances with respect to H$_2$O (derived from the pure rotational transitions) are compared to the ice abundances in Fig.~\ref{fig:B1-c_L1448-mm_column_ratio}. 
It is directly evident that the abundances between gas and ice are remarkably similar. It is important to note that the gas-phase column densities and their ratios for B1-c are derived from absorption lines and therefore more reliable than those derived for sources showing emission lines. Moreover, the velocity of all absorbing species is consistent with the $v_{\rm lsr}=6$~\kms, therefore excluding an outflow origin since its outflow is not in the plane of the sky \citep[e.g.,][]{Jorgensen2006,Hatchell2007,Tychoniec2021}. In particular the ratios of CO$_2$ (i.e., the $^{13}$CO$_2$ ratio multiplied by 70) and NH$_3$ are in very good agreement with {these} in the ices, showing that the gas-phase lines are directly tracing the composition of the sublimated ices. For CH$_4$, the gas-phase abundance is on the higher side but still within the uncertainties of typical CH$_4$ ice abundances \citep[][]{Oberg2008,Boogert2015,Rocha2024,Chen2024}. The only species that significantly deviates is SO$_2$ for which the gas-phase abundance is about an order of magnitude higher than in the ices \citep[][]{Rocha2024,Chen2024}. This indicates that for SO$_2$ (additional) gas-phase chemistry may enhance its abundance above that of the ices \citep[e.g.,][]{Charnley1997,Garrod2022}. 

Similarly, for some other sources a hot core origin seems to be the most plausible. 
In Ser-SMM1B, CO$_2$ is detected in emission and the upper limits {for} other species such as SO$_2$, and NH$_3$ are all still consistent with ices. An ice origin is further supported by the velocity of the CO$_2$ that is redshifted by $\sim10$~\kms with respect to the $v_{\rm lsr}$, likely coming from infalling material. However, only the $Q$-branch of CO$_2$ is detected and therefore the number of CO$_2$ molecules may be underestimated. 
SVS4-5 is another source showing deep ice absorption features \citep[][]{Pontoppidan2004,Perotti2020}. The gas-phase ratios of CO$_2$ and CH$_4$ with respect to H$_2$O also point toward a hot core origin, but the lines are shifted by $-80$~\kms and $-40$~\kms with respect to the $v_{\rm lsr}$ and are more consistent with outflowing material \citep[i.e., molecules in the spatially-extended outflow of the nearby Class~0 source Ser-SMM4 seen in absorption against the IR continuum of SVS4-5;][]{Pontoppidan2004}.
% although only CO$_2$ and CH$_4$ have been detected with ratios a factor of few lower than in the ices. However, only their possibly optically thick $Q$-branch are detected, meaning that their abundance could be underestimated. The upper limit on CO$_2$ derived from $^{13}$CO$_2$ is in good agreement with the ice abundance. Furthermore, the high blueshifted velocities are more consistent with outflowing material. For all other species, only upper limits could be derived which are consistent with the ice abundances. 

\subsubsection{Outflows and disk winds}
A completely different case to the hot core in B1-c is present in L1448-mm, see Fig.~\ref{fig:L1448-mm_column_ratio}. Here, no strong agreement with the ices is seen for many species. Whereas the ratio of CO$_2$ is in agreement when IR pumping is not considered, its abundance with respect to H$_2$O may be as low as $10^{-5}$ when IR pumping through its 4~\mum band is taken into account. Similarly, the upper limit for NH$_3$ suggests a lower gas-phase abundance than in the ices. On the other hand, the abundances of CH$_4$ and SO$_2$ are significantly higher than the ices. 
{Given that the derived excitation temperatures of CH$_4$ and SO$_2$ ($130\pm10$~K and $140\pm20$~K, respectively) are higher than the assumed vibrational temperatures (119~K and 122~K), their excitation is assumed not to be dominated by IR pumping, hence the latter cannot explain the high abundances of both species.} 
Likewise {to L1448-mm}, the abundance ratios in BHR71-IRS1 do not agree with those in the ices. 

% This suggests that the molecular features are not directly tracing ice sublimation but rather regions where their abundances are dominated by gas-phase chemistry. This can happen either in the hot core itself \cite[e.g.,][]{Charnley1992,Charnley1997} or in shocks in the outflow or disk wind close to the source \citep[e.g.,][]{Caselli1997,Gusdorf2008_a,Gusdorf2008_b}.
This suggests that the molecular features are not directly tracing thermal ice sublimation but rather regions where their abundances are dominated by other effects. One possibility is that the molecular abundances are altered by gas-phase chemistry following thermal ice sublimation in a hot core \cite[e.g.,][]{Charnley1992,Charnley1997,Garrod2022}. Alternatively, the molecular features could originate {from} shocks in either the outflow or disk wind close to the source, where they can be formed through high-temperature gas-phase chemistry \citep[e.g.,][]{Caselli1997,Gusdorf2008_a,Gusdorf2008_b}. 
% {One well-known shock tracer is SO$_2$, which indeed shows a higher abundance with respect to H$_2$O than the ices. Similarly, CH$_4$ shows an even higher abundance with respect to H$_2$O but is not commonly associated with gas-phase chemistry in shocks.} 
Otherwise, the ices can be sputtered {off} the dust grains in such shocks, although this seems to be limited only to shock velocities of up to $\lesssim$15~\kms \citep[e.g.,][]{Suutarinen2014,vanDishoeck2021}. 
% (i.e., within the spatial resolution of the MRS).

\begin{figure*}
    \centering
    \includegraphics[width=\linewidth]{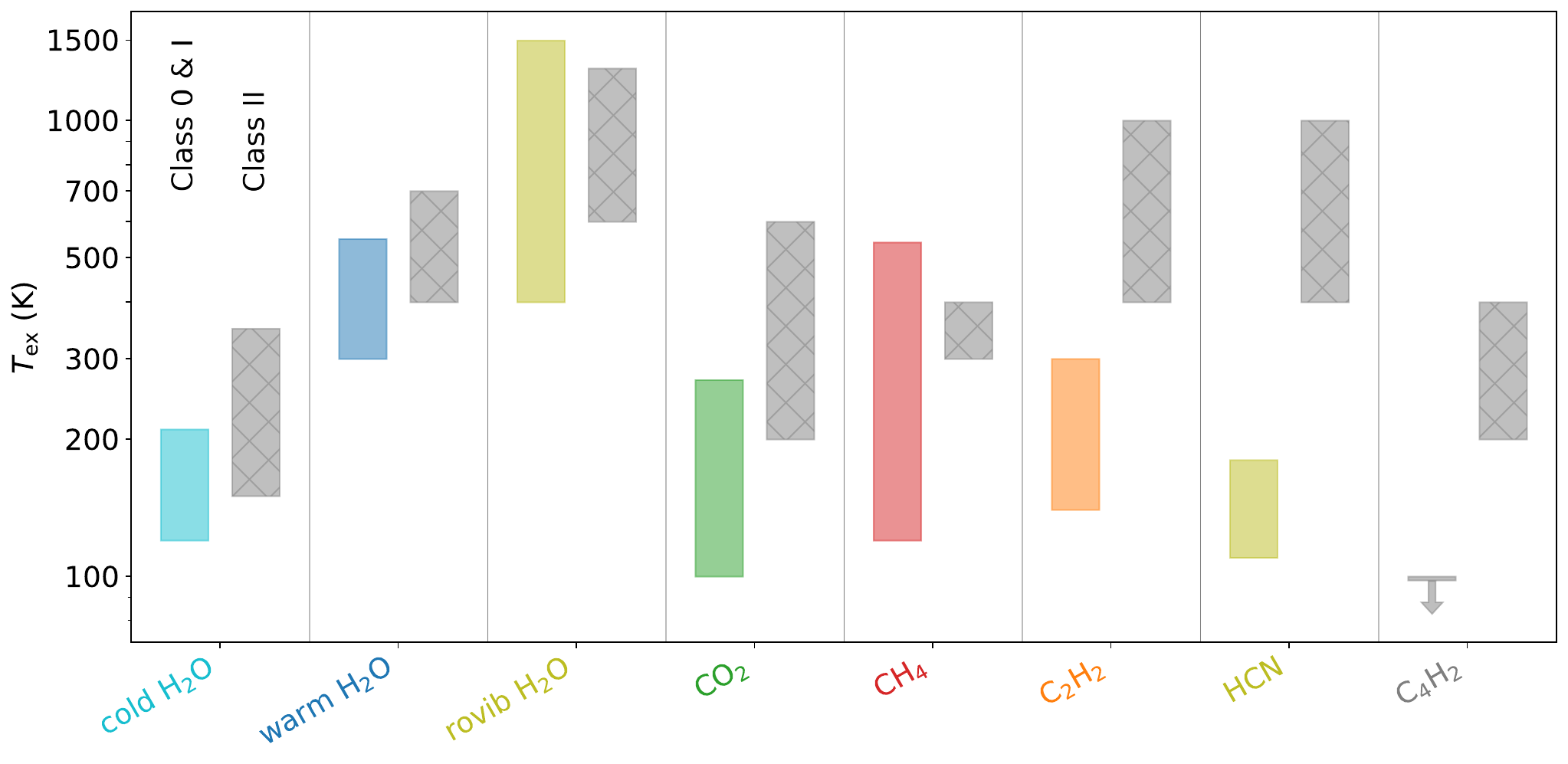}
    \caption{
    Range in excitation temperatures (colored shaded areas) for all detected species toward the embedded low-mass protostellar systems in the JOYS sample. For each species, the measured range of more evolved Class~II sources is presented in the gray shaded areas \citep[][]{Salyk2011,Grant2023,Gasman2023,Tabone2023,Temmink2024_H2O,Schwarz2024,Arabhavi2024}. {Since no data are available for SiO, SO$_2$, CS, and NH$_3$ in Class~II sources, these are not displayed}. It is evident that the molecular features in Class~0$\&$I sources are tracing colder regions than in Class~II sources, although the number of detections in embedded sources remains limited.
    }
    \label{fig:Temp_dist_Classes}
\end{figure*}

An (unresolved) outflow or disk wind origin is supported by the fact that the mid-IR lines are shifted by {$-10$ up to $-45$~\kms} with respect to the $v_{\rm lsr}$ in BHR71-IRS1. Toward L1448-mm, the velocity shift is smaller ($\sim$-25~\kms) and only directly evident at shorter wavelengths for the rovibrational lines of H$_2$O and SiO (see Table~\ref{tab:LTE_results_L1448-mm}). The emission lines of other molecules such as CO$_2$, C$_2$H$_2$, and HCN do not show significant velocity offsets up to $-20$~\kms. Moreover, especially CO$_2$ clearly shows spatially extended emission consistent with outflowing material (Navarro et al. in prep.). Furthermore, SiO is detected at mid-IR wavelengths toward both sources. Silicon monoxide is commonly observed at millimeter wavelengths in the bullets of molecular jets \citep[e.g.,][]{Podio2016,Podio2021,Lee2017,Tychoniec2019,Tychoniec2021} and indeed also detected on larger scales in the jets of both L1448-mm \citep[][]{Guilloteau1992,Jimenez-Serra2011,Toledano-Juarez2023,Nazari2024_L1448mm} and BHR71-IRS1 \citep[and IRS2;][]{Gusdorf2011,Gusdorf2015,Gavino2024}. In both cases, the mid-IR SiO emission and absorption is spatially unresolved, but showing similar velocity offsets as the other molecules and atomic species (Tychoniec et al. in prep., Navarro et al. in prep.). 
This is in agreement with results from {\it Herschel} comparing the H$_2$O velocity profiles with those of SiO \citep[e.g.,][]{Nisini2013,Leurini2014,vanDishoeck2021}.
Moreover, the derived temperatures of $300-400$~K are consistent with those in jet shocks \citep[e.g.,][]{Caselli1997,Gusdorf2008_a,Gusdorf2008_b} and not with thermal sublimation of silicate grains ($\sim1500$~K). Toward both sources, emission from other molecules such as CO$_2$ is also present in the MIRI-MRS data on larger scales in the outflows, but this will be presented in a separate paper.

For many other sources, there is also a strong indication that especially molecular emission features are tracing unresolved outflows. This is especially evident for CO$_2$, for which temperatures of $\sim100-300$~K are measured for the majority of the sources, consistent with it being present on larger scales and not in the hot inner regions of an embedded disk (see Sect.~\ref{subsec:embedded_disks}). On the other hand, not all sources show velocity offsets in their CO$_2$ with respect to the $v_{\rm lsr}$ (i.e., Ser-SMM3, TMC1A). 
Toward SVS4-5, the CO$_2$ emission is modeled at a velocity of $-80$~\kms with respect to the $v_{\rm lsr}$, likely associated to the outflow of the nearby Class~0 source Ser-SMM4. For B1-a-NS, on the other hand, CO$_2$ is seen in absorption at $+35$~\kms with respect to the $v_{\rm lsr}$, but also weak spatially extended emission of CO$_2$ is present toward the south. Overall, CO$_2$ appears to be very commonly present on larger scales and not necessarily solely tracing ice sublimation. 

\subsection{Embedded disks}
\label{subsec:embedded_disks}
\subsubsection{Temperatures from Class 0 and I to Class II}
Figure~\ref{fig:Temp_dist_Classes} compares the ranges of derived excitation temperatures from Class~0 and I sources in JOYS to those derived for more evolved Class~II protoplanetary disks.
% Molecular emission at mid-IR wavelengths is commonly detected in more evolved Class~II protoplanetary disks \citep[e.g.,][]{Salyk2011,Grant2023}. 
Typical temperatures in Class~II disks are $T\gtrsim500$~K \citep[e.g.,][]{Grant2023,Banzatti2023_JWST,Gasman2023,Temmink2024_H2O,Temmink2024_CO}, suggesting that the emission arises either from the inner disk or a warm disk surface layer. In contrast, for many molecules, the derived temperatures ($100-300$~K) are significantly lower toward the protostars in the JOYS sample. As discussed in Sect.~\ref{subsec:hotcore_outflow}, in many cases it is evident that the molecular features are not tracing a disk-like structure but are rather present in the hot core or warm inner envelope (e.g., B1-c, Ser-SMM1B) or are located in the disk wind or outflow (e.g., L1448-mm, BHR71-IRS1). However, in a few embedded sources, the rovibrational lines of H$_2$O show temperatures of up to 1200~K, suggesting that these may be tracing embedded disks.

Within the JOYS sample, several sources are known to host rotationally supported embedded disks based on resolved millimeter observations \citep[e.g., TMC1A, TMC1, L1527;][]{Tobin2012,Harsono2014,Tychoniec2021,Ohashi2023}. Most notably, toward TMC1-W, hot H$_2$O ($1280\pm20$~K) is detected through its rovibrational lines between 5.5~\mum and 7.5~\mum, see Fig.~\ref{fig:specfit_TMC1-W_H2O-rovib}. Given the high temperature and the small emitting area predicted by the LTE models ($0.05\pm0.01$~au), the rovibrational H$_2$O lines are clearly originating from hot compact material that likely resides within its inner embedded disk. Interestingly, toward its companion TMC1-E, the H$_2$O lines are in absorption and not as hot ($580\pm20$~K), but still consistent with a cooler H$_2$O component in the disk \citep[][]{Gasman2023,Temmink2024_H2O}. Moreover, given that the lines are shifted by $-10$~\kms with respect to the $v_{\rm lsr}$, this H$_2$O may also originate from a disk wind rather than the embedded disk \citep[][]{Tychoniec2024}. Similarly hot H$_2$O as in TMC1-W is detected toward Ser-SMM1B ($980\pm20$~K), but for this source the presence of an embedded disks has yet to be confirmed at millimeter wavelengths. However, given the similar temperature as seen toward TMC1-W, an embedded disk origin for the hot H$_2$O is plausible. 
Toward SVS4-5, also hot H$_2$O ($1040\pm20$~K) is detected, which in contrast to CO$_2$ and CH$_4$ is not blueshifted but consistent with the systemic velocity. This indicates that it is likely originating from the background Class~I/II source that is located behind or inside the envelope and outflow of the nearby Class~0 source Ser-SMM4 \citep[][]{Pontoppidan2004}.
% Toward SVS4-5, also hot H$_2$O ($1040\pm20$~K) is detected, which in contrast to CO$_2$ and CH$_4$ may be coming from a disk in the background Class~I/II source behind the envelope and outflow of Ser-SMM4.  

\begin{figure*}
    \centering
    \includegraphics[width=\linewidth]{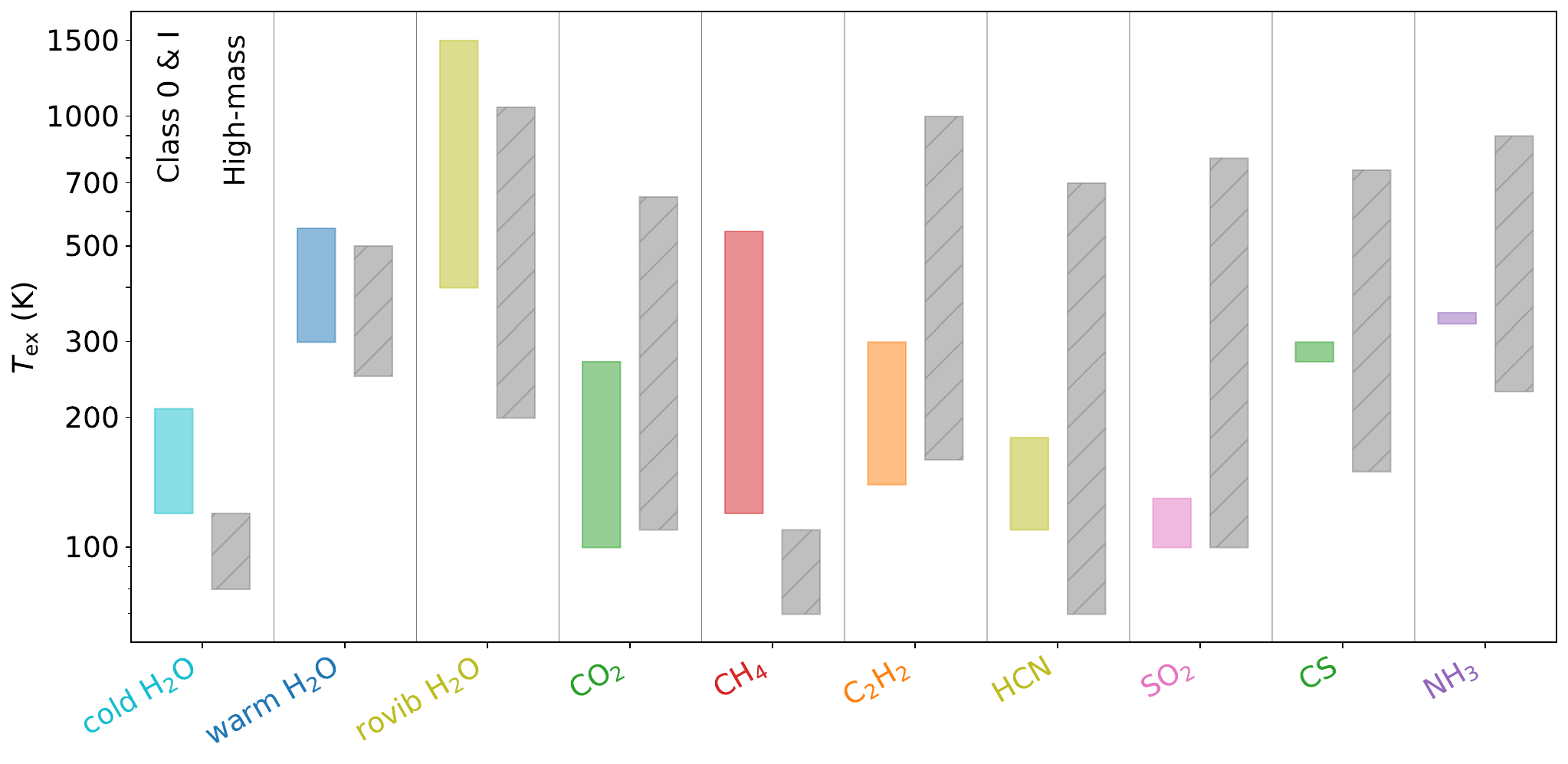}
    \caption{
    Range in excitation temperatures (colored shaded areas) for all detected species measured toward the embedded low-mass protostellar systems in the JOYS sample. For each species, the range measured toward high-mass sources is presented in the gray shaded areas \citep[][]{Boogert1998,Lahuis2000,Keane2001,Boonman2003_Orion,Knez2009,Karska2014,Dungee2018,Barr2018,Barr2020,Nickerson2023,Francis2024}. {Since for no data are available for SiO and C$_4$H$_2$ in high-mass sources, these are not displayed.}
    }
    \label{fig:Temp_dist_Highmass}
\end{figure*}

On the other hand, several sources hosting embedded disks do not show such hot temperatures in H$_2$O. TMC1A shows rovibrational H$_2$O in absorption at $440\pm10$~K, but it is not directly evident whether this is tracing the embedded disk or whether it is originating from its disk wind since the velocity {with respect to the $v_{\rm lsr}$} needed to fit the data is $-10$~\kms \citep[e.g.,][]{Herczeg2011,Bjerkeli2016}. Similarly, B1-a-NS \citep[displaying weak rotation in $^{13}$CO at millimeter wavelengths;][]{Tobin2018} shows the rovibrational lines in emission at $405\pm15$~K at $-10$~\kms offset for which neither an embedded disk nor an outflow contribution can be neglected.

\subsubsection{Absence of molecular features}
A major surprise is the non-detection of clear embedded disk tracers at high temperatures in other species than H$_2$O. As is evident from Fig.~\ref{fig:Temp_dist_Classes}, the temperatures of especially CO$_2$, C$_2$H$_2$, and HCN are systematically lower ($\sim100-300$~K) than typically seen in Class~II sources \citep[$\gtrsim500$~K; e.g.,][]{Grant2023,Tabone2023,Schwarz2024,Arabhavi2024}. 
% Only CH$_4$ appears at similar temperatures as in Class~II disks, especially for B1-c ($620\pm10$~K) and SVS4-5 ($460\pm80$~K). However, as discussed in Sect.~\ref{subsec:hotcore_outflow}, CH$_4$ likely traces the hot core in B1-c and an embedded disk has not been detected down to 10~au scales. For SVS4-5, the CH$_4$ emission is located at a $-60$~\kms offset of the $v_{\rm lsr}$, implying that it is located in outflowing material.
% Other secies like CO$_2$, C$_2$H$_2$, and HCN are systematically tracing colder temperatures than in Class~II disks . 
Moreover, several embedded sources such as B1-b, Per-emb~8 and Ser-SMM3 \citep[{the latter two} hosting rotating disk-like structures;][]{Tobin2018,Tychoniec2021}, and L1527 \citep[hosting an edge-on rotationally-supported disk;][]{Tobin2012} show no or hardly any emission or absorption features despite bright mid-IR continuum. For edge-on disks like L1527 this could easily be explained by the outer disk absorbing the inner disk, but for the other sources this cannot explain the lack of hot molecular gas. No clear trend between the amount of detected molecules and their physical properties is visible with either the bolometric luminosity or bolometric temperature.

One possible explanation could be that the inner regions of embedded disks are not as warm as more evolved Class~II disks. However, this would be rather surprising given that embedded disks are generally warmer than their Class~II counterparts \citep[e.g.,][]{van'tHoff2018,van'tHoff2020_Taurus,van'tHoff2020_IRAS16293,Podio2020}. Moreover, young protostars are still actively accreting and therefore generally more luminous \citep[][]{Evans2009,Fischer2017}. The presence of molecular features does not increase with luminosity since two of the more luminous sources in our sample, Ser-SMM1A and Ser-SMM3, hardly show any molecular emission and absorption. The absence of hot molecular features from the inner regions of embedded disks is thus likely not a temperature effect.

A different scenario is where the emission (or absorption) from the inner embedded disks is completely extincted. 
% This extinction cannot occur solely in the envelope since then the continuum and molecular emission and absorption features attributed to the inner envelope (hot core) or outflow should also be extincted (i.e., the cases of IRAS~4B and Ser-S68N-S). 
In more evolved Class~II disks, the larger dust grains are suggested to have settled to the midplane of the disk \citep[e.g.,][]{Dullemond2005,Dullemond2007}. In the disk surface layers, often gas-to-dust ratios of $>1000$ are needed in order to explain the observed molecular emission and high line-to-continuum ratios \citep[e.g.,][]{Meijerink2009,Woitke2018,Greenwood2019}. However, in embedded disks, the dust may not have fully grown and settled yet and can therefore severely extinct the emission from the inner disk. Indeed, recent observations of the embedded disks from the Early Planet Formation in Embedded Disks (eDisk) program show little settling of dust grains in several Class~I sources \citep[][]{Lin2023,Encalada2024,Gavino2024}. Moreover, the embedded disks are continuously being replenished in small dust grains by their envelopes \citep[][]{Visser2009,Visser2011,Cridland2022,Gupta2023}. The non-settled dust grains in the upper layers of the disk can easily fully extinct the hot molecular features emitted from lower layers of the disk.

Alternatively, substructures in the disk (i.e., gaps, cavities, rings) can also reduce the amount of gas-phase molecules in the inner disk \citep[][]{Kalyaan2021,Kalyaan2023,Vlasblom2024}. However, embedded disks show very little substructures at millimeter wavelengths in the eDisk program \citep[][]{Ohashi2023}, although some embedded sources are known to show subtructures \citep[e.g., HL~Tau, Oph-IRS63;][]{ALMA2015,Sheehan2020,Segura-Cox2020,Flores2023}. BHR71-IRS1 (and IRS2) were also part of the eDisk program and lack clear substructures in the continuum \citep[][]{Gavino2024}, but clear disk related molecular emission or absorption are also absent. This suggests that substructures alone cannot explain the lack of gas-phase molecular features in embedded disks. Extinction by non-settled dust in grains in the upper layers of the disks is therefore a more likely explanation.

\subsection{Comparison to high-mass protostars}
Many of the molecules detected in this work have also been observed toward embedded high-mass sources. With previous space-based telescopes such as ISO and {\it Spitzer}, species such as H$_2$O, CO$_2$, and C$_2$H$_2$ were commonly detected \citep[e.g.,][]{Helmich1996,Boonman2003_Orion,vanDishoeck2004,Sonnentrucker2006,Sonnentrucker2007}. Moreover, also ground-based instruments such as the VLT and airborne telescopes such as SOFIA were able to observe gas-phase molecular features toward high-mass sources at very high spectral resolution \citep[][]{Evans1991,Dungee2018,Indriolo2020,Barr2020,Barr2022}. More recently, also MIRI-MRS observations of a high-mass protostellar region were presented \citep[][]{Francis2024}, showing molecular emission and absorption features. Now that these species are also detected toward low-mass sources, it is interesting to determine whether similar components are traced between low-mass and high-mass systems.

In Fig.~\ref{fig:Temp_dist_Highmass}, the range of measured excitation temperatures is compared to those measured in high-mass sources similarly to Fig.~\ref{fig:Temp_dist_Classes}. The temperatures measured toward low-mass sources all overlap with those in high-mass sources, but are on the lower side of the distribution. An exception is CH$_4$ for which higher temperatures are measured for low-mass sources, although it must be noted that both the low-mass and high-mass range are based on two sources \citep[][]{Boogert1998} and that higher temperatures are suggested from a tentative detection in IRAS~23385+6053 \citep[][]{Francis2024}. The high temperatures of {several 100}~K for C$_2$H$_2$ and HCN were suggested to originate from ice sublimation, perhaps followed by high-temperature gas-phase chemistry \citep[][]{Lahuis2000}, which would mean that they are tracing hot core regions similar to B1-c, but can also trace hot surface layers of a viscously heated disk \citep[][]{Barr2020}. On the other hand, lower temperatures on the order of 100-300~K were measured for CO$_2$ and toward several other high-mass sources and linked to either sputtering the from ices and/or via high-temperature gas-phase chemistry in shocks \citep[][]{Boonman2003_Orion,Sonnentrucker2006,Sonnentrucker2007}. More recently, lower excitation temperatures were also linked to the warm disk surface layers, {but at larger radii of $>100$~au} \citep[][]{vanDishoeck2023,Francis2024}. In many of our low-mass sources, the low temperatures are linked to outflow shock activity (e.g., Ser-SMM3, L1448-mm) based on their blueshifted velocities and disagreement between gas-phase and ice abundance ratios.

Three molecules that were only detected toward low-mass sources for the first time with JWST, SO$_2$, CS, and NH$_3$ \citep[][this work]{vanGelder2024}, were previously detected toward high-mass sources \citep[e.g.,][]{Keane2001,Knez2009,Dungee2018,Barr2018,Barr2020,Nickerson2023}. For SO$_2$, the measured excitation temperatures toward high-mass sources are mostly similar to those in low-mass sources, on the order of $100-300$~K \citep[][]{Dungee2018,Nickerson2023}, suggesting that they are likely either tracing the hot cores or outflow shocks. On the other hand, also higher temperatures of up to 700~K have been seen for high-mass sources \citep[][]{Keane2001} which are more consistent with hot disk surface layers. 
For CS, a similarly large range in temperatures is measured toward high-mass sources \citep[][]{Knez2009,Barr2018,Barr2020,Nickerson2023} and the derived temperature of B1-c lies right in the middle of the high-mass range. 
{The derived temperature of NH$_3$ of $340\pm10$~K in the hot core of B1-c is on the lower side of the high-mass range \citep[$250-900$~K;][]{Barr2020,Nickerson2023}, which is consistent with high-mass hot cores. Higher temperature of up to 900~K for NH$_3$ have only been seen in hot disk surface layers \citep[][]{Barr2020}.}
% A different picture is seen for NH$_3$ where only higher temperatures of $300-900$~K have been observed related to hot disk surface layers \citep[][]{Barr2020}, whereas the only detection toward a low-mass source is consistent with the hot core (B1-c, see Sect.~\ref{subsec:hotcore_outflow} and Fig.~\ref{fig:B1-c_column_ratio}). 
Both toward low-mass and high-mass source molecular features can thus trace various components ranging from the hot core to disk surface layers and (outflow) shocked regions.

\section{Summary}
\label{sec:conclusions}
This paper presents an overview of molecular emission and absorption features detected in JWST/MIRI-MRS observations of the sample of 18 low-mass protostars in the JOYS program. The spectra of the central regions of 16 sources where mid-IR continuum was detected were fitted using LTE slab models to determine the physical conditions in which the molecules reside. {This includes molecules such as CO$_2$, C$_2$H$_2$, and CH$_4$ that cannot be studied at millimeter wavelengths.} Abundance ratios were derived with respect to H$_2$O, which together with the derived excitation conditions were used to determine the origin of the molecular emission and absorption features. Based on the aforementioned analysis, our main conclusions are as follows:
\begin{itemize}
    \item The spectra show an unprecedented richness in molecular emission and absorption features. Molecular features can be either in absorption against the bright embedded IR continuum (e.g., B1-c) or in emission (e.g., L1448-mm). 
    \item Water is the most commonly detected species in the MIRI-MRS observations, being present in 12/16 sources. Almost similarly present are CO$_2$ (11/16), CO (8/16), and OH (7/16). All other molecules are detected in at most three sources (predominantly B1-c and L1448-mm) and include $^{13}$CO$_2$, C$_2$H$_2$, $^{13}$CCH$_2$, HCN, C$_4$H$_2$, CH$_4$, and SO$_2$. 
    \item The data yield the surprising detection of SiO toward L1448-mm (emission) and BHR71-IRS1 (absorption) and for the first time CS and NH$_3$ at mid-IR wavelengths toward a low-mass protostellar source (B1-c).
    \item The excitation temperatures derived toward the majority of the species lie in the $100-300$~K range, consistent with those derived toward high-mass sources but significantly lower than typically observed toward more evolved Class~II disks. 
    % In a few sources, the rovibrational lines of H$_2$O show significantly higher temperatures up to 1200~K, likely tracing the embedded disks.
    \item For sources showing molecules in emission, IR pumping has to be taken into account in deriving accurate column densities and total number of molecules.
    \item Molecular emission and absorption features in embedded protostellar systems can trace various components. The abundance ratios with respect to H$_2$O are fully consistent with ice sublimation toward the famous hot core source B1-c and to a lesser extent in several other sources (e.g., Ser-SMM1B). On the other hand, the discrepancy between gas and ice abundances with respect to H$_2$O together with the detection of SiO suggest outflow shock origins toward L1448-mm and BHR71-IRS1. Also the cold and velocity-shifted CO$_2$ emission toward several other sources (e.g., Ser-SMM3, Ser-S68N-N) is consistent with outflowing material.
    \item The JOYS data yield surprisingly few detections of molecular gas in the hot inner regions of the disk.  TMC1-W is the only source with both a confirmed embedded disk and detected hot ($\sim1200$~K) molecular emission (H$_2$O), with only two other sources showing such hot H$_2$O features. This is in strong contrast with the more evolved Class~II sources where hot molecular features are detected for many species (CO$_2$, C$_2$H$_2$, HCN). The non-detection of a broader range of molecular features from the embedded disks likely originates from extinction due to non-settled dust grains in the upper layers of the disk.
\end{itemize}

\noindent The JOYS data have shown the great potential of detecting and characterizing molecular emission and absorption features toward embedded protostellar systems. The unprecedented spatial and spectral resolution and sensitivity of JWST/MIRI-MRS will certainly yield more beautiful and line-rich data of embedded protostellar systems, further enlightening us about the processes {that take place} on disk and outflow scales. Furthermore, additional work on modeling the structure of embedded systems, including the disk and outflow, as well as their evolution will further shed light on what the molecular features detected at mid-IR wavelengths are tracing and what the chemical composition is during the onset of planet formation.

\begin{acknowledgements}

We would like to thank {the anonymous referee for their constructive comments on the manuscript and} John Black for fruitful discussions on non-LTE effects and infrared pumping.

This work is based on observations made with the NASA/ESA/CSA James Webb Space Telescope. The data were obtained from the Mikulski Archive for Space Telescopes at the Space Telescope Science Institute, which is operated by the Association of Universities for Research in Astronomy, Inc., under NASA contract NAS 5-03127 for JWST. These observations are associated with program \#1290.
The following National and International Funding Agencies funded and supported the MIRI development: NASA; ESA; Belgian Science Policy Office (BELSPO); Centre Nationale d’Études Spatiales (CNES); Danish National Space Centre; Deutsches Zentrum fur Luftund Raumfahrt (DLR); Enterprise Ireland; Ministerio De Economiá y Competividad; The Netherlands Research School for Astronomy (NOVA); The Netherlands Organisation for Scientific Research (NWO); Science and Technology Facilities Council; Swiss Space Office; Swedish National Space Agency; and UK Space Agency. 

MvG, LF, EvD, YC, KS, and WR acknowledge support from ERC Advanced grant 101019751 MOLDISK, TOP-1 grant 614.001.751 from the Dutch Research Council (NWO), The Netherlands Research School for Astronomy (NOVA), the Danish National Research Foundation through the Center of Excellence “InterCat” (DNRF150), and DFG-grant 325594231, FOR 2634/2.

T.P.R. acknowledges support from ERC grant 743029 EASY. 

H.B. acknowledges support from the Deutsche Forschungsgemeinschaft in the Collaborative Research Center (SFB 881) “The Milky Way System” (subproject B1)

A.C.G. acknowledges support from PRIN-MUR 2022 20228JPA3A “The path to star and planet formation in the JWST era (PATH)” funded by NextGeneration EU and by INAF-GoG 2022 “NIR-dark Accretion Outbursts in Massive Young stellar objects (NAOMY)” and Large Grant INAF 2022 “YSOs Outflows, Disks and Accretion: towards a global framework for the evolution of planet forming systems (YODA)”. 

KJ acknowledges the support from the Swedish National Space Agency.

P.J.K. acknowledges support from the Science Foundation Ireland/Irish Research Council Pathway programme under Grant Number 21/PATH-S/9360.

P.N. acknowledges support from the ESO Fellowship and IAU Gruber Foundation Fellowship programs.

This research made use of NumPy \citep[][]{Harris2020}; Astropy, a community-developed core Python package for Astronomy \citep[][]{AstropyCollaboration2022}; Matplotlib \citep[][]{Hunter2007}.

\end{acknowledgements}

\bibliographystyle{aa}
\bibliography{refs}

\begin{appendix}

\onecolumn
\section{Sample properties and spectral extraction}
\label{app:specextr}
\begin{multicols}{2}
% \nolinenumbers
\noindent The JOYS sample consists of 20 low-mass protostellar systems, see summary of their main properties in Table~\ref{tab:sample}. The observations of NGC~1333~IRAS~4A and Ser-emb~8(N) do not cover the central protostellar position itself and are therefore excluded from the analysis of this paper, leaving a sample of 18 protostars. Table~\ref{tab:sample} also indicates the total exposure time divided over the three MIRI-MRS gratings, as well as the number of dither positions. It is important to note that several observations included multiple pointings (i.e., B1-c, L1448-mm, IRAS~4B, BHR71-IRS1) covering parts of their extended blueshifted outflows. Furthermore, also toward some sources with a single pointing the center of the FOV was placed toward the blueshifted outflow (i.e., TMC1A). However, given that only the spectra from the central protostellar position are covered here, only the pointings containing the central protostellar system are listed in Table~\ref{tab:sample}.

\begin{figure*}[h]
    \centering
    \includegraphics[width=0.85\linewidth]{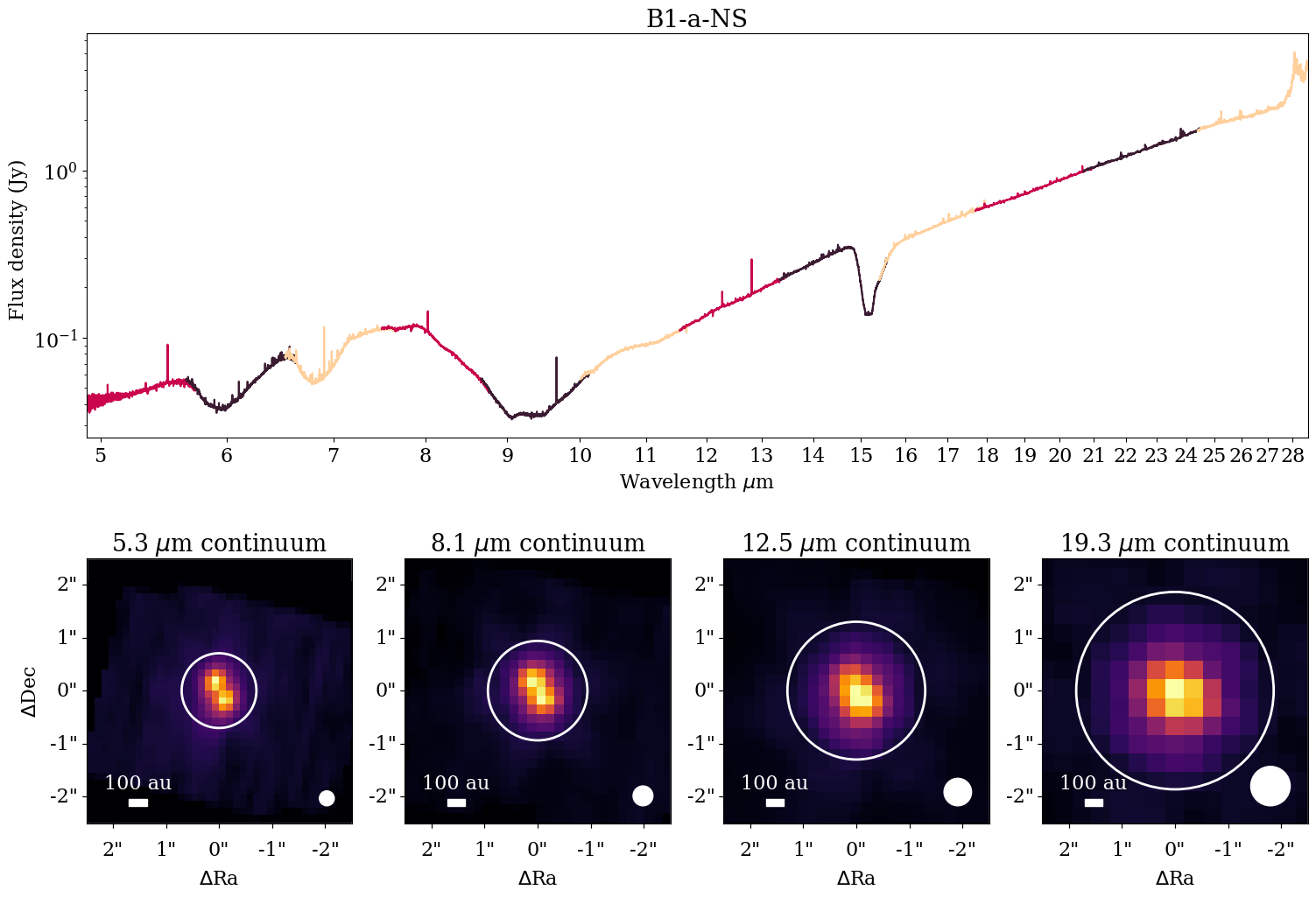}
    \caption{Extracted spectrum (top panel) and continuum images (bottom row) for B1-a-NS. The 12 subbands in the spectrum are not stitched,  and each color represent one of the three MIRI-MRS gratings: A (red), B (black), and C (yellow). The bottom row shows from left to right the mid-IR continuum images around 5.3~\mum (channel 1 short), 8.1~\mum (channel 2 short), 12.5~\mum (channel 3 short), and 19.3~\mum (channel 4 short). The continuum images are scaled with a {\tt sqrt} stretch to enhance fainter features without saturating brighter emission. A scale bar is displayed in the bottom right and the FWHM of the PSF is shown in the bottom right of each panel.}
    \label{fig:specext_B1-a-NS}
\end{figure*}

\begin{figure*}[h]
    \centering
    \includegraphics[width=0.85\linewidth]{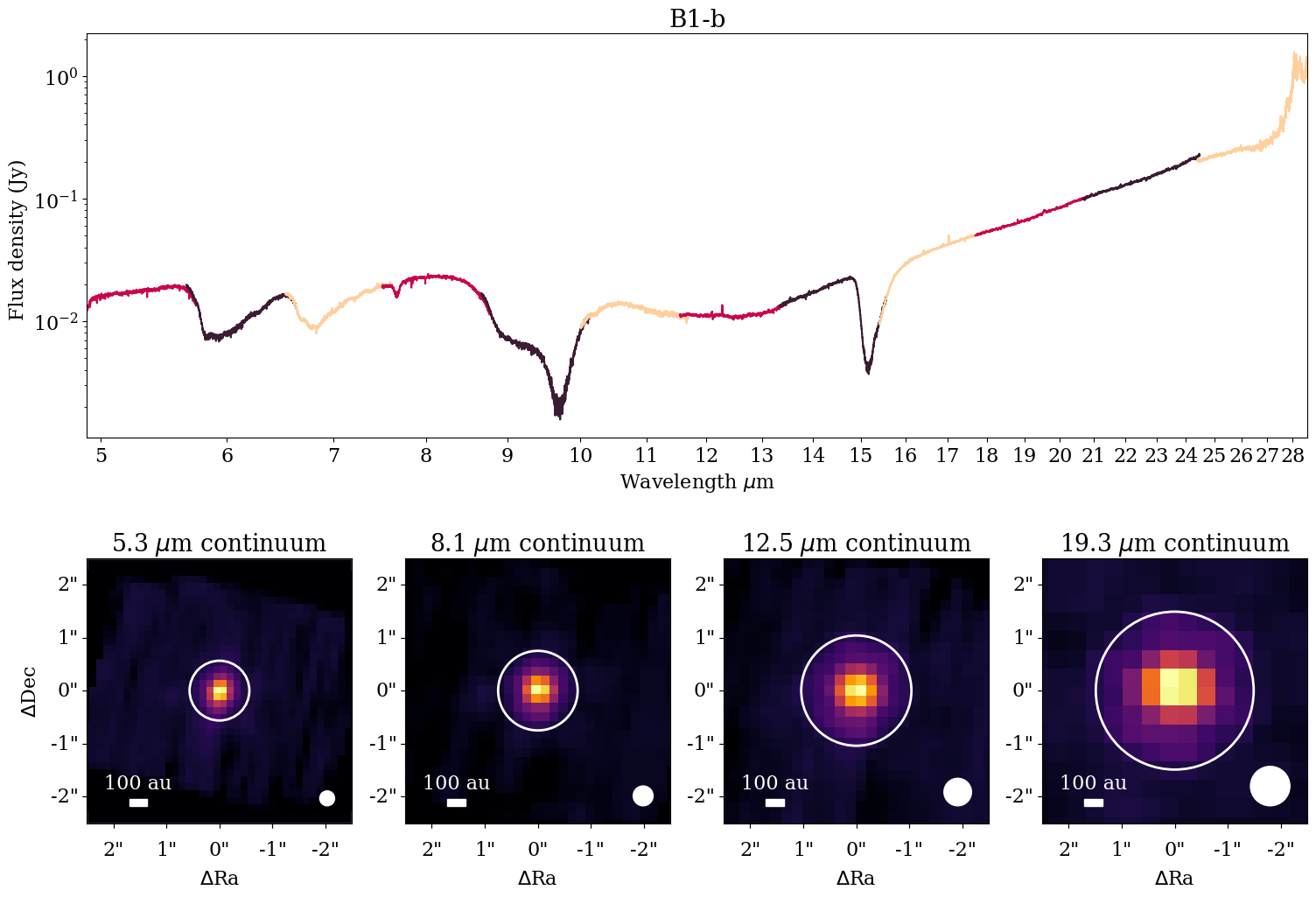}
    \caption{Same as Fig.~\ref{fig:specext_B1-a-NS} but for B1-b.}
    \label{fig:specext_B1-b}
\end{figure*}

\begin{figure*}
    \centering
    \includegraphics[width=0.92\linewidth]{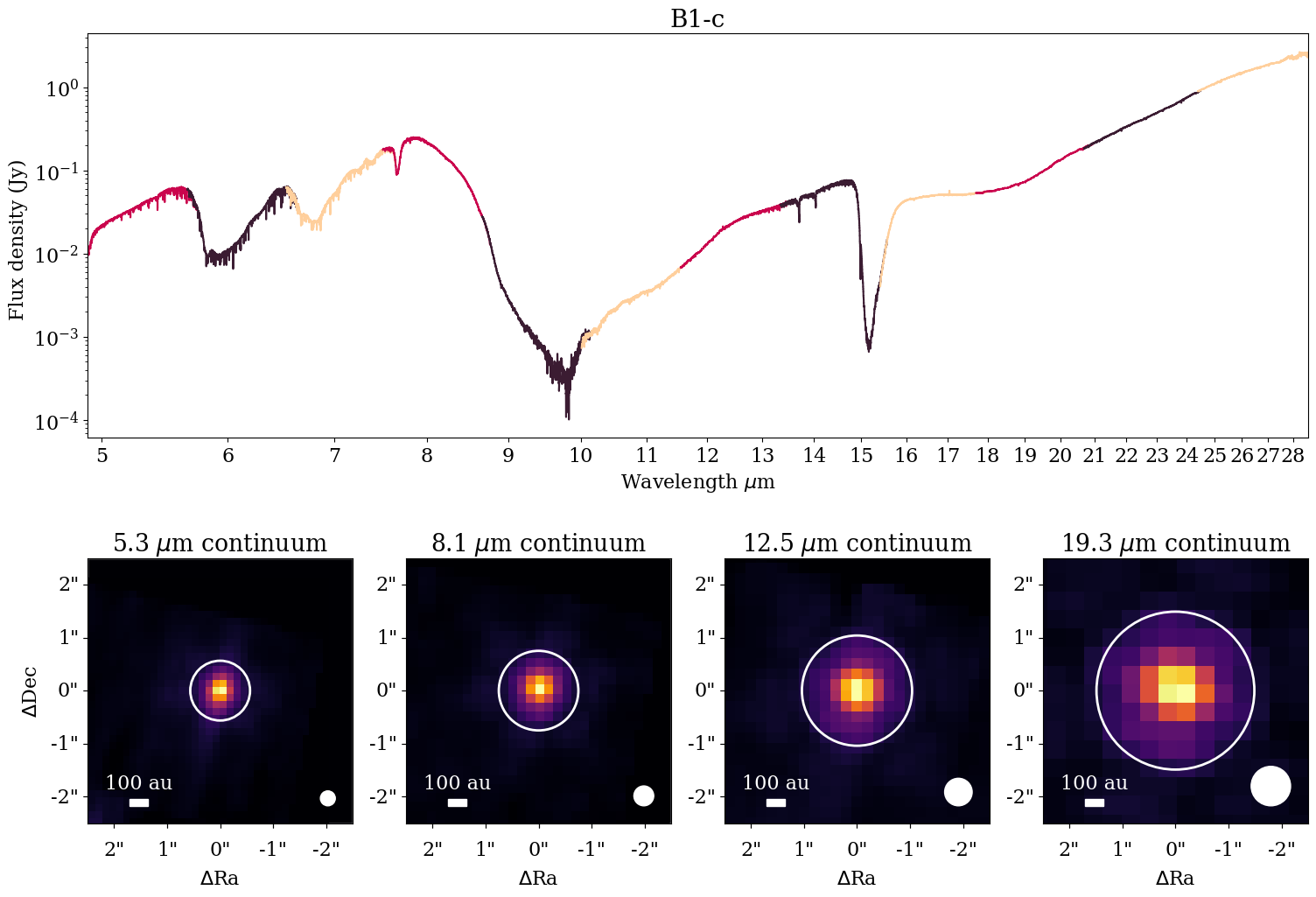}
    \caption{
    Same as Fig.~\ref{fig:specext_B1-a-NS} but for B1-c.
    % Extracted spectrum (top panel) and continuum images (bottom row) for B1-c. The 12 subbands in the spectrum are not stitched,  and each color represents one of the three MIRI-MRS gratings: A (red), B (black), and C (yellow). The bottom row shows from left to right the mid-IR continuum images around 5.3~\mum (channel 1 short), 8.1~\mum (channel 2 short), 12.5~\mum (channel 3 short), and 19.3~\mum (channel 4 short). The continuum images are scaled with a {\tt sqrt} stretch to enhance fainter features without saturating brighter emission. A 100~au scale bar is displayed in the bottom right and the FWHM of the PSF is shown in the bottom right of each panel.
    }
    \label{fig:specext_B1-c}
\end{figure*}

\begin{figure*}[h]
    \centering
    \includegraphics[width=0.92\linewidth]{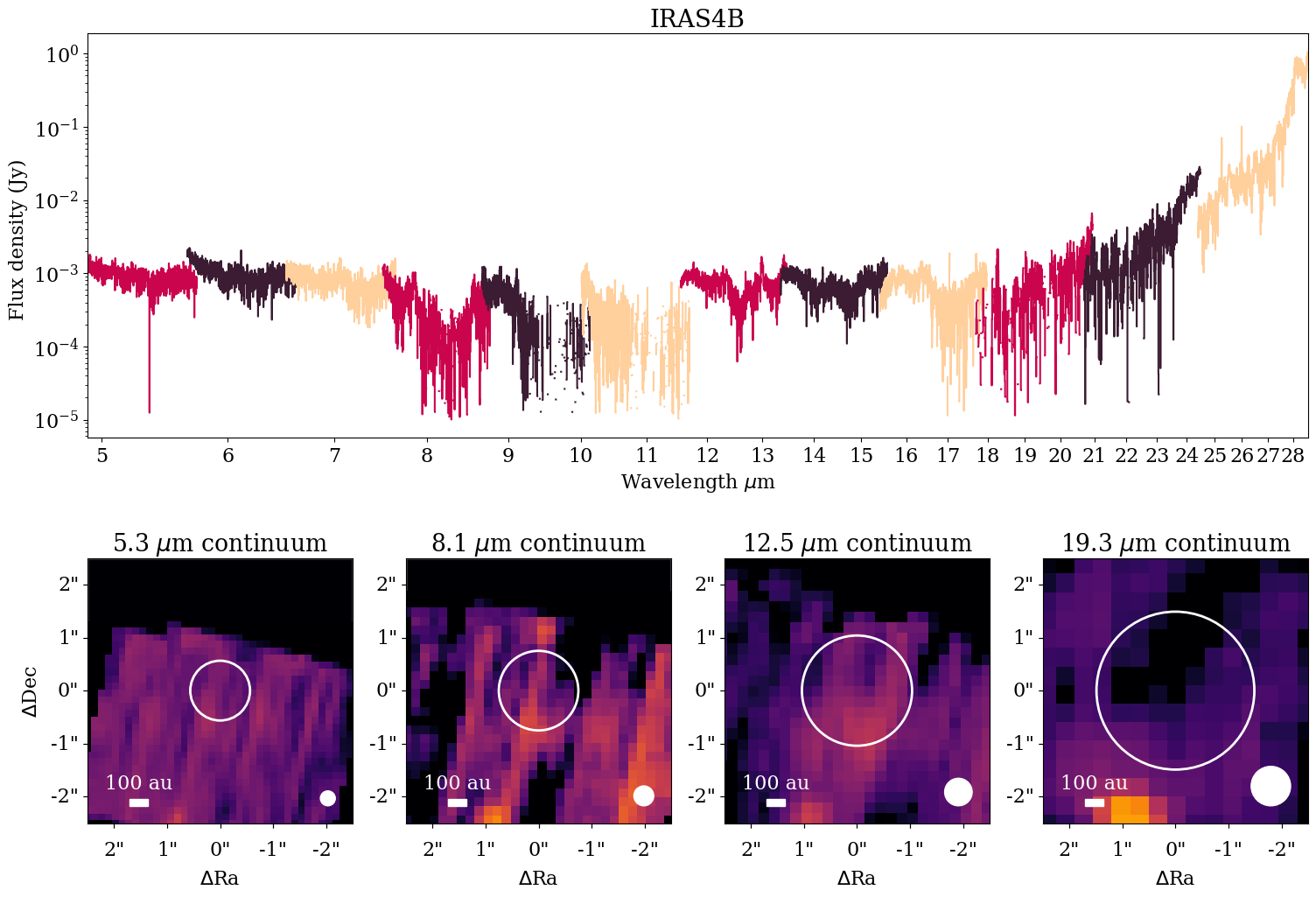}
    \caption{Same as Fig.~\ref{fig:specext_B1-a-NS} but for IRAS~4B.}
    \label{fig:specext_IRAS4B}
\end{figure*}

\begin{figure*}[h]
    \centering
    \includegraphics[width=0.92\linewidth]{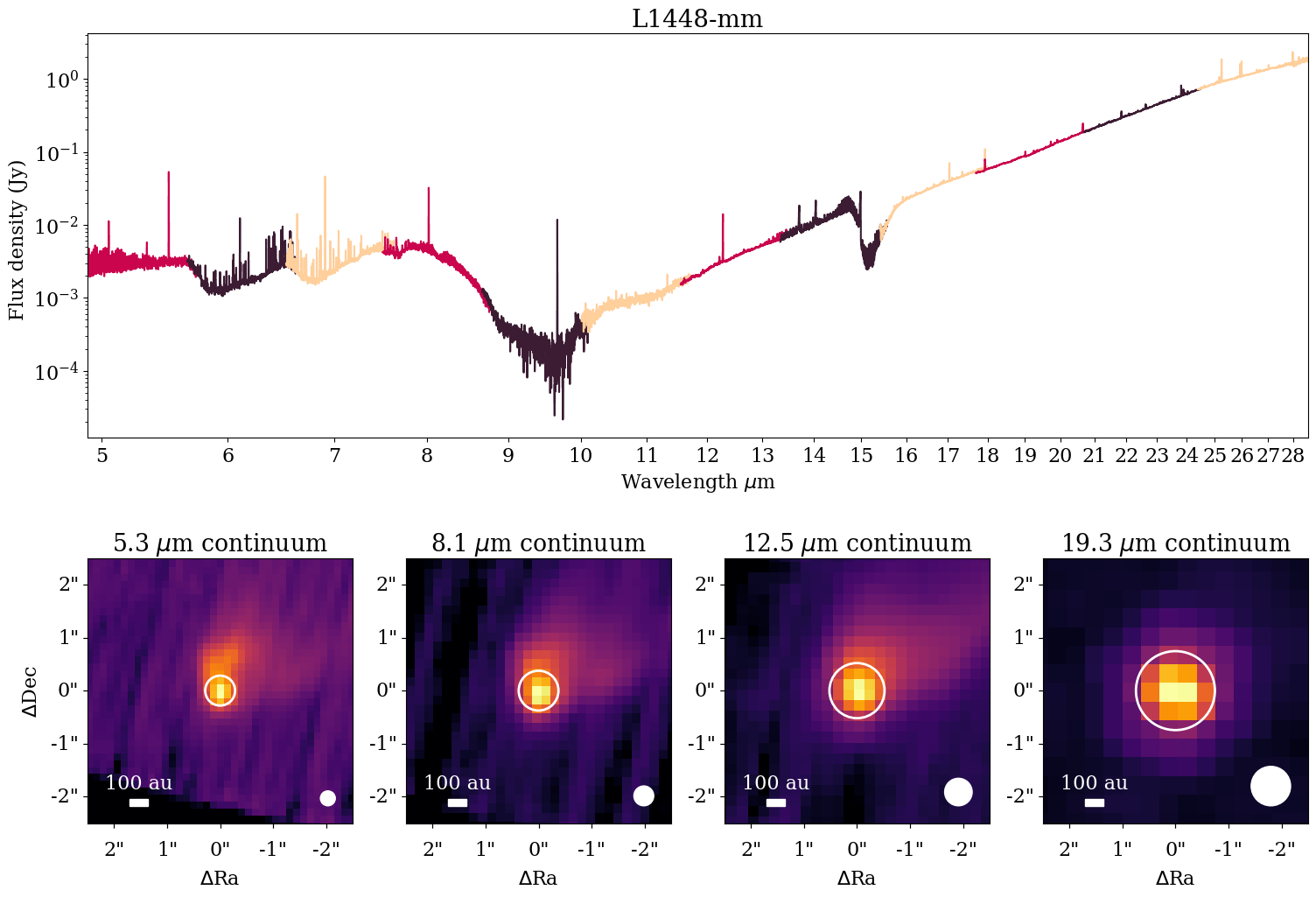}
    \caption{Same as Fig.~\ref{fig:specext_B1-a-NS} but for L1448-mm.}
    \label{fig:specext_L1448-mm}
\end{figure*}

\begin{figure*}[h]
    \centering
    \includegraphics[width=0.92\linewidth]{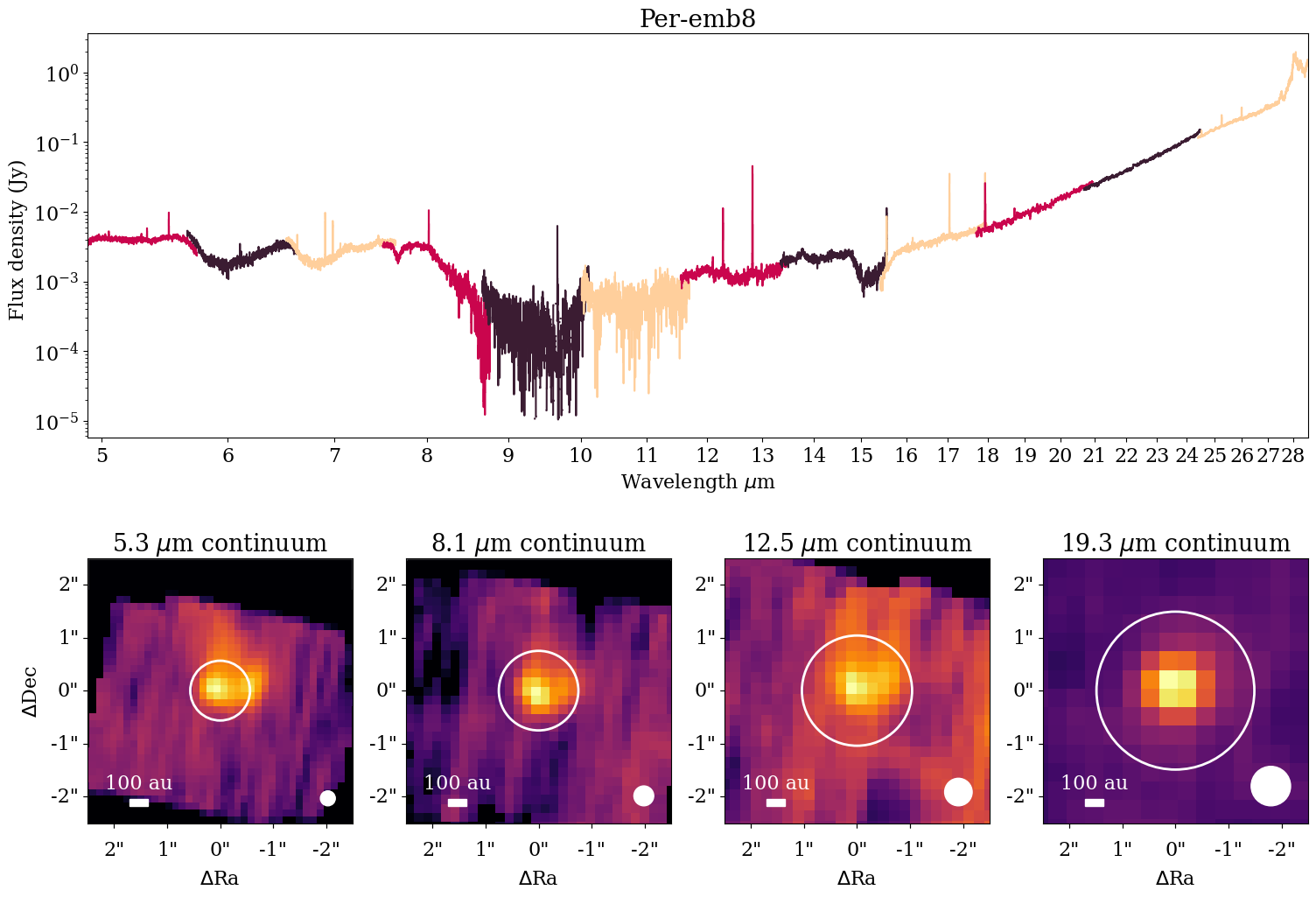}
    \caption{Same as Fig.~\ref{fig:specext_B1-a-NS} but for Per-emb~8.}
    \label{fig:specext_Per-emb8}
\end{figure*}

\begin{figure*}[h]
    \centering
    \includegraphics[width=0.92\linewidth]{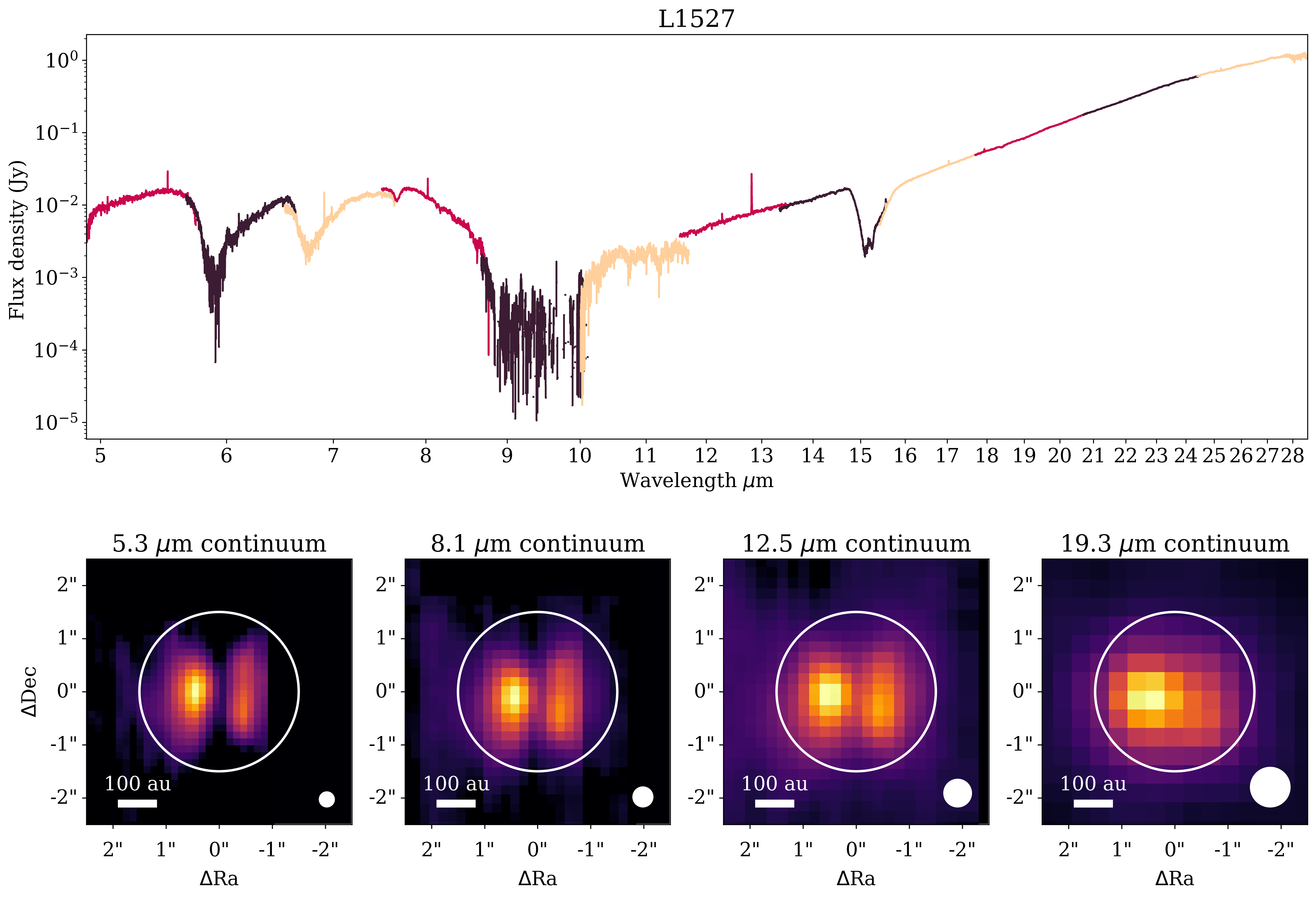}
    \caption{Same as Fig.~\ref{fig:specext_B1-a-NS} but for L1527.}
    \label{fig:specext_L1527}
\end{figure*}

\begin{figure*}[h]
    \centering
    \includegraphics[width=0.92\linewidth]{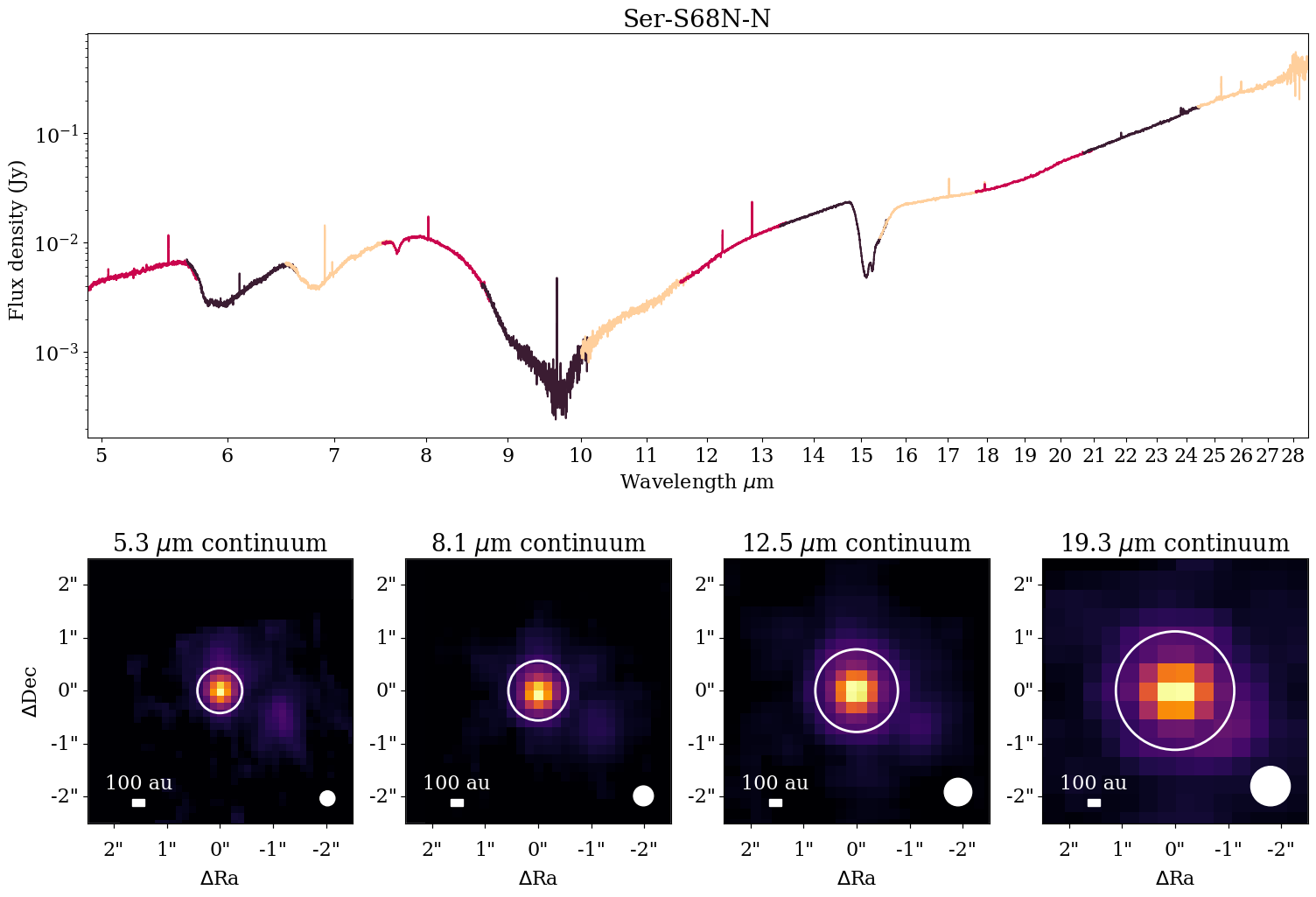}
    \caption{Same as Fig.~\ref{fig:specext_B1-a-NS} but for Ser-S68N-N.}
    \label{fig:specext_Ser-S68N-N}
\end{figure*}

\begin{figure*}[h]
    \centering
    \includegraphics[width=0.92\linewidth]{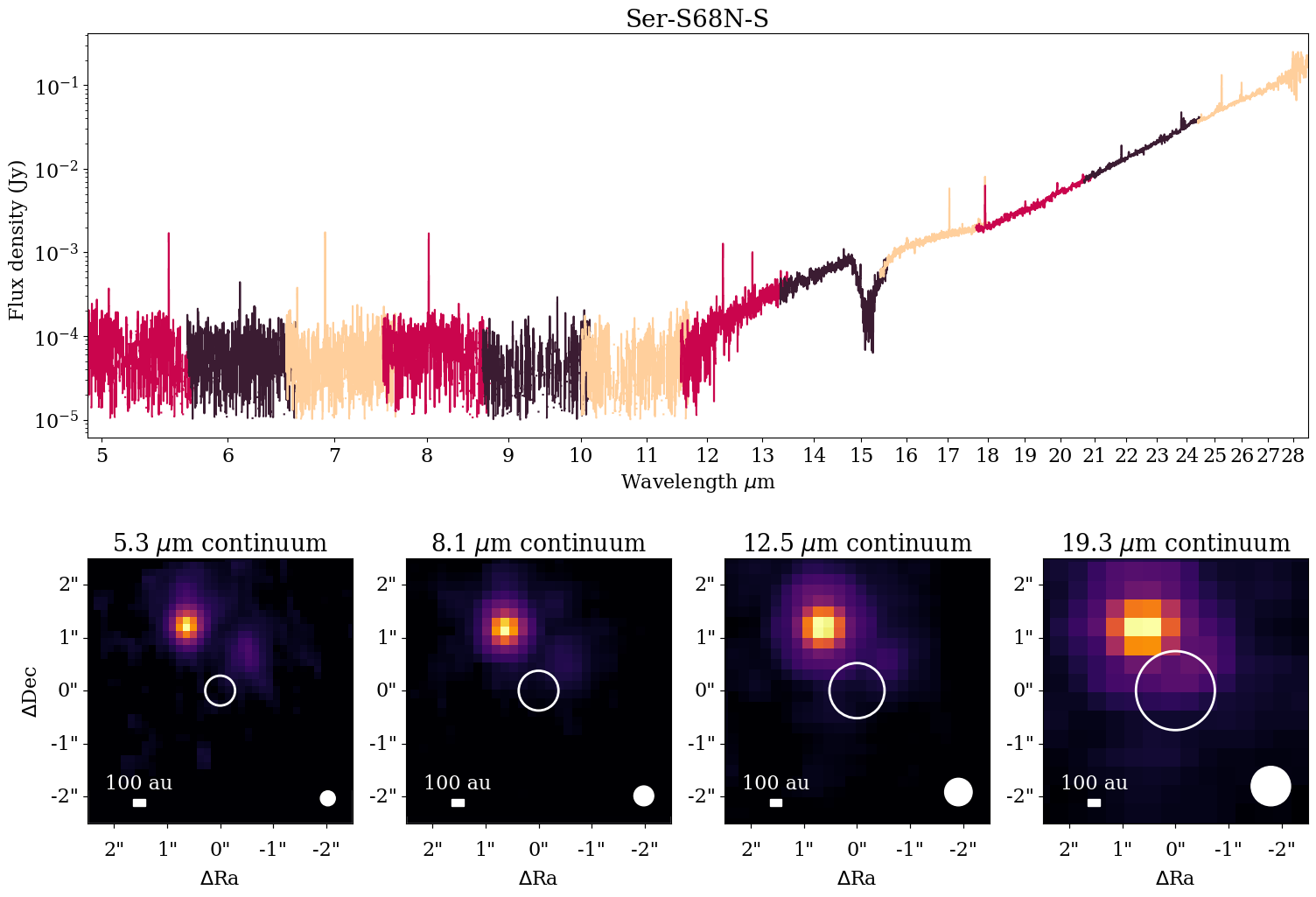}
    \caption{Same as Fig.~\ref{fig:specext_B1-a-NS} but for Ser-S68N-S.}
    \label{fig:specext_Ser-S68N-S}
\end{figure*}

\begin{figure*}[h]
    \centering
    \includegraphics[width=0.92\linewidth]{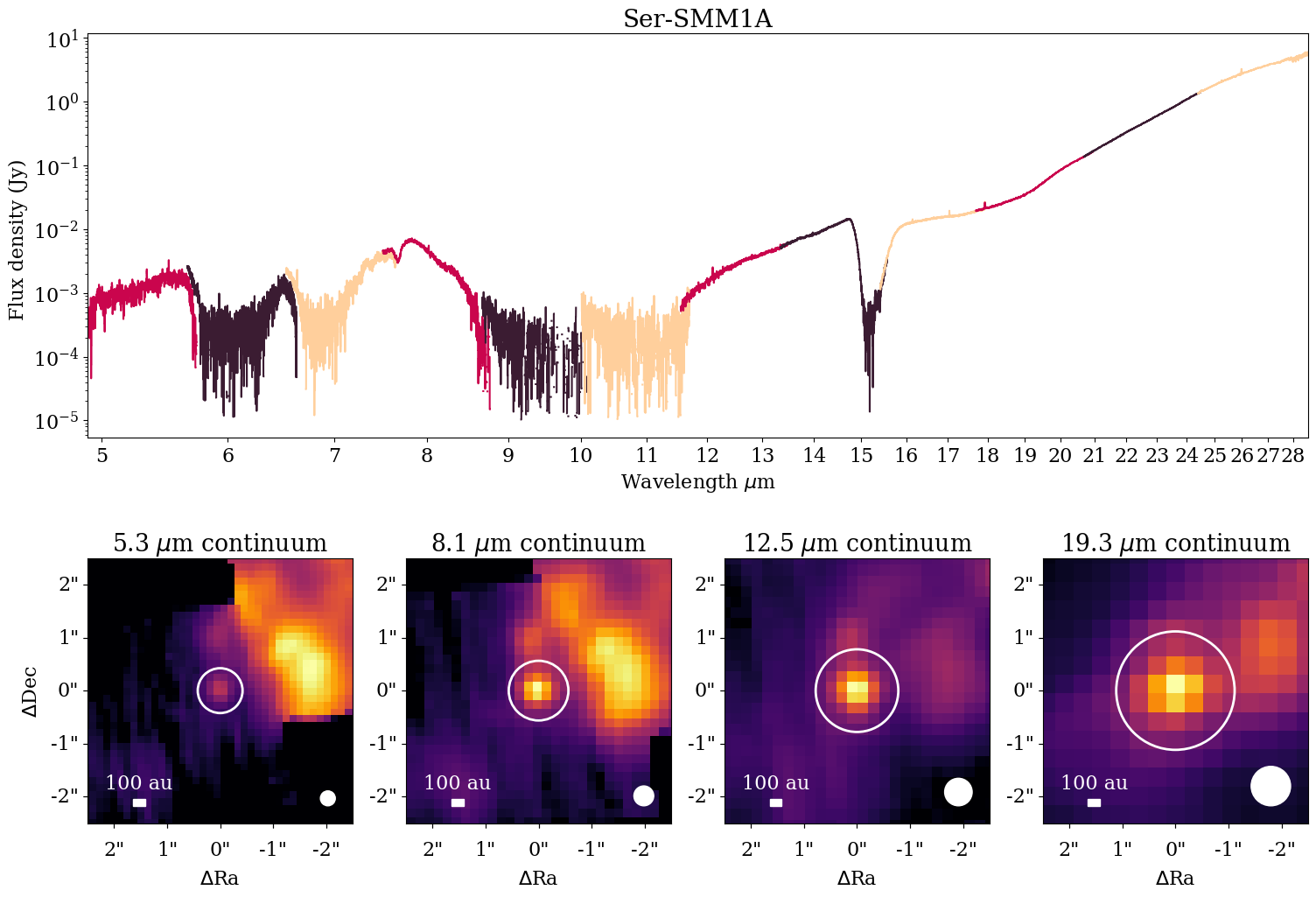}
    \caption{Same as Fig.~\ref{fig:specext_B1-a-NS} but for Ser-SMM1A.}
    \label{fig:specext_Ser-SMM1A}
\end{figure*}

\begin{figure*}[h]
    \centering
    \includegraphics[width=0.92\linewidth]{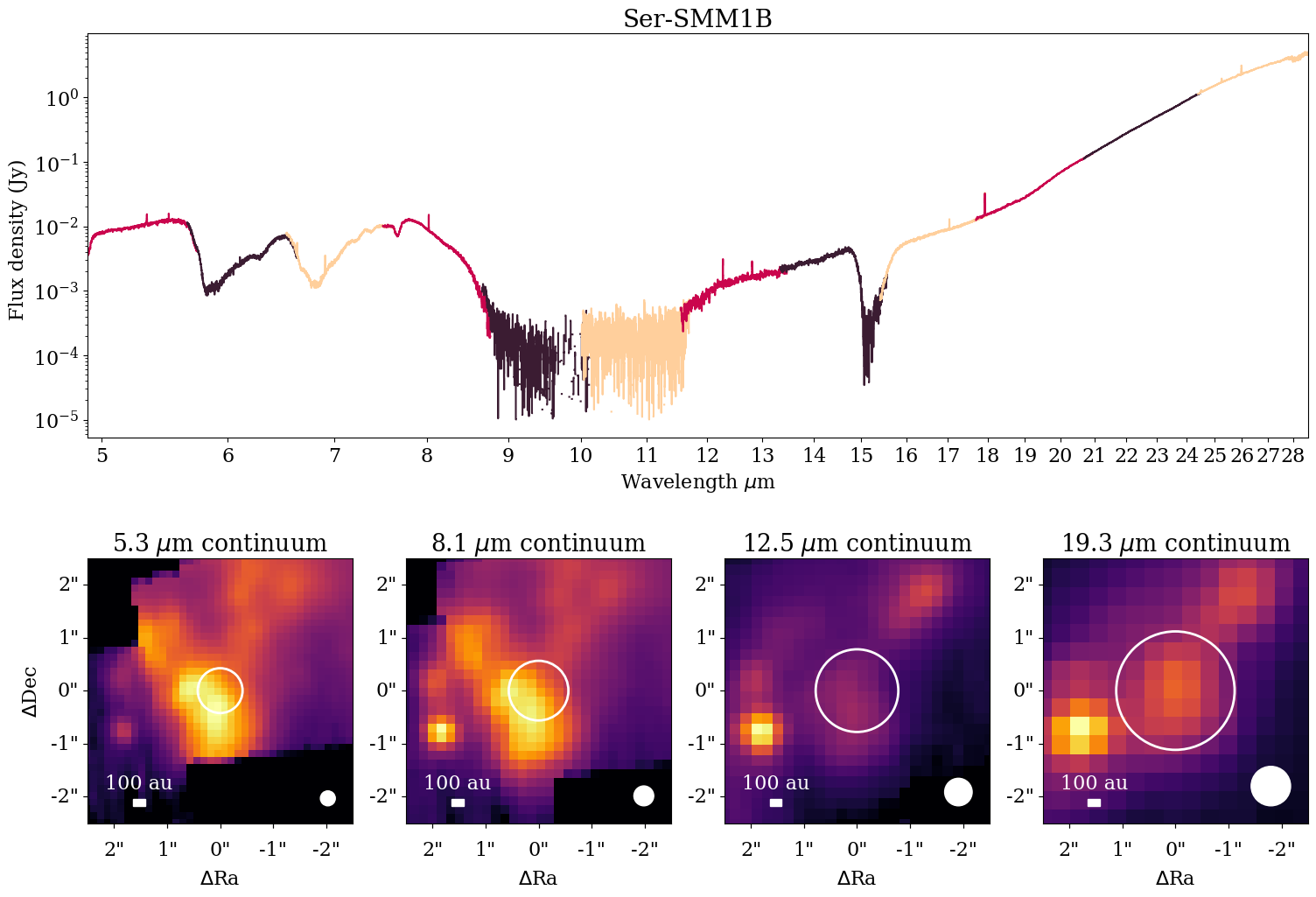}
    \caption{Same as Fig.~\ref{fig:specext_B1-a-NS} but for Ser-SMM1B.}
    \label{fig:specext_Ser-SMM1B}
\end{figure*}

\begin{figure*}[h]
    \centering
    \includegraphics[width=0.92\linewidth]{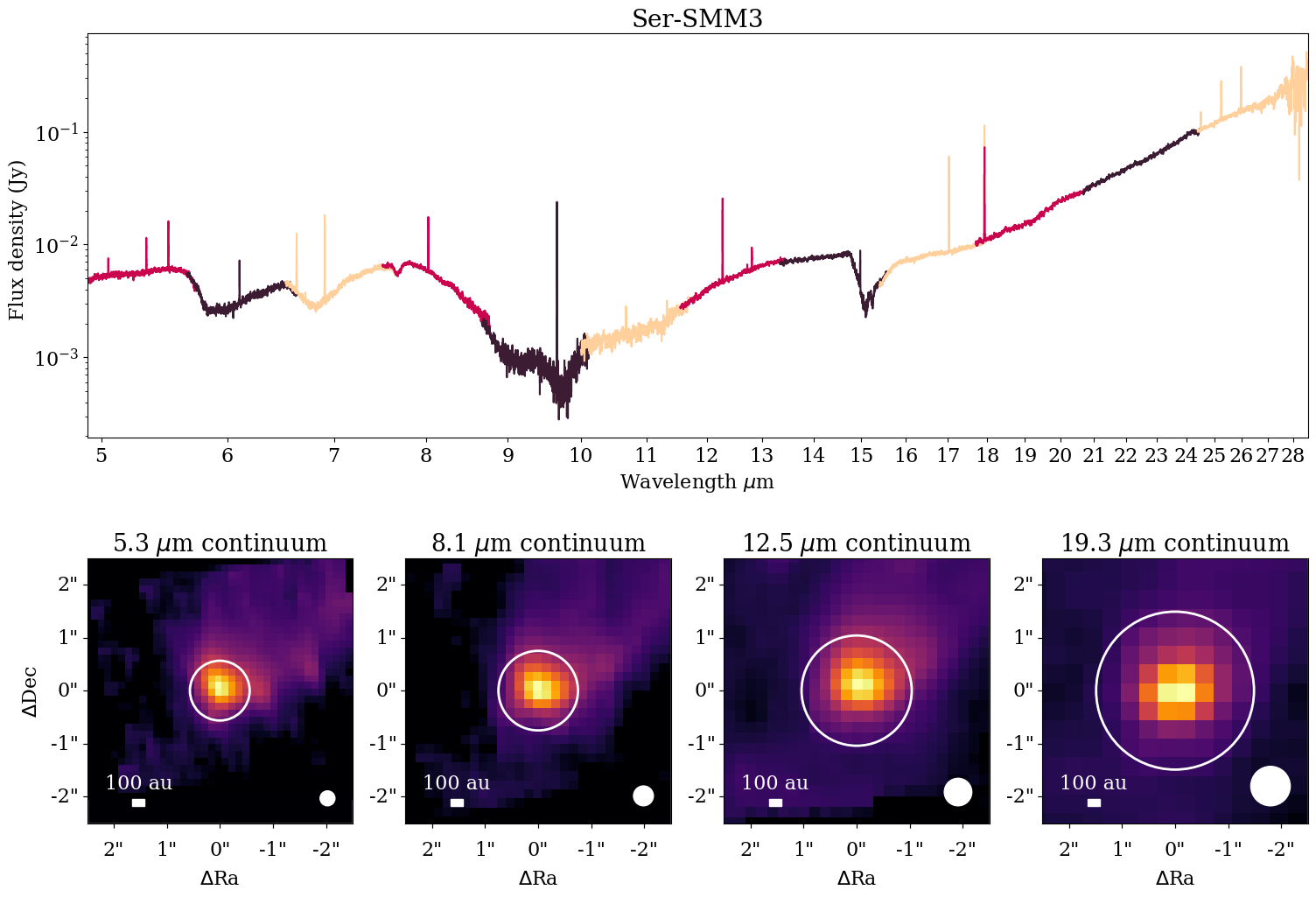}
    \caption{Same as Fig.~\ref{fig:specext_B1-a-NS} but for Ser-SMM3.}
    \label{fig:specext_Ser-SMM3}
\end{figure*}

\begin{figure*}[h]
    \centering
    \includegraphics[width=0.91\linewidth]{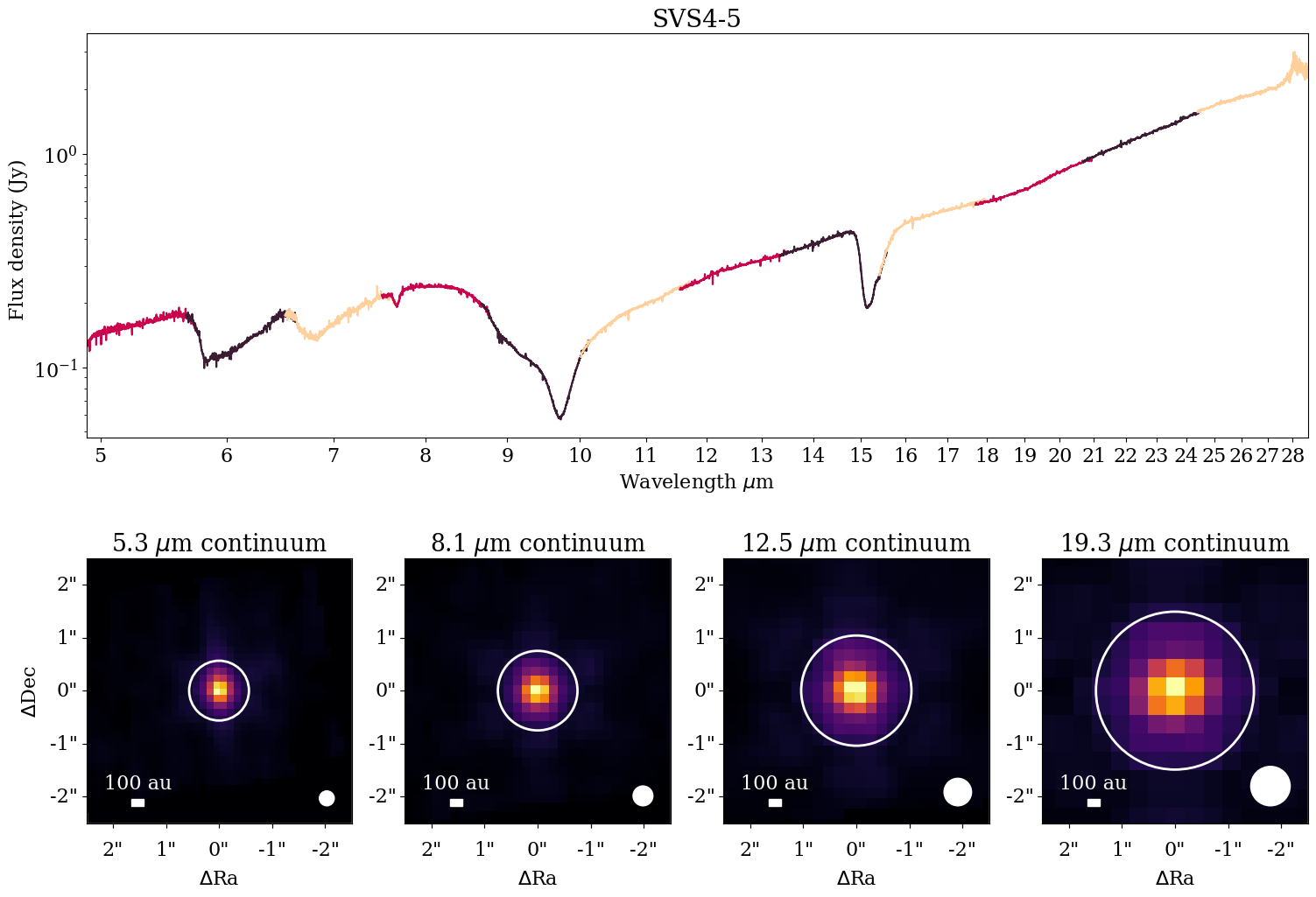}
    \caption{Same as Fig.~\ref{fig:specext_B1-a-NS} but for SVS4-5.}
    \label{fig:specext_SVS4-5}
\end{figure*}

\begin{figure*}[h]
    \centering
    \includegraphics[width=0.91\linewidth]{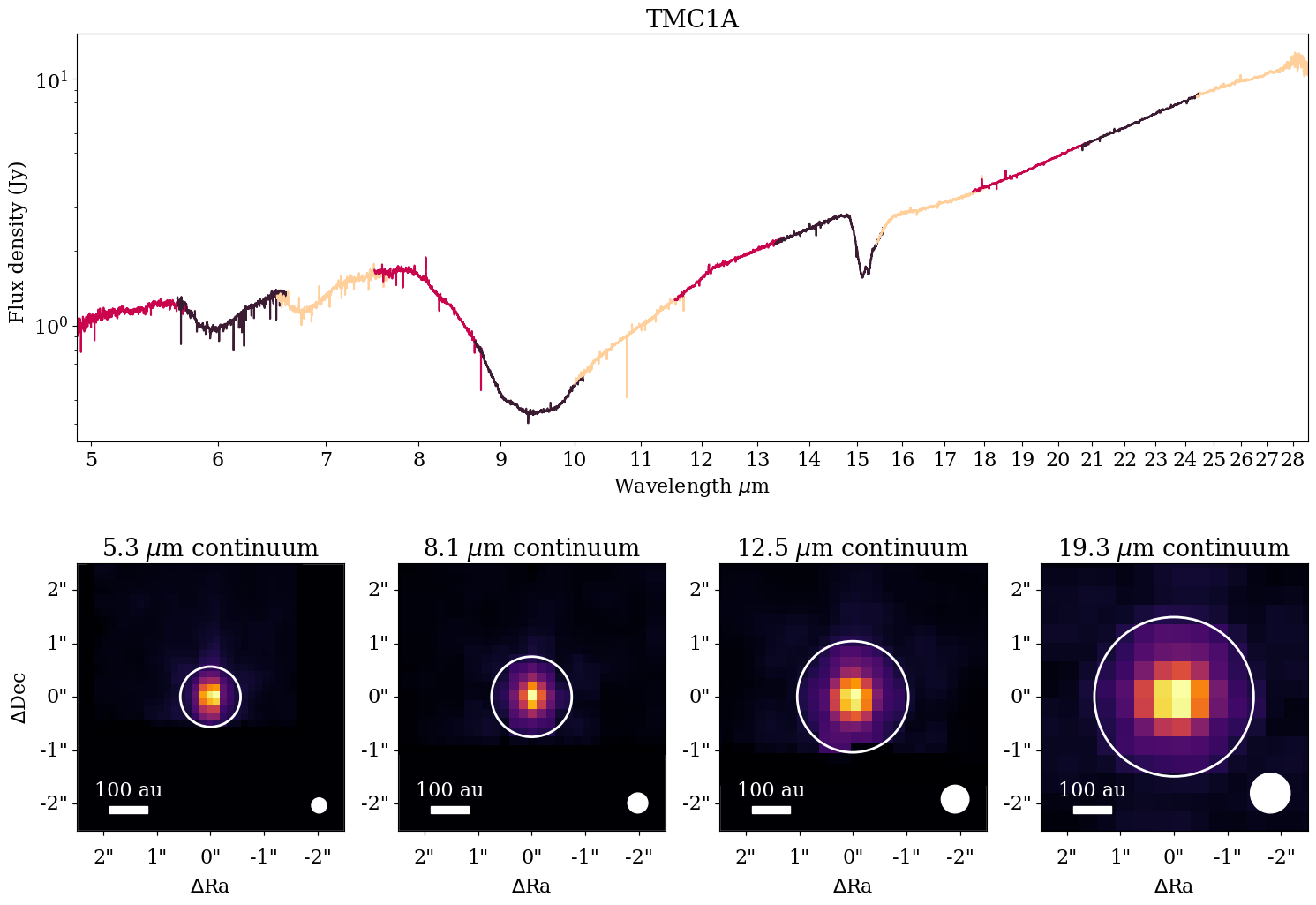}
    \caption{Same as Fig.~\ref{fig:specext_B1-a-NS} but for TMC1A.}
    \label{fig:specext_TMC1A}
\end{figure*}

\begin{figure*}[h]
    \centering
    \includegraphics[width=0.92\linewidth]{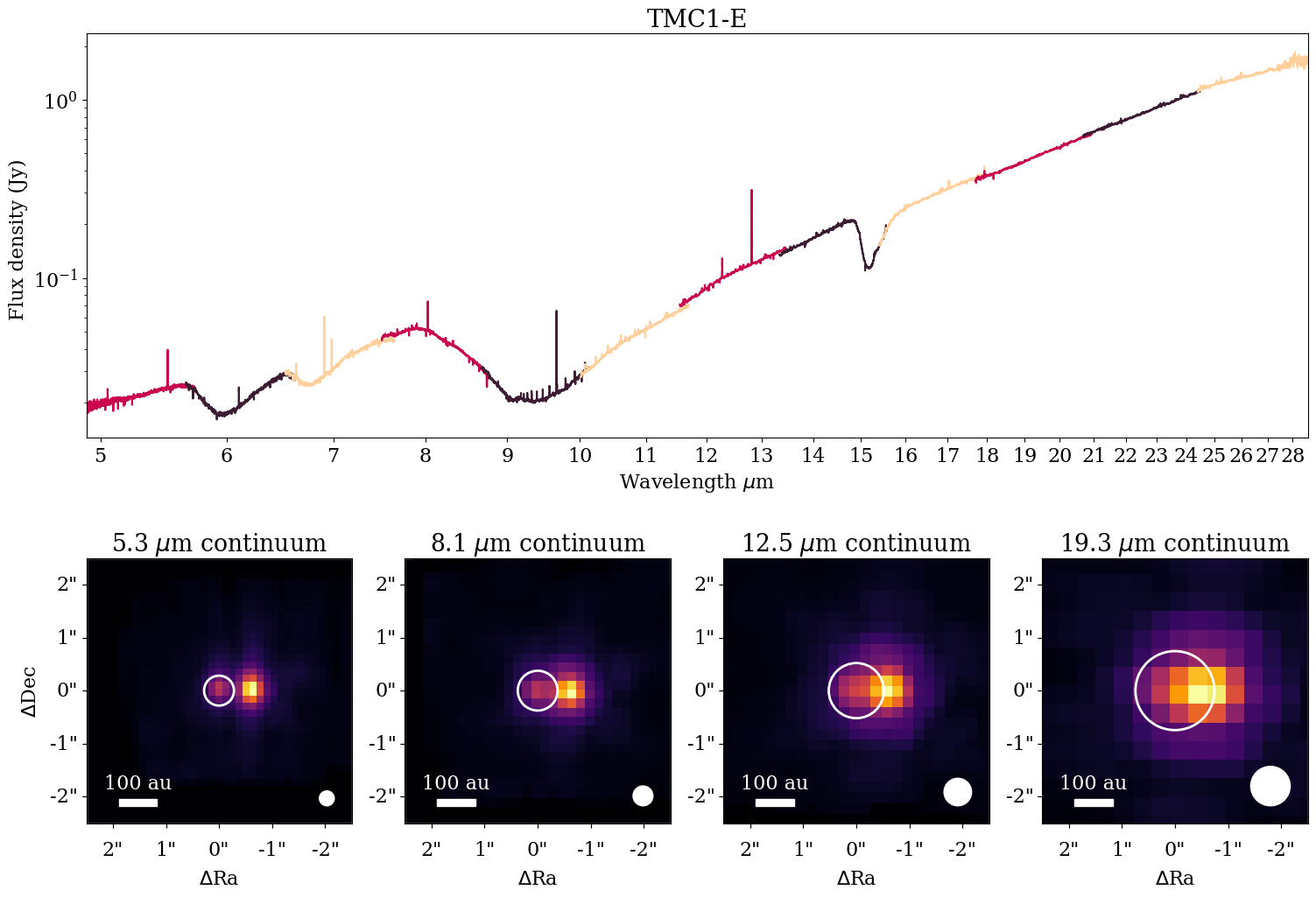}
    \caption{Same as Fig.~\ref{fig:specext_B1-a-NS} but for TMC1-E.}
    \label{fig:specext_TMC1-E}
\end{figure*}

\begin{figure*}[h]
    \centering
    \includegraphics[width=0.92\linewidth]{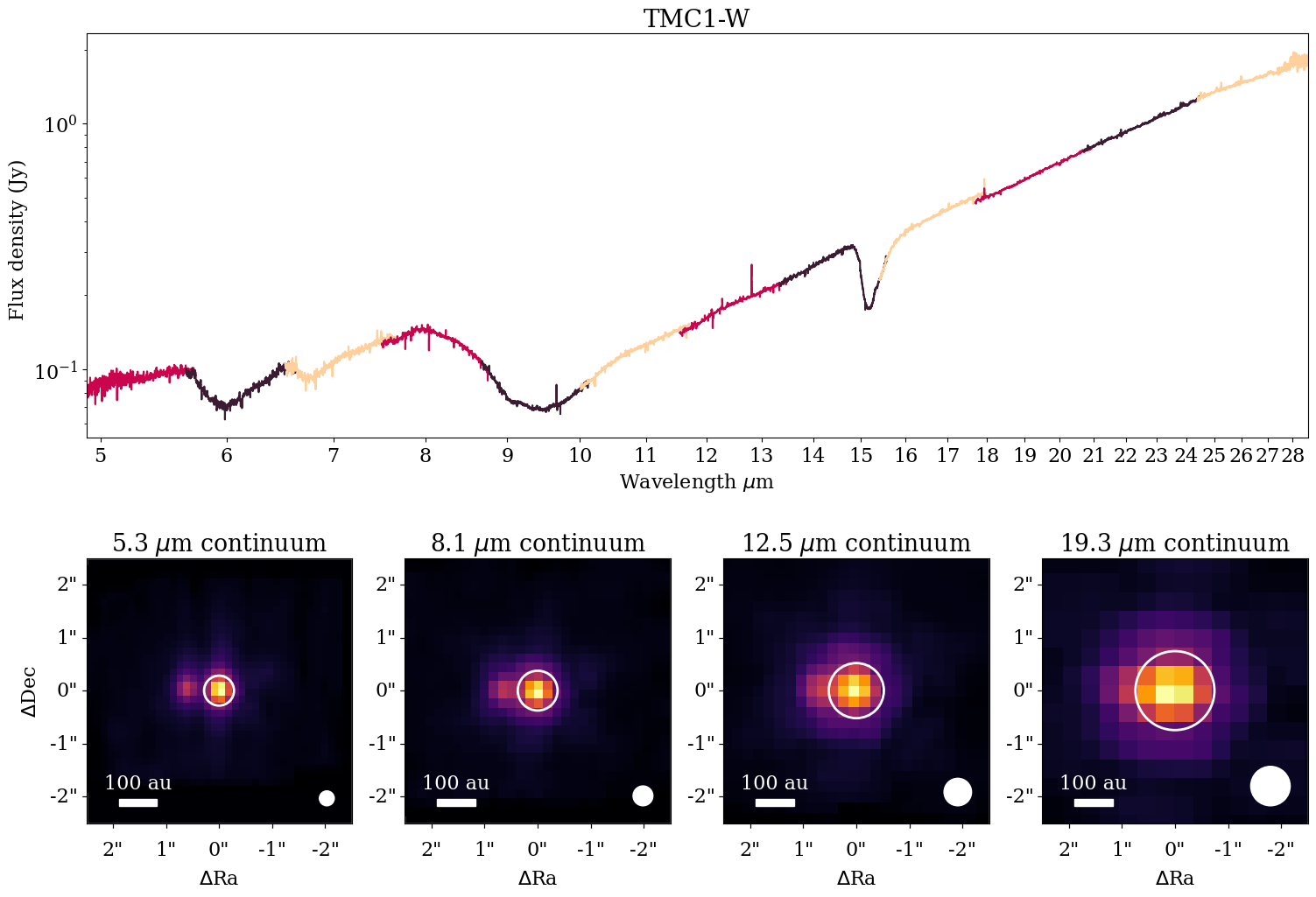}
    \caption{Same as Fig.~\ref{fig:specext_B1-a-NS} but for TMC1-W.}
    \label{fig:specext_TMC1-W}
\end{figure*}

\begin{figure*}[h]
    \centering
    \includegraphics[width=0.92\linewidth]{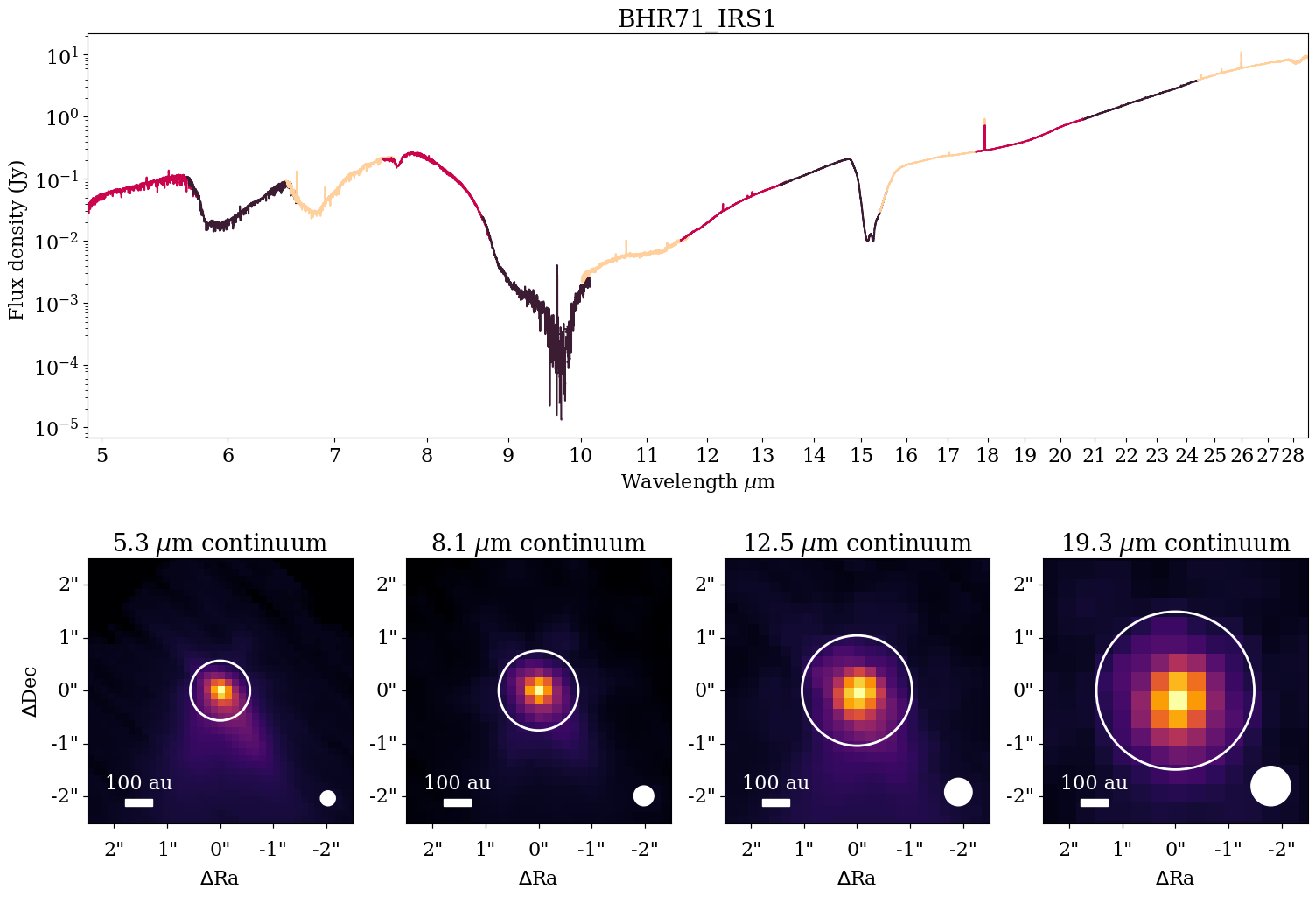}
    \caption{Same as Fig.~\ref{fig:specext_B1-a-NS} but for BHR71-IRS1.}
    \label{fig:specext_BHR71-IRS1}
\end{figure*}

\begin{figure*}[h]
    \centering
    \includegraphics[width=0.92\linewidth]{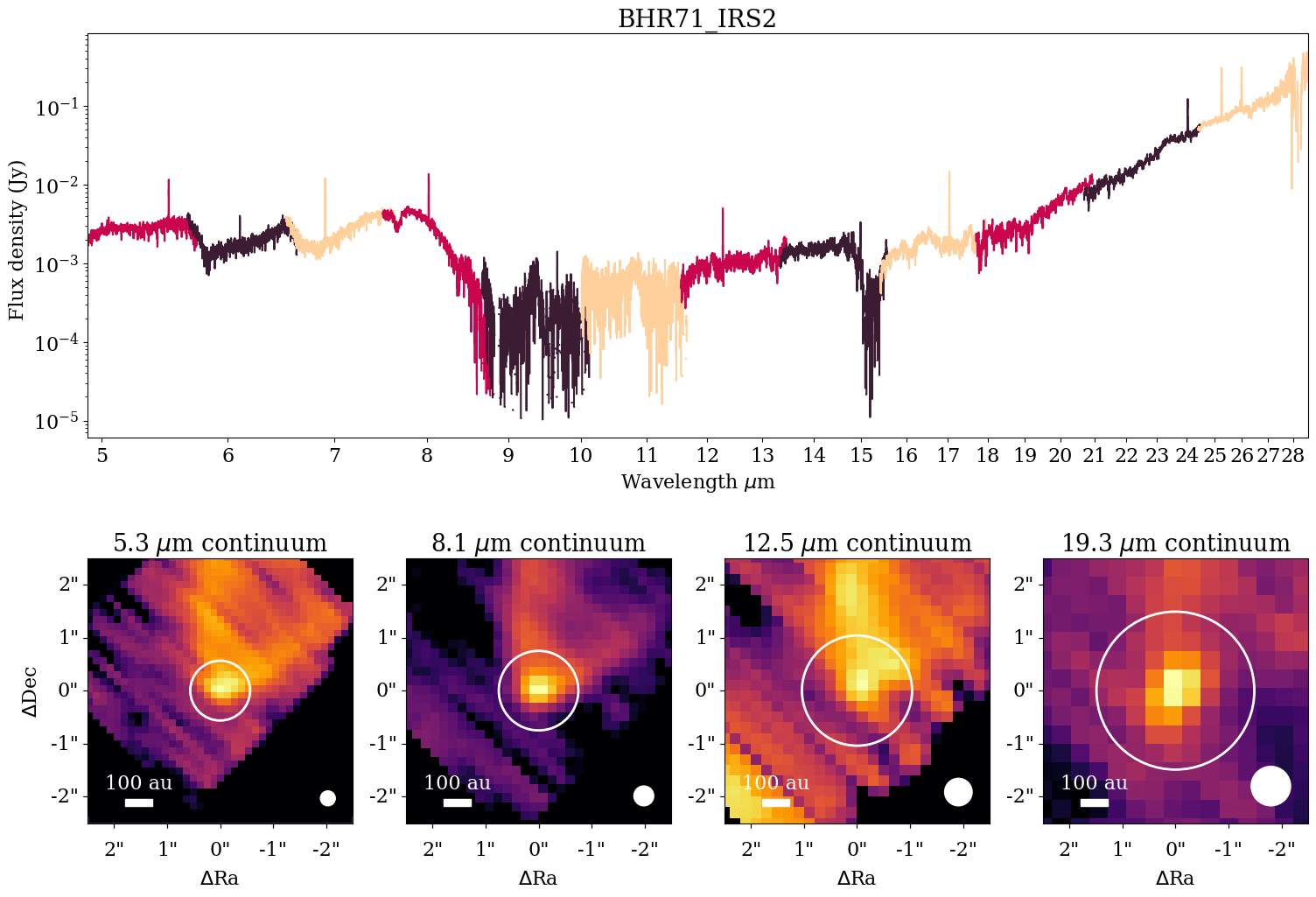}
    \caption{Same as Fig.~\ref{fig:specext_B1-a-NS} but for BHR71-IRS2.}
    \label{fig:specext_BHR71-IRS2}
\end{figure*}

The list of coordinates from which apertures were extracted is also listed in Table~\ref{tab:sample}. Two sources, NGC~1333~IRAS~4B and Ser-S68N-S, do not show any MIR continuum and therefore spectra were extracted from the millimeter continuum positions \citep[][]{vanGelder2020,Tychoniec2021}. The full extracted spectra and continuum images at 5.3~\mum (channel 1 short), 8.1~\mum (channel 2 short), 12.5~\mum (channel 3 short), and 19.3~\mum (channel 4 short) are shown in Figs.~\ref{fig:specext_B1-a-NS}-\ref{fig:specext_BHR71-IRS2} for all sources.
\end{multicols}

\renewcommand{\arraystretch}{1.2}
\begin{table*}[h]
    \centering
    \caption{Sample of low-mass JOYS sources and their main properties in the PID 1290 JWST GTO program. }
    \begin{tabular}{lccccccccc}
    \hline\hline
    Source & RA & Dec & $d$ & $L_{\rm bol}$ & $T_{\rm bol}$ & Class & Dithers & Exposure time & Refs\\
     & J2000 & J2000 & pc & L$_\odot$ & K & & & s &  \\ 
    \hline
    B1-a-NS & 03:33:16.69 & +31:07:55.10 & 293 & 2.3 & 115 & I & 2 & 600 & 1,2 \\
    B1-b & 03:33:20.35 & +31:07:21.35 & 293 & 0.23 & 157 & I & 2 & 600 & 1,3  \\
    B1-c & 03:33:17.90 & +31:09:31.86 & 293 & 5.0 & 48 & 0 & 4 & 8000 & 1,2 \\
    Per-emb 8 & 03:44:43.99 & +32:01:35.56 & 321 & 4.5 & 45 & 0 & 2 & 600 & 1,2  \\
    L1448-mm & 03:25:38.88 & +30:44:05.64 & 293 & 8.5 & 49 & 0 & 2 & 600 & 1,2 \\
    IRAS 4A\tablefootmark{(1)} & -- & -- &  293 & 14.1 & 34 & 0 & 2 & 600 & 1,2 \\
    IRAS 4B & 03:29:12.02\tablefootmark{(2)} & +31:13:08.03\tablefootmark{(2)} & 293 & 6.8 & 28 & 0 & 2 & 600 & 1,2 \\
    \hline
    Ser-emb 8(N)\tablefootmark{(1)} & -- & -- & 436 & 1.8 & -- & 0 & 2 & 600 & 3,4 \\
    Ser-S68N-N & 18:29:48.13 & +01:16:44.51 & 436 & 6.90 & 58 & 0 & 2 & 600 & 3,5  \\
    Ser-S68N-S & 18:29:48.09\tablefootmark{(2)} & +01:16:43.30\tablefootmark{(2)} & 436 & 6.90 & 58 & 0 & 2 & 600 & 3,5 \\
    Ser-SMM1-A & 18:29:49.79 & +01:15:20.37 & 436 & 109 & 39 & 0 & 4 & 600 & 3,5 \\
    Ser-SMM1-B & 18:29:49.66 & +01:15:21.16 & 436 & 109 & 39 & 0 & 2 & 600 & 3,5  \\
    Ser-SMM3 & 18:29:59.31 & +01:14:00.65 & 436 & 28 & 38 & 0 & 2 & 600 & 3,5 \\
    SVS 4-5\tablefootmark{(3)} & 18:29:57.61 & +01:13:00.12 & 436 & 38 & -- & I/II & 2 & 600 & 5,6 \\
    \hline
    TMC1-E & 04:41:12.73 & +25:46:34.64 & 142 & 0.7 & 161 & I & 2 & 600 & 2,7 \\
    TMC1-W & 04:41:12.69 & +25:46:34.63 & 142 & 0.7 & 161 & I & 2 & 600 & 2,7 \\
    TMC1A & 04:39:35.21 & +25:41:43.76 & 142 & 2.7 & 189 & I & 2\tablefootmark{(4)} & 600 & 2,7 \\
    L1527 & 04:39:53.84 & +26:03:10.18 & 142 & 1.6 & 60 & 0/I & 2 & 3000 & 2,7 \\
    \hline
    BHR71-IRS1 & 12:01:36.50 & -65:08:49.40 & 200 & 14.6 & 68 & 0 & 2 & 600 & 8,9 \\
    BHR71-IRS2 & 12:01:33.96 & -65:08:47.78 & 200 & 1.7 & 38 & 0 & 2 & 600 & 8,9 \\
    \hline
    \end{tabular}
    \label{tab:sample}
    \tablefoot{The coordinates are those of the mid-IR continuum peak. The exposure time is the total exposure time divided over the A,B,C gratings for the pointing from which the spectrum was extracted. \tablefoottext{1}{The pointing of IRAS~4A and Ser-emb8(N) was not centered on the protostar itself but offset on blueshifted outflow.} 
    \tablefoottext{2}{Coordinates for the millimeter continuum peak since mid-IR continuum is not detected.}
    \tablefoottext{3}{SVS4-5 is a Class I/II sources that is behind or inside the envelope and outflow of the Class~0 source Ser-SMM4.} 
    \tablefoottext{4}{Only one dither is used for channels 1 and 2 due to source lying at the edge of the FOV.}
    \\
    {\bf References.} 1: \citet{Ortiz-Leon2018}; 2: \citet{Karska2018}; 3: \citet{Enoch2009}. 4: \citet{Podio2021}. 5: \citet{Ortiz-Leon2017}. 6: \citet{Pontoppidan2004}. 7: \citet{Krolikowski2021}. 8: \citet{Seidensticker1989}. 9: \citet{Tobin2019}.}
\end{table*}
\renewcommand{\arraystretch}{1.0}


\onecolumn
\begin{multicols}{2}
% \nolinenumbers
\section{Spectral information tables}
In Table~\ref{tab:sigma_data}, the noise estimates in units of mJy are presented. The noise {level} is estimated from line-free {wavelength regions} for each sub-band separately. For most sources, the noise level is about $\sim0.1-0.5$~mJy in channels~$1-3$, whereas in channel~4 the noise levels increase toward several~mJy in channels~4A and 4B and up to $\sim10$~mJy in channel~4C. One major outlier is TMC1A where the noise levels are significantly higher. At the shorter wavelengths, this likely originates from the source lying on the very edge of the FOV, which hampers the sensitivity. At longer wavelengths, the noise levels are likely dominated by residual fringes at the 1\% level of the continuum, which is several Jy in channels 3 and 4.
\end{multicols}

\renewcommand{\arraystretch}{1.2}
\begin{table}[h]
    \centering
    \caption{Noise level estimates ($\sigma$) in mJy derived from line-free {wavelength regions} per MIRI-MRS sub-band for all sources.}
    \begin{tabular}{lcccccccccccc}
\hline \hline
Source & $\sigma_{\rm 1A}$ & $\sigma_{\rm 1B}$ & $\sigma_{\rm 1C}$ & $\sigma_{\rm 2A}$ & $\sigma_{\rm 2B}$ & $\sigma_{\rm 2C}$ & $\sigma_{\rm 3A}$ & $\sigma_{\rm 3B}$ & $\sigma_{\rm 3C}$ & $\sigma_{\rm 4A}$ & $\sigma_{\rm 4B}$ & $\sigma_{\rm 4C}$ \\
 & mJy & mJy & mJy & mJy & mJy & mJy & mJy & mJy & mJy & mJy & mJy & mJy \\
\hline
B1-a-NS & 0.36 & 0.40 & 0.40 & 0.49 & 0.37 & 0.60 & 0.52 & 0.73 & 0.69 & 2.49 & 4.33 & 11.85 \\
B1-b & 0.16 & 0.20 & 0.15 & 0.21 & 0.19 & 0.26 & 0.16 & 0.21 & 0.29 & 0.59 & 2.48 & 2.76 \\
B1-c & 0.40 & 0.17 & 0.49 & 0.05 & 0.05 & 0.05 & 0.07 & 0.05 & 0.08 & 0.30 & 2.43 & 2.33 \\
L1448-mm & 0.07 & 0.06 & 0.07 & 0.06 & 0.06 & 0.06 & 0.04 & 0.05 & 0.14 & 0.32 & 1.64 & 2.68 \\
Per-emb8 & 0.17 & 0.18 & 0.14 & 0.17 & 0.13 & 0.16 & 0.10 & 0.13 & 0.15 & 0.42 & 0.95 & 2.20 \\
Ser-S68N-N & 0.11 & 0.11 & 0.08 & 0.12 & 0.07 & 0.11 & 0.08 & 0.11 & 0.15 & 0.39 & 0.79 & 1.76 \\
Ser-SMM1A & 0.14 & 0.12 & 0.13 & 0.12 & 0.12 & 0.12 & 0.10 & 0.10 & 0.12 & 0.29 & 0.85 & 4.74 \\
Ser-SMM1B & 0.09 & 0.06 & 0.02 & 0.07 & 0.06 & 0.10 & 0.08 & 0.07 & 0.11 & 0.25 & 1.11 & 2.88 \\
Ser-SMM3 & 0.14 & 0.12 & 0.06 & 0.12 & 0.10 & 0.11 & 0.12 & 0.11 & 0.13 & 0.33 & 1.12 & 2.57 \\
SVS4-5 & 0.32 & 0.14 & 0.28 & 0.27 & 0.25 & 0.31 & 0.63 & 0.78 & 0.61 & 1.82 & 2.61 & 4.10 \\
L1527 & 0.35 & 0.22 & 0.18 & 0.22 & 0.24 & 0.24 & 0.14 & 0.13 & 0.14 & 0.46 & 0.95 & 1.52 \\
TMC1A & 14.5 & 3.68 & 7.53 & 6.58 & 3.71 & 3.68 & 8.03 & 9.10 & 7.16 & 8.66 & 16.9 & 30.0 \\
TMC1-E & 0.15 & 0.13 & 0.13 & 0.22 & 0.15 & 0.16 & 0.26 & 0.38 & 0.53 & 1.61 & 2.82 & 6.05 \\
TMC1-W & 0.31 & 0.36 & 0.33 & 0.26 & 0.49 & 0.34 & 0.34 & 0.35 & 0.71 & 1.66 & 2.02 & 5.46 \\
BHR71-IRS1 & 0.69 & 0.28 & 0.42 & 0.16 & 0.16 & 0.15 & 0.10 & 0.21 & 0.29 & 1.49 & 3.43 & 9.69 \\
BHR71-IRS2 & 0.12 & 0.18 & 0.13 & 0.13 & 0.16 & 0.20 & 0.15 & 0.13 & 0.15 & 0.55 & 1.12 & 2.04 \\
\hline
    \end{tabular}
    \tablefoot{No noise estimates are provided for IRAS~4B and Ser-S68N-S since no continuum source is detected.}
    \label{tab:sigma_data}
\end{table}
\renewcommand{\arraystretch}{1.0}

\vspace{-0.6cm}
\begin{multicols}{2}
% \nolinenumbers
\section{Infrared pumping}
\label{app:IR_pumping}
The effect of infrared pumping is very dependent on both the rotational temperature $T_{\rm rot}$ and vibrational temperature $T_{\rm vib}$. This is evident through the correction factor applied in Eq.~\eqref{eq:IR_pump}, with both $T_{\rm rot}$ and $T_{\rm vib}$ present in an exponential. Moreover, also the wavelength (or frequency $\nu$) through which the infrared pumping is thought to occur is very important. 

\begin{figure*}[h]
    \centering
    \includegraphics[width=0.33\linewidth]{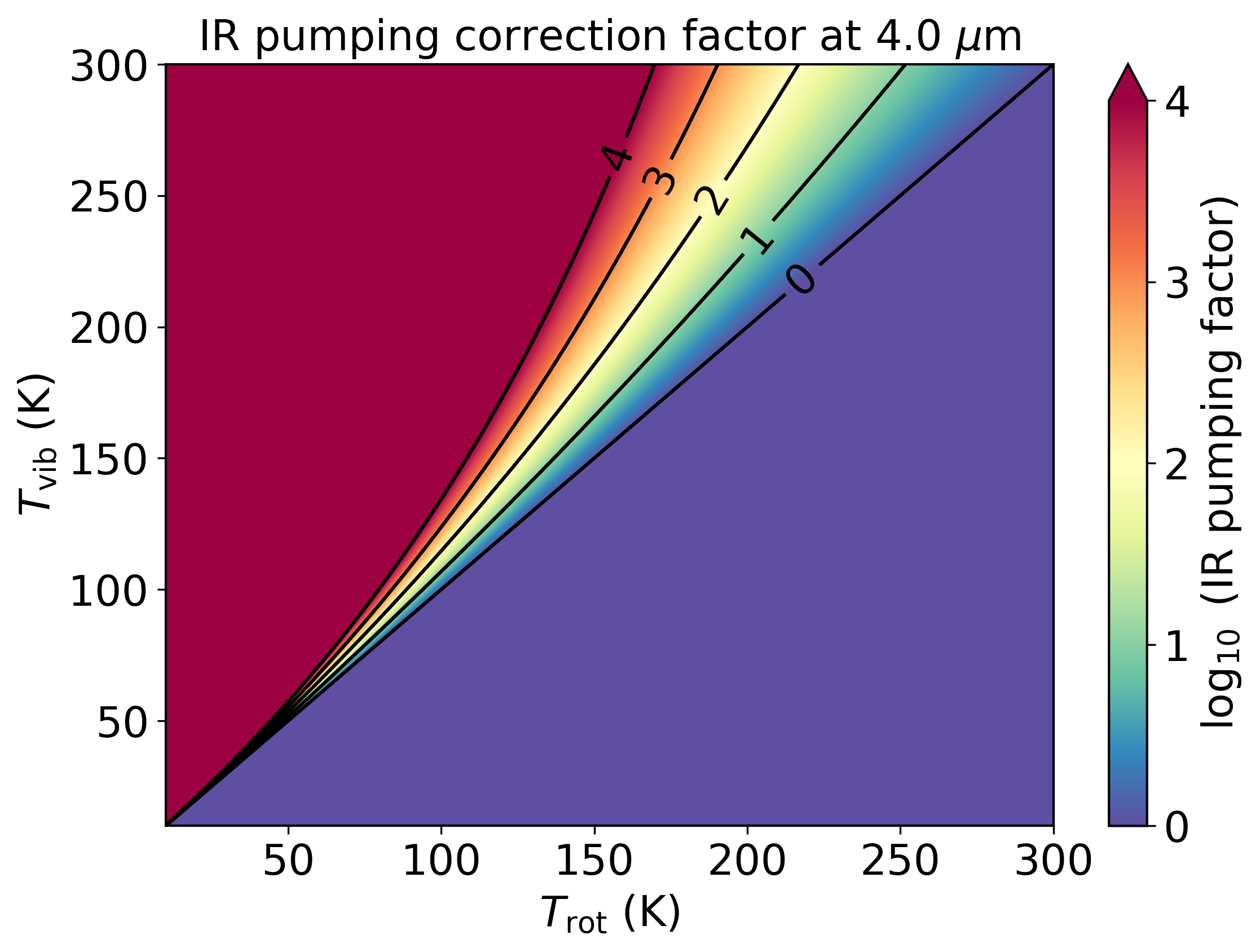}
    \includegraphics[width=0.33\linewidth]{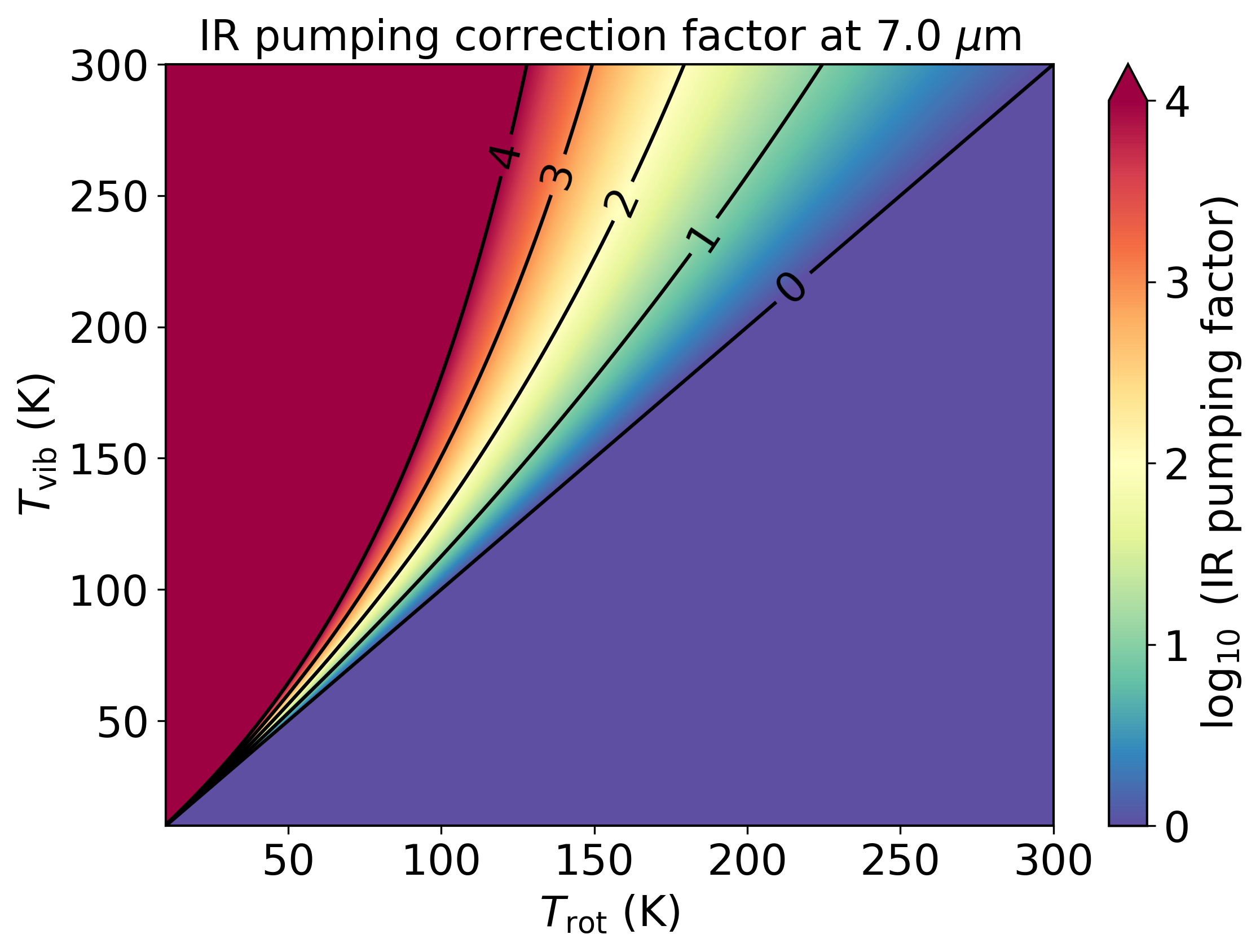}
    \includegraphics[width=0.33\linewidth]{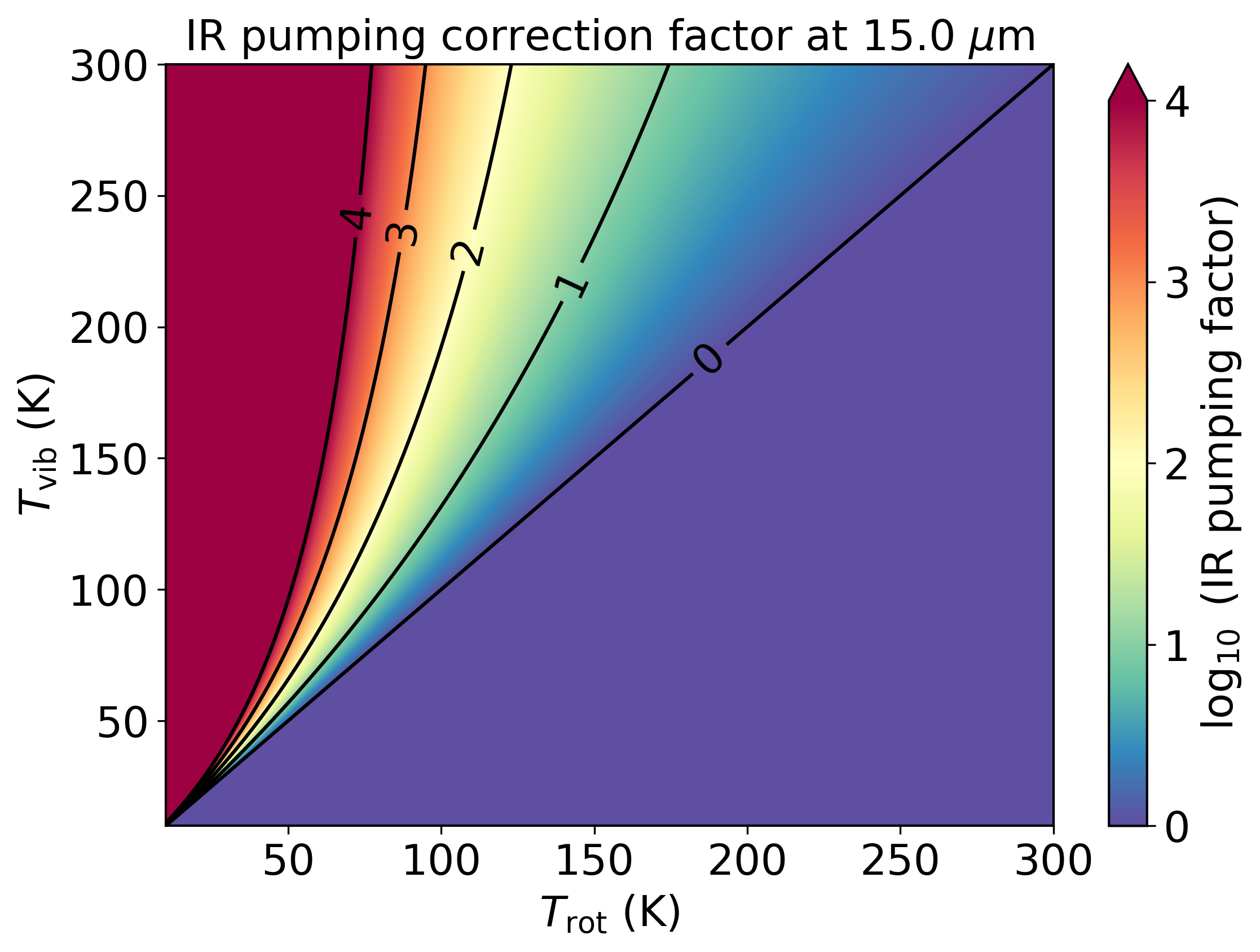}
    \caption{Infrared pumping correction factor of Eq.~\eqref{eq:IR_pump} as function of $T_{\rm rot}$ and $T_{\rm vib}$ at 4.0~\mum (left), 7.0~\mum (middle), and 15.0~\mum (right). The correction factor is presented on a logarithmic scale, with the black contours indicating orders of magnitude (up to 10$^4$). For the lower right of each panel where $T_{\rm vib} < T_{\rm rot}$, the correction factor is assumed to be 1 (i.e., no correction).}
    \label{fig:IR_pump}
\end{figure*}

In Fig.~\ref{fig:IR_pump}, the correction factor (on logarithmic scale) to the number of molecules is presented as function of both $T_{\rm rot}$ and $T_{\rm vib}$ for three different wavelengths. It is clear that even small differences in $T_{\rm vib}$, which is estimated from the infrared brightness temperature via Eq.~\eqref{eq:T_IR}, can lead to several orders of magnitude differences in the correction factor to the number of molecules. This effect is strongest at shorter wavelengths and gets less pronounced when going to longer wavelengths, yet still the correction can be more than an order of magnitude. This thus results in large uncertainties on the number of molecules for species exhibiting emission with $T_{\rm vib} > T_{\rm rot}$. When $T_{\rm vib} < T_{\rm rot}$, no correction is applied since infrared pumping is likely not dominating the excitation. 
\end{multicols}

\onecolumn
\begin{multicols}{2}
% \nolinenumbers
\section{Importance of fitting optically thin and unblended lines}
\label{app:line_optical_depth}
It is important to consider the optical depth of emission or absorption features in determining their physical parameters. The LTE slab models adopted in this paper do take line optical depth into account in determining the line strength \citep[e.g.,][see also Eq.~\eqref{eq:chi2_abs}]{Tabone2023}. Moreover, also the blending of multiple lines that remain unresolved at the spectral resolution of MIRI-MRS is taken into account. However, despite taking all these effects into account in LTE models, it remains important to fit weak optically thin transitions to derive the physical parameters such as the excitation temperature and column density.

Recently, \citet{Li2024} showed the importance of fitting only weak and unblended H$_2$O absorption features in JWST/MIRI-MRS data. Focusing mostly on absorption models of the rovibrational H$_2$O lines between 5~\mum and 8~\mum, they showed that the column density derived from these optically thick and often blended lines can lead to under- or overestimating the column density by two orders of magnitude. For some of our sources, the fits between rotational and rovibrational lines agree very well (i.e., B1-c), but for others the difference in column densities is as large as two orders of magnitude despite similar excitation temperatures (B1-a-NS). This can also originate from IR pumping (higher $N$ derived from rovibrational lines than from rotational lines) and or subthermal excitation (lower $N$ derived from rovibrational lines than from rotational lines) but blending of rovibrational lines cannot be excluded. However, in this paper the main H$_2$O column used for abundance ratios is derived from pure rotational lines of H$_2$O at longer wavelengths for which line blending is less of an issue than for the rovibrational lines. Line blending is therefore not affecting the total H$_2$O column densities derived in this work.

\begin{figure*}[h]
    \centering
    \includegraphics[width=\linewidth]{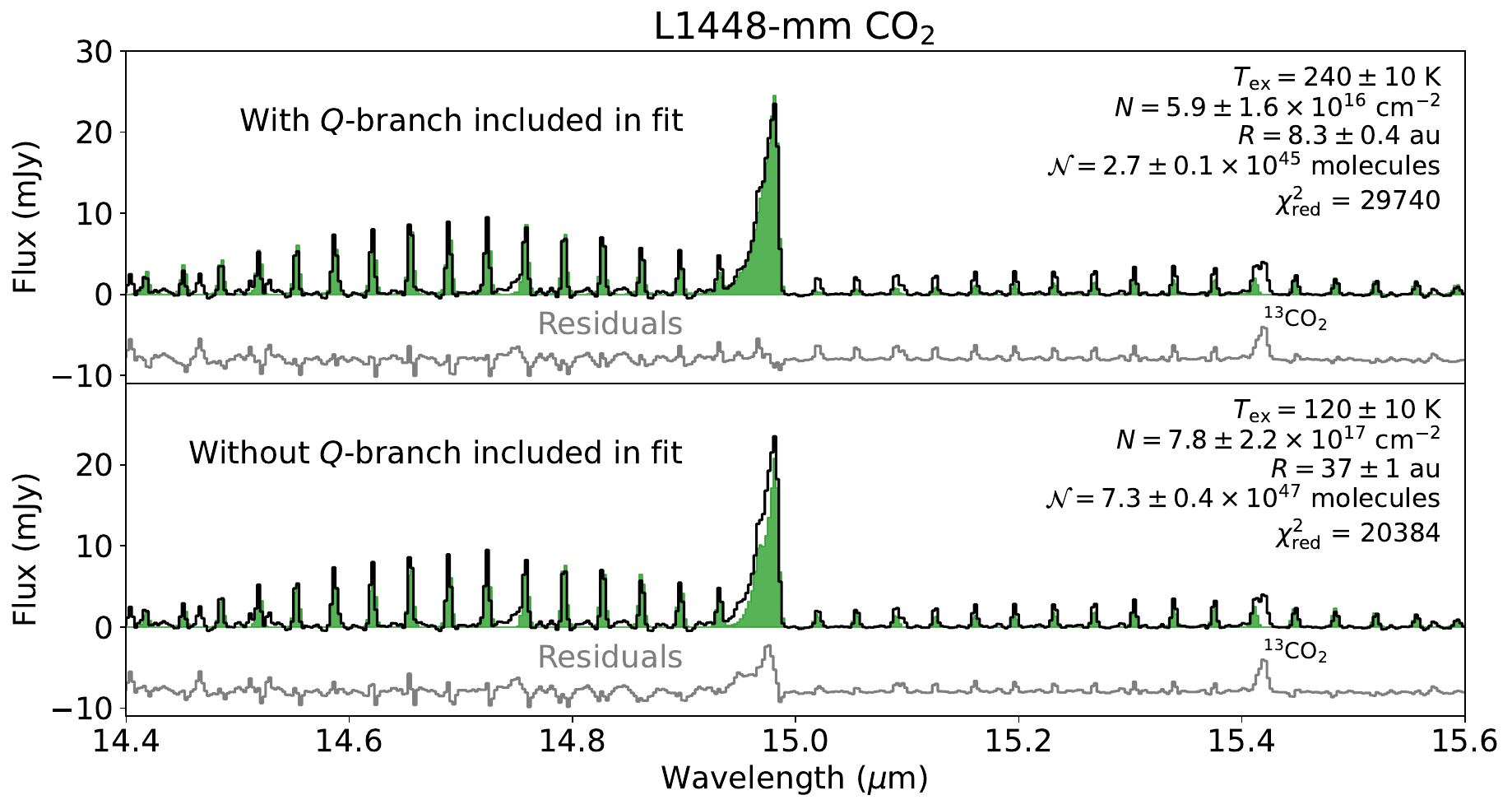}
    \caption{The best fit model of CO$_2$ (green) to the observed spectrum of L1448-mm (black) when the $Q$-branch around 15~\mum is included in the fit (top panel) and when it is excluded (bottom panel). {The residuals of the fit are shown in gray in each panel.} The best-fit parameters for both cases are presented in the top right of each panel. The best-fit parameters can differ significantly depending on whether the optically thick and blended $Q$-branch is taken into account in the fit. Excluding the $Q$-branch leads to more reliable derived quantities, see main text.}
    \label{fig:L1448-mm_Q-vs-withoutQ}
\end{figure*}

For other species, it is important to take line blending into account in fitting. This is most important for molecules showing emission or absorption through a $Q$-branch, which is the superposition of several transitions with $\Delta J = 0$ in a small wavelength region. These lines are spectrally unresolved at the spectral resolution of MIRI-MRS \citep[$R\sim3500-1000$;][]{Labiano2021,Jones2023}. In Fig.~\ref{fig:L1448-mm_Q-vs-withoutQ}, the best-fit LTE models to CO$_2$ in L1448-mm are presented for both cases of taking into account the $Q$-branch in the fit (top panel) and excluding it (bottom panel). Taking into account the $Q$-branch in the fit results in a significantly higher derived excitation temperature and an order of magnitude lower column density and number of molecules. However, the best-fit model depends highly on the shape of the $Q$-branch, which is very sensitive to the excitation temperature, but is also severely affected by blending of optically thick lines. In the bottom panel, the best-fit model excluding the $Q$-branch shows a much better fit to the $R$ and $P$-branches and only underestimates the emission in the $Q$-branch by at most 30\%. Moreover, {the $\chi^2_{\rm red}$ is significantly lower when the $Q$-branch is excluded and} the derived excitation temperature is in perfect agreement with that of C$_2$H$_2$, HCN, and CH$_4$ (see Table~\ref{tab:LTE_Tex}), further supporting that the fit including the $Q$-branch is not correct. Furthermore, the number of molecules derived from CO$_2$ excluding the $Q$-branch is in perfect agreement with that derived from the $^{13}$CO$_2$ isotopologue multiplied by the $^{12}$C/$^{13}$C ratio of the local ISM \citep[70;][]{Milam2005}. This thus shows the importance of fitting unblended $R$ and $P$-branch lines (if detected) in determining accurate excitation temperatures and column densities. 
\end{multicols}

\onecolumn
\begin{multicols}{2}
% \nolinenumbers
\section{Residuals of LTE fits to B1-c}
\label{app:B1-c_residuals}
\begin{figure*}[h]
    \centering
    \includegraphics[width=0.95\linewidth]{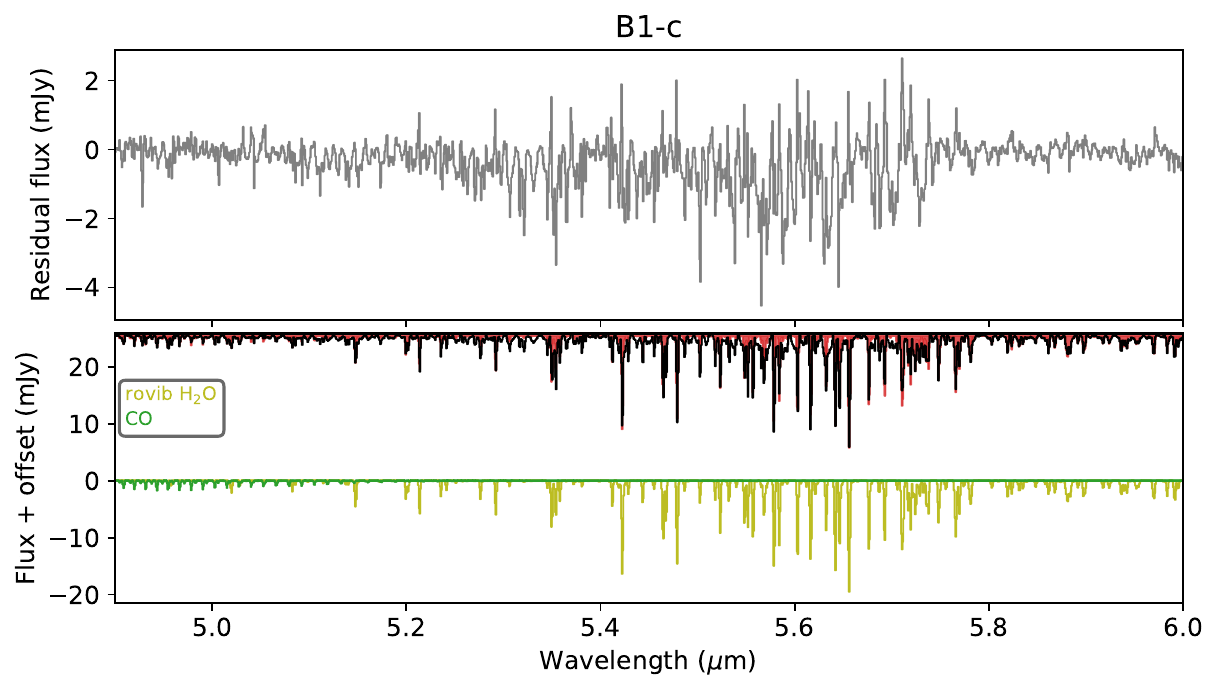}
    \includegraphics[width=0.95\linewidth]{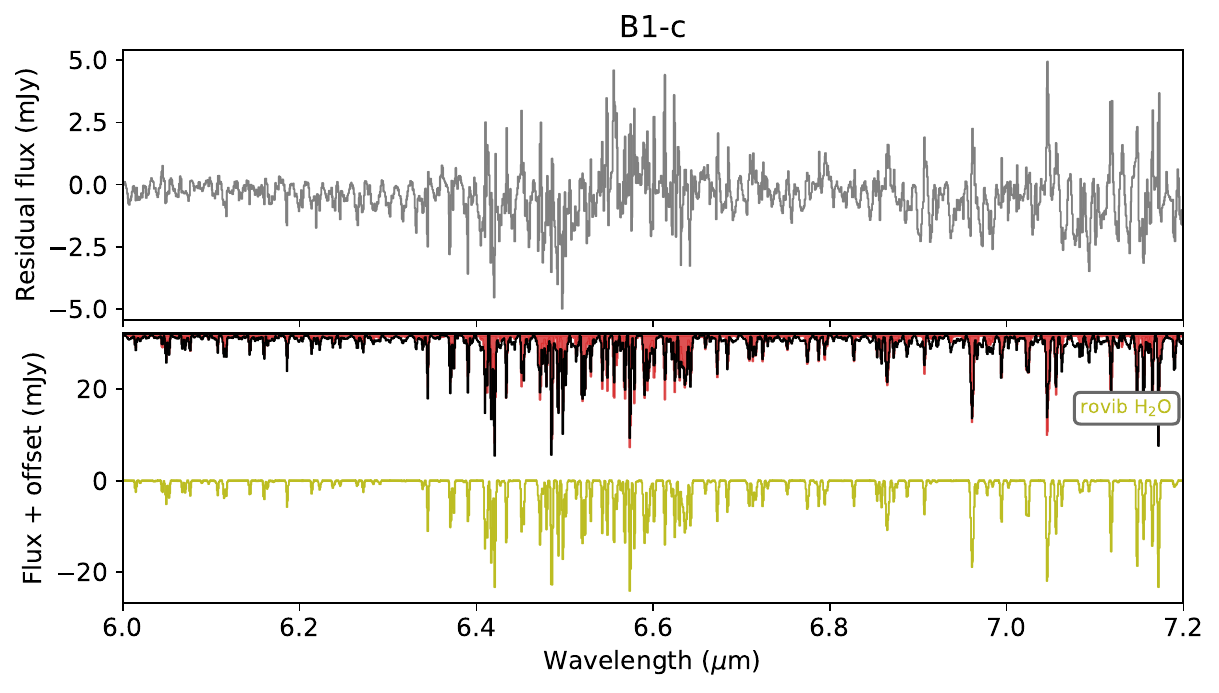}
    \caption{{Baseline-subtracted spectrum (black) and best-fit LTE models (overlayed in red) for B1c in the $4.9-7.2$~\mum range. The residual to the fit is presented in gray in the panel above each spectrum, where it is important to note that the scale of the y-axis is different (smaller). Each molecule contributing to this wavelength range is shown at an arbitrary constant offset in the bottom of each panel.}}
    \label{fig:B1-c_residuals_pt1}
\end{figure*}

\begin{figure*}[h]
    \centering
    \includegraphics[width=\linewidth]{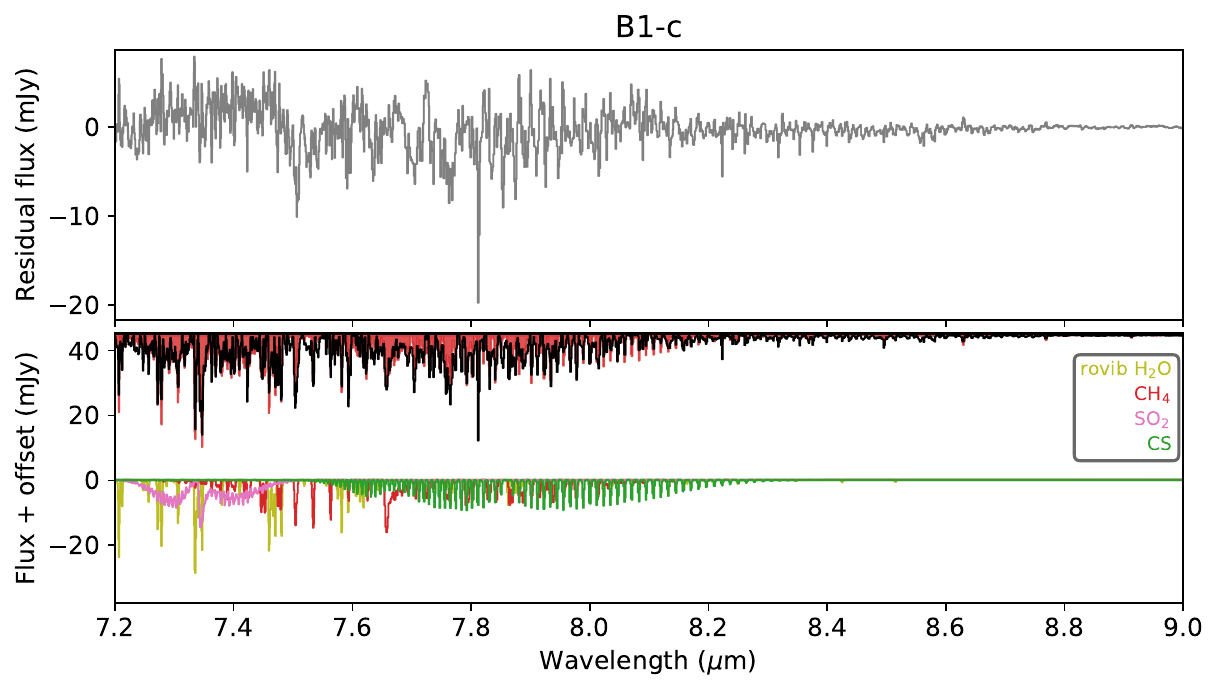}
    \includegraphics[width=\linewidth]{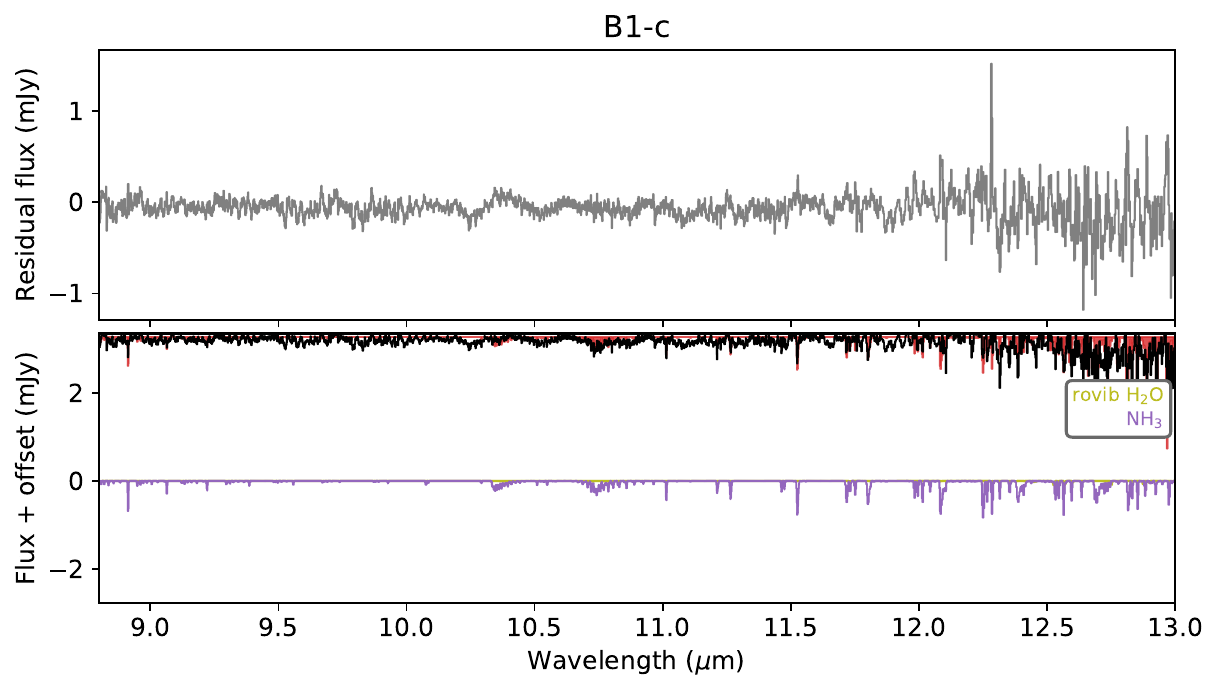}
    \caption{{Same as Fig.~\ref{fig:B1-c_residuals_pt1} but in the $7.2-13.0$~\mum range.}}
    \label{fig:B1-c_residuals_pt2}
\end{figure*}

\begin{figure*}[h]
    \centering
    \includegraphics[width=\linewidth]{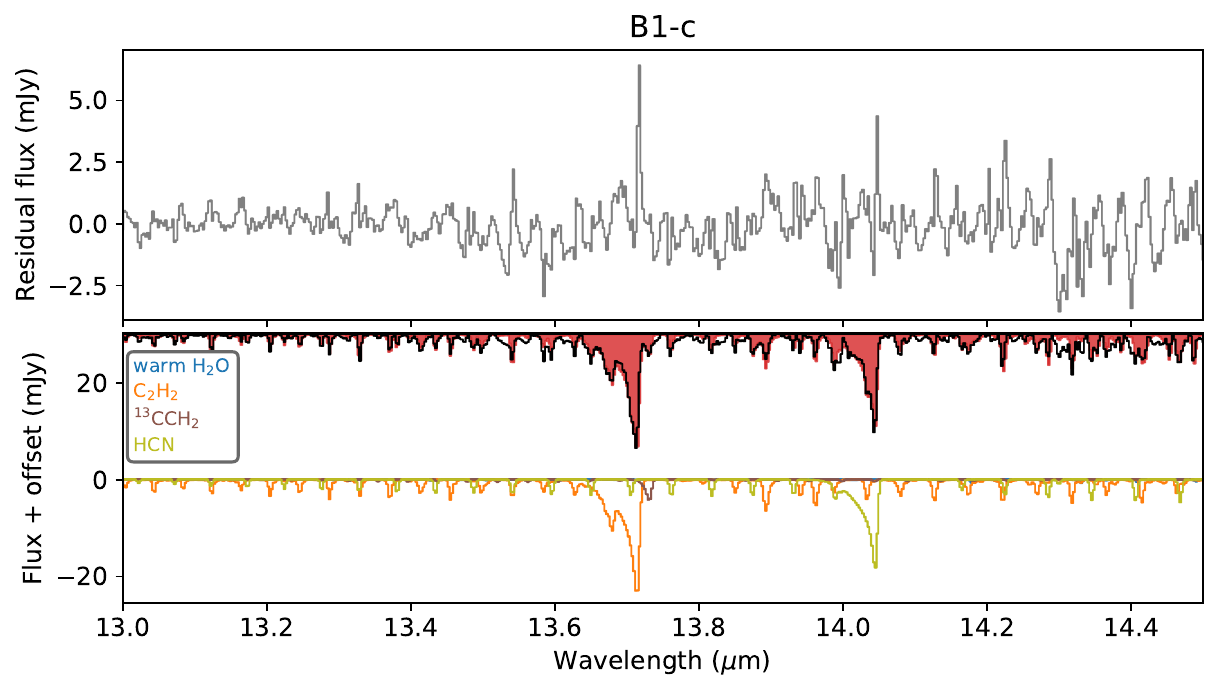}
    \includegraphics[width=\linewidth]{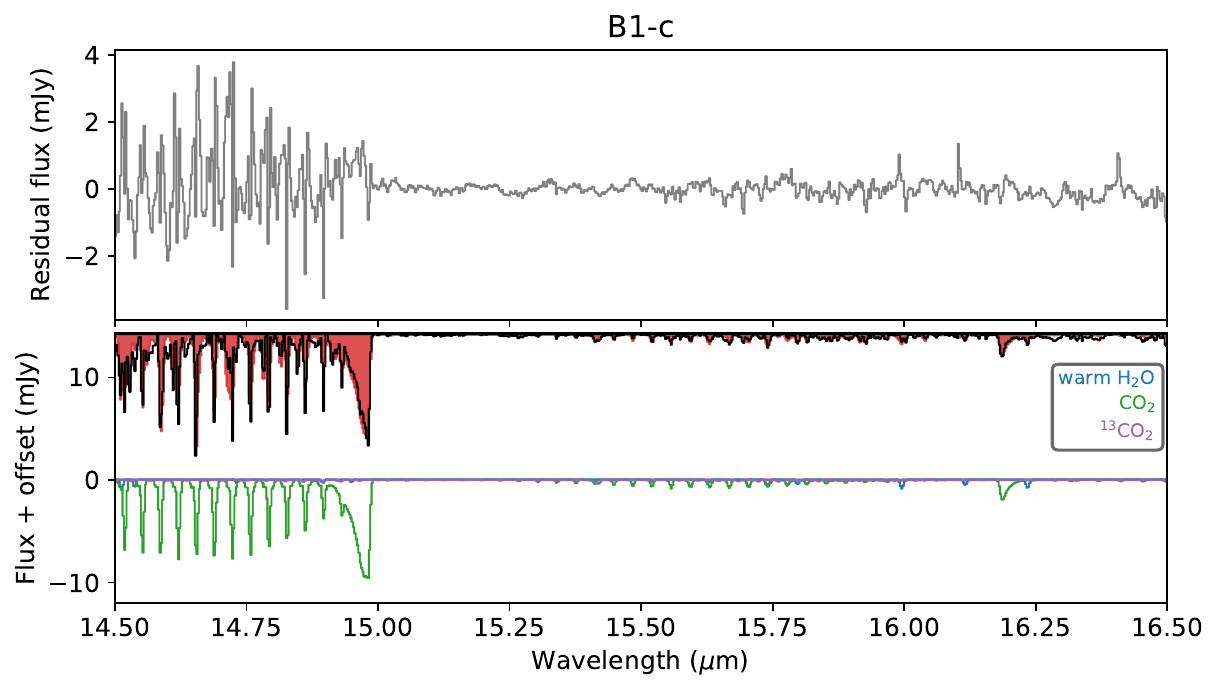}
    \caption{{Same as Fig.~\ref{fig:B1-c_residuals_pt1} but in the $13.0-16.5$~\mum range.}}
    \label{fig:B1-c_residuals_pt3}
\end{figure*}

\begin{figure*}[h]
    \centering
    \includegraphics[width=\linewidth]{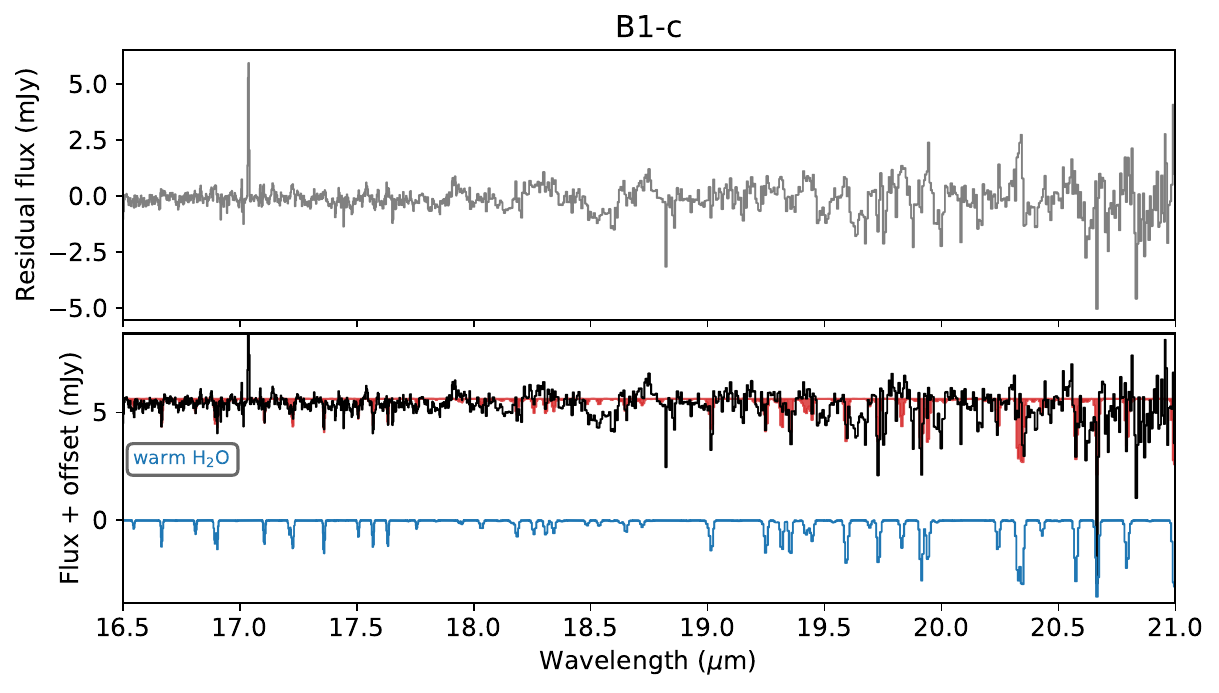}
    \includegraphics[width=\linewidth]{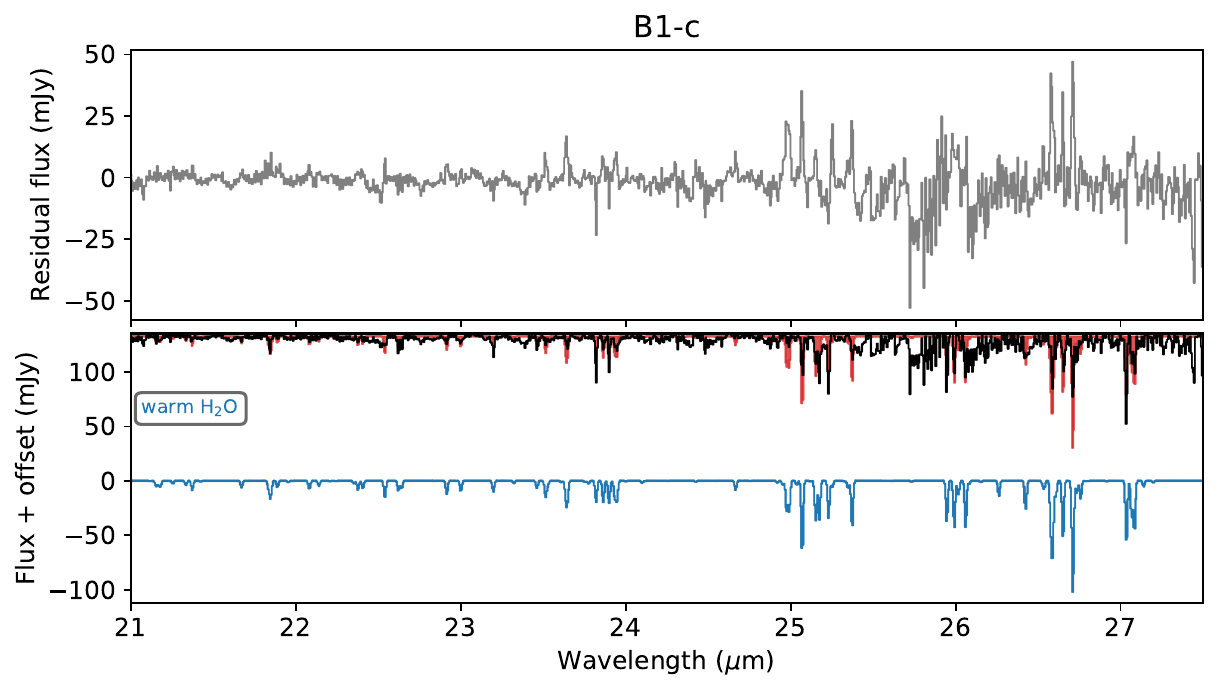}
    \caption{{Same as Fig.~\ref{fig:B1-c_residuals_pt1} but in the $16.5-27.5$~\mum range.}}
    \label{fig:B1-c_residuals_pt4}
\end{figure*}

{In Figs.~\ref{fig:B1-c_residuals_pt1}-\ref{fig:B1-c_residuals_pt4}, the residuals to the best-fit LTE models are presented for our most line-rich source, B1-c. The residuals are plotted using a smaller scale on the y-axis, but are typically on the level of 10-20\%. In some cases the residuals indicate that a second temperature component could be present for some species. Alternatively, the residual lines could also originate from species not explored in this work.}
\end{multicols}

\onecolumn
\begin{multicols}{2}
% \nolinenumbers
\section{LTE model fits to full spectra}
\label{app:specfit}
\begin{figure*}[h]
    \centering
    \includegraphics[width=0.9\linewidth]{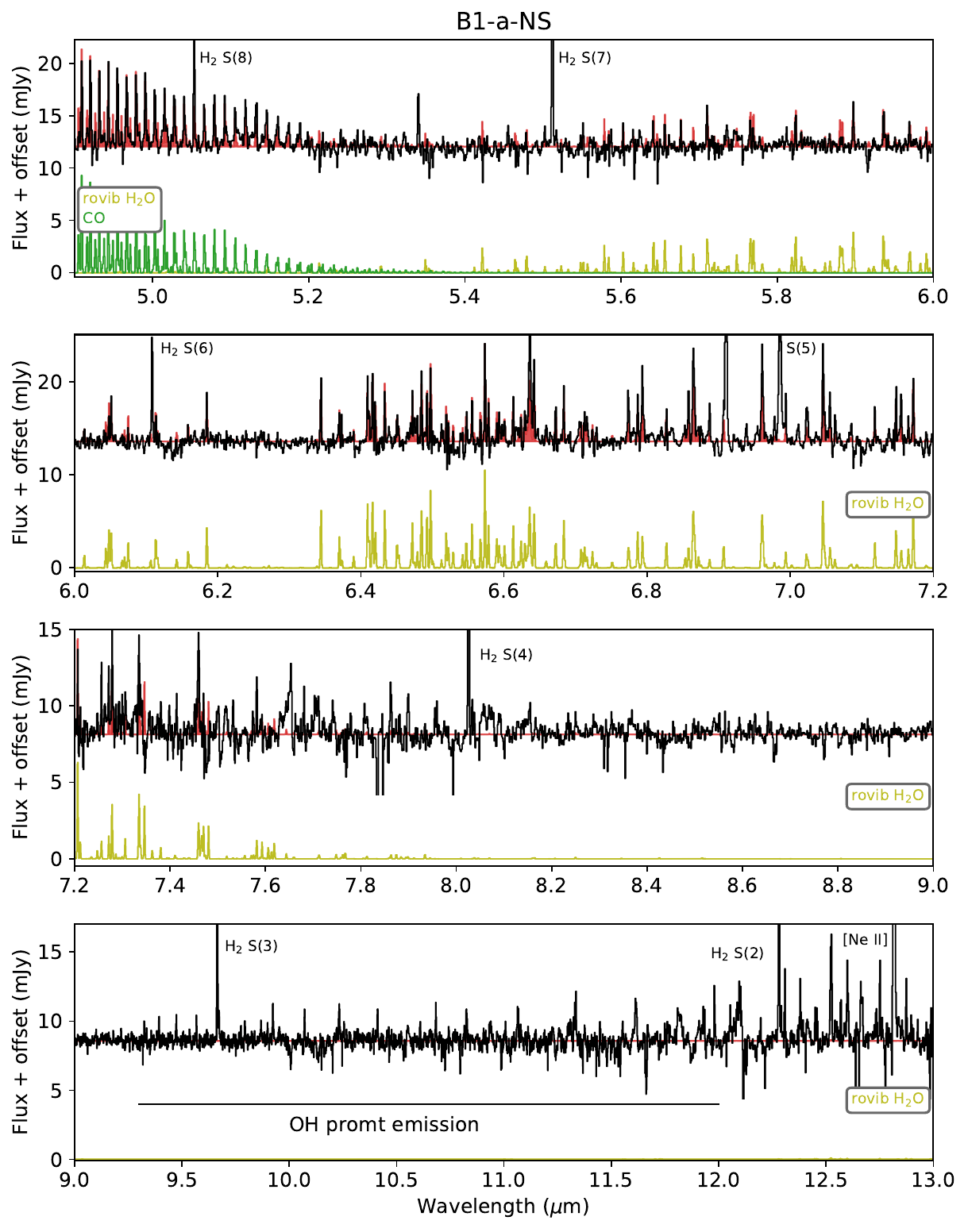}
    \caption{{Baseline-subtracted} spectrum (black) and best-fit LTE models for B1-a-NS in the $4.9-13$~\mum range. In each panel, the red shaded spectrum overlaid on top of the observed spectrum is the full best-fit LTE model. Each molecule contributing to this wavelength range is shown at an arbitrary constant offset in the bottom of each panel. 
    {Strong H$_2$ and atomic emission lines are labeled.}
    % Prompt emission lines between 9.0--13~\mum are present in the bottom panel. 
    }
    \label{fig:specfitfull_B1-a-NS_4.9-13.0}
\end{figure*}

\begin{figure*}[h]
    \centering
    \includegraphics[width=\linewidth]{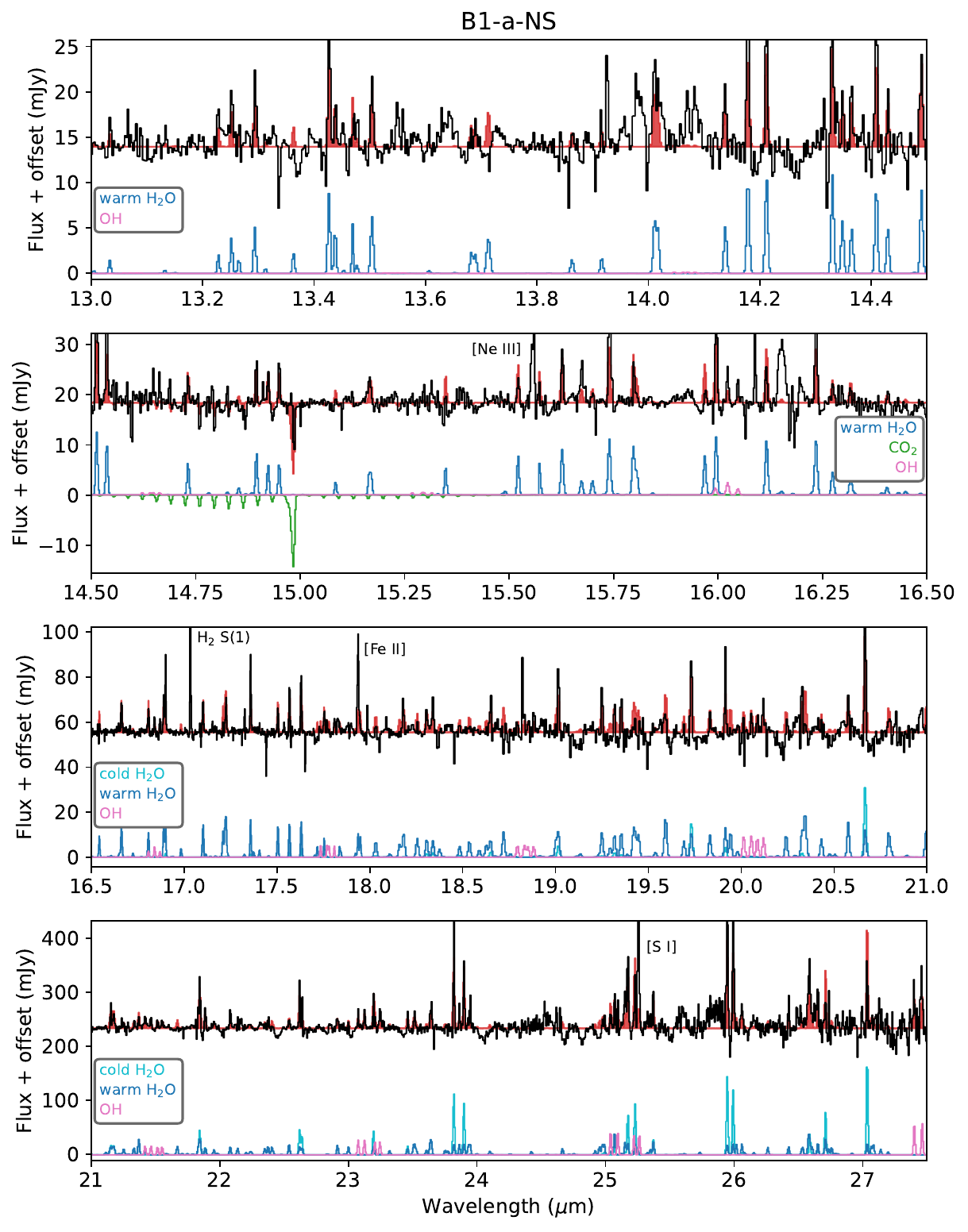}
    \caption{Same as Fig.~\ref{fig:specfitfull_B1-a-NS_4.9-13.0} but now for the $13-27.5$~\mum range.}
    \label{fig:specfitfull_B1-a-NS_13.0-27.5}
\end{figure*}

\begin{figure*}[h]
    \centering
    \includegraphics[width=\linewidth]{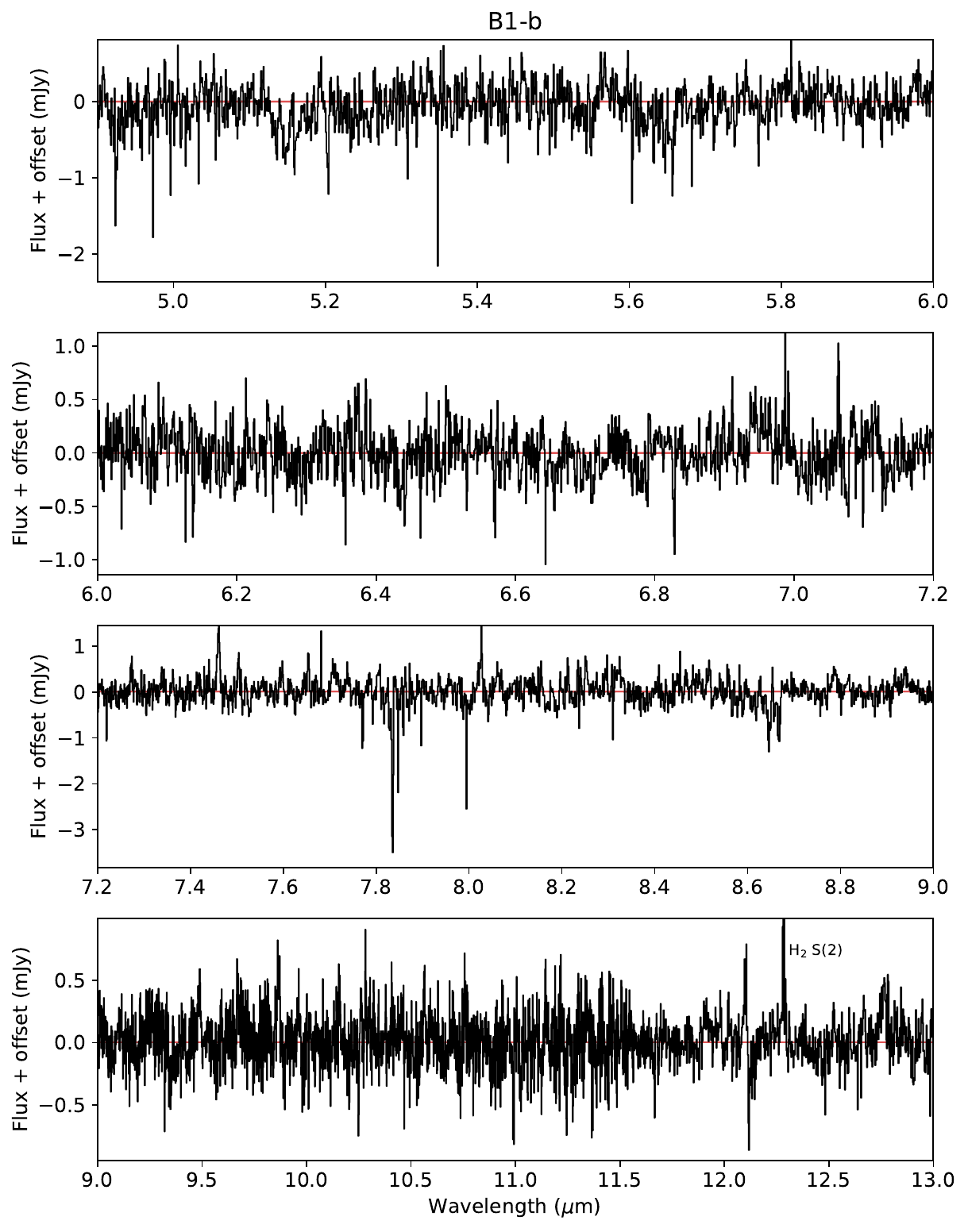}
    \caption{{Baseline-subtracted} spectrum (black) and best-fit LTE models for B1-b in the $4.9-13$~\mum range. In each panel, the red shaded spectrum overlaid on top of the observed spectrum is the full best-fit LTE model. Each molecule contributing to this wavelength range is shown at an arbitrary constant offset in the bottom of each panel. {Strong H$_2$ and atomic emission lines are labeled.}}
    \label{fig:specfitfull_B1-b_4.9-13.0}
\end{figure*}

\begin{figure*}[h]
    \centering
    \includegraphics[width=\linewidth]{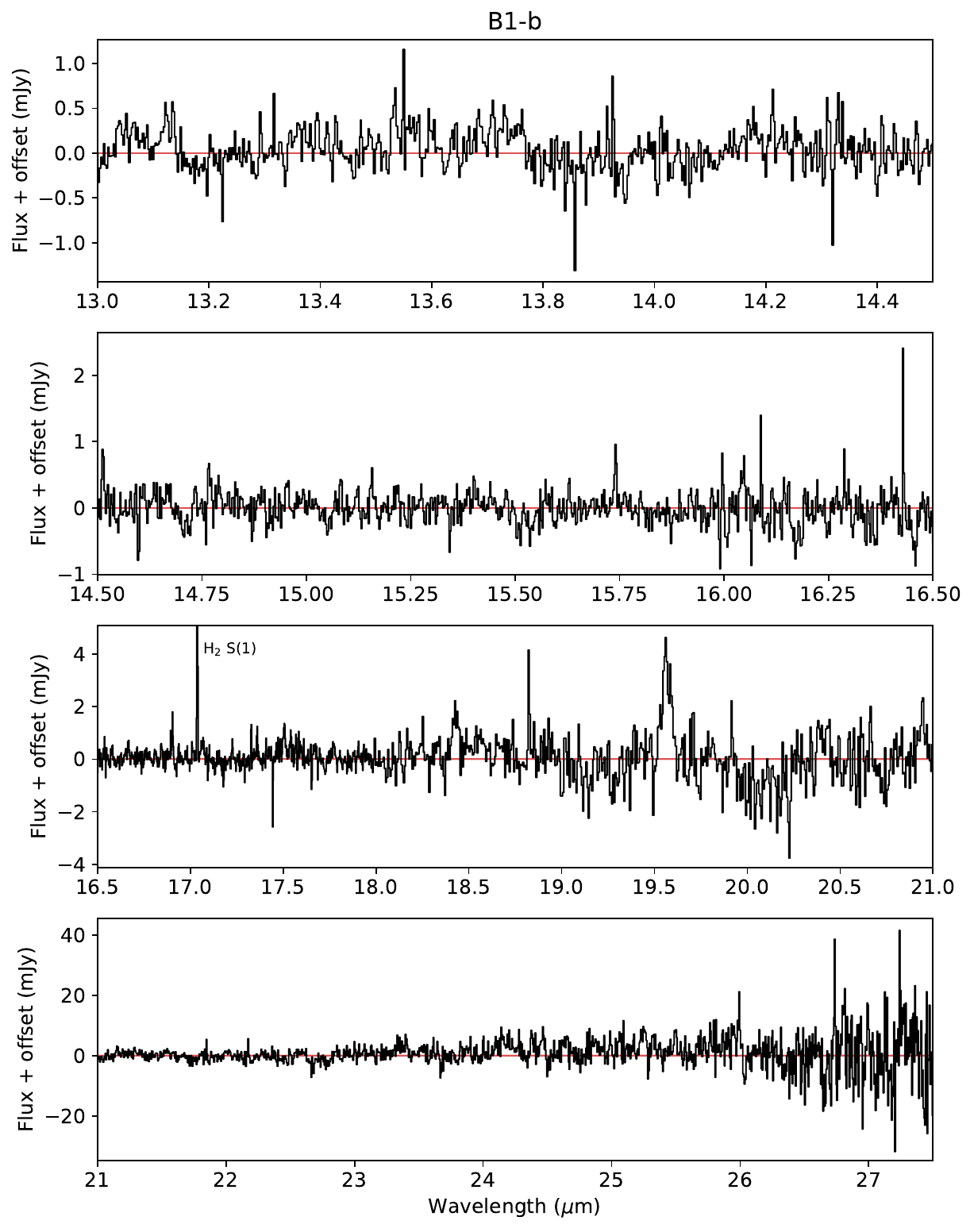}
    \caption{Same as Fig.~\ref{fig:specfitfull_B1-b_4.9-13.0} but now for the $13-27.5$~\mum range.}
    \label{fig:specfitfull_B1-b_13.0-27.5}
\end{figure*}

\begin{figure*}[h]
    \centering
    \includegraphics[width=\linewidth]{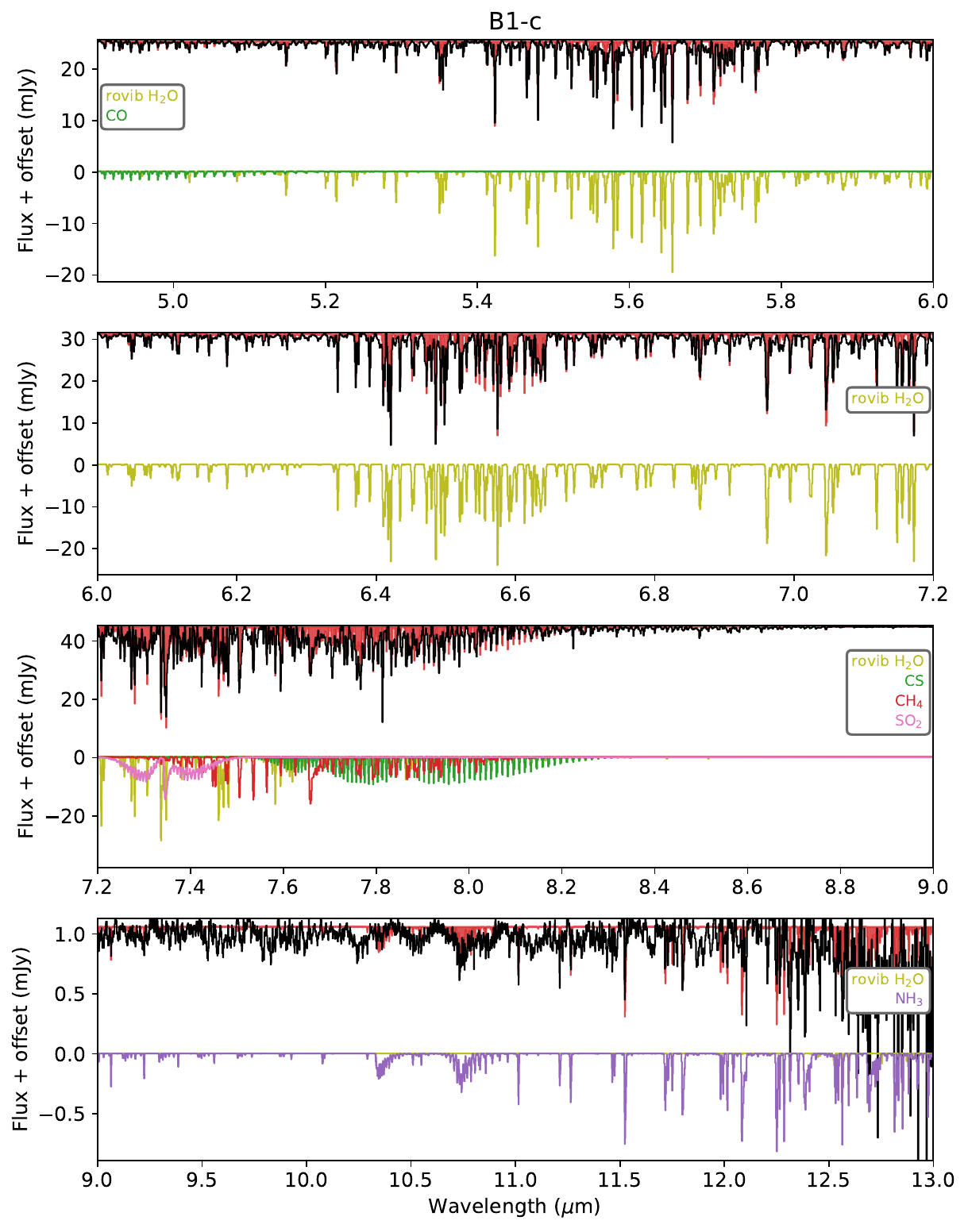}
    \caption{{Baseline-subtracted} spectrum (black) and best-fit LTE models for B1-c in the $4.9-13$~\mum range. In each panel, the red shaded spectrum overlaid on top of the observed spectrum is the full best-fit LTE model. Each molecule contributing to this wavelength range is shown at an arbitrary constant offset in the bottom of each panel. {Strong H$_2$ and atomic emission lines are labeled.}}
    \label{fig:specfitfull_B1-c_4.9-13.0}
\end{figure*}

\begin{figure*}[h]
    \centering
    \includegraphics[width=\linewidth]{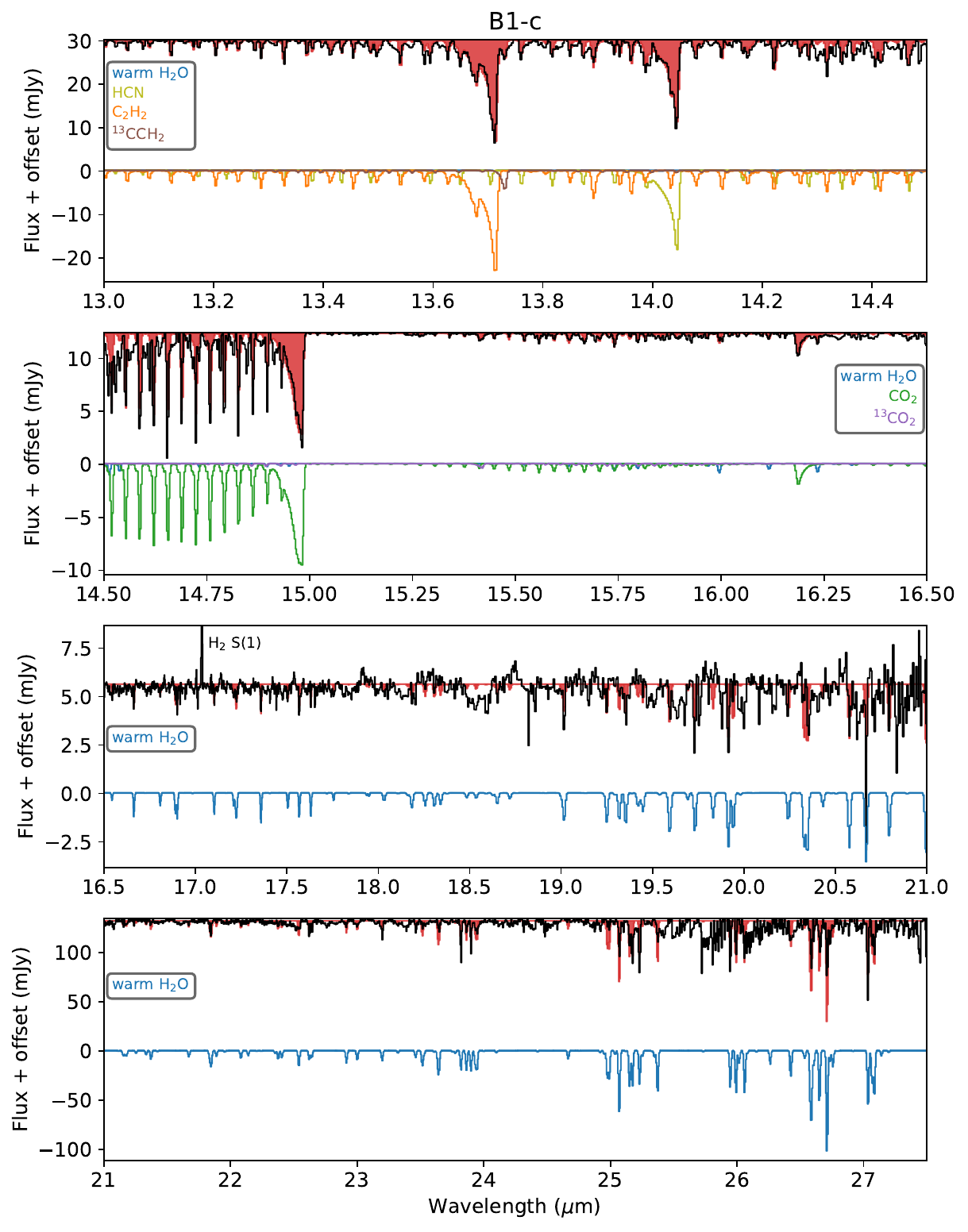}
    \caption{Same as Fig.~\ref{fig:specfitfull_B1-c_4.9-13.0} but now for the $13-27.5$~\mum range.}
    \label{fig:specfitfull_B1-c_13.0-27.5}
\end{figure*}

\begin{figure*}[h]
    \centering
    \includegraphics[width=\linewidth]{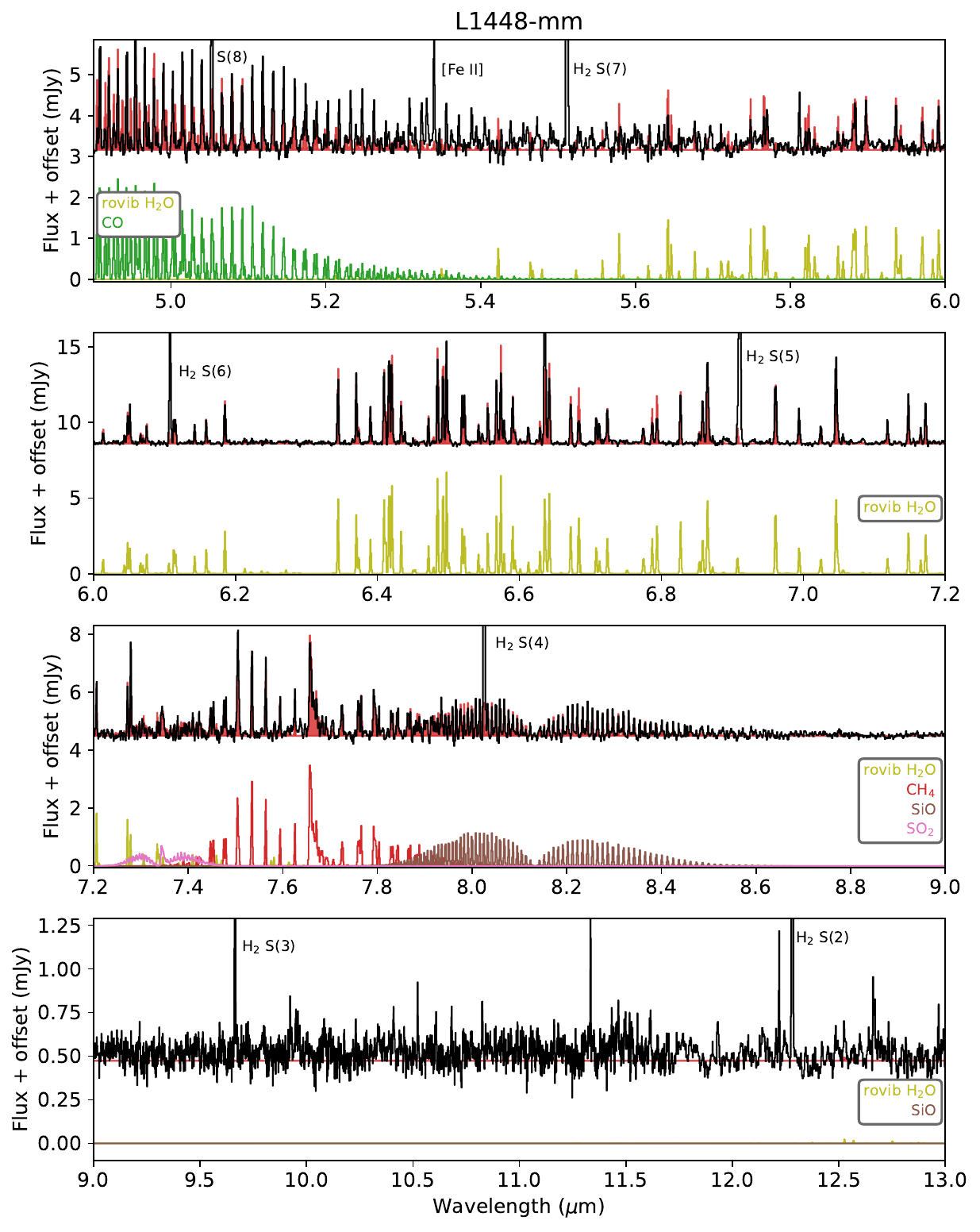}
    \caption{{Baseline-subtracted} spectrum (black) and best-fit LTE models for L1448-mm in the $4.9-13$~\mum range. In each panel, the red shaded spectrum overlaid on top of the observed spectrum is the full best-fit LTE model. Each molecule contributing to this wavelength range is shown at an arbitrary constant offset in the bottom of each panel. {Strong H$_2$ and atomic emission lines are labeled.}}
    \label{fig:specfitfull_L1448-mm_4.9-13.0}
\end{figure*}

\begin{figure*}[h]
    \centering
    \includegraphics[width=\linewidth]{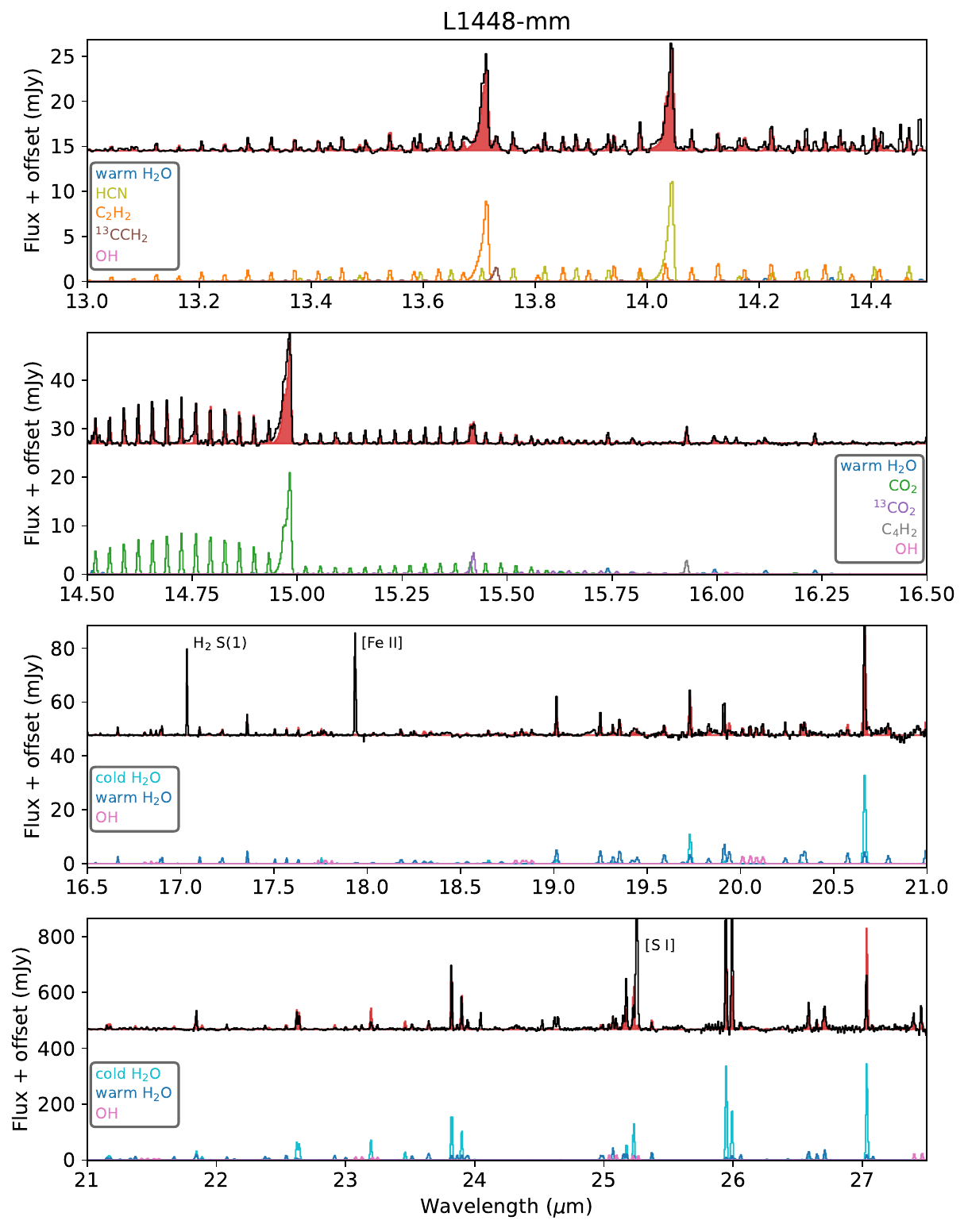}
    \caption{Same as Fig.~\ref{fig:specfitfull_L1448-mm_4.9-13.0} but now for the $13-27.5$~\mum range.}
    \label{fig:specfitfull_L1448-mm_13.0-27.5}
\end{figure*}

\begin{figure*}[h]
    \centering
    \includegraphics[width=\linewidth]{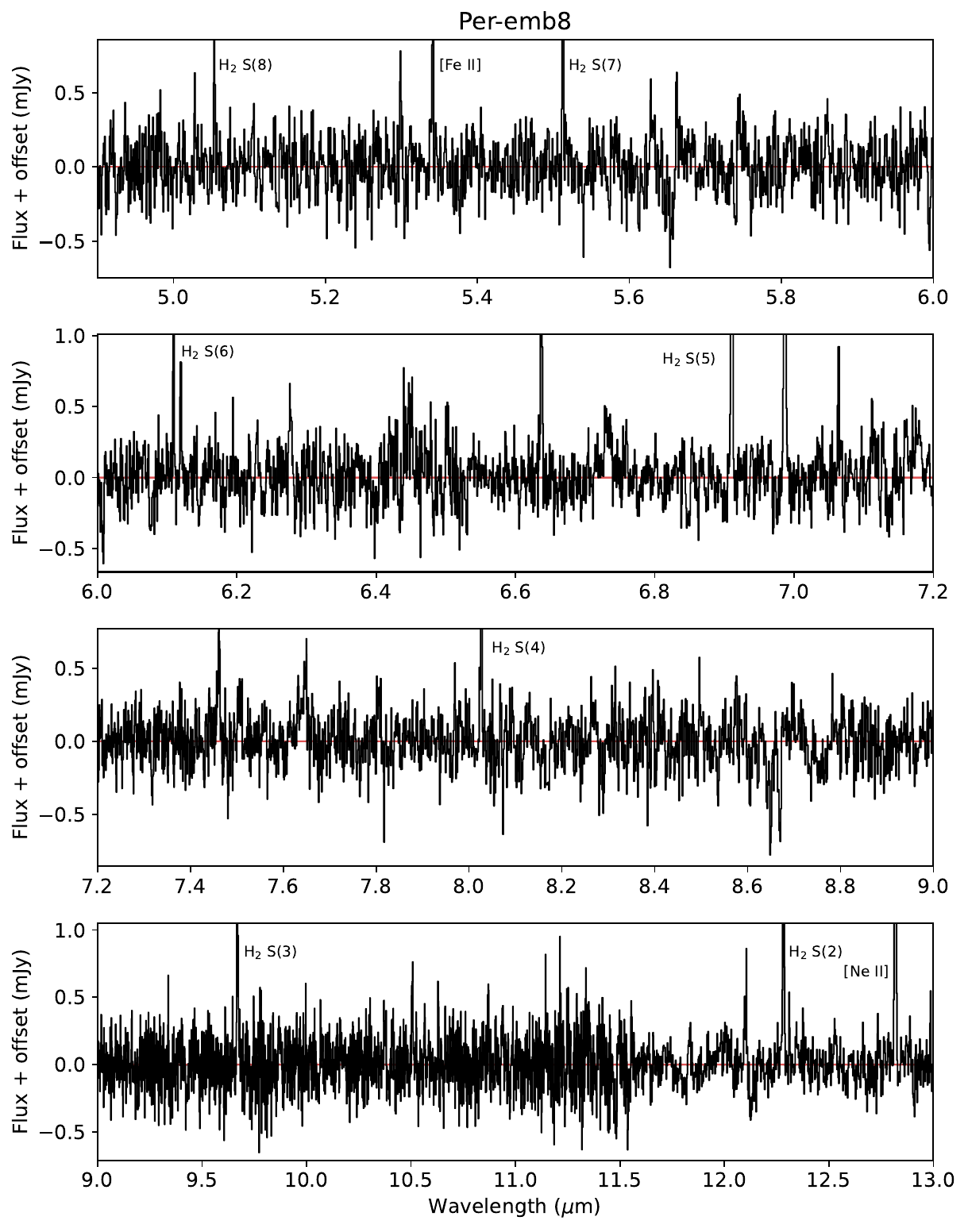}
    \caption{{Baseline-subtracted} spectrum (black) and best-fit LTE models for Per-emb~8 in the $4.9-13$~\mum range. In each panel, the red shaded spectrum overlaid on top of the observed spectrum is the full best-fit LTE model. Each molecule contributing to this wavelength range is shown at an arbitrary constant offset in the bottom of each panel. {Strong H$_2$ and atomic emission lines are labeled.}}
    \label{fig:specfitfull_Per-emb8_4.9-13.0}
\end{figure*}

\begin{figure*}[h]
    \centering
    \includegraphics[width=\linewidth]{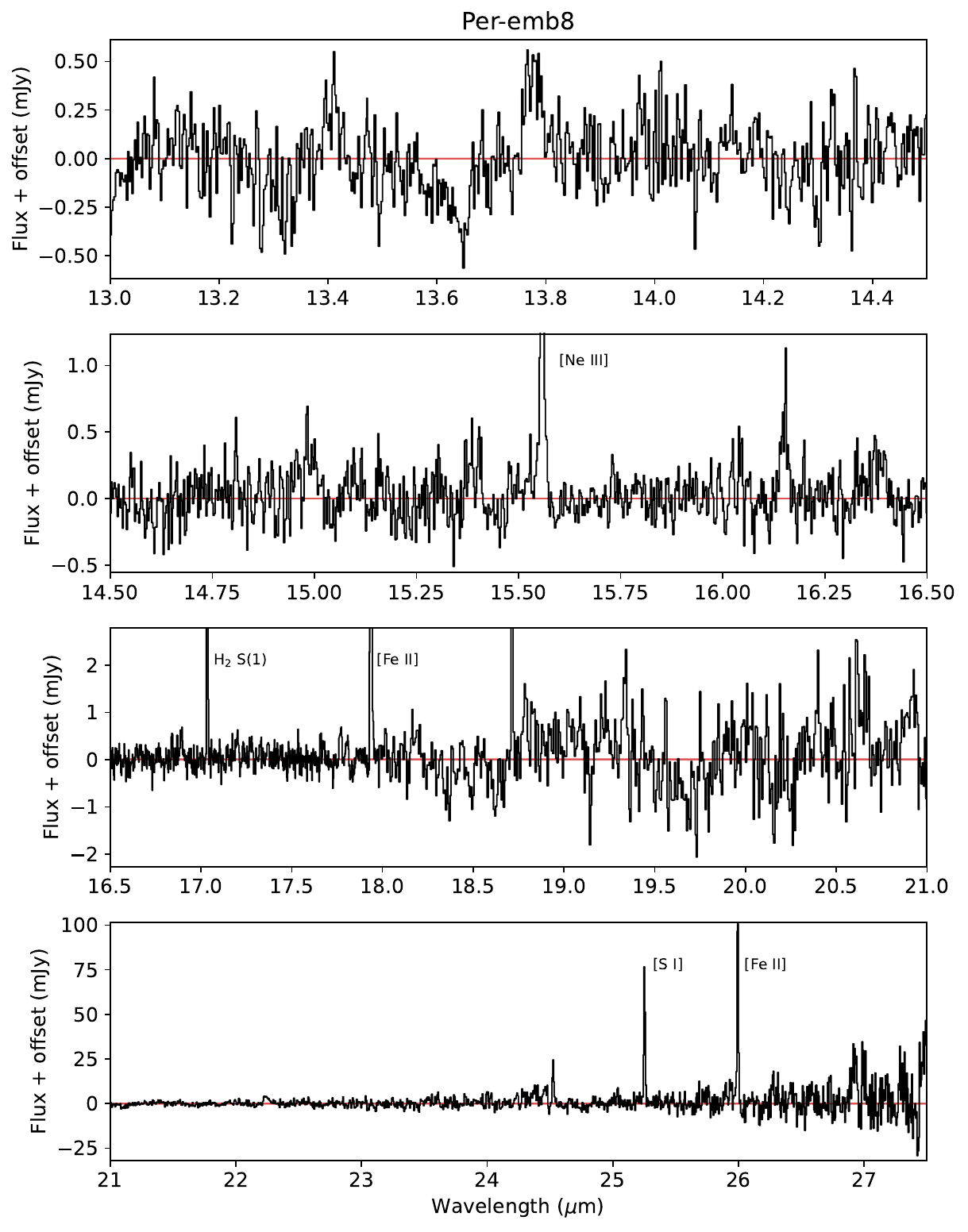}
    \caption{Same as Fig.~\ref{fig:specfitfull_Per-emb8_4.9-13.0} but now for the $13-27.5$~\mum range.}
    \label{fig:specfitfull_Per-emb8_13.0-27.5}
\end{figure*}

\begin{figure*}[h]
    \centering
    \includegraphics[width=\linewidth]{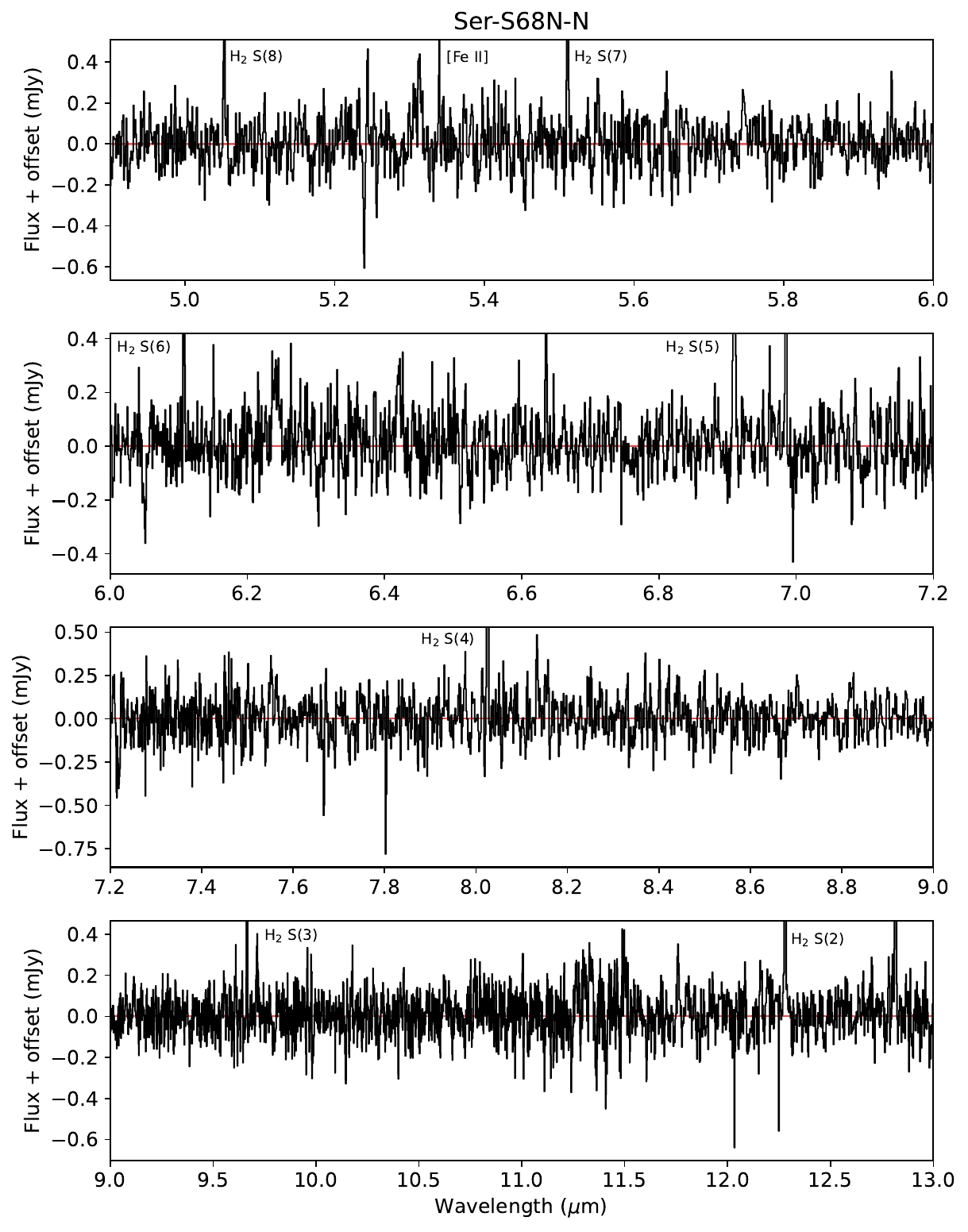}
    \caption{{Baseline-subtracted} spectrum (black) and best-fit LTE models for Ser-S68N-N in the $4.9-13$~\mum range. In each panel, the red shaded spectrum overlaid on top of the observed spectrum is the full best-fit LTE model. Each molecule contributing to this wavelength range is shown at an arbitrary constant offset in the bottom of each panel. {Strong H$_2$ and atomic emission lines are labeled.}}
    \label{fig:specfitfull_Ser-S68N-N_4.9-13.0}
\end{figure*}

\begin{figure*}[h]
    \centering
    \includegraphics[width=\linewidth]{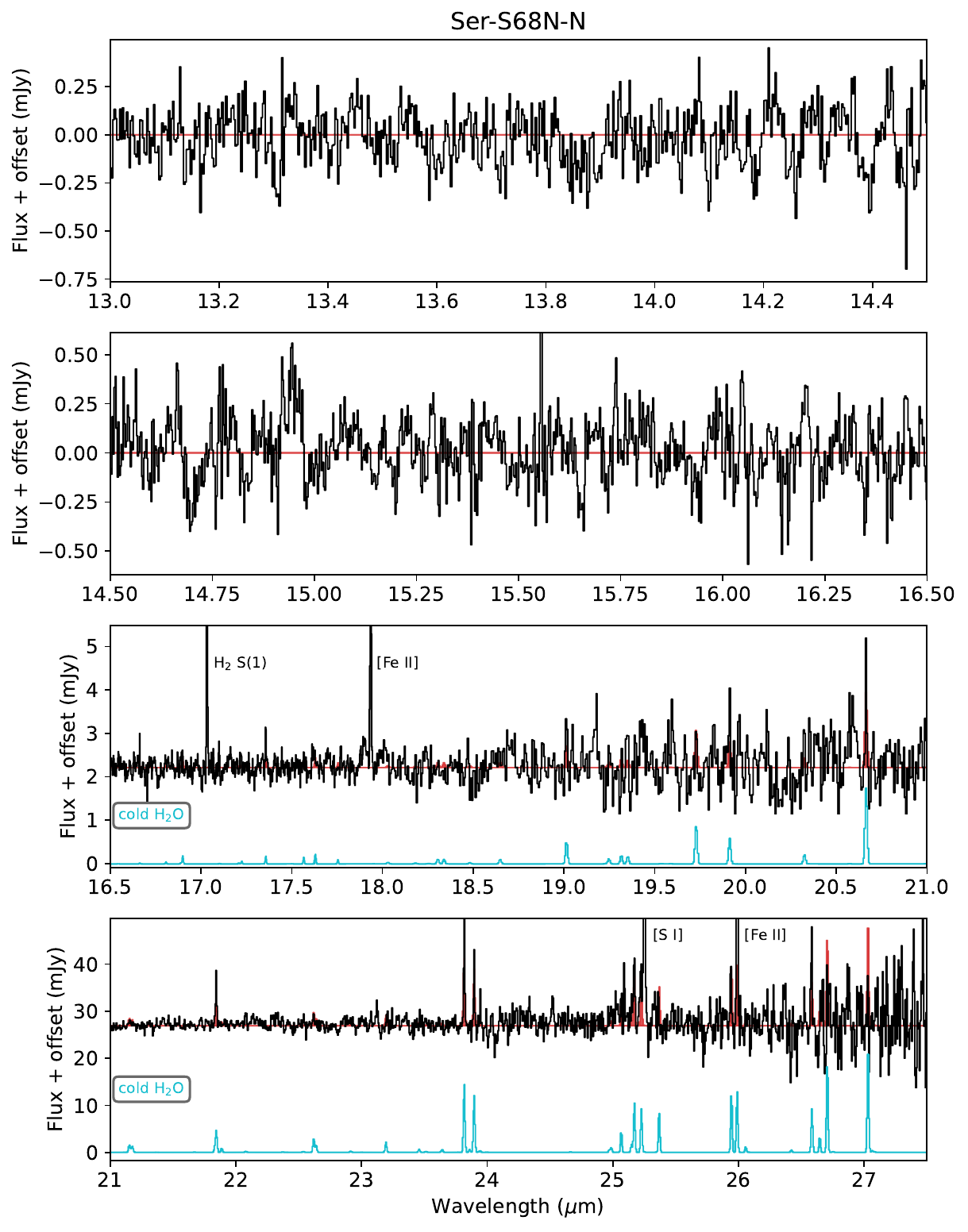}
    \caption{Same as Fig.~\ref{fig:specfitfull_Ser-S68N-N_4.9-13.0} but now for the $13-27.5$~\mum range.}
    \label{fig:specfitfull_Ser-S68N-N_13.0-27.5}
\end{figure*}

\begin{figure*}[h]
    \centering
    \includegraphics[width=\linewidth]{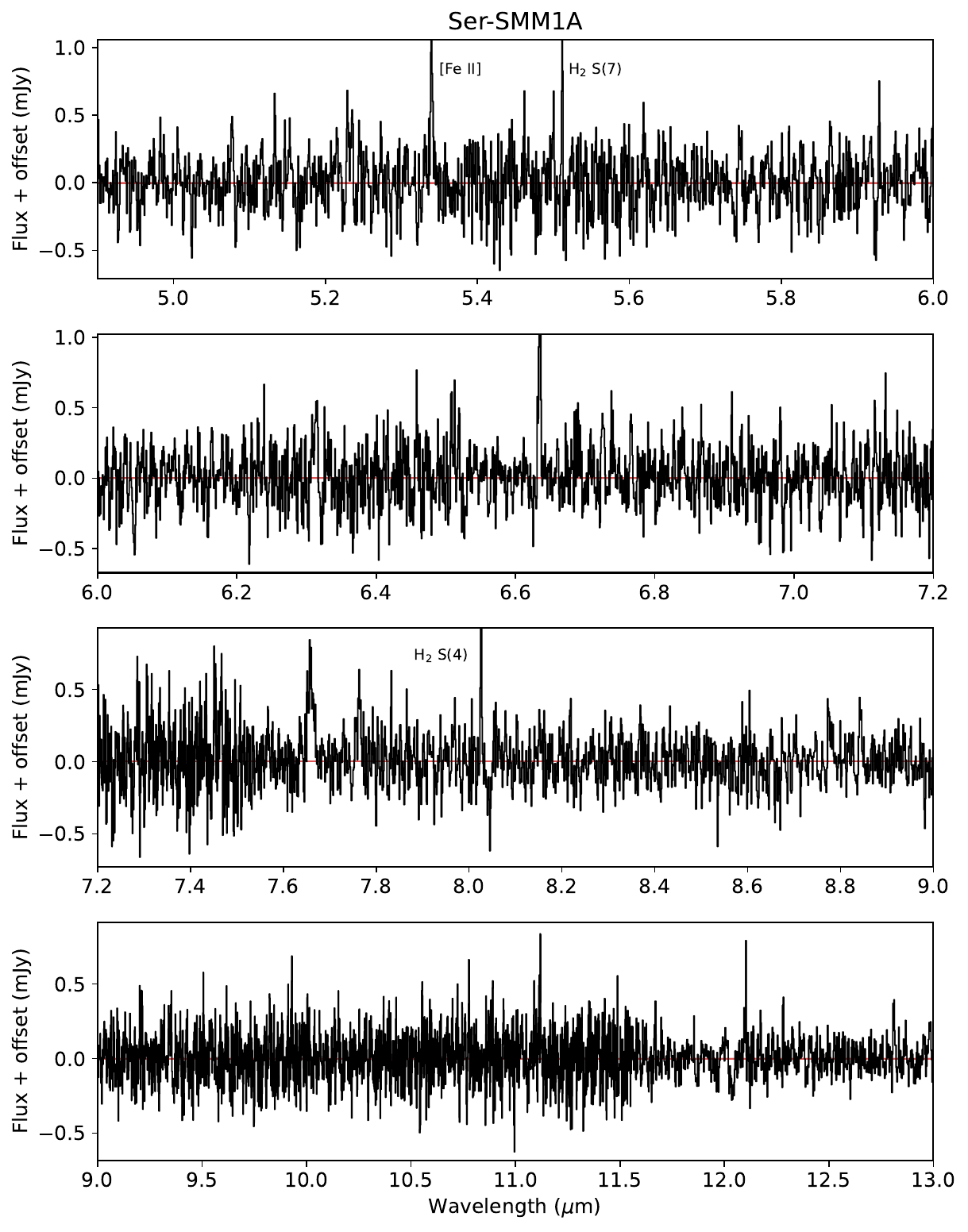}
    \caption{{Baseline-subtracted} spectrum (black) and best-fit LTE models for Ser-SMM1A in the $4.9-13$~\mum range. In each panel, the red shaded spectrum overlaid on top of the observed spectrum is the full best-fit LTE model. Each molecule contributing to this wavelength range is shown at an arbitrary constant offset in the bottom of each panel. {Strong H$_2$ and atomic emission lines are labeled.}}
    \label{fig:specfitfull_Ser-SMM1A_4.9-13.0}
\end{figure*}

\begin{figure*}[h]
    \centering
    \includegraphics[width=\linewidth]{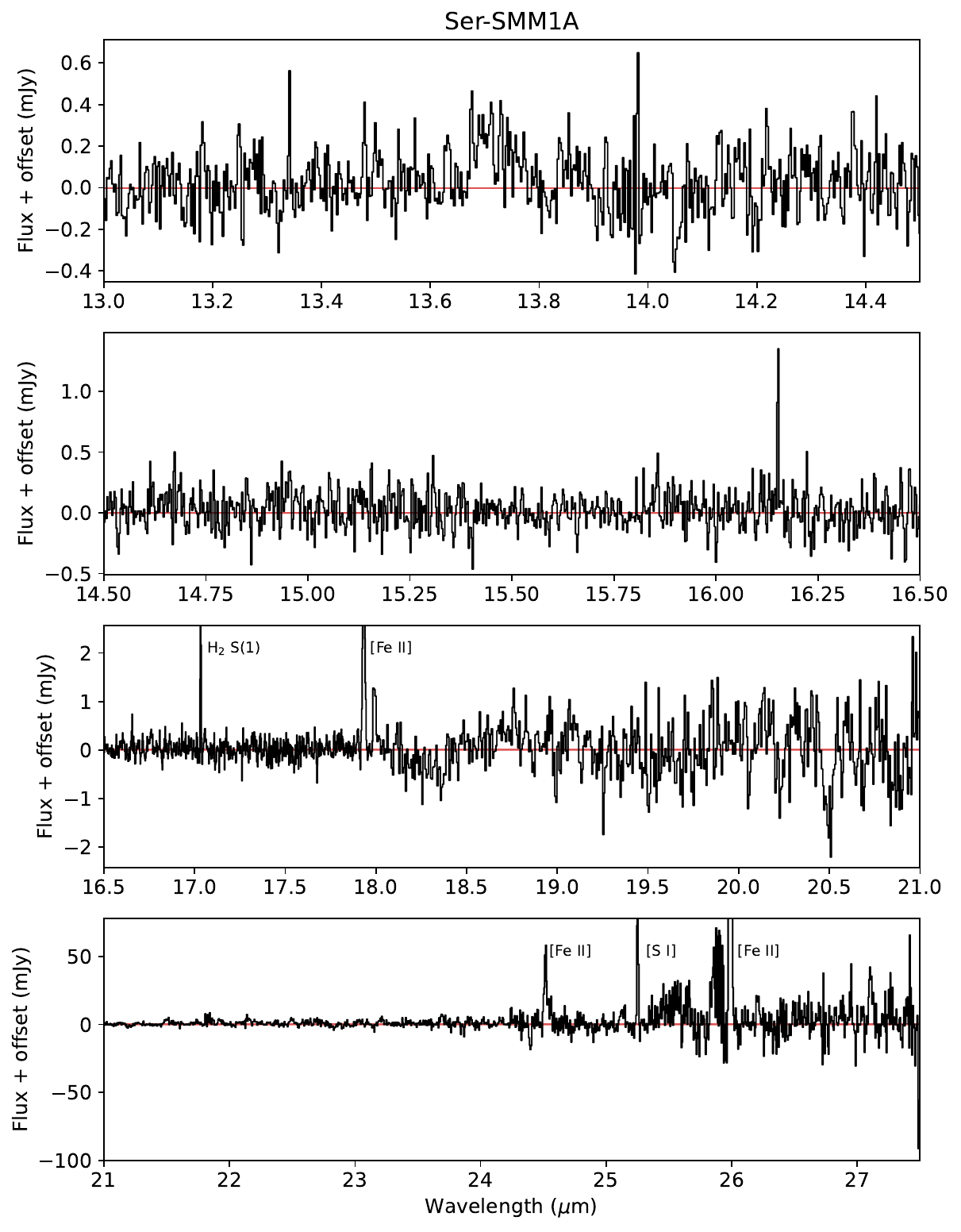}
    \caption{Same as Fig.~\ref{fig:specfitfull_Ser-SMM1A_4.9-13.0} but now for the $13-27.5$~\mum range.}
    \label{fig:specfitfull_Ser-SMM1A_13.0-27.5}
\end{figure*}

\begin{figure*}[h]
    \centering
    \includegraphics[width=\linewidth]{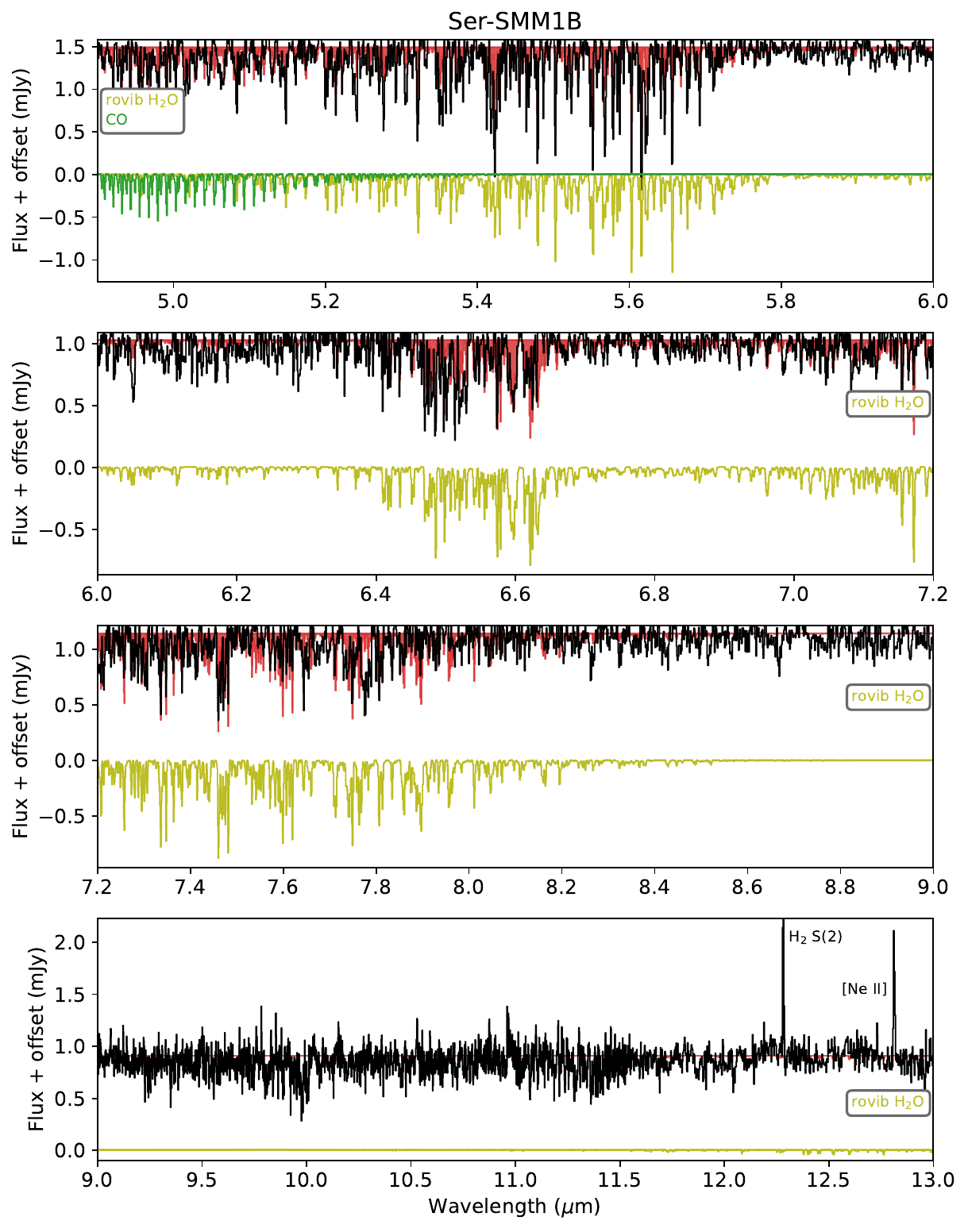}
    \caption{{Baseline-subtracted} spectrum (black) and best-fit LTE models for Ser-SMM1B in the $4.9-13$~\mum range. In each panel, the red shaded spectrum overlaid on top of the observed spectrum is the full best-fit LTE model. Each molecule contributing to this wavelength range is shown at an arbitrary constant offset in the bottom of each panel. {Strong H$_2$ and atomic emission lines are labeled.}}
    \label{fig:specfitfull_Ser-SMM1B_4.9-13.0}
\end{figure*}

\begin{figure*}[h]
    \centering
    \includegraphics[width=\linewidth]{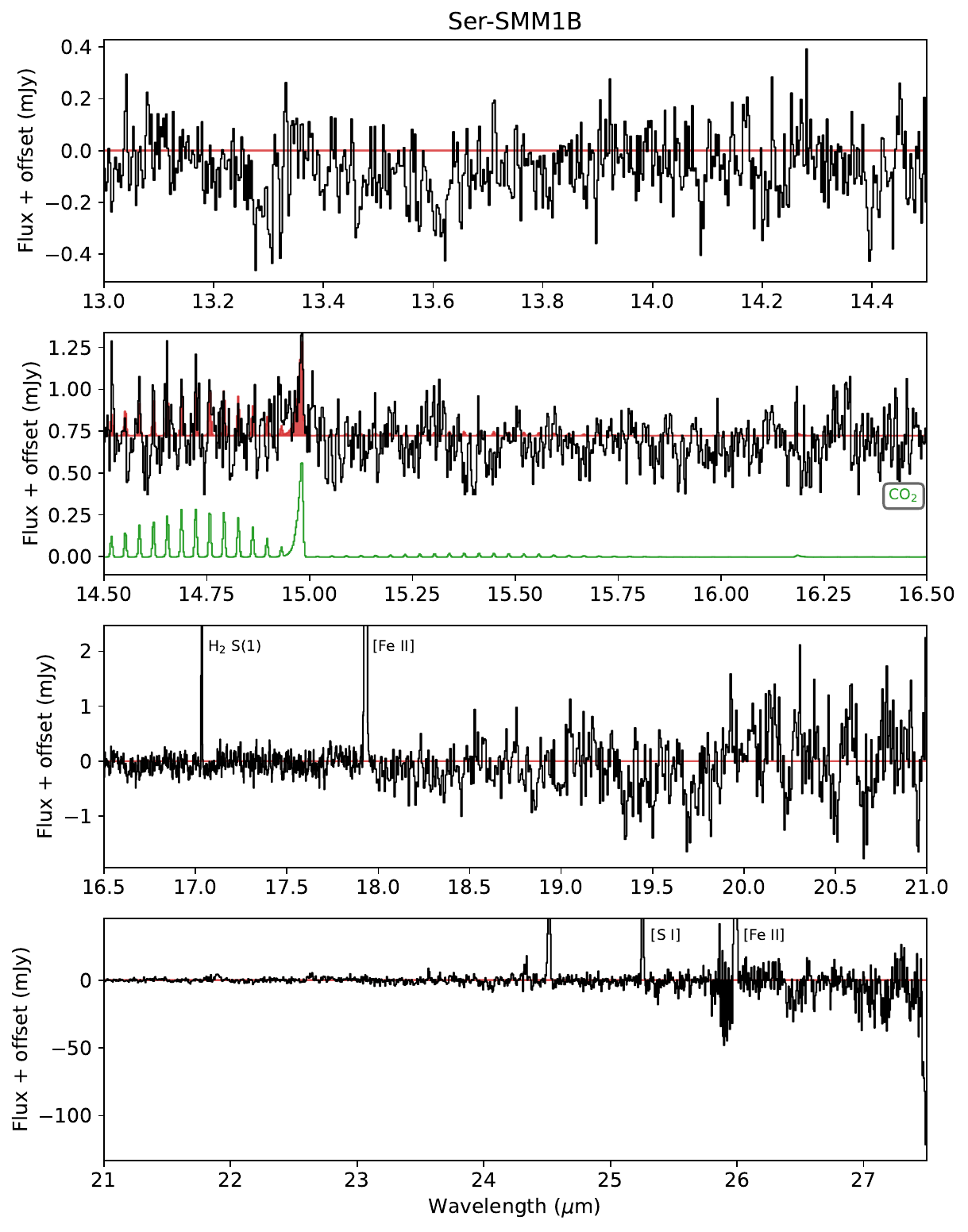}
    \caption{Same as Fig.~\ref{fig:specfitfull_Ser-SMM1B_4.9-13.0} but now for the $13-27.5$~\mum range.}
    \label{fig:specfitfull_Ser-SMM1B_13.0-27.5}
\end{figure*}

\begin{figure*}[h]
    \centering
    \includegraphics[width=\linewidth]{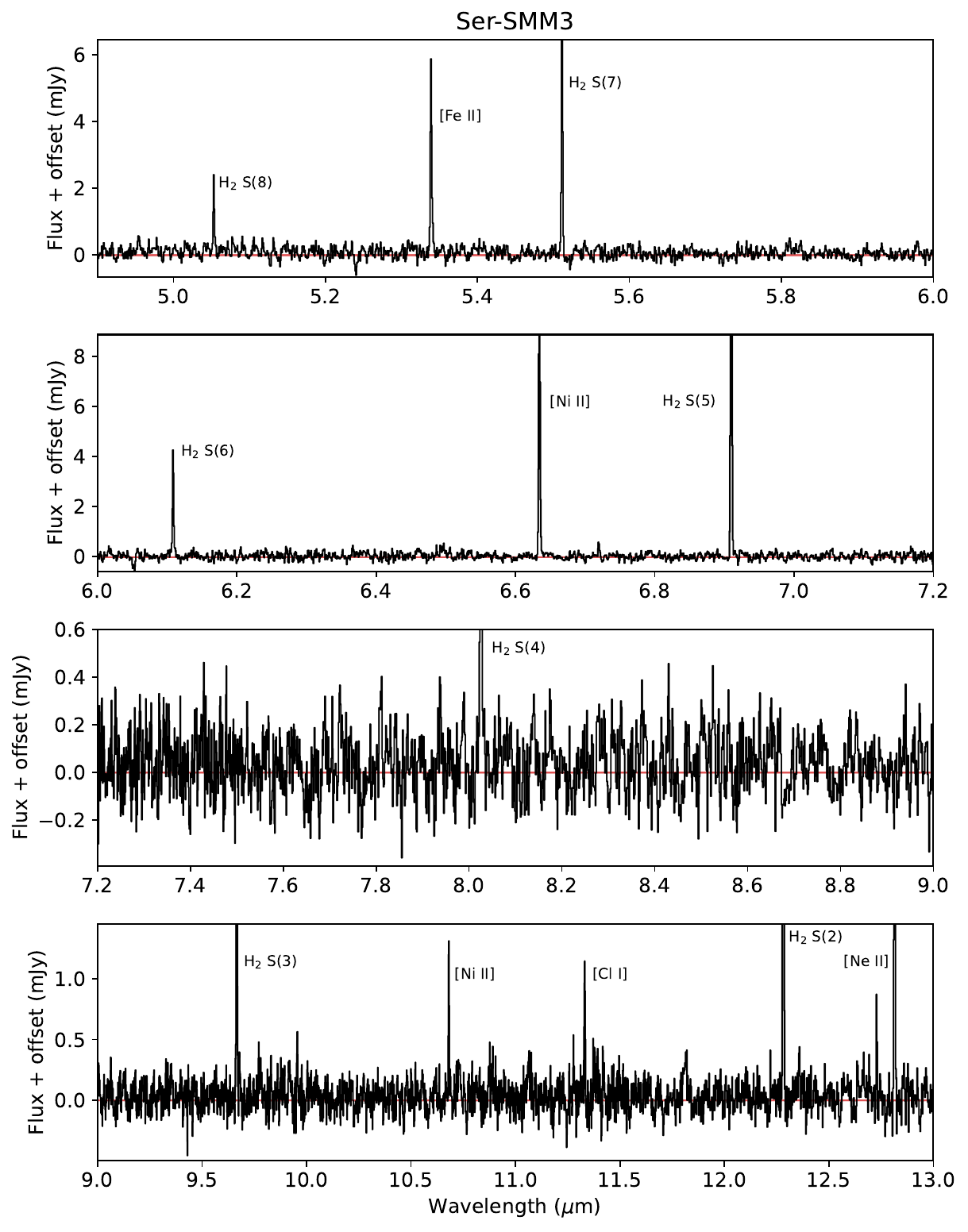}
    \caption{{Baseline-subtracted} spectrum (black) and best-fit LTE models for Ser-SMM3 in the $4.9-13$~\mum range. In each panel, the red shaded spectrum overlaid on top of the observed spectrum is the full best-fit LTE model. Each molecule contributing to this wavelength range is shown at an arbitrary constant offset in the bottom of each panel. {Strong H$_2$ and atomic emission lines are labeled.}}
    \label{fig:specfitfull_Ser-SMM3_4.9-13.0}
\end{figure*}

\begin{figure*}[h]
    \centering
    \includegraphics[width=\linewidth]{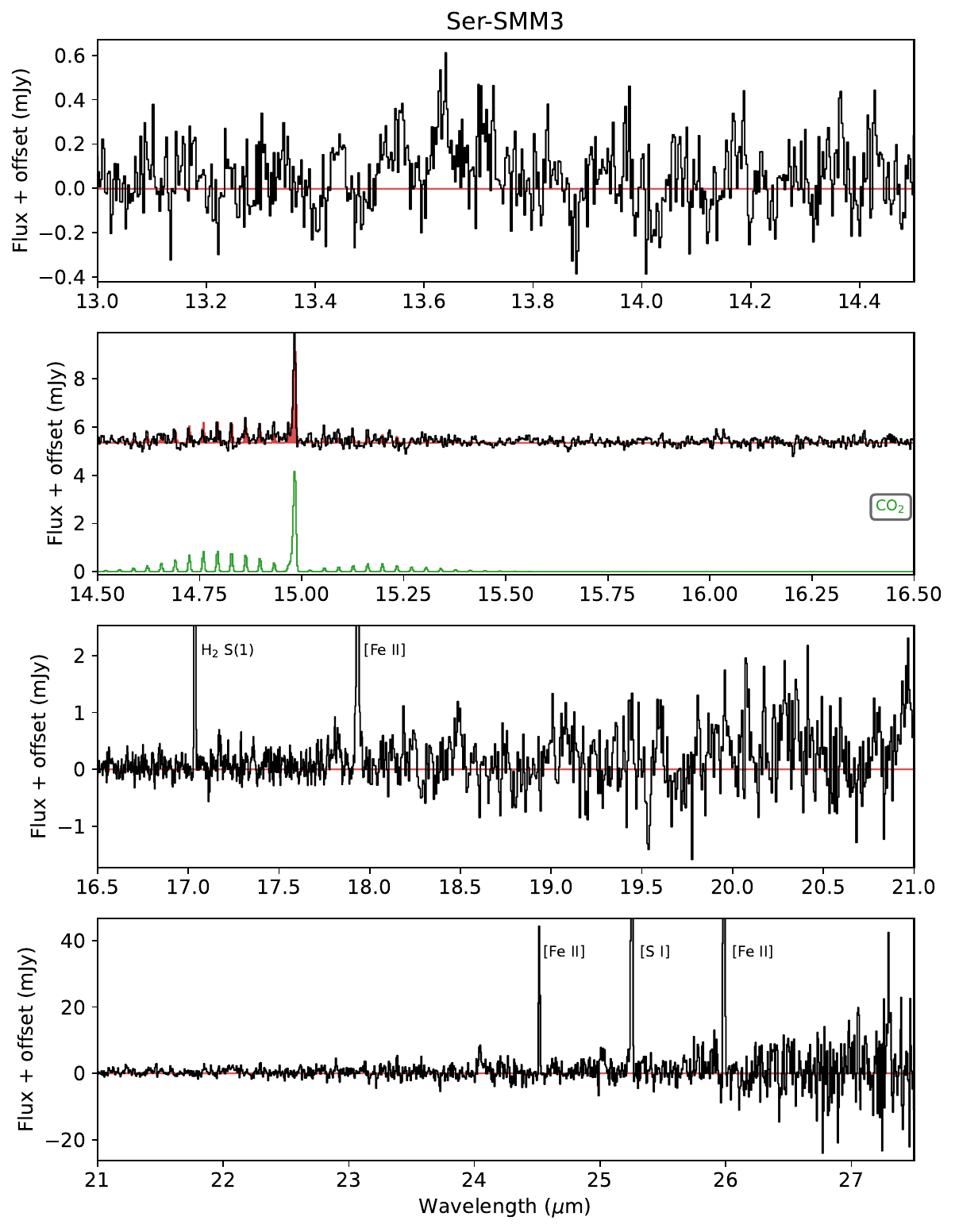}
    \caption{Same as Fig.~\ref{fig:specfitfull_Ser-SMM3_4.9-13.0} but now for the $13-27.5$~\mum range.}
    \label{fig:specfitfull_Ser-SMM3_13.0-27.5}
\end{figure*}

\begin{figure*}[h]
    \centering
    \includegraphics[width=\linewidth]{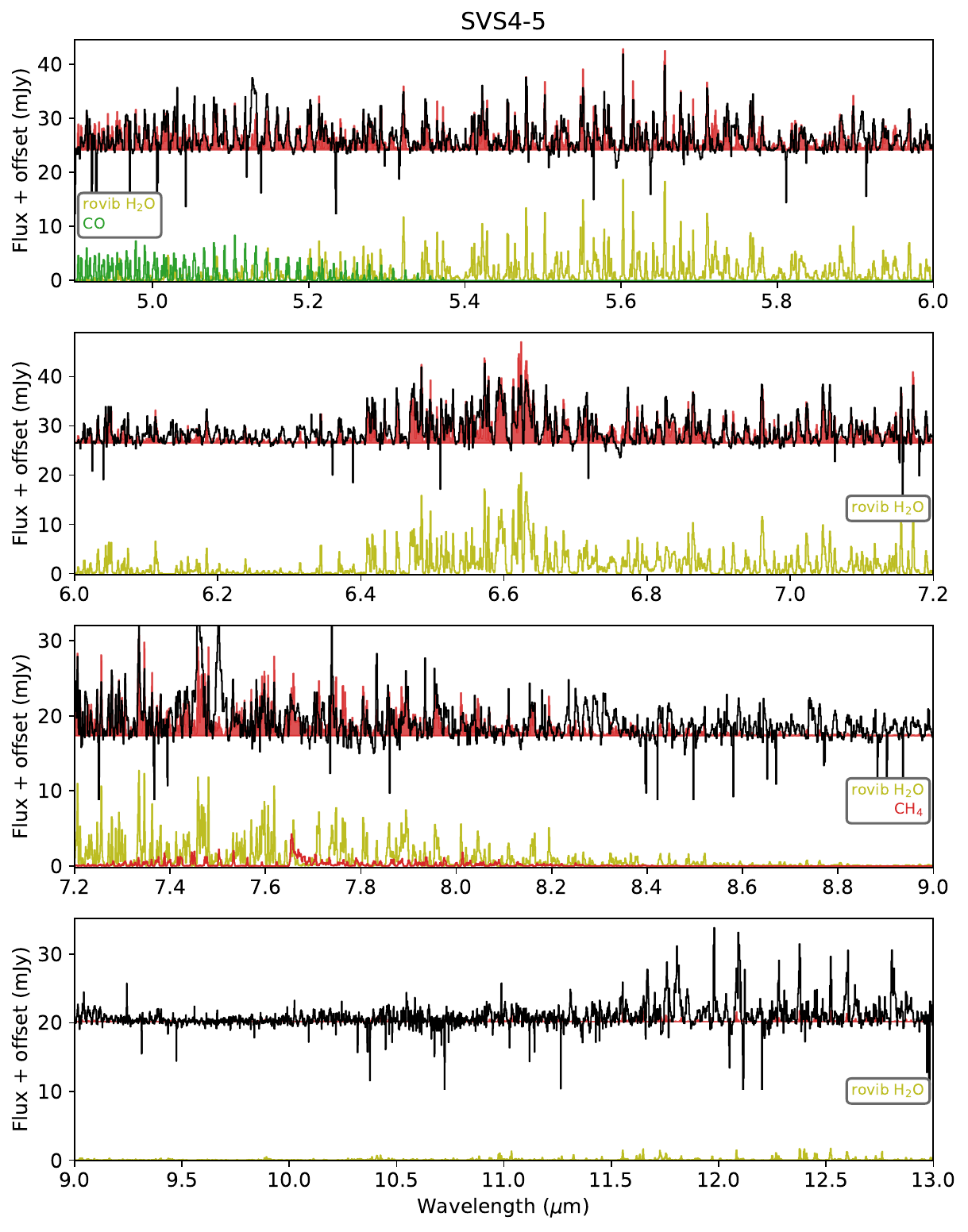}
    \caption{{Baseline-subtracted} spectrum (black) and best-fit LTE models for SVS4-5 in the $4.9-13$~\mum range. In each panel, the red shaded spectrum overlaid on top of the observed spectrum is the full best-fit LTE model. Each molecule contributing to this wavelength range is shown at an arbitrary constant offset in the bottom of each panel. {Strong H$_2$ and atomic emission lines are labeled.}}
    \label{fig:specfitfull_SVS4-5_4.9-13.0}
\end{figure*}

\begin{figure*}[h]
    \centering
    \includegraphics[width=\linewidth]{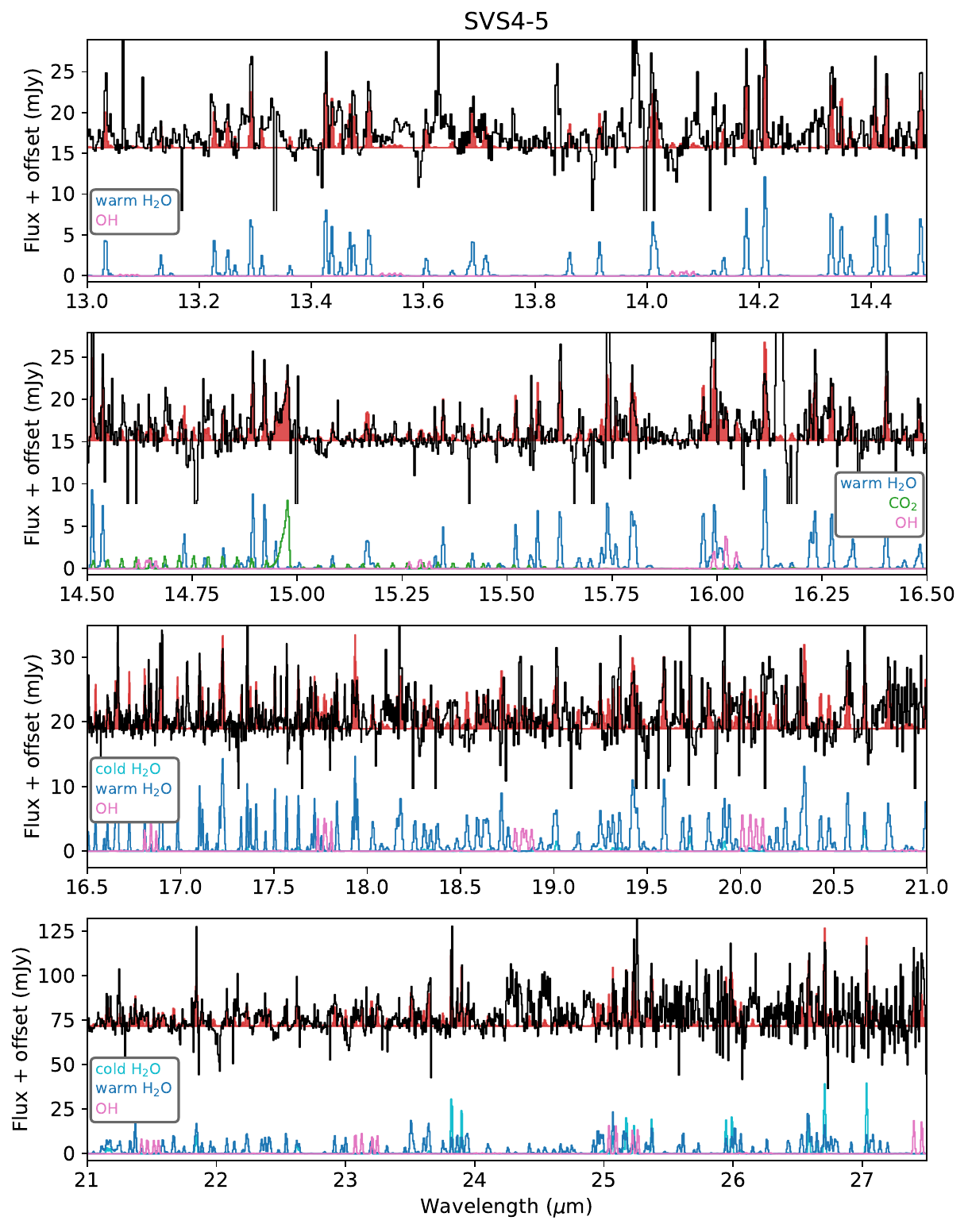}
    \caption{Same as Fig.~\ref{fig:specfitfull_SVS4-5_4.9-13.0} but now for the $13-27.5$~\mum range.}
    \label{fig:specfitfull_SVS4-5_13.0-27.5}
\end{figure*}

\begin{figure*}[h]
    \centering
    \includegraphics[width=\linewidth]{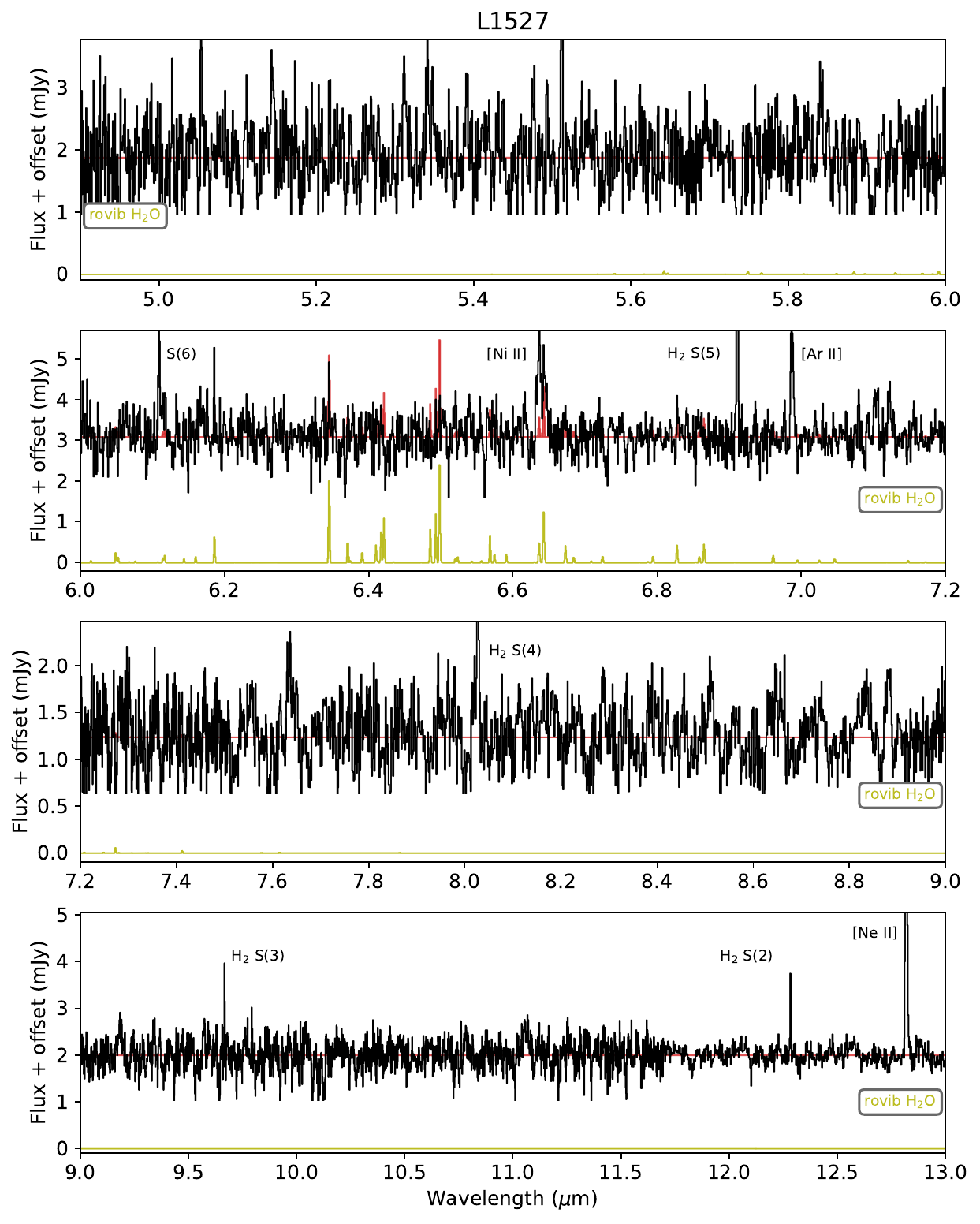}
    \caption{{Baseline-subtracted} spectrum (black) and best-fit LTE models for L1527 in the $4.9-13$~\mum range. In each panel, the red shaded spectrum overlaid on top of the observed spectrum is the full best-fit LTE model. Each molecule contributing to this wavelength range is shown at an arbitrary constant offset in the bottom of each panel. {Strong H$_2$ and atomic emission lines are labeled.}}
    \label{fig:specfitfull_L1527_4.9-13.0}
\end{figure*}

\begin{figure*}[h]
    \centering
    \includegraphics[width=\linewidth]{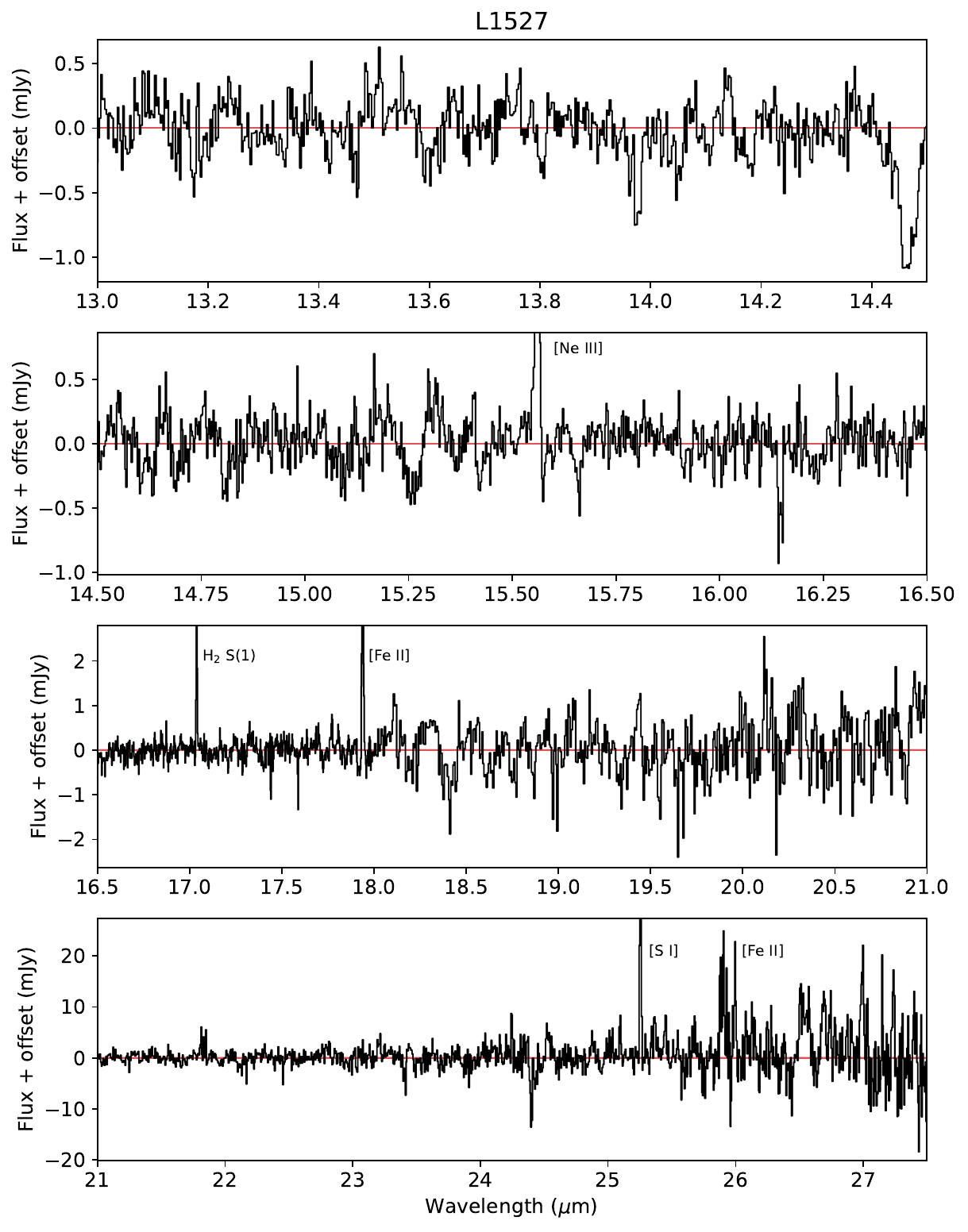}
    \caption{Same as Fig.~\ref{fig:specfitfull_L1527_4.9-13.0} but now for the $13-27.5$~\mum range.}
    \label{fig:specfitfull_L1527_13.0-27.5}
\end{figure*}

\begin{figure*}[h]
    \centering
    \includegraphics[width=\linewidth]{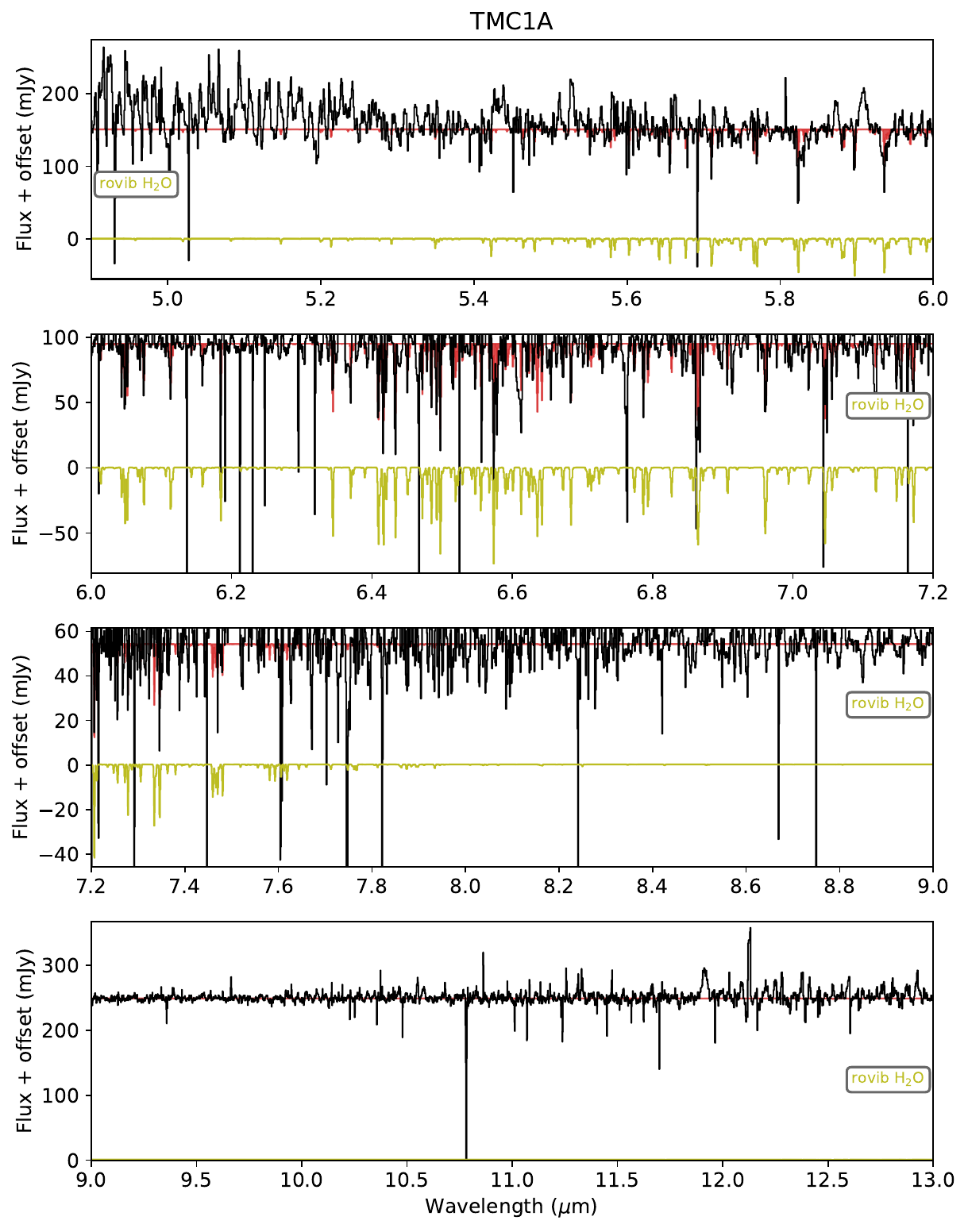}
    \caption{{Baseline-subtracted} spectrum (black) and best-fit LTE models for TMC1A in the $4.9-13$~\mum range. In each panel, the red shaded spectrum overlaid on top of the observed spectrum is the full best-fit LTE model. Each molecule contributing to this wavelength range is shown at an arbitrary constant offset in the bottom of each panel. {Strong H$_2$ and atomic emission lines are labeled.}}
    \label{fig:specfitfull_TMC1A_4.9-13.0}
\end{figure*}

\begin{figure*}[h]
    \centering
    \includegraphics[width=\linewidth]{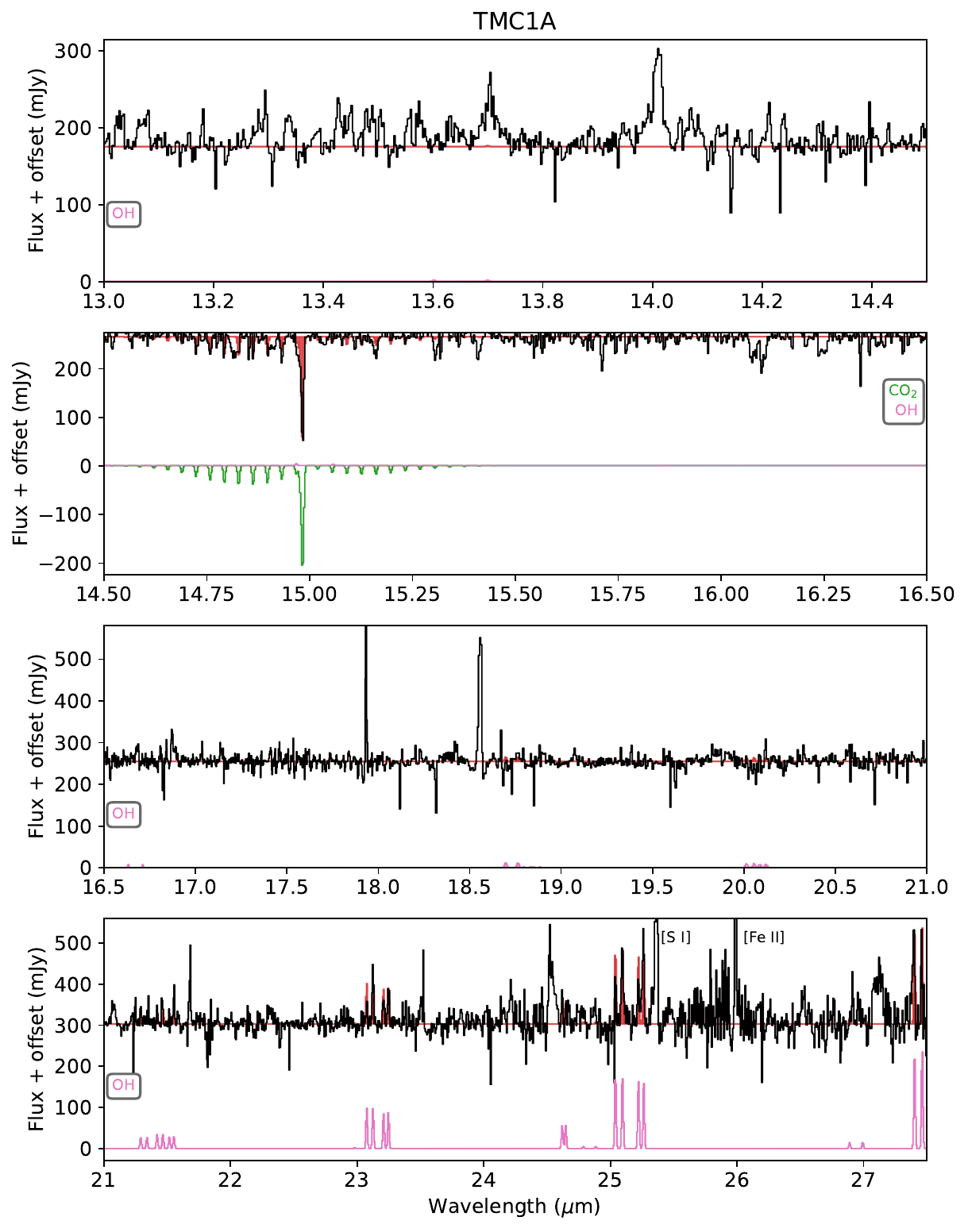}
    \caption{Same as Fig.~\ref{fig:specfitfull_TMC1A_4.9-13.0} but now for the $13-27.5$~\mum range.}
    \label{fig:specfitfull_TMC1A_13.0-27.5}
\end{figure*}

\begin{figure*}[h]
    \centering
    \includegraphics[width=\linewidth]{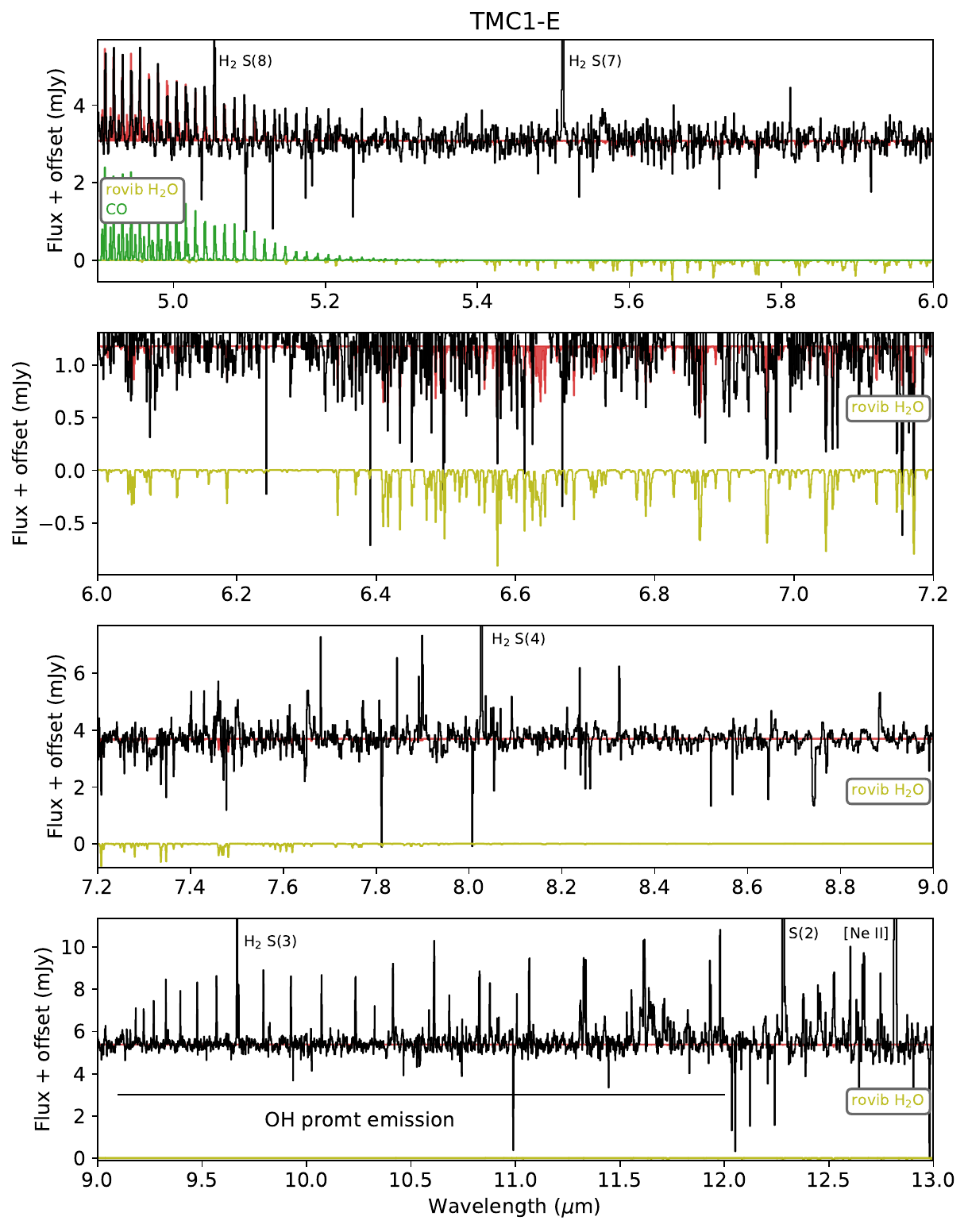}
    \caption{{Baseline-subtracted} spectrum (black) and best-fit LTE models for TMC1-E in the $4.9-13$~\mum range. In each panel, the red shaded spectrum overlaid on top of the observed spectrum is the full best-fit LTE model. Each molecule contributing to this wavelength range is shown at an arbitrary constant offset in the bottom of each panel. {Strong H$_2$ and atomic emission lines are labeled.}}
    \label{fig:specfitfull_TMC1-E_4.9-13.0}
\end{figure*}

\begin{figure*}[h]
    \centering
    \includegraphics[width=\linewidth]{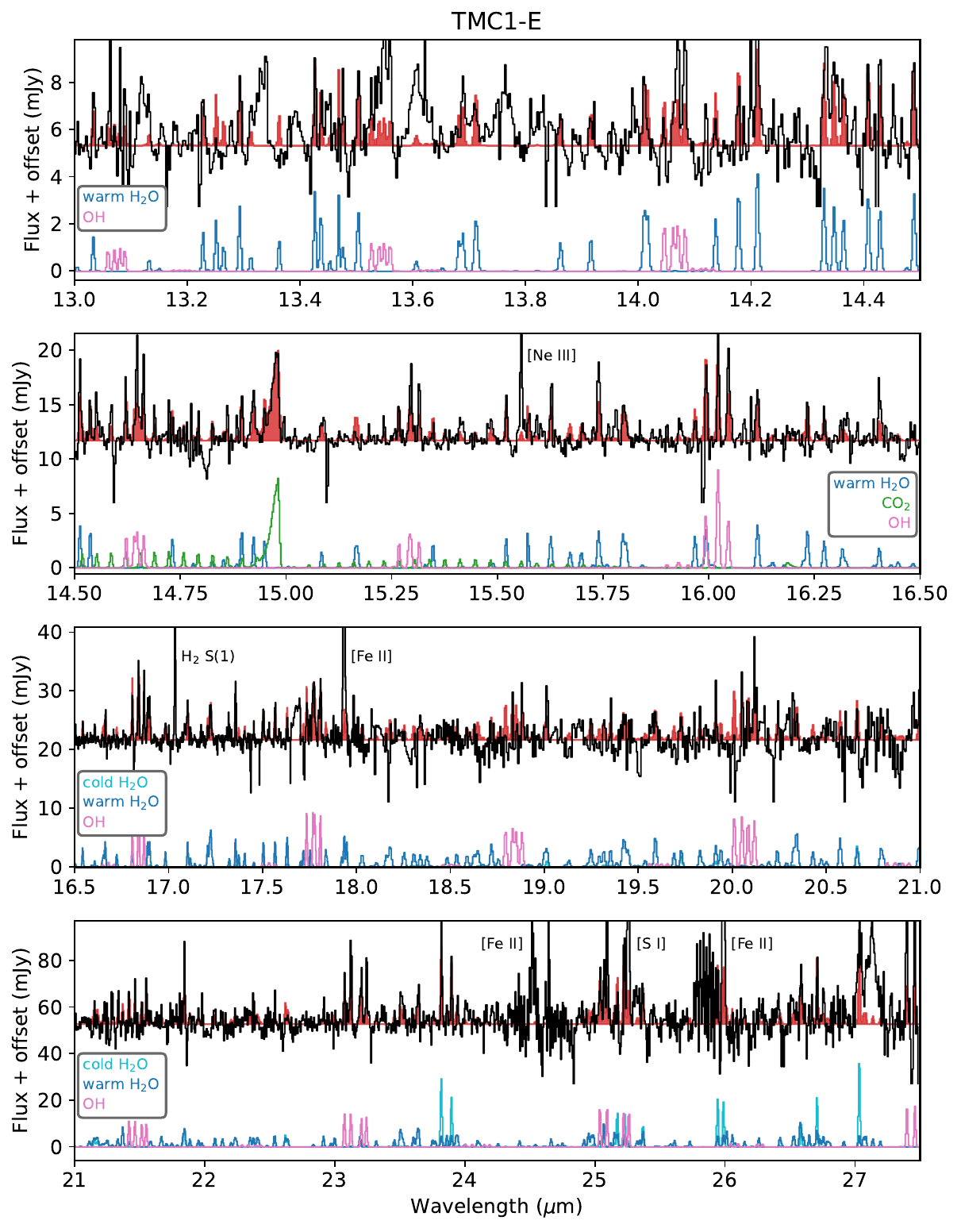}
    \caption{Same as Fig.~\ref{fig:specfitfull_TMC1-E_4.9-13.0} but now for the $13-27.5$~\mum range.}
    \label{fig:specfitfull_TMC1-E_13.0-27.5}
\end{figure*}

\begin{figure*}[h]
    \centering
    \includegraphics[width=\linewidth]{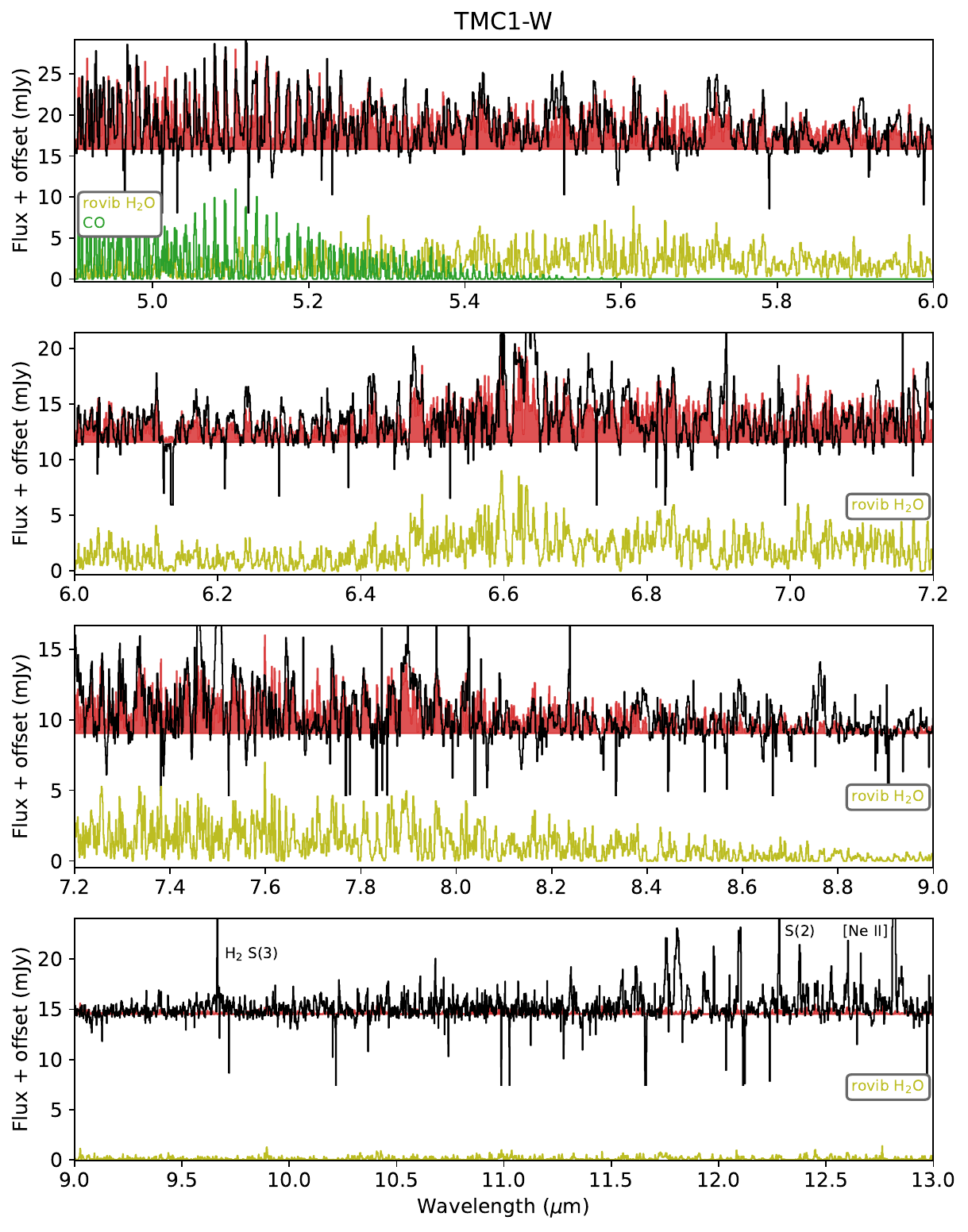}
    \caption{{Baseline-subtracted} spectrum (black) and best-fit LTE models for TMC1-W in the $4.9-13$~\mum range. In each panel, the red shaded spectrum overlaid on top of the observed spectrum is the full best-fit LTE model. Each molecule contributing to this wavelength range is shown at an arbitrary constant offset in the bottom of each panel. {Strong H$_2$ and atomic emission lines are labeled.}}
    \label{fig:specfitfull_TMC1-W_4.9-13.0}
\end{figure*}

\begin{figure*}[h]
    \centering
    \includegraphics[width=\linewidth]{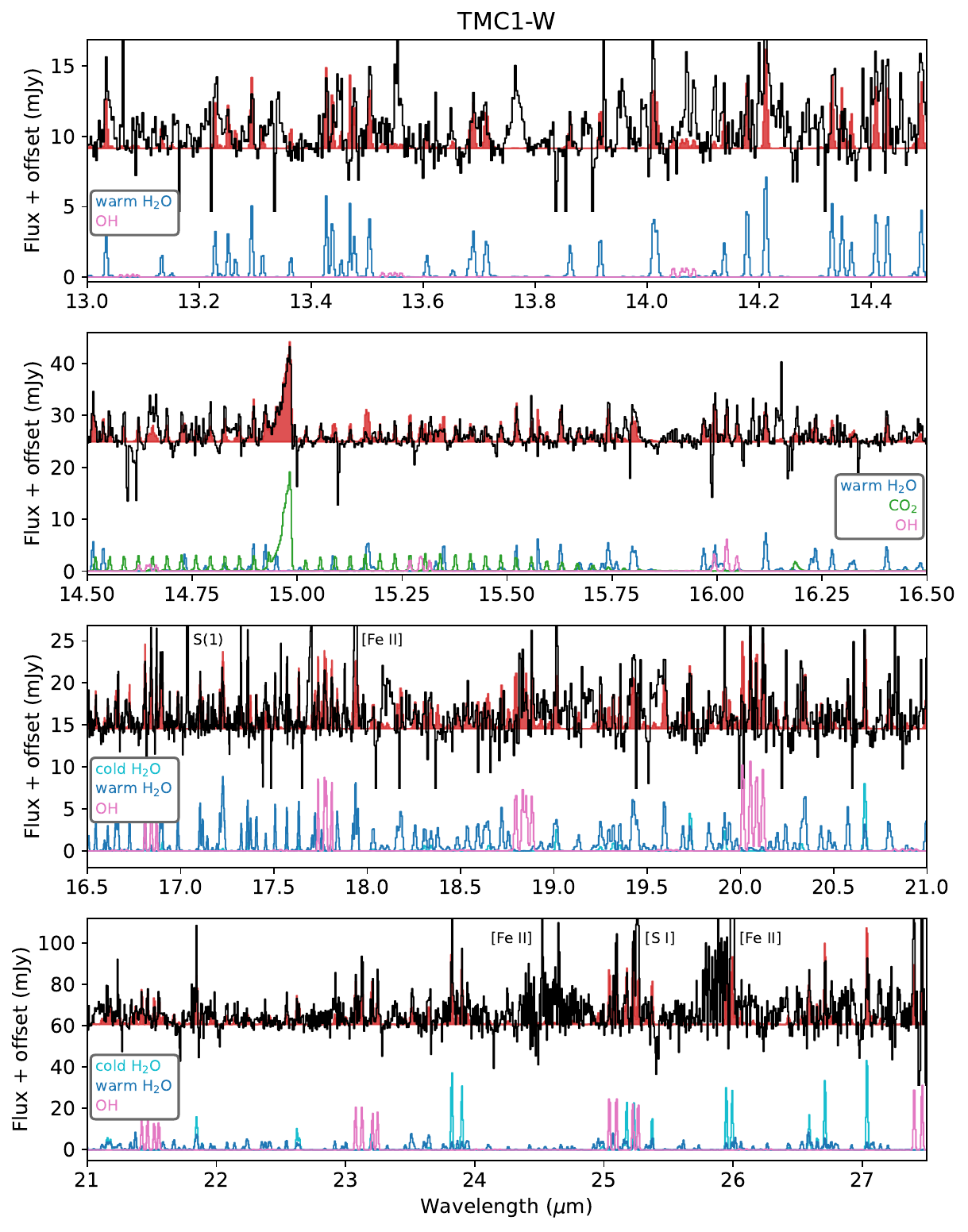}
    \caption{Same as Fig.~\ref{fig:specfitfull_TMC1-W_4.9-13.0} but now for the $13-27.5$~\mum range.}
    \label{fig:specfitfull_TMC1-W_13.0-27.5}
\end{figure*}

\begin{figure*}[h]
    \centering
    \includegraphics[width=\linewidth]{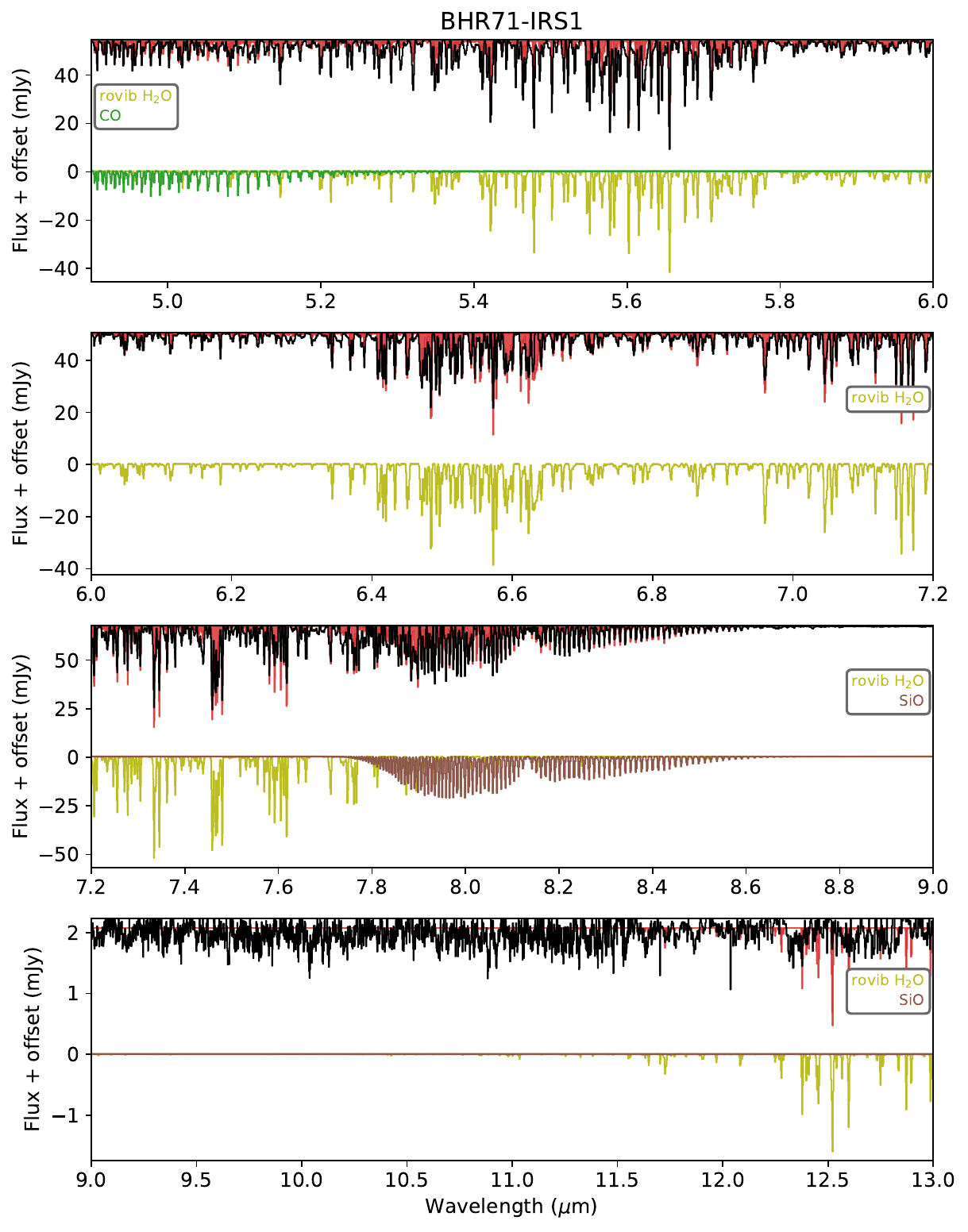}
    \caption{{Baseline-subtracted} spectrum (black) and best-fit LTE models for BHR71-IRS1 in the $4.9-13$~\mum range. In each panel, the red shaded spectrum overlaid on top of the observed spectrum is the full best-fit LTE model. Each molecule contributing to this wavelength range is shown at an arbitrary constant offset in the bottom of each panel. {Strong H$_2$ and atomic emission lines are labeled.}}
    \label{fig:specfitfull_BHR71-IRS1_4.9-13.0}
\end{figure*}

\begin{figure*}[h]
    \centering
    \includegraphics[width=\linewidth]{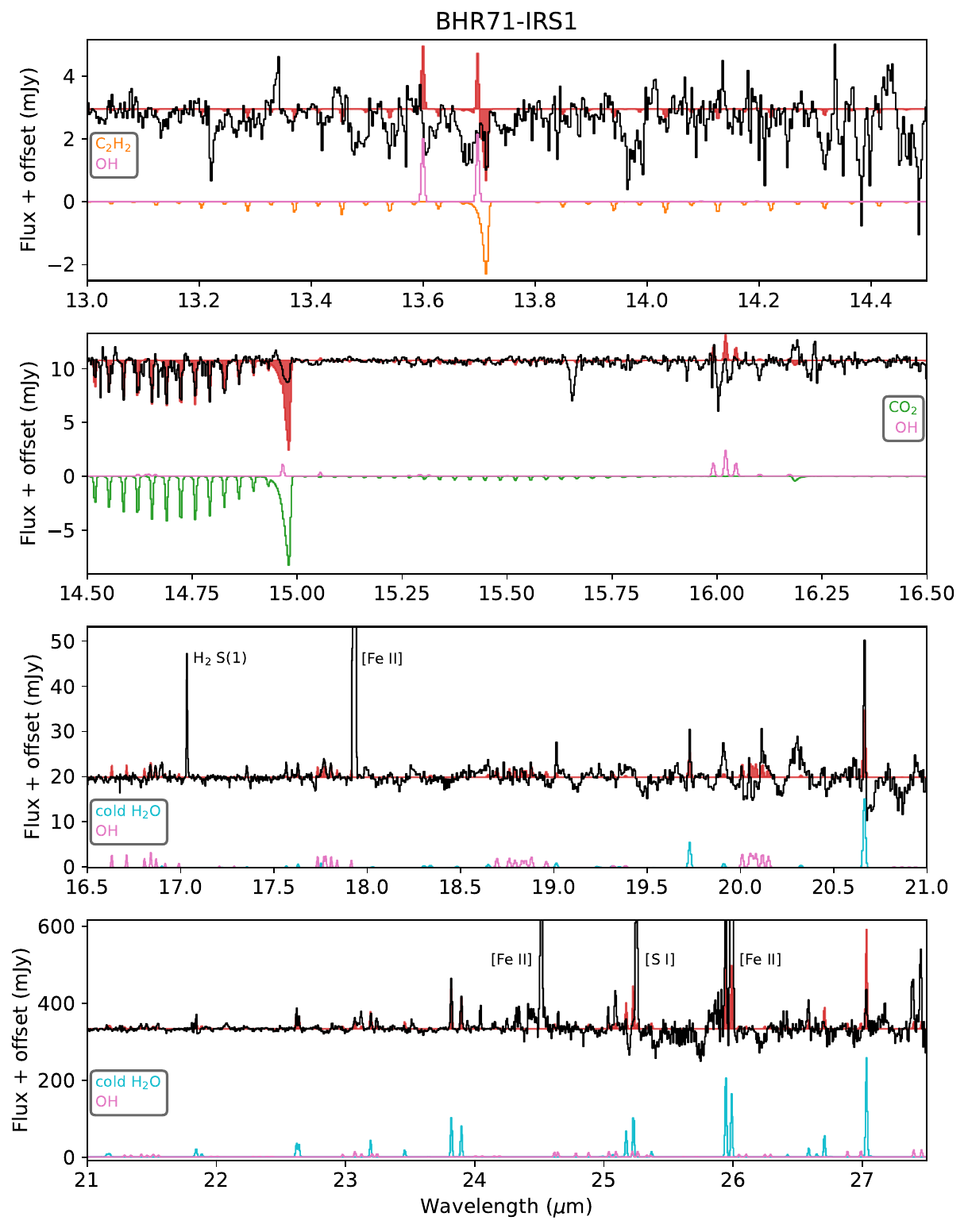}
    \caption{Same as Fig.~\ref{fig:specfitfull_BHR71-IRS1_4.9-13.0} but now for the $13-27.5$~\mum range.}
    \label{fig:specfitfull_BHR71-IRS1_13.0-27.5}
\end{figure*}

\begin{figure*}[h]
    \centering
    \includegraphics[width=\linewidth]{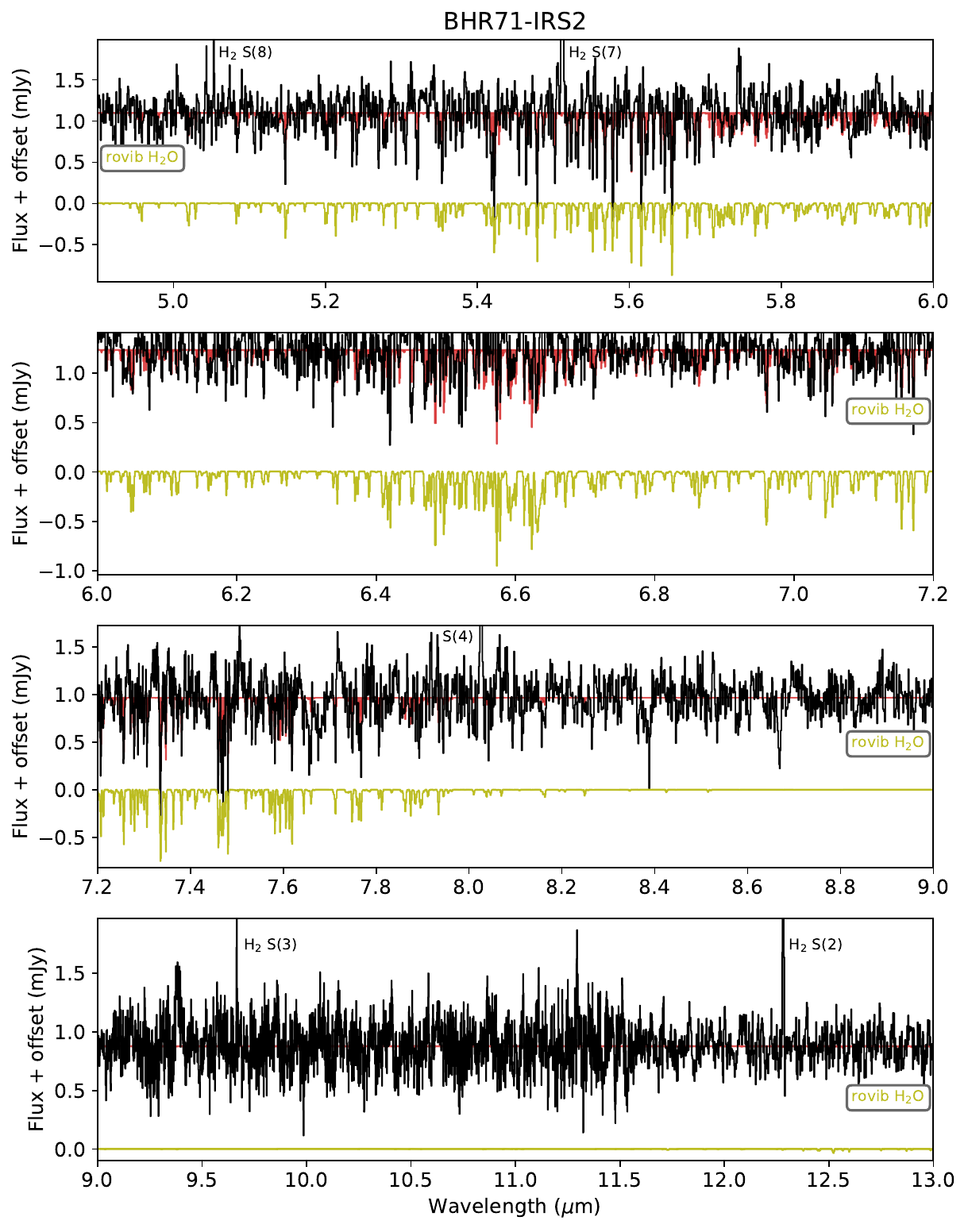}
    \caption{{Baseline-subtracted} spectrum (black) and best-fit LTE models for BHR71-IRS2 in the $4.9-13$~\mum range. In each panel, the red shaded spectrum overlaid on top of the observed spectrum is the full best-fit LTE model. Each molecule contributing to this wavelength range is shown at an arbitrary constant offset in the bottom of each panel. {Strong H$_2$ and atomic emission lines are labeled.}}
    \label{fig:specfitfull_BHR71-IRS2_4.9-13.0}
\end{figure*}

\begin{figure*}[h]
    \centering
    \includegraphics[width=\linewidth]{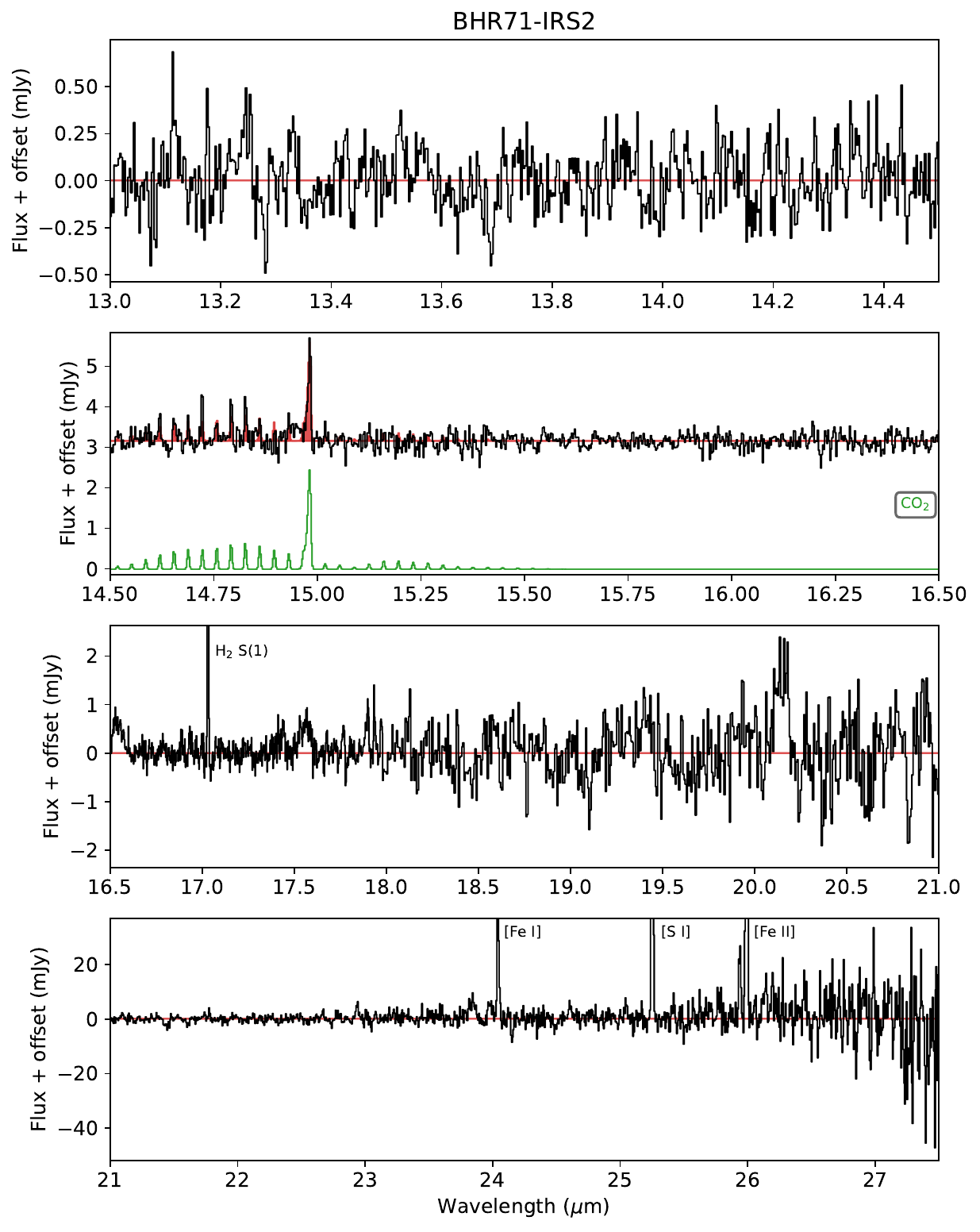}
    \caption{Same as Fig.~\ref{fig:specfitfull_BHR71-IRS2_4.9-13.0} but now for the $13-27.5$~\mum range.}
    \label{fig:specfitfull_BHR71-IRS2_13.0-27.5}
\end{figure*}

{In Figs.~\ref{fig:specfitfull_B1-a-NS_4.9-13.0}-\ref{fig:specfitfull_BHR71-IRS2_13.0-27.5}, the full baseline-subtracted spectra are presented for all sources. The full best-fit LTE model is overlayed as the red shaded region and each species contributing at the LTE model is presented below the observed spectrum. The physical parameters derived from the LTE fits are tabulated in Appendix~\ref{app:LTE_fit_results}.}
\end{multicols}
















% \onecolumn
% \clearpage
\onecolumn
\begin{multicols}{2}
% \nolinenumbers
\section{LTE fit results}
\label{app:LTE_fit_results}
\renewcommand{\arraystretch}{1.2}
Tables~\ref{tab:LTE_results_B1-a-NS}-\ref{tab:LTE_results_BHR71-IRS2} present the derived excitation temperatures $T_{\rm ex}$ and column densities $N$ of the best-fit LTE models, as well as the emitting area ($R_{\rm disk}$) and number of molecules $\mathcal{N}_{\rm mol}$ for emission models. For the latter, also the range $\mathcal{N}_{\rm corr}$ is given when IR pumping is taken into account. Upper limits are presented for molecules that are not detected. 
\end{multicols}

\begin{table*}[h]
    \centering
    \caption{LTE slab model results for B1-a-NS.}
    \scriptsize
    \begin{tabular}{lcccccccccc}
\hline \hline
Species & Mode & Detected & $v_{\rm line}$ & $\Delta V$ & $T_{\rm ex}$ & $N$ & $R_{\rm disk}$ & $\mathcal{N}_{\rm mol}$ & $T_{\rm IR}$ &$\mathcal{N}_{\rm corr}$\\
 & & & (km s$^{-1}$) & (km s$^{-1}$) & (K) & (cm$^{-2}$) & (au) & & (K) & \\
\hline
cold H$_2$O & E & Y & 0 & 4.71 & $160 \pm 10$ & $1.9\pm0.5 \times 10^{18}$ & $25\pm1$ & $7.9\pm0.4 \times 10^{47}$ & -- & -- \\
warm H$_2$O & E & Y & 0 & 4.71 & $370 \pm 10$ & $4.4\pm1.2 \times 10^{19}$ & $2.3\pm0.1$ & $1.6\pm0.1 \times 10^{47}$ & -- & -- \\
rovib H$_2$O & E & Y & -15 & 4.71 & $405 \pm 15$ & $1.9\pm0.5 \times 10^{17}$ & $2.6\pm0.1$ & $7.2\pm1.1 \times 10^{44}$ & 153 & -- \\
CO$_2$ & A & Y & 35 & 2.0 & $115 \pm 25$ & $7.8\pm2.2 \times 10^{15}$ & -- & -- & -- & -- \\
$^{13}$CO$_2$ & A & N & -- & 4.71 & [150] & $<1.3 \times 10^{15}$ & -- & -- & -- & -- \\
C$_2$H$_2$ & A & N & -- & 4.71 & [150] & $<1.0 \times 10^{15}$ & -- & -- & -- & -- \\
$^{13}$CCH$_2$ & A & N & -- & 4.71 & [150] & $<1.8 \times 10^{14}$ & -- & -- & -- & -- \\
HCN & E & N & -- & 4.71 & [150] & $<1.3 \times 10^{16}$ & [$10$] & $<9.4 \times 10^{44}$ & 137 & -- \\
C$_4$H$_2$ & E & N & -- & 4.71 & [150] & $<1.8 \times 10^{15}$ & [$10$] & $<1.3 \times 10^{44}$ & 124 & -- \\
CH$_4$ & E & N & -- & 4.71 & [150] & $<3.2 \times 10^{17}$ & [$10$] & $<2.2 \times 10^{46}$ & 128 & -- \\
SO$_2$ & E & N & -- & 4.71 & [150] & $<1.0 \times 10^{21}$ & [$10$] & $<7.0 \times 10^{49}$ & 132 & -- \\
CS & E & N & -- & 4.71 & [150] & $<1.0 \times 10^{21}$ & [$10$] & $<7.0 \times 10^{49}$ & 124 & -- \\
SiO & E & N & -- & 4.71 & [150] & $<1.0 \times 10^{21}$ & [$10$] & $<7.0 \times 10^{49}$ & 122 & -- \\
NH$_3$ & E & N & -- & 4.71 & [150] & $<5.6 \times 10^{16}$ & [$10$] & $<4.0 \times 10^{45}$ & 151 & $<3.2 \times 10^{45}$ \\
% CO & E & Y & 0 & 4.71 & $1560 \pm 60$ & $<1.5 \times 10^{17}$ & $8.5\pm8.1$ & $2.0\pm0.3 \times 10^{43}$ & 178 & -- \\
% OH & E & Y & 0 & 4.71 & $1120 \pm 60$ & $<3.2 \times 10^{15}$ & $12\pm9$ & $2.9\pm0.7 \times 10^{43}$ & 56 & -- \\
\hline
\end{tabular}
\tablefoot{Square brackets indicate that the parameter was fixed. The second column indicates whether the molecular features were fitted using an absorption (A) or emission (E) model. The velocity $v_{\rm line}$ {was determined through visual inspection and} is with respect to the $v_{\rm lsr}=7.4$~\kms of B1-a-NS. The last column shows the range in total number of molecules corrected for IR pumping using Eq.~\eqref{eq:IR_pump} when $T_{\rm IR} > T_{\rm ex}$.}
\label{tab:LTE_results_B1-a-NS}
\end{table*}

\begin{table*}[h]
    \centering
    \caption{LTE slab model results for B1-b.}
    \scriptsize
    \begin{tabular}{lcccccccccc}
\hline \hline
Species & Mode & Detected & $v_{\rm line}$ & $\Delta V$ & $T_{\rm ex}$ & $N$ & $R_{\rm disk}$ & $\mathcal{N}_{\rm mol}$ & $T_{\rm IR}$ &$\mathcal{N}_{\rm corr}$\\
 & & & (km s$^{-1}$) & (km s$^{-1}$) & (K) & (cm$^{-2}$) & (au) & & (K) & \\
\hline
cold H$_2$O & E & N & -- & 4.71 & [150] & $<3.2 \times 10^{17}$ & [$10$] & $<2.2 \times 10^{46}$ & -- & -- \\
warm H$_2$O & E & N & -- & 4.71 & [450] & $<3.2 \times 10^{15}$ & [$10$] & $<2.2 \times 10^{44}$ & -- & -- \\
rovib H$_2$O & E & N & -- & 4.71 & [150] & $<5.6 \times 10^{20}$ & [$10$] & $<4.0 \times 10^{49}$ & 146 & -- \\
CO$_2$ & E & N & -- & 4.71 & [150] & $<3.2 \times 10^{15}$ & [$10$] & $<2.2 \times 10^{44}$ & 195 & $<1.2 \times 10^{42}$ \\
$^{13}$CO$_2$ & E & N & -- & 4.71 & [150] & $<5.6 \times 10^{15}$ & [$10$] & $<4.0 \times 10^{44}$ & 195 & $<2.2 \times 10^{42}$ \\
C$_2$H$_2$ & E & N & -- & 4.71 & [150] & $<1.8 \times 10^{15}$ & [$10$] & $<1.3 \times 10^{44}$ & 118 & -- \\
$^{13}$CCH$_2$ & E & N & -- & 4.71 & [150] & $<2.4 \times 10^{15}$ & [$10$] & $<1.7 \times 10^{44}$ & 118 & -- \\
HCN & E & N & -- & 4.71 & [150] & $<2.4 \times 10^{15}$ & [$10$] & $<1.7 \times 10^{44}$ & 128 & -- \\
C$_4$H$_2$ & E & N & -- & 4.71 & [150] & $<4.2 \times 10^{14}$ & [$10$] & $<3.0 \times 10^{43}$ & 114 & -- \\
CH$_4$ & E & N & -- & 4.71 & [150] & $<3.2 \times 10^{17}$ & [$10$] & $<2.2 \times 10^{46}$ & 119 & -- \\
SO$_2$ & E & N & -- & 4.71 & [150] & $<1.3 \times 10^{17}$ & [$10$] & $<9.4 \times 10^{45}$ & 123 & -- \\
CS & E & N & -- & 4.71 & [150] & $<4.2 \times 10^{18}$ & [$10$] & $<3.0 \times 10^{47}$ & 115 & -- \\
SiO & E & N & -- & 4.71 & [150] & $<1.0 \times 10^{18}$ & [$10$] & $<7.0 \times 10^{46}$ & 113 & -- \\
NH$_3$ & E & N & -- & 4.71 & [150] & $<1.8 \times 10^{16}$ & [$10$] & $<1.3 \times 10^{45}$ & 144 & -- \\
% CO & E & N & -- & 4.71 & [1500] & $<2.2 \times 10^{14}$ & [$1$] & $<1.5 \times 10^{41}$ & 174 & -- \\
% OH & E & N & -- & 4.71 & [1500] & $<6.8 \times 10^{14}$ & [$1$] & $<4.8 \times 10^{41}$ & 47 & -- \\
\hline
\end{tabular}
\tablefoot{Square brackets indicate that the parameter was fixed. The second column indicates whether the molecular features were fitted using an absorption (A) or emission (E) model. The velocity $v_{\rm line}$ {was determined through visual inspection and} is with respect to the $v_{\rm lsr}=6.0$~\kms of B1-b. The last column shows the range in total number of molecules corrected for IR pumping using Eq.~\eqref{eq:IR_pump} when $T_{\rm IR} > T_{\rm ex}$.}
\label{tab:LTE_results_B1-b}
\end{table*}

\begin{table*}[h]
    \centering
    \caption{LTE slab model results for B1-c.}
    \scriptsize
    \begin{tabular}{lcccccccccc}
\hline \hline
Species & Mode & Detected & $v_{\rm line}$ & $\Delta V$ & $T_{\rm ex}$ & $N$ & $R_{\rm disk}$ & $\mathcal{N}_{\rm mol}$ & $T_{\rm IR}$ &$\mathcal{N}_{\rm corr}$\\
 & & & (km s$^{-1}$) & (km s$^{-1}$) & (K) & (cm$^{-2}$) & (au) & & (K) & \\
\hline
cold H$_2$O & A & N & -- & 4.71 & [150] & $<2.4 \times 10^{17}$ & -- & -- & -- & -- \\
warm H$_2$O & A & Y & $6\pm4$ & 2.0 & $325 \pm 15$ & $4.4\pm1.2 \times 10^{18}$ & -- & -- & -- & -- \\
rovib H$_2$O & A & Y & $-1\pm4$ & 10.0 & $300 \pm 10$ & $7.8\pm2.2 \times 10^{18}$ & -- & -- & -- & -- \\
CO$_2$ & A & Y & $5\pm4$ & 10.0 & $330 \pm 10$ & $3.3\pm0.9 \times 10^{17}$ & -- & -- & -- & -- \\
$^{13}$CO$_2$ & A & Y & 0 & 4.71 & $225 \pm 35$ & $1.4\pm0.4 \times 10^{16}$ & -- & -- & -- & -- \\
C$_2$H$_2$ & A & Y & 0 & 4.71 & $285 \pm 25$ & $1.9\pm0.5 \times 10^{17}$ & -- & -- & -- & -- \\
$^{13}$CCH$_2$ & A & Y & 0 & 2.0 & $100 \pm 20$ & $5.9\pm1.6 \times 10^{15}$ & -- & -- & -- & -- \\
HCN & A & Y & 0 & 4.71 & $180 \pm 10$ & $3.3\pm0.9 \times 10^{17}$ & -- & -- & -- & -- \\
C$_4$H$_2$ & A & N & -- & 4.71 & [150] & $<5.6 \times 10^{14}$ & -- & -- & -- & -- \\
CH$_4$ & A & Y & 0 & 2.0 & $200 \pm 10$ & $4.4\pm1.2 \times 10^{17}$ & -- & -- & -- & -- \\
SO$_2$ & A & Y & 0 & 2.0 & $150 \pm 10$ & $5.9\pm1.6 \times 10^{16}$ & -- & -- & -- & -- \\
CS & A & Y & 0 & 2.0 & $285 \pm 15$ & $2.5\pm0.7 \times 10^{17}$ & -- & -- & -- & -- \\
SiO & A & N & -- & 4.71 & [150] & $<2.4 \times 10^{16}$ & -- & -- & -- & -- \\
NH$_3$ & A & Y & 0 & 4.71 & $340 \pm 10$ & $1.9\pm0.5 \times 10^{17}$ & -- & -- & -- & -- \\
% CO & A & Y & 0 & 20.0 & $1220 \pm 60$ & $1.1\pm0.4 \times 10^{18}$ & -- & -- & -- & -- \\
% OH & A & N & -- & 2.0 & $2350 \pm 150$ & $<4.6 \times 10^{15}$ & -- & -- & -- & -- \\
\hline
\end{tabular}
\tablefoot{Square brackets indicate that the parameter was fixed. The second column indicates whether the molecular features were fitted using an absorption (A) or emission (E) model. The velocity $v_{\rm line}$ {was computed using Gaussian models to selected unblended lines for H$_2$O and CO$_2$ and} is with respect to the $v_{\rm lsr}=6.0$~\kms of B1-c. {For all other species,  $v_{\rm line}$ was determined through visual inspection. } The last column shows the range in total number of molecules corrected for IR pumping using Eq.~\eqref{eq:IR_pump} when $T_{\rm IR} > T_{\rm ex}$.}
\label{tab:LTE_results_B1-c}
\end{table*}

\begin{table*}[h]
    \centering
    \caption{LTE slab model results for L1448-mm.}
    \scriptsize
    \begin{tabular}{lcccccccccc}
\hline \hline
Species & Mode & Detected & $v_{\rm line}$ & $\Delta V$ & $T_{\rm ex}$ & $N$ & $R_{\rm disk}$ & $\mathcal{N}_{\rm mol}$ & $T_{\rm IR}$ &$\mathcal{N}_{\rm corr}$\\
 & & & (km s$^{-1}$) & (km s$^{-1}$) & (K) & (cm$^{-2}$) & (au) & & (K) & \\
\hline
cold H$_2$O & E & Y & $-25\pm3$ & 4.71 & $130 \pm 10$ & $2.5\pm0.7 \times 10^{18}$ & $44\pm2$ & $3.3\pm0.2 \times 10^{48}$ & -- & -- \\
warm H$_2$O & E & Y & $-25\pm3$ & 4.71 & $390 \pm 10$ & $7.8\pm2.2 \times 10^{17}$ & $2.3\pm0.1$ & $2.4\pm0.2 \times 10^{45}$ & -- & -- \\
rovib H$_2$O & E & Y & $-12\pm3$ & 4.71 & $180 \pm 10$ & $4.4\pm1.2 \times 10^{17}$ & $59\pm2$ & $1.1\pm0.1 \times 10^{48}$ & 143 & -- \\
CO$_2$ & E & Y & $2\pm2$ & 4.71 & $120 \pm 10$ & $7.8\pm2.2 \times 10^{17}$ & $37\pm1$ & $7.3\pm0.4 \times 10^{47}$ & 187 & $3.1 \times 10^{43}$--$7.3 \times 10^{47}$ \\
$^{13}$CO$_2$ & E & Y & 0 & 4.71 & [120] & $<1.8 \times 10^{15}$ & $244\pm150$ & $1.1\pm0.1 \times 10^{46}$ & 187 & $4.7 \times 10^{41}$--$1.1 \times 10^{46}$ \\
C$_2$H$_2$ & E & Y & 0 & 4.71 & $110 \pm 10$ & $1.4\pm0.4 \times 10^{17}$ & $49\pm11$ & $2.5\pm1.1 \times 10^{47}$ & 118 & $4.3 \times 10^{46}$--$3.6 \times 10^{47}$ \\
$^{13}$CCH$_2$ & E & Y & 0 & 4.71 & [110] & $<3.2 \times 10^{15}$ & $128\pm87$ & $3.5\pm0.2 \times 10^{45}$ & 118 & $1.0 \times 10^{45}$--$3.6 \times 10^{45}$ \\
HCN & E & Y & 0 & 4.71 & $110 \pm 10$ & $1.4\pm0.4 \times 10^{16}$ & $83\pm4$ & $6.5\pm0.3 \times 10^{46}$ & 127 & $5.3 \times 10^{45}$--$6.5 \times 10^{46}$ \\
C$_4$H$_2$ & E & Y & 0 & 4.71 & [100] & $<2.4 \times 10^{15}$ & $112\pm72$ & $2.5\pm0.1 \times 10^{45}$ & 114 & $2.4 \times 10^{44}$--$2.7 \times 10^{45}$ \\
CH$_4$ & E & Y & 0 & 4.71 & $130 \pm 10$ & $7.8\pm2.2 \times 10^{16}$ & $213\pm10$ & $2.1\pm0.1 \times 10^{48}$ & 119 & -- \\
SO$_2$ & E & Y & -25 & 4.71 & $140 \pm 20$ & $<1.0 \times 10^{17}$ & $1.5\pm1.4 \times 10^{3}$ & $3.6\pm3.5 \times 10^{47}$ & 122 & -- \\
CS & E & N & -- & 4.71 & [150] & $<1.3 \times 10^{19}$ & [$10$] & $<9.4 \times 10^{47}$ & 115 & -- \\
SiO & E & Y & $-24\pm2$ & 4.71 & $310 \pm 10$ & $3.3\pm0.9 \times 10^{17}$ & $1.9\pm0.3$ & $6.9\pm1.5 \times 10^{44}$ & 113 & -- \\
NH$_3$ & E & N & -- & 4.71 & [150] & $<1.8 \times 10^{17}$ & [$10$] & $<1.3 \times 10^{46}$ & 141 & -- \\
% CO & E & Y & -25 & 4.71 & $1910 \pm 50$ & $3.4\pm1.2 \times 10^{17}$ & $0.2\pm0.0$ & $5.3\pm0.6 \times 10^{42}$ & 168 & -- \\
% OH & E & Y & 0 & 4.71 & $1110 \pm 70$ & $<4.6 \times 10^{15}$ & $7.0\pm5.3$ & $8.7\pm2.5 \times 10^{42}$ & 58 & -- \\
\hline
\end{tabular}
\tablefoot{Square brackets indicate that the parameter was fixed. The second column indicates whether the molecular features were fitted using an absorption (A) or emission (E) model. The velocity $v_{\rm line}$ {was computed using Gaussian models to selected unblended lines for H$_2$O, CO$_2$, and SiO and} is with respect to the $v_{\rm lsr}=5.4$~\kms of L1448-mm. {For all other species,  $v_{\rm line}$ was determined through visual inspection.} The last column shows the range in total number of molecules corrected for IR pumping using Eq.~\eqref{eq:IR_pump} when $T_{\rm IR} > T_{\rm ex}$.}
\label{tab:LTE_results_L1448-mm}
\end{table*}

\begin{table*}[h]
    \centering
    \caption{LTE slab model results for Per-emb~8.}
    \scriptsize
    \begin{tabular}{lcccccccccc}
\hline \hline
Species & Mode & Detected & $v_{\rm line}$ & $\Delta V$ & $T_{\rm ex}$ & $N$ & $R_{\rm disk}$ & $\mathcal{N}_{\rm mol}$ & $T_{\rm IR}$ &$\mathcal{N}_{\rm corr}$\\
 & & & (km s$^{-1}$) & (km s$^{-1}$) & (K) & (cm$^{-2}$) & (au) & & (K) & \\
\hline
cold H$_2$O & E & N & -- & 4.71 & [150] & $<1.0 \times 10^{18}$ & [$10$] & $<7.0 \times 10^{46}$ & -- & -- \\
warm H$_2$O & E & N & -- & 4.71 & [450] & $<5.6 \times 10^{15}$ & [$10$] & $<4.0 \times 10^{44}$ & -- & -- \\
rovib H$_2$O & E & N & -- & 4.71 & [150] & $<1.0 \times 10^{21}$ & [$10$] & $<7.0 \times 10^{49}$ & 132 & -- \\
CO$_2$ & E & N & -- & 4.71 & [150] & $<2.4 \times 10^{16}$ & [$10$] & $<1.7 \times 10^{45}$ & 181 & $<3.4 \times 10^{43}$ \\
$^{13}$CO$_2$ & E & N & -- & 4.71 & [150] & $<1.0 \times 10^{16}$ & [$10$] & $<7.0 \times 10^{44}$ & 181 & $<1.4 \times 10^{43}$ \\
C$_2$H$_2$ & E & N & -- & 4.71 & [150] & $<1.3 \times 10^{15}$ & [$10$] & $<9.4 \times 10^{43}$ & 105 & -- \\
$^{13}$CCH$_2$ & E & N & -- & 4.71 & [150] & $<1.8 \times 10^{15}$ & [$10$] & $<1.3 \times 10^{44}$ & 105 & -- \\
HCN & E & N & -- & 4.71 & [150] & $<7.5 \times 10^{15}$ & [$10$] & $<5.3 \times 10^{44}$ & 114 & -- \\
C$_4$H$_2$ & E & N & -- & 4.71 & [150] & $<1.8 \times 10^{15}$ & [$10$] & $<1.3 \times 10^{44}$ & 101 & -- \\
CH$_4$ & E & N & -- & 4.71 & [150] & $<3.2 \times 10^{17}$ & [$10$] & $<2.2 \times 10^{46}$ & 106 & -- \\
SO$_2$ & E & N & -- & 4.71 & [150] & $<7.5 \times 10^{16}$ & [$10$] & $<5.3 \times 10^{45}$ & 110 & -- \\
CS & E & N & -- & 4.71 & [150] & $<1.0 \times 10^{18}$ & [$10$] & $<7.0 \times 10^{46}$ & 102 & -- \\
SiO & E & N & -- & 4.71 & [150] & $<1.0 \times 10^{18}$ & [$10$] & $<7.0 \times 10^{46}$ & 100 & -- \\
NH$_3$ & E & N & -- & 4.71 & [150] & $<1.0 \times 10^{17}$ & [$10$] & $<7.0 \times 10^{45}$ & 130 & -- \\
% CO & E & N & -- & 4.71 & [1500] & $<1.0 \times 10^{15}$ & [$1$] & $<7.0 \times 10^{41}$ & 160 & -- \\
% OH & E & N & -- & 4.71 & [1500] & $<6.8 \times 10^{15}$ & [$1$] & $<4.8 \times 10^{42}$ & 49 & -- \\
\hline
\end{tabular}
\tablefoot{Square brackets indicate that the parameter was fixed. The second column indicates whether the molecular features were fitted using an absorption (A) or emission (E) model. The velocity $v_{\rm line}$ {was determined through visual inspection and} is with respect to the $v_{\rm lsr}=11.0$~\kms of Per-emb8. The last column shows the range in total number of molecules corrected for IR pumping using Eq.~\eqref{eq:IR_pump} when $T_{\rm IR} > T_{\rm ex}$.}
\label{tab:LTE_results_Per-emb8}
\end{table*}

\begin{table*}[h]
    \centering
    \caption{LTE slab model results for Ser-S68N-N.}
    \scriptsize
    \begin{tabular}{lcccccccccc}
\hline \hline
Species & Mode & Detected & $v_{\rm line}$ & $\Delta V$ & $T_{\rm ex}$ & $N$ & $R_{\rm disk}$ & $\mathcal{N}_{\rm mol}$ & $T_{\rm IR}$ &$\mathcal{N}_{\rm corr}$\\
 & & & (km s$^{-1}$) & (km s$^{-1}$) & (K) & (cm$^{-2}$) & (au) & & (K) & \\
\hline
cold H$_2$O & E & Y & 0 & 4.71 & $200 \pm 10$ & $<4.2 \times 10^{17}$ & $15\pm4$ & $3.1\pm0.5 \times 10^{46}$ & -- & -- \\
warm H$_2$O & E & N & -- & 4.71 & [450] & $<5.6 \times 10^{15}$ & [$10$] & $<4.0 \times 10^{44}$ & -- & -- \\
rovib H$_2$O & E & N & -- & 4.71 & [150] & $<1.0 \times 10^{21}$ & [$10$] & $<7.0 \times 10^{49}$ & 143 & -- \\
CO$_2$ & E & N & -- & 4.71 & [150] & $<1.8 \times 10^{15}$ & [$10$] & $<1.3 \times 10^{44}$ & 178 & $<3.6 \times 10^{42}$ \\
$^{13}$CO$_2$ & E & N & -- & 4.71 & [150] & $<5.6 \times 10^{15}$ & [$10$] & $<4.0 \times 10^{44}$ & 178 & $<1.1 \times 10^{43}$ \\
C$_2$H$_2$ & E & N & -- & 4.71 & [150] & $<5.6 \times 10^{14}$ & [$10$] & $<4.0 \times 10^{43}$ & 117 & -- \\
$^{13}$CCH$_2$ & E & N & -- & 4.71 & [150] & $<7.5 \times 10^{14}$ & [$10$] & $<5.3 \times 10^{43}$ & 117 & -- \\
HCN & E & N & -- & 4.71 & [150] & $<1.8 \times 10^{15}$ & [$10$] & $<1.3 \times 10^{44}$ & 126 & -- \\
C$_4$H$_2$ & E & N & -- & 4.71 & [150] & $<4.2 \times 10^{14}$ & [$10$] & $<3.0 \times 10^{43}$ & 113 & -- \\
CH$_4$ & E & N & -- & 4.71 & [150] & $<1.0 \times 10^{18}$ & [$10$] & $<7.0 \times 10^{46}$ & 118 & -- \\
SO$_2$ & E & N & -- & 4.71 & [150] & $<2.4 \times 10^{17}$ & [$10$] & $<1.7 \times 10^{46}$ & 122 & -- \\
CS & E & N & -- & 4.71 & [150] & $<4.2 \times 10^{20}$ & [$10$] & $<3.0 \times 10^{49}$ & 114 & -- \\
SiO & E & N & -- & 4.71 & [150] & $<1.0 \times 10^{19}$ & [$10$] & $<7.0 \times 10^{47}$ & 112 & -- \\
NH$_3$ & E & N & -- & 4.71 & [150] & $<1.8 \times 10^{17}$ & [$10$] & $<1.3 \times 10^{46}$ & 141 & -- \\
% CO & E & N & -- & 4.71 & [1500] & $<1.0 \times 10^{15}$ & [$1$] & $<7.0 \times 10^{41}$ & 166 & -- \\
% OH & E & N & -- & 4.71 & [1500] & $<3.2 \times 10^{15}$ & [$1$] & $<2.2 \times 10^{42}$ & 51 & -- \\
\hline
\end{tabular}
\tablefoot{Square brackets indicate that the parameter was fixed. The second column indicates whether the molecular features were fitted using an absorption (A) or emission (E) model. The velocity $v_{\rm line}$ {was determined through visual inspection and} is with respect to the $v_{\rm lsr}=8.5$~\kms of Ser-S68N-N. The last column shows the range in total number of molecules corrected for IR pumping using Eq.~\eqref{eq:IR_pump} when $T_{\rm IR} > T_{\rm ex}$.}
\label{tab:LTE_results_Ser-S68N-N}
\end{table*}

\begin{table*}[h]
    \centering
    \caption{LTE slab model results for Ser-SMM1A.}
    \scriptsize
    \begin{tabular}{lcccccccccc}
\hline \hline
Species & Mode & Detected & $v_{\rm line}$ & $\Delta V$ & $T_{\rm ex}$ & $N$ & $R_{\rm disk}$ & $\mathcal{N}_{\rm mol}$ & $T_{\rm IR}$ &$\mathcal{N}_{\rm corr}$\\
 & & & (km s$^{-1}$) & (km s$^{-1}$) & (K) & (cm$^{-2}$) & (au) & & (K) & \\
\hline
cold H$_2$O & E & N & -- & 4.71 & [150] & $<2.4 \times 10^{19}$ & [$10$] & $<1.7 \times 10^{48}$ & -- & -- \\
warm H$_2$O & E & N & -- & 4.71 & [450] & $<7.5 \times 10^{15}$ & [$10$] & $<5.3 \times 10^{44}$ & -- & -- \\
rovib H$_2$O & E & N & -- & 4.71 & [150] & $<1.0 \times 10^{21}$ & [$10$] & $<7.0 \times 10^{49}$ & 134 & -- \\
CO$_2$ & E & N & -- & 4.71 & [150] & $<7.5 \times 10^{16}$ & [$10$] & $<5.3 \times 10^{45}$ & 15 & -- \\
$^{13}$CO$_2$ & E & N & -- & 4.71 & [150] & $<2.4 \times 10^{17}$ & [$10$] & $<1.7 \times 10^{46}$ & 15 & -- \\
C$_2$H$_2$ & E & N & -- & 4.71 & [150] & $<2.4 \times 10^{16}$ & [$10$] & $<1.7 \times 10^{45}$ & 113 & -- \\
$^{13}$CCH$_2$ & E & N & -- & 4.71 & [150] & $<1.3 \times 10^{16}$ & [$10$] & $<9.4 \times 10^{44}$ & 113 & -- \\
HCN & E & N & -- & 4.71 & [150] & $<1.3 \times 10^{16}$ & [$10$] & $<9.4 \times 10^{44}$ & 121 & -- \\
C$_4$H$_2$ & E & N & -- & 4.71 & [150] & $<2.4 \times 10^{15}$ & [$10$] & $<1.7 \times 10^{44}$ & 110 & -- \\
CH$_4$ & E & N & -- & 4.71 & [150] & $<1.0 \times 10^{21}$ & [$10$] & $<7.0 \times 10^{49}$ & 114 & -- \\
SO$_2$ & E & N & -- & 4.71 & [150] & $<4.2 \times 10^{18}$ & [$10$] & $<3.0 \times 10^{47}$ & 117 & -- \\
CS & E & N & -- & 4.71 & [150] & $<1.0 \times 10^{21}$ & [$10$] & $<7.0 \times 10^{49}$ & 110 & -- \\
SiO & E & N & -- & 4.71 & [150] & $<1.0 \times 10^{21}$ & [$10$] & $<7.0 \times 10^{49}$ & 109 & -- \\
NH$_3$ & E & N & -- & 4.71 & [150] & $<1.0 \times 10^{21}$ & [$10$] & $<7.0 \times 10^{49}$ & 133 & -- \\
% CO & E & N & -- & 4.71 & [1500] & $<1.0 \times 10^{15}$ & [$1$] & $<7.0 \times 10^{41}$ & 149 & -- \\
% OH & E & N & -- & 4.71 & [1500] & $<1.0 \times 10^{16}$ & [$1$] & $<7.0 \times 10^{42}$ & 62 & -- \\
\hline
\end{tabular}
\tablefoot{Square brackets indicate that the parameter was fixed. The second column indicates whether the molecular features were fitted using an absorption (A) or emission (E) model. The velocity $v_{\rm line}$ {was determined through visual inspection and} is with respect to the $v_{\rm lsr}=8.5$~\kms of Ser-SMM1A. The last column shows the range in total number of molecules corrected for IR pumping using Eq.~\eqref{eq:IR_pump} when $T_{\rm IR} > T_{\rm ex}$.}
\label{tab:LTE_results_Ser-SMM1A}
\end{table*}

\begin{table*}[h]
    \centering
    \caption{LTE slab model results for Ser-SMM1B.}
    \scriptsize
    \begin{tabular}{lcccccccccc}
\hline \hline
Species & Mode & Detected & $v_{\rm line}$ & $\Delta V$ & $T_{\rm ex}$ & $N$ & $R_{\rm disk}$ & $\mathcal{N}_{\rm mol}$ & $T_{\rm IR}$ &$\mathcal{N}_{\rm corr}$\\
 & & & (km s$^{-1}$) & (km s$^{-1}$) & (K) & (cm$^{-2}$) & (au) & & (K) & \\
\hline
cold H$_2$O & E & N & -- & 4.71 & [150] & $<7.5 \times 10^{16}$ & [$10$] & $<5.3 \times 10^{45}$ & -- & -- \\
warm H$_2$O & E & N & -- & 4.71 & [450] & $<3.2 \times 10^{15}$ & [$10$] & $<2.2 \times 10^{44}$ & -- & -- \\
rovib H$_2$O & A & Y & 10 & 2.0 & $980 \pm 20$ & $2.3\pm0.9 \times 10^{18}$ & -- & -- & -- & -- \\
CO$_2$ & E & Y & 10 & 4.71 & $205 \pm 25$ & $<1.8 \times 10^{16}$ & $110\pm97$ & $2.2\pm1.2 \times 10^{45}$ & 15 & -- \\
$^{13}$CO$_2$ & E & N & -- & 4.71 & [150] & $<4.2 \times 10^{16}$ & [$10$] & $<3.0 \times 10^{45}$ & 15 & -- \\
C$_2$H$_2$ & E & N & -- & 4.71 & [150] & $<4.2 \times 10^{15}$ & [$10$] & $<3.0 \times 10^{44}$ & 118 & -- \\
$^{13}$CCH$_2$ & E & N & -- & 4.71 & [150] & $<3.2 \times 10^{15}$ & [$10$] & $<2.2 \times 10^{44}$ & 118 & -- \\
HCN & E & N & -- & 4.71 & [150] & $<7.5 \times 10^{15}$ & [$10$] & $<5.3 \times 10^{44}$ & 129 & -- \\
C$_4$H$_2$ & E & N & -- & 4.71 & [150] & $<3.2 \times 10^{15}$ & [$10$] & $<2.2 \times 10^{44}$ & 114 & -- \\
CH$_4$ & A & N & -- & 4.71 & [150] & $<3.2 \times 10^{15}$ & -- & -- & -- & -- \\
SO$_2$ & E & N & -- & 4.71 & [150] & $<5.6 \times 10^{15}$ & [$10$] & $<4.0 \times 10^{44}$ & 124 & -- \\
CS & A & N & -- & 4.71 & [150] & $<3.2 \times 10^{16}$ & -- & -- & -- & -- \\
SiO & E & N & -- & 4.71 & [150] & $<2.4 \times 10^{17}$ & [$10$] & $<1.7 \times 10^{46}$ & 112 & -- \\
NH$_3$ & E & N & -- & 4.71 & [150] & $<7.5 \times 10^{17}$ & [$10$] & $<5.3 \times 10^{46}$ & 146 & -- \\
% CO & A & Y & 10 & 20.0 & $1810 \pm 130$ & $7.3\pm2.7 \times 10^{17}$ & -- & -- & -- & -- \\
% OH & E & N & -- & 4.71 & [1500] & $<1.0 \times 10^{16}$ & [$1$] & $<7.0 \times 10^{42}$ & 60 & -- \\
\hline
\end{tabular}
\tablefoot{Square brackets indicate that the parameter was fixed. The second column indicates whether the molecular features were fitted using an absorption (A) or emission (E) model. The velocity $v_{\rm line}$ {was determined through visual inspection and} is with respect to the $v_{\rm lsr}=8.5$~\kms of Ser-SMM1B. The last column shows the range in total number of molecules corrected for IR pumping using Eq.~\eqref{eq:IR_pump} when $T_{\rm IR} > T_{\rm ex}$.}
\label{tab:LTE_results_Ser-SMM1B}
\end{table*}

\begin{table*}[h]
    \centering
    \caption{LTE slab model results for Ser-SMM3.}
    \scriptsize
    \begin{tabular}{lcccccccccc}
\hline \hline
Species & Mode & Detected & $v_{\rm line}$ & $\Delta V$ & $T_{\rm ex}$ & $N$ & $R_{\rm disk}$ & $\mathcal{N}_{\rm mol}$ & $T_{\rm IR}$ &$\mathcal{N}_{\rm corr}$\\
 & & & (km s$^{-1}$) & (km s$^{-1}$) & (K) & (cm$^{-2}$) & (au) & & (K) & \\
\hline
cold H$_2$O & E & N & -- & 4.71 & [150] & $<7.5 \times 10^{17}$ & [$10$] & $<5.3 \times 10^{46}$ & -- & -- \\
warm H$_2$O & E & N & -- & 4.71 & [450] & $<7.5 \times 10^{15}$ & [$10$] & $<5.3 \times 10^{44}$ & -- & -- \\
rovib H$_2$O & E & N & -- & 4.71 & [150] & $<1.0 \times 10^{21}$ & [$10$] & $<7.0 \times 10^{49}$ & 136 & -- \\
CO$_2$ & E & Y & 0 & 4.71 & $90 \pm 10$ & $<7.5 \times 10^{15}$ & $1.1\pm0.9 \times 10^{3}$ & $2.9\pm0.2 \times 10^{47}$ & 180 & $2.3 \times 10^{39}$--$3.1 \times 10^{47}$ \\
$^{13}$CO$_2$ & E & N & -- & 4.71 & [150] & $<7.5 \times 10^{15}$ & [$10$] & $<5.3 \times 10^{44}$ & 180 & $<1.2 \times 10^{43}$ \\
C$_2$H$_2$ & E & N & -- & 4.71 & [150] & $<3.2 \times 10^{15}$ & [$10$] & $<2.2 \times 10^{44}$ & 109 & -- \\
$^{13}$CCH$_2$ & E & N & -- & 4.71 & [150] & $<2.4 \times 10^{15}$ & [$10$] & $<1.7 \times 10^{44}$ & 109 & -- \\
HCN & E & N & -- & 4.71 & [150] & $<2.4 \times 10^{15}$ & [$10$] & $<1.7 \times 10^{44}$ & 119 & -- \\
C$_4$H$_2$ & E & N & -- & 4.71 & [150] & $<7.5 \times 10^{14}$ & [$10$] & $<5.3 \times 10^{43}$ & 106 & -- \\
CH$_4$ & E & N & -- & 4.71 & [150] & $<2.4 \times 10^{18}$ & [$10$] & $<1.7 \times 10^{47}$ & 111 & -- \\
SO$_2$ & E & N & -- & 4.71 & [150] & $<4.2 \times 10^{17}$ & [$10$] & $<3.0 \times 10^{46}$ & 115 & -- \\
CS & E & N & -- & 4.71 & [150] & $<1.0 \times 10^{19}$ & [$10$] & $<7.0 \times 10^{47}$ & 106 & -- \\
SiO & E & N & -- & 4.71 & [150] & $<1.0 \times 10^{21}$ & [$10$] & $<7.0 \times 10^{49}$ & 105 & -- \\
NH$_3$ & E & N & -- & 4.71 & [150] & $<1.8 \times 10^{17}$ & [$10$] & $<1.3 \times 10^{46}$ & 134 & -- \\
% CO & E & N & -- & 4.71 & [1500] & $<3.2 \times 10^{15}$ & [$1$] & $<2.2 \times 10^{42}$ & 162 & -- \\
% OH & E & N & -- & 4.71 & [1500] & $<6.8 \times 10^{15}$ & [$1$] & $<4.8 \times 10^{42}$ & 47 & -- \\
\hline
\end{tabular}
\tablefoot{Square brackets indicate that the parameter was fixed. The second column indicates whether the molecular features were fitted using an absorption (A) or emission (E) model. The velocity $v_{\rm line}$ {was determined through visual inspection and} is with respect to the $v_{\rm lsr}=8.5$~\kms of Ser-SMM3. The last column shows the range in total number of molecules corrected for IR pumping using Eq.~\eqref{eq:IR_pump} when $T_{\rm IR} > T_{\rm ex}$.}
\label{tab:LTE_results_Ser-SMM3}
\end{table*}

\begin{table*}[h]
    \centering
    \caption{LTE slab model results for SVS4-5.}
    \scriptsize
    \begin{tabular}{lcccccccccc}
\hline \hline
Species & Mode & Detected & $v_{\rm line}$ & $\Delta V$ & $T_{\rm ex}$ & $N$ & $R_{\rm disk}$ & $\mathcal{N}_{\rm mol}$ & $T_{\rm IR}$ &$\mathcal{N}_{\rm corr}$\\
 & & & (km s$^{-1}$) & (km s$^{-1}$) & (K) & (cm$^{-2}$) & (au) & & (K) & \\
\hline
cold H$_2$O & E & Y & 0 & 4.71 & $210 \pm 10$ & $<3.2 \times 10^{17}$ & $332\pm318$ & $3.3\pm1.1 \times 10^{46}$ & -- & -- \\
warm H$_2$O & E & Y & 0 & 4.71 & $515 \pm 15$ & $1.9\pm0.5 \times 10^{19}$ & $1.8\pm0.1$ & $4.9\pm0.8 \times 10^{46}$ & -- & -- \\
rovib H$_2$O & E & Y & 0 & 4.71 & $1040 \pm 20$ & $2.3\pm0.9 \times 10^{18}$ & $0.5\pm0.1$ & $4.3\pm0.2 \times 10^{44}$ & 170 & -- \\
CO$_2$ & E & Y & -80 & 4.71 & [200] & $<4.2 \times 10^{17}$ & $5.3\pm1.7$ & $2.4\pm1.5 \times 10^{45}$ & 220 & $1.7 \times 10^{44}$--$3.9 \times 10^{45}$ \\
$^{13}$CO$_2$ & E & N & -- & 4.71 & [150] & $<2.4 \times 10^{16}$ & [$10$] & $<1.7 \times 10^{45}$ & 220 & $<1.3 \times 10^{42}$ \\
C$_2$H$_2$ & E & N & -- & 4.71 & [150] & $<3.2 \times 10^{16}$ & [$10$] & $<2.2 \times 10^{45}$ & 138 & -- \\
$^{13}$CCH$_2$ & E & N & -- & 4.71 & [150] & $<1.0 \times 10^{16}$ & [$10$] & $<7.0 \times 10^{44}$ & 138 & -- \\
HCN & E & N & -- & 4.71 & [150] & $<1.0 \times 10^{16}$ & [$10$] & $<7.0 \times 10^{44}$ & 149 & -- \\
C$_4$H$_2$ & E & N & -- & 4.71 & [150] & $<1.0 \times 10^{15}$ & [$10$] & $<7.0 \times 10^{43}$ & 134 & -- \\
CH$_4$ & E & Y & -40 & 4.71 & $460 \pm 80$ & $<5.6 \times 10^{16}$ & $44\pm42$ & $3.5\pm1.9 \times 10^{44}$ & 140 & -- \\
SO$_2$ & E & N & -- & 4.71 & [150] & $<1.0 \times 10^{21}$ & [$10$] & $<7.0 \times 10^{49}$ & 144 & -- \\
CS & E & N & -- & 4.71 & [150] & $<1.0 \times 10^{21}$ & [$10$] & $<7.0 \times 10^{49}$ & 133 & -- \\
SiO & E & N & -- & 4.71 & [150] & $<1.0 \times 10^{21}$ & [$10$] & $<7.0 \times 10^{49}$ & 133 & -- \\
NH$_3$ & E & N & -- & 4.71 & [150] & $<1.3 \times 10^{17}$ & [$10$] & $<9.4 \times 10^{45}$ & 167 & $<1.8 \times 10^{45}$ \\
% CO & E & Y & 0 & 4.71 & $500 \pm 20$ & $8.4\pm1.6 \times 10^{23}$ & $1.2\pm0.1$ & $9.5\pm0.5 \times 10^{50}$ & 193 & -- \\
% OH & E & Y & 0 & 4.71 & $1630 \pm 190$ & $<6.8 \times 10^{15}$ & $8.5\pm6.7$ & $1.3\pm0.5 \times 10^{43}$ & 58 & -- \\
\hline
\end{tabular}
\tablefoot{Square brackets indicate that the parameter was fixed. The second column indicates whether the molecular features were fitted using an absorption (A) or emission (E) model. The velocity $v_{\rm line}$ {was determined through visual inspection and} is with respect to the $v_{\rm lsr}=8.5$~\kms of SVS4-5. The last column shows the range in total number of molecules corrected for IR pumping using Eq.~\eqref{eq:IR_pump} when $T_{\rm IR} > T_{\rm ex}$.}
\label{tab:LTE_results_SVS4-5}
\end{table*}

\begin{table*}[h]
    \centering
    \caption{LTE slab model results for L1527.}
    \scriptsize
    \begin{tabular}{lcccccccccc}
\hline \hline
Species & Mode & Detected & $v_{\rm line}$ & $\Delta V$ & $T_{\rm ex}$ & $N$ & $R_{\rm disk}$ & $\mathcal{N}_{\rm mol}$ & $T_{\rm IR}$ &$\mathcal{N}_{\rm corr}$\\
 & & & (km s$^{-1}$) & (km s$^{-1}$) & (K) & (cm$^{-2}$) & (au) & & (K) & \\
\hline
cold H$_2$O & E & N & -- & 4.71 & [150] & $<5.6 \times 10^{16}$ & [$10$] & $<4.0 \times 10^{45}$ & -- & -- \\
warm H$_2$O & E & N & -- & 4.71 & [450] & $<1.8 \times 10^{14}$ & [$10$] & $<1.3 \times 10^{43}$ & -- & -- \\
rovib H$_2$O & E & Y & 0 & 4.71 & $90 \pm 10$ & $<4.2 \times 10^{16}$ & $1.3\pm1.2 \times 10^{5}$ & $7.1\pm7.0 \times 10^{52}$ & 132 & $8.2 \times 10^{46}$--$1.4 \times 10^{53}$ \\
CO$_2$ & E & N & -- & 4.71 & [150] & $<1.0 \times 10^{15}$ & [$10$] & $<7.0 \times 10^{43}$ & 15 & -- \\
$^{13}$CO$_2$ & E & N & -- & 4.71 & [150] & $<5.6 \times 10^{14}$ & [$10$] & $<4.0 \times 10^{43}$ & 15 & -- \\
C$_2$H$_2$ & E & N & -- & 4.71 & [150] & $<1.3 \times 10^{14}$ & [$10$] & $<9.4 \times 10^{42}$ & 106 & -- \\
$^{13}$CCH$_2$ & E & N & -- & 4.71 & [150] & $<2.4 \times 10^{14}$ & [$10$] & $<1.7 \times 10^{43}$ & 106 & -- \\
HCN & E & N & -- & 4.71 & [150] & $<2.4 \times 10^{14}$ & [$10$] & $<1.7 \times 10^{43}$ & 115 & -- \\
C$_4$H$_2$ & E & N & -- & 4.71 & [150] & $<1.0 \times 10^{14}$ & [$10$] & $<7.0 \times 10^{42}$ & 102 & -- \\
CH$_4$ & E & N & -- & 4.71 & [150] & $<7.5 \times 10^{16}$ & [$10$] & $<5.3 \times 10^{45}$ & 107 & -- \\
SO$_2$ & E & N & -- & 4.71 & [150] & $<4.2 \times 10^{16}$ & [$10$] & $<3.0 \times 10^{45}$ & 111 & -- \\
CS & E & N & -- & 4.71 & [150] & $<7.5 \times 10^{16}$ & [$10$] & $<5.3 \times 10^{45}$ & 103 & -- \\
SiO & E & N & -- & 4.71 & [150] & $<1.3 \times 10^{17}$ & [$10$] & $<9.4 \times 10^{45}$ & 101 & -- \\
NH$_3$ & E & N & -- & 4.71 & [150] & $<1.8 \times 10^{16}$ & [$10$] & $<1.3 \times 10^{45}$ & 130 & -- \\
% CO & E & N & -- & 4.71 & [1500] & $<4.6 \times 10^{14}$ & [$1$] & $<3.3 \times 10^{41}$ & 148 & -- \\
% OH & E & N & -- & 4.71 & [1500] & $<3.2 \times 10^{14}$ & [$1$] & $<2.2 \times 10^{41}$ & 53 & -- \\
\hline
\end{tabular}
\tablefoot{Square brackets indicate that the parameter was fixed. The second column indicates whether the molecular features were fitted using an absorption (A) or emission (E) model. The velocity $v_{\rm line}$ {was determined through visual inspection and} is with respect to the $v_{\rm lsr}=5.9$~\kms of L1527. The last column shows the range in total number of molecules corrected for IR pumping using Eq.~\eqref{eq:IR_pump} when $T_{\rm IR} > T_{\rm ex}$.}
\label{tab:LTE_results_L1527}
\end{table*}

\begin{table*}[h]
    \centering
    \caption{LTE slab model results for TMC1A.}
    \scriptsize
    \begin{tabular}{lcccccccccc}
\hline \hline
Species & Mode & Detected & $v_{\rm line}$ & $\Delta V$ & $T_{\rm ex}$ & $N$ & $R_{\rm disk}$ & $\mathcal{N}_{\rm mol}$ & $T_{\rm IR}$ &$\mathcal{N}_{\rm corr}$\\
 & & & (km s$^{-1}$) & (km s$^{-1}$) & (K) & (cm$^{-2}$) & (au) & & (K) & \\
\hline
cold H$_2$O & A & N & -- & 4.71 & [150] & $<1.0 \times 10^{17}$ & -- & -- & -- & -- \\
warm H$_2$O & A & N & -- & 4.71 & [450] & $<1.8 \times 10^{16}$ & -- & -- & -- & -- \\
rovib H$_2$O & A & Y & -20 & 20.0 & $440 \pm 10$ & $1.4\pm0.4 \times 10^{17}$ & -- & -- & -- & -- \\
CO$_2$ & A & Y & 0 & 2.0 & $60 \pm 10$ & $1.4\pm0.4 \times 10^{16}$ & -- & -- & -- & -- \\
$^{13}$CO$_2$ & A & N & -- & 4.71 & [150] & $<1.3 \times 10^{15}$ & -- & -- & -- & -- \\
C$_2$H$_2$ & A & N & -- & 4.71 & [150] & $<7.5 \times 10^{14}$ & -- & -- & -- & -- \\
$^{13}$CCH$_2$ & A & N & -- & 4.71 & [150] & $<2.4 \times 10^{14}$ & -- & -- & -- & -- \\
HCN & A & N & -- & 4.71 & [150] & $<7.5 \times 10^{14}$ & -- & -- & -- & -- \\
C$_4$H$_2$ & A & N & -- & 4.71 & [150] & $<2.4 \times 10^{14}$ & -- & -- & -- & -- \\
CH$_4$ & A & N & -- & 4.71 & [150] & $<5.6 \times 10^{15}$ & -- & -- & -- & -- \\
SO$_2$ & A & N & -- & 4.71 & [150] & $<3.2 \times 10^{15}$ & -- & -- & -- & -- \\
CS & A & N & -- & 4.71 & [150] & $<2.4 \times 10^{15}$ & -- & -- & -- & -- \\
SiO & A & N & -- & 4.71 & [150] & $<4.2 \times 10^{15}$ & -- & -- & -- & -- \\
NH$_3$ & A & N & -- & 4.71 & [150] & $<2.4 \times 10^{15}$ & -- & -- & -- & -- \\
% CO & E & N & -- & 4.71 & $1760 \pm 740$ & $<1.0 \times 10^{24}$ & $0.2\pm0.1$ & $<7.5 \times 10^{49}$ & 224 & -- \\
% OH & E & Y & 0 & 4.71 & $530 \pm 90$ & $<6.8 \times 10^{17}$ & $66\pm63$ & $2.0\pm1.7 \times 10^{45}$ & 69 & -- \\
\hline
\end{tabular}
\tablefoot{Square brackets indicate that the parameter was fixed. The second column indicates whether the molecular features were fitted using an absorption (A) or emission (E) model. The velocity $v_{\rm line}$ {was determined through visual inspection and} is with respect to the $v_{\rm lsr}=6.6$~\kms of TMC1A. The last column shows the range in total number of molecules corrected for IR pumping using Eq.~\eqref{eq:IR_pump} when $T_{\rm IR} > T_{\rm ex}$.}
\label{tab:LTE_results_TMC1A}
\end{table*}

\begin{table*}[h]
    \centering
    \caption{LTE slab model results for TMC1-E.}
    \scriptsize
    \begin{tabular}{lcccccccccc}
\hline \hline
Species & Mode & Detected & $v_{\rm line}$ & $\Delta V$ & $T_{\rm ex}$ & $N$ & $R_{\rm disk}$ & $\mathcal{N}_{\rm mol}$ & $T_{\rm IR}$ &$\mathcal{N}_{\rm corr}$\\
 & & & (km s$^{-1}$) & (km s$^{-1}$) & (K) & (cm$^{-2}$) & (au) & & (K) & \\
\hline
cold H$_2$O & E & Y & 0 & 4.71 & $185 \pm 15$ & $<4.2 \times 10^{17}$ & $155\pm150$ & $8.4\pm5.1 \times 10^{45}$ & -- & -- \\
warm H$_2$O & E & Y & 0 & 4.71 & $405 \pm 15$ & $1.0\pm0.3 \times 10^{20}$ & $0.5\pm0.1$ & $1.5\pm0.5 \times 10^{46}$ & -- & -- \\
rovib H$_2$O & A & Y & -20 & 2.0 & $580 \pm 20$ & $1.0\pm0.3 \times 10^{17}$ & -- & -- & -- & -- \\
CO$_2$ & E & Y & 0 & 4.71 & $315 \pm 25$ & $<1.3 \times 10^{17}$ & $0.8\pm0.2$ & $2.9\pm1.3 \times 10^{43}$ & 212 & -- \\
$^{13}$CO$_2$ & E & N & -- & 4.71 & [150] & $<1.0 \times 10^{15}$ & [$10$] & $<7.0 \times 10^{43}$ & 212 & $<1.0 \times 10^{41}$ \\
C$_2$H$_2$ & E & N & -- & 4.71 & [150] & $<3.2 \times 10^{14}$ & [$10$] & $<2.2 \times 10^{43}$ & 137 & -- \\
$^{13}$CCH$_2$ & E & N & -- & 4.71 & [150] & $<5.6 \times 10^{14}$ & [$10$] & $<4.0 \times 10^{43}$ & 137 & -- \\
HCN & E & N & -- & 4.71 & [150] & $<3.2 \times 10^{14}$ & [$10$] & $<2.2 \times 10^{43}$ & 146 & -- \\
C$_4$H$_2$ & E & N & -- & 4.71 & [150] & $<1.3 \times 10^{14}$ & [$10$] & $<9.4 \times 10^{42}$ & 133 & -- \\
CH$_4$ & E & N & -- & 4.71 & [150] & $<1.3 \times 10^{17}$ & [$10$] & $<9.4 \times 10^{45}$ & 138 & -- \\
SO$_2$ & E & N & -- & 4.71 & [150] & $<7.5 \times 10^{15}$ & [$10$] & $<5.3 \times 10^{44}$ & 142 & -- \\
CS & E & N & -- & 4.71 & [150] & $<3.2 \times 10^{15}$ & [$10$] & $<2.2 \times 10^{44}$ & 133 & -- \\
SiO & E & N & -- & 4.71 & [150] & $<1.3 \times 10^{17}$ & [$10$] & $<9.4 \times 10^{45}$ & 132 & -- \\
NH$_3$ & E & N & -- & 4.71 & [150] & $<2.4 \times 10^{15}$ & [$10$] & $<1.7 \times 10^{44}$ & 162 & $<5.0 \times 10^{43}$ \\
% CO & E & Y & 10 & 4.71 & $1070 \pm 90$ & $<1.5 \times 10^{18}$ & $0.10\pm0.10$ & $7.7\pm3.5 \times 10^{42}$ & 191 & -- \\
% OH & E & Y & 0 & 4.71 & $2040 \pm 200$ & $<1.0 \times 10^{16}$ & $2.4\pm2.0$ & $1.3\pm0.3 \times 10^{42}$ & 64 & -- \\
\hline
\end{tabular}
\tablefoot{Square brackets indicate that the parameter was fixed. The second column indicates whether the molecular features were fitted using an absorption (A) or emission (E) model. The velocity $v_{\rm line}$ {was determined through visual inspection and} is with respect to the $v_{\rm lsr}=5.4$~\kms of TMC1-E. The last column shows the range in total number of molecules corrected for IR pumping using Eq.~\eqref{eq:IR_pump} when $T_{\rm IR} > T_{\rm ex}$.}
\label{tab:LTE_results_TMC1-E}
\end{table*}

\begin{table*}[h]
    \centering
    \caption{LTE slab model results for TMC1-W.}
    \scriptsize
    \begin{tabular}{lcccccccccc}
\hline \hline
Species & Mode & Detected & $v_{\rm line}$ & $\Delta V$ & $T_{\rm ex}$ & $N$ & $R_{\rm disk}$ & $\mathcal{N}_{\rm mol}$ & $T_{\rm IR}$ &$\mathcal{N}_{\rm corr}$\\
 & & & (km s$^{-1}$) & (km s$^{-1}$) & (K) & (cm$^{-2}$) & (au) & & (K) & \\
\hline
cold H$_2$O & E & Y & 0 & 4.71 & $190 \pm 10$ & $<5.6 \times 10^{17}$ & $6.4\pm1.9$ & $5.7\pm2.1 \times 10^{45}$ & -- & -- \\
warm H$_2$O & E & Y & 0 & 4.71 & $485 \pm 15$ & $4.4\pm1.2 \times 10^{19}$ & $0.3\pm0.1$ & $2.6\pm0.6 \times 10^{45}$ & -- & -- \\
rovib H$_2$O & E & Y & 0 & 4.71 & $1180 \pm 20$ & $1.1\pm0.4 \times 10^{20}$ & $0.06\pm0.10$ & $2.3\pm0.1 \times 10^{44}$ & 180 & -- \\
CO$_2$ & E & Y & 0 & 4.71 & $190 \pm 10$ & $1.9\pm0.5 \times 10^{18}$ & $1.1\pm0.1$ & $1.8\pm0.9 \times 10^{45}$ & 236 & $2.9 \times 10^{43}$--$2.7 \times 10^{45}$ \\
$^{13}$CO$_2$ & E & N & -- & 4.71 & [150] & $<4.2 \times 10^{14}$ & [$10$] & $<3.0 \times 10^{43}$ & 236 & $<8.3 \times 10^{39}$ \\
C$_2$H$_2$ & E & N & -- & 4.71 & [150] & $<2.4 \times 10^{14}$ & [$10$] & $<1.7 \times 10^{43}$ & 148 & -- \\
$^{13}$CCH$_2$ & E & N & -- & 4.71 & [150] & $<1.3 \times 10^{14}$ & [$10$] & $<9.4 \times 10^{42}$ & 148 & -- \\
HCN & E & N & -- & 4.71 & [150] & $<3.2 \times 10^{14}$ & [$10$] & $<2.2 \times 10^{43}$ & 159 & $<9.8 \times 10^{42}$ \\
C$_4$H$_2$ & E & N & -- & 4.71 & [150] & $<1.3 \times 10^{14}$ & [$10$] & $<9.4 \times 10^{42}$ & 143 & -- \\
CH$_4$ & E & N & -- & 4.71 & [150] & $<1.3 \times 10^{18}$ & [$10$] & $<9.4 \times 10^{46}$ & 149 & -- \\
SO$_2$ & E & N & -- & 4.71 & [150] & $<1.0 \times 10^{17}$ & [$10$] & $<7.0 \times 10^{45}$ & 154 & $<4.8 \times 10^{45}$ \\
CS & E & N & -- & 4.71 & [150] & $<5.6 \times 10^{17}$ & [$10$] & $<4.0 \times 10^{46}$ & 144 & -- \\
SiO & E & N & -- & 4.71 & [150] & $<1.8 \times 10^{17}$ & [$10$] & $<1.3 \times 10^{46}$ & 142 & -- \\
NH$_3$ & E & N & -- & 4.71 & [150] & $<3.2 \times 10^{15}$ & [$10$] & $<2.2 \times 10^{44}$ & 177 & $<1.9 \times 10^{43}$ \\
% CO & E & Y & 10 & 4.71 & $2280 \pm 200$ & $7.3\pm2.7 \times 10^{18}$ & $0.06\pm0.10$ & $1.3\pm0.4 \times 10^{43}$ & 205 & -- \\
% OH & E & Y & 0 & 4.71 & $1360 \pm 60$ & $<1.5 \times 10^{15}$ & $3.5\pm2.1$ & $1.9\pm0.3 \times 10^{42}$ & 65 & -- \\
\hline
\end{tabular}
\tablefoot{Square brackets indicate that the parameter was fixed. The second column indicates whether the molecular features were fitted using an absorption (A) or emission (E) model. The velocity $v_{\rm line}$ {was determined through visual inspection and} is with respect to the $v_{\rm lsr}=5.4$~\kms of TMC1-W. The last column shows the range in total number of molecules corrected for IR pumping using Eq.~\eqref{eq:IR_pump} when $T_{\rm IR} > T_{\rm ex}$.}
\label{tab:LTE_results_TMC1-W}
\end{table*}

\begin{table*}[h]
    \centering
    \caption{LTE slab model results for BHR71-IRS1.}
    \scriptsize
    \begin{tabular}{lcccccccccc}
\hline \hline
Species & Mode & Detected & $v_{\rm line}$ & $\Delta V$ & $T_{\rm ex}$ & $N$ & $R_{\rm disk}$ & $\mathcal{N}_{\rm mol}$ & $T_{\rm IR}$ &$\mathcal{N}_{\rm corr}$\\
 & & & (km s$^{-1}$) & (km s$^{-1}$) & (K) & (cm$^{-2}$) & (au) & & (K) & \\
\hline
cold H$_2$O & E & Y & $-42\pm3$ & 4.71 & $140 \pm 10$ & $3.3\pm0.9 \times 10^{18}$ & $24\pm1$ & $1.3\pm0.1 \times 10^{48}$ & -- & -- \\
warm H$_2$O & E & N & -- & 4.71 & [450] & $<4.2 \times 10^{15}$ & [$10$] & $<3.0 \times 10^{44}$ & -- & -- \\
rovib H$_2$O & A & Y & $-44\pm9$ & 10.0 & $560 \pm 10$ & $5.9\pm1.6 \times 10^{18}$ & -- & -- & -- & -- \\
CO$_2$ & A & Y & $-10\pm4$ & 10.0 & $260 \pm 10$ & $3.3\pm0.9 \times 10^{16}$ & -- & -- & -- & -- \\
$^{13}$CO$_2$ & A & N & -- & 10.0 & [150] & $<3.2 \times 10^{14}$ & -- & -- & -- & -- \\
C$_2$H$_2$ & A & Y & -5 & 10.0 & [150] & $1.0\pm0.3 \times 10^{15}$ & -- & -- & -- & -- \\
$^{13}$CCH$_2$ & A & N & -- & 10.0 & [150] & $<3.2 \times 10^{14}$ & -- & -- & -- & -- \\
HCN & A & N & -- & 10.0 & [150] & $<1.0 \times 10^{15}$ & -- & -- & -- & -- \\
C$_4$H$_2$ & A & N & -- & 10.0 & [150] & $<3.2 \times 10^{14}$ & -- & -- & -- & -- \\
CH$_4$ & A & N & -- & 10.0 & [150] & $<4.2 \times 10^{16}$ & -- & -- & -- & -- \\
SO$_2$ & A & N & -- & 10.0 & [150] & $<1.8 \times 10^{16}$ & -- & -- & -- & -- \\
CS & A & N & -- & 10.0 & [150] & $<2.4 \times 10^{16}$ & -- & -- & -- & -- \\
SiO & A & Y & $-23\pm2$ & 4.71 & $400 \pm 20$ & $1.9\pm0.5 \times 10^{18}$ & -- & -- & -- & -- \\
NH$_3$ & A & N & -- & 10.0 & [150] & $<3.2 \times 10^{16}$ & -- & -- & -- & -- \\
% CO & A & Y & -45 & 10.0 & $1400 \pm 20$ & $5.0\pm1.8 \times 10^{18}$ & -- & -- & -- & -- \\
% OH & E & Y & -85 & 4.71 & $380 \pm 20$ & $8.4\pm1.6 \times 10^{23}$ & $1.0\pm0.1$ & $6.8\pm0.6 \times 10^{50}$ & 65 & -- \\
\hline
\end{tabular}
\tablefoot{Square brackets indicate that the parameter was fixed. The second column indicates whether the molecular features were fitted using an absorption (A) or emission (E) model. The velocity $v_{\rm line}$ {was computed using Gaussian models to selected unblended lines for H$_2$O, CO$_2$, and SiO and} is with respect to the $v_{\rm lsr}=-4.7$~\kms of BHR71-IRS1. {For all other species, $v_{\rm line}$ was determined through visual inspection.}  The last column shows the range in total number of molecules corrected for IR pumping using Eq.~\eqref{eq:IR_pump} when $T_{\rm IR} > T_{\rm ex}$.}
\label{tab:LTE_results_BHR71-IRS1}
\end{table*}

\begin{table*}[h]
    \centering
    \caption{LTE slab model results for BHR71-IRS2.}
    \scriptsize
    \begin{tabular}{lcccccccccc}
\hline \hline
Species & Mode & Detected & $v_{\rm line}$ & $\Delta V$ & $T_{\rm ex}$ & $N$ & $R_{\rm disk}$ & $\mathcal{N}_{\rm mol}$ & $T_{\rm IR}$ &$\mathcal{N}_{\rm corr}$\\
 & & & (km s$^{-1}$) & (km s$^{-1}$) & (K) & (cm$^{-2}$) & (au) & & (K) & \\
\hline
cold H$_2$O & E & N & -- & 4.71 & [150] & $<4.2 \times 10^{17}$ & [$10$] & $<3.0 \times 10^{46}$ & -- & -- \\
warm H$_2$O & E & N & -- & 4.71 & [450] & $<5.6 \times 10^{15}$ & [$10$] & $<4.0 \times 10^{44}$ & -- & -- \\
rovib H$_2$O & A & Y & -45 & 4.71 & $440 \pm 40$ & $1.4\pm0.4 \times 10^{19}$ & -- & -- & -- & -- \\
CO$_2$ & E & Y & 0 & 4.71 & $75 \pm 15$ & $3.3\pm2.3 \times 10^{17}$ & $541\pm466$ & $1.8\pm1.7 \times 10^{50}$ & 173 & $4.4 \times 10^{36}$--$3.6 \times 10^{50}$ \\
$^{13}$CO$_2$ & E & N & -- & 4.71 & [150] & $<1.8 \times 10^{16}$ & [$10$] & $<1.3 \times 10^{45}$ & 173 & $<6.1 \times 10^{43}$ \\
C$_2$H$_2$ & E & N & -- & 4.71 & [150] & $<1.3 \times 10^{15}$ & [$10$] & $<9.4 \times 10^{43}$ & 107 & -- \\
$^{13}$CCH$_2$ & E & N & -- & 4.71 & [150] & $<1.3 \times 10^{15}$ & [$10$] & $<9.4 \times 10^{43}$ & 107 & -- \\
HCN & E & N & -- & 4.71 & [150] & $<5.6 \times 10^{15}$ & [$10$] & $<4.0 \times 10^{44}$ & 116 & -- \\
C$_4$H$_2$ & E & N & -- & 4.71 & [150] & $<5.6 \times 10^{14}$ & [$10$] & $<4.0 \times 10^{43}$ & 103 & -- \\
CH$_4$ & E & N & -- & 4.71 & [150] & $<5.6 \times 10^{16}$ & [$10$] & $<4.0 \times 10^{45}$ & 108 & -- \\
SO$_2$ & E & N & -- & 4.71 & [150] & $<1.8 \times 10^{17}$ & [$10$] & $<1.3 \times 10^{46}$ & 112 & -- \\
CS & E & N & -- & 4.71 & [150] & $<5.6 \times 10^{16}$ & [$10$] & $<4.0 \times 10^{45}$ & 104 & -- \\
SiO & E & N & -- & 4.71 & [150] & $<3.2 \times 10^{16}$ & [$10$] & $<2.2 \times 10^{45}$ & 102 & -- \\
NH$_3$ & E & N & -- & 4.71 & [150] & $<1.8 \times 10^{16}$ & [$10$] & $<1.3 \times 10^{45}$ & 130 & -- \\
% CO & E & N & -- & 4.71 & [1500] & $<1.5 \times 10^{14}$ & [$1$] & $<1.0 \times 10^{41}$ & 156 & -- \\
% OH & E & N & -- & 4.71 & $<2500$ & $<1.0 \times 10^{24}$ & $5.0\pm4.9 \times 10^{4}$ & $<7.0 \times 10^{60}$ & 46 & -- \\
\hline
\end{tabular}
\tablefoot{Square brackets indicate that the parameter was fixed. The second column indicates whether the molecular features were fitted using an absorption (A) or emission (E) model. The velocity $v_{\rm line}$ {was determined through visual inspection and} is with respect to the $v_{\rm lsr}=-4.7$~\kms of BHR71-IRS2. The last column shows the range in total number of molecules corrected for IR pumping using Eq.~\eqref{eq:IR_pump} when $T_{\rm IR} > T_{\rm ex}$.}
\label{tab:LTE_results_BHR71-IRS2}
\end{table*}

\renewcommand{\arraystretch}{1.0}
\end{appendix}

\end{document}